\documentclass[a4paper, twoside, openright, 12pt]{report}
\usepackage[T1]{fontenc}
\usepackage[english]{babel}
\usepackage{leqno}
\usepackage{url}
\usepackage{array,multirow}
\usepackage{graphicx}
\usepackage[compress]{cite}
\usepackage{float}
\usepackage{amsmath}
\usepackage{amssymb}
\usepackage{dcolumn}
\usepackage{bm}
\usepackage{tabularx}
\usepackage{lipsum}
\usepackage{geometry}
\usepackage{afterpage}
\usepackage{color}

\usepackage{fancyhdr}

\usepackage[colorlinks,
            linkcolor=black,
            anchorcolor=black,
            citecolor=blue,
            urlcolor=black,
]{hyperref}

\newcolumntype{L}[1]{>{\raggedright\let\newline\\\arraybackslash\hspace{0pt}}m{#1}}
\newcolumntype{C}[1]{>{\centering\let\newline\\\arraybackslash\hspace{0pt}}m{#1}}
\newcolumntype{R}[1]{>{\raggedleft\let\newline\\\arraybackslash\hspace{0pt}}m{#1}}

\makeatletter
\newcommand*{\rom}[1]{\expandafter\@slowromancap\romannumeral #1@}
\makeatother

\selectfont

\DeclareMathOperator\erf{erf}

\begin{document}

\pagenumbering{gobble}
\author{Hongfeng Ma}

\thispagestyle{fancy}
\lhead{} 
\chead{\includegraphics[width=0.4\textwidth]{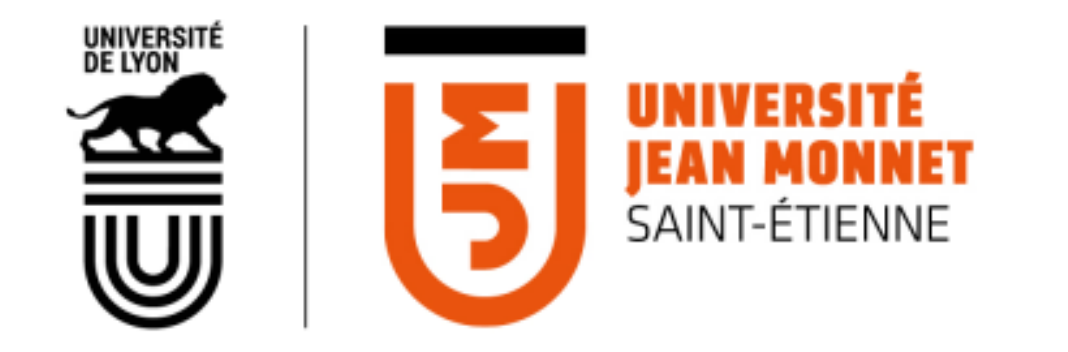}} 
\rhead{}

\renewcommand{\headrulewidth}{0.0pt}

 \begin{center}

\noindent {\textbf{\large{THESE de DOCTORAT DE L'UNIVERSITE DE LYON}}} \\
\noindent {op\'{e}r\'{e}e au sein de} \\
\noindent {\textbf{\large{Laboratoire Hubert Curien}}} \\
\vspace*{0.4cm}

\noindent {\textbf{\large{Ecole Doctorale N$^{\circ}$ 488}}} \\
\noindent {\textbf{\large{Science Ing\'{e}nierie Sant\'{e}}}} \\
\vspace*{0.4cm}

\noindent {\textbf{\large{Sp\'{e}cialit\'{e} de doctorat: Optique, Photonique}}} \\
\noindent {\textbf{\large{Discipline: Laser interaction avec des mat\'{e}riaux}}} \\
\vspace*{0.4cm}

\noindent {Soutenue publiquement le 15/01/2020, par} \\

\noindent \textbf{\large Hongfeng \textsc{Ma}} \\
\vspace*{0.4cm}

\noindent {\large\textbf{\'{E}tude num\'{e}rique de la micro et nano structuration laser de mat\'{e}riaux poreux nanocomposites }}\\

\vspace*{0.4cm}

\leftline{ \noindent \large{Devant le jury compos\'{e} de:}}

\vspace*{0.1cm}
\end{center}
\begin{center}
\noindent 
\begin{tabular}{C{6cm}L{6cm}L{4cm}} 

\shortstack{\textsc{N. Del Fatti} \\ Professeur } &Institut Lumi\`{e}re Mati\`{e}re (iLM) UMR CNRS 5306, Lyon, France \bigskip & Pr\'{e}sidente \\
\shortstack{\textsc{E. Mariotti} \\ Professeur }  & Universit\'{a} degli Studi di Siena , Siena, Italie &  Rapporteur\\
\shortstack{\textsc{G. Duchateau} \\ Chercheur au CELIA-CEA, HdR} & CEA CELIA, Bordeaux, France & Rapporteur \\
\shortstack{\textsc{A. Kabashin } \\  Directeur de Recherche au CNRS}& Laboratoire Lasers, Plasmas et Proc\`{e}d\`{e}s Photoniques (LP3), UMR CNRS 7341, Marseille, France \bigskip & Examinateur \\
\shortstack{\textsc{N. Destouches  } \\ Professeur } & Laboratoire Hubert Curien, UJM/UMR CNRS 5516, Saint-\'{E}tienne, France \bigskip & Examinatrice \\
\shortstack{\textsc{F. Goutaland} \\ Ma\^{i}tre de conf\`{e}rences} & Laboratoire Hubert Curien, UJM/CNRS UMR 5516, Saint-\'{E}tienne, France \bigskip & Examinateur \\
\shortstack{\textsc{T. E. Itina} \\ Directeur de Recherche au CNRS} & Laboratoire Hubert Curien, UJM/UMR CNRS 5516, Saint-\'{E}tienne, France & Directrice de th\'{e}se \\

\end{tabular}
\end{center}

\newpage\null\thispagestyle{empty}\newpage

\author{Hongfeng Ma}
 \begin{center}
\noindent {\large{Doctoral thesis University of Lyon}} \\
\noindent {\large{Jean Monnet University - Saint-\'{E}tienne}} \\
\vspace*{0.4cm}

\noindent {\large\textbf{Numerical study of laser micro- and nano-processing of nanocomposite porous materials}} \\

\vspace*{0.4cm}
\noindent \large{by} \\
\vspace*{0.4cm}
\noindent \LARGE Hongfeng \textsc{Ma} \\
\vspace*{0.4cm}

\noindent \large {Thesis presented in partial fulfillment of the requirements} \\
\noindent \large{for the degree of Doctor of Philosophy} \\
\noindent \large{in the subject of Optics, photonics and hyperfrequencies}\\
\vspace*{0.4cm}

\noindent \large {before the committee:}\\
\vspace*{0.2cm}
\end{center}
\begin{center}
\noindent 
\begin{tabular}{C{6cm}L{6cm}L{3cm}}

\shortstack{\textsc{N. Del Fatti} \\ Professor } &Institut Lumi\`{e}re Mati\`{e}re (iLM) UMR CNRS 5306, Lyon, France \bigskip & President \\			
\shortstack{\textsc{E. Mariotti} \\ Professor }  & Universit\'{a} degli Studi di Siena , Siena, Italie &  Reporter\\
\shortstack{\textsc{G. Duchateau} \\ CEA Researcher, HdR} & CEA CELIA, Bordeaux, France & Reporter \\
\shortstack{\textsc{A. Kabashin } \\ CNRS Research Director}& Laboratoire Lasers, Plasmas et Proc\`{e}d\`{e}s Photoniques (LP3), UMR CNRS 7341, Marseille, France \bigskip & Examiner \\
\shortstack{\textsc{N. Destouches  } \\ Professor} & Laboratoire Hubert Curien, UJM/UMR CNRS 5516, Saint-\'{E}tienne, France \bigskip & Examiner \\
\shortstack{\textsc{F. Goutaland} \\ Ma\^{i}tre de conf\`{e}rences } & Laboratoire Hubert Curien, UJM/CNRS UMR 5516, Saint-\'{E}tienne, France \bigskip & Examiner \\
\shortstack{\textsc{T. E. Itina} \\ CNRS Research Director} & Laboratoire Hubert Curien, UJM/UMR CNRS 5516, Saint-\'{E}tienne, France \bigskip  & Superviser\\

\end{tabular}
\end{center}

\newpage\null\thispagestyle{empty}\newpage

\pagenumbering{roman}

\newpage
\begin{center} \textbf{R\'{e}sum\'{e}} \end{center}

Cette th\`{e}se porte sur les simulations num\'{e}riques de l'interaction laser avec des mat\'{e}riaux poreux. Une possibilit\'{e} de traitement bien contr\^{o}l\'{e} est particuli\`{e}rement importante pour la microstructuration laser du verre poreux et le nano-usinage de mat\'{e}riaux poreux semi-conducteurs en pr\'{e}sence de nanoparticules m\'{e}talliques. La mod\'{e}lisation auto-coh\'{e}rente se concentre donc sur une \'{e}tude d\'{e}taill\'{e}e des processus impliqu\'{e}s. En particulier, pour comprendre les structures des micro-vides p\'{e}riodiques produits \`{a} l'int\'{e}rieur du verre poreux par des impulsions laser femtoseconde, une analyse thermodynamique num\'{e}rique d\'{e}taill\'{e}e a \'{e}t\'{e} r\'{e}alis\'{e}e. Les r\'{e}sultats des calculs montrent la possibilit\'{e} de contr\^{o}ler le micro-usinage laser en volume de SiO$_2$. De plus, les dimensions des structures densifi\'{e}es par laser sont examin\'{e}es pour diff\'{e}rentes conditions de focalisation \`{a} de faibles \'{e}nergies d'impulsion. Les dimensions caract\'{e}ristiques obtenues \`{a} partir des structures sont corr\'{e}l\'{e}es avec les r\'{e}sultats exp\'{e}rimentaux. Compar\'{e}s au verre poreux, les films m\'{e}soporeux TiO$_2$ charg\'{e}s d'ions Ag et de nanoparticules supportent des r\'{e}sonances plasmoniques localis\'{e}es. Les films nanocomposites obtenus sont capables de transf\'{e}rer des \'{e}lectrons libres et d'absorber l'\'{e}nergie laser de mani\`{e}re r\'{e}sonnante, offrant des possibilit\'{e}s suppl\'{e}mentaires pour contr\^{o}ler la taille des nanoparticules d'Ag. Pour identifier les param\`{e}tres optimaux du laser \`{a} onde continue, un mod\`{e}le multi-physique prenant en compte la croissance des nanoparticules d'Ag, photo-oxydation, r\'{e}duction a \'{e}t\'{e} d\'{e}velopp\'{e}. Les simulations r\'{e}alis\'{e}es montrent que la vitesse d'\'{e}criture laser contr\^{o}le la taille des nanoparticules d'Ag. Les calculs ont \'{e}galement repr\'{e}sent\'{e} une nouvelle vision selon laquelle les nanoparticules d'Ag se d\'{e}veloppent devant le centre du faisceau laser du fait de la diffusion de chaleur. Il a \'{e}t\'{e} d\'{e}montr\'{e} que la croissance rapide activ\'{e}e thermiquement suivie d'une photo-oxydation est la principale raison du changement de taille et de temp\'{e}rature en fonction de la vitesse d'\'{e}criture. Un mod\`{e}le tridimensionnel a \'{e}t\'{e} d\'{e}velopp\'{e} et reproduit les lignes \'{e}crites au laser.

L'\'{e}criture de films m\'{e}soporeux TiO$_2$ charg\'{e}s de nanoparticules d'Ag par un laser puls\'{e} promet \'{e}galement d'offrir des possibilit\'{e}s suppl\'{e}mentaires dans la g\'{e}n\'{e}ration de deux types de nanostructures: les rainures de surface p\'{e}riodiques induites par laser (LIPSS) et les nanogratings Ag \`{a} l'int\'{e}rieur du film TiO$_2$. Pour mieux comprendre les effets d'un laser puls\'{e}, deux mod\`{e}les multi-impulsions - un semi-analytique et un autre bas\'{e} sur une m\'{e}thode par \'{e}l\'{e}ments finis (FEM) - sont d\'{e}velopp\'{e}s pour simuler la croissance des nanoparticules d'Ag. Le mod\`{e}le FEM s'av\`{e}re pr\'{e}cis car il traite mieux la diffusion de la chaleur \`{a} l'int\'{e}rieur des films minces TiO$_2$. Le mod\`{e}le pourrait \^{e}tre \'{e}tendu \`{a} l'avenir pour comprendre la formation de nanogratings LIPSS et Ag dans de tels milieux en les couplant avec les migrations de nanoparticules, la fusion de surface et l'hydrodynamique. Les r\'{e}sultats obtenus ont ouvert de nouvelles perspectives sur le microtraitement laser des mat\'{e}riaux poreux et un meilleur contr\^{o}le laser sur la nanostructuration dans les films semi-conducteurs poreux charg\'{e}s de nanoparticules m\'{e}talliques.

\newpage
\begin{center} \textbf{Abstract} \end{center}

This thesis is focused on numerical simulations of the laser interaction with porous materials. A possibility of well-controlled processing is particularly important for the laser based micro-structuring of porous glass and nano-machining of semiconducting porous materials in the presence of metallic nanoparticles. The self-consistent modeling is, therefore, focused on a detailed investigation of the involved processes. Particularly, to understand the periodic micro-void structures produced inside porous glass by femtosecond laser pulses, a detailed numerical thermodynamic analysis was performed. The calculation results show the possibility to control laser micro-machining in volume of SiO$_2$. Furthermore, the dimensions of laser-densified structures are examined for different focusing conditions at low pulse energies. The obtained characteristic dimensions of the structures correlate with the experimental results. Comparing to the porous glass, the mesoporous TiO$_2$ films loaded by Ag ions and nanoparticles support localized plasmon resonances. The resulted nanocomposite films are capable to transfer free electrons and to resonantly absorb laser energy providing additional possibilities in controlling Ag nanoparticle size. To identify the optimum parameters of the continuous-wave laser, a multi-physical model considering Ag nanoparticle growth, photo-oxidation, reduction was developed. The performed simulations show that the laser writing speed controls the Ag nanoparticles size. The calculations also depicted a novel view that Ag nanoparticles grow ahead of the laser beam center due to the heat diffusion. The thermally activated fast growth followed by the photo-oxidation was found to be the main reason for the writing speed dependent size change and temperature rises. A three-dimensional model was developed and reproduced the laser written lines.

Writing of mesoporous TiO$_2$ films loaded with Ag nanoparticles by a pulsed laser is, furthermore, promising to provide additional possibilities in the generation of two kinds of nanostructures: laser induced periodic surface grooves (LIPSS) and Ag nanogratings inside the TiO$_2$ film. To better understand the effects of a pulsed laser, two multi-pulses models - one semi-analytic and another one based on a finite element method (FEM) - are developed to simulate the Ag nanoparticle growth. The FEM model is shown to be precise because it better treats heat diffusion inside the TiO$_2$ thin films. The model could be extended in future to understand the formation of LIPSS and Ag nanogratings in such media by coupling with nanoparticle migrations, surface melting and hydrodynamics. The obtained results provided new insights into laser micro-processing of porous material and better laser controlling over nanostructuring in porous semiconducting films loaded with metallic nanoparticles.

\tableofcontents

\newpage
\begin{center} \textbf{Acknowledgments} \end{center}
\hspace{0.5cm}

I would like to express my sincere gratitude to my adviser, Dr. Tatiana E. Itina, for her continuous supports by providing infinite original ideas for my study, for her patience and kind help that make my Ph.D life easier in France.

My sincere thanks also goes to Prof. Nathalie Destouches, for her guidance, insightful comments and encouragement that broaden my research in many aspects.

I am very grateful to my thesis reporters, Prof. Emilio Mariotti and Dr. Guillaume Duchateau for their rigorous analysis of the manuscript, as well as useful advices and comments in order to improve its quality. I am also thankful to examiners Prof. Natalia Del Fatti, Prof. Andrei Kabashin and Prof. Fran\c{c}ois Goutaland for taking time to attend my defense and asking interesting questions. 

I would like to thank Florence Garrelie, Director of Laboratoire Hubert Curien, Julie Debiesse, Head of administrative staff, for providing me the excellent working environment and all necessary facilities to carry out this work. 

I would like to acknowledge the financial supports by the French Ministry of Science and Education, Jean Monnet University and Laboratoire Hubert Curien.

I am grateful to all the co-authors, including Roman Zakoldaev, Maksim Sergeev, Yaroslava Andreeva, Vadim P. Veiko, Nipun Sharma, Francis Vocanson, Said Bakhti, Daniel S. Slaughter, and others for their contribution. Additionally, I would like to thank particularly Razvan Stoian, Jean-Philippe Colombier, Ciro D'Amico, St\'{e}phan Mottin, Anton Rudenko, Florent Bourquard, Cyril Mauclair, Christophe Hubert,  as well as Francis Vocanson and Youcef Ouerdane for stimulating discussions during my research work.

My special thanks goes to all those who have made my stay memorable and who helped me along the long way to obtain this degree: Andrey Voloshko, Emile B\'{e}villon, Adriana Morana, Sedao, Guanghua Cheng, Zeming Liu, Elena Silaeva, Ayoub Ladaci, Chen Wu, Florent Bourquard, Daiwei Zhang, Anthony Abou Saleh, Madhura Somayaji, Nipun Sharma, Bobin Varghese, Guodong Zhang, Muzi Yang, Xuan Qi, Wenxi Wang, Doan Le, Yao Lu, Kevin Delmares, Jian Cao, Omar Qawasmeh, Su Luo, Nian Liu, Rim Faraj, Dat Nguyen, Wissam Farhat, Chen He, Sixiang Xu, Shaobo Zhang, Nicolas Dalloz, Yannick Bleu, Shenlin Tian, and many others. 

I would like to thank also staffs from our lab: Emmanuel Marin, Aziz Boukenter, Nicolas Faur\'{e}, Sedao, Eric Sigronde, Thomas Gautrais, as well as visiting professors Guanghua Cheng, Konstantin Khishchenko for supplementary support and participation.

Finally, I would like to thank all my family members, especially, my brother and my parents for their unconditional support, understanding and encouragement to do a PhD in France.

\newpage

\chapter{Introduction}
\pagenumbering{arabic}
\hspace{0.5cm} Material processing by laser interactions has attracted a wide interest both in scientific and engineering communities. The possibilities in micro- and nano-structuring utilizing laser energy density to make material changes or damages \cite{nolte1997ablation} enable a tremendous number of applications, such as optical data storage\cite{glezer1996three,watanabe1998three,juodkazis2003recording,canioni2008three}, waveguides \cite{szameit2010discrete,davis1996writing,miura1997photowritten,nolte2003femtosecond,eaton2011high}, Bragg gratings \cite{gattass2008femtosecond,marshall2006direct,zhang2007single,marshall2010point, zhang2018efficient}, volume gratings \cite{klyukin2017volume,klyukin2018thermal}, photonic quantum circuits \cite{marshall2009laser,matthews2009manipulation,aspuru2012photonic}, micro-welding \cite{tamaki2005welding,horn2008investigations}, metasurfaces for light field manipulation \cite{drevinskas2017ultrafast, wang2016optically}, micromechanics \cite{maruo2003submicron}, microfludic devices \cite{bellouard2004fabrication,wu2009femtosecond,amato2012integrated}, and biological channels \cite{wu2015channel,kim2009single,sugioka2012femtosecond}. These expanded applications stem from the fundamental progresses and improvements in high power lasers. In 1987, the finding of laser ablation without heat-affected zone by ultrafast lasers whose pulse duration was less than a few picoseconds had an important impact on the researches working in the field of laser interactions with materials \cite{srinivasan1987ablation,sugioka2014ultrafast,kuper1987femtosecond}. Since then, and, importantly, because of the invention of the chirped-pulse amplification in 1985 by D. Strickland and G. Mourou \cite{strickland1985compression} and further development of this technique \cite{rudd1993chirped}, high quality micro- and nano-fabrication attracted lots of attentions for the applications in semiconductors and dielectrics. The pioneering work for producing optical waveguides inside a glass via ultrafast lasers was reported in 1996 by K. Davis et al. \cite{davis1996writing}. In fact, at attempt to fabricate such optical elements by ultra-violet(UV) light was first performed much earlier, in the 1970s \cite{hill1978photosensitivity}. However, it was until the development of the high energy density pulsed lasers that a plausibility to greatly increase the refractive-index changes appeared due to multiphoton absorption. Other improvements were proposed to increase the spatial resolutions in the three-dimensional micro-fabrications. Nowadays, the two-photon polymerization enables the three-dimensional fabrications of objects with dimensions down to sub-100 nm \cite{serbin2003femtosecond,emons2012two,sun1999three}. 

Furthermore, the laser-induced periodic surface structures (LIPSS) emerged as another useful surface nano-structuring technique. The effect was firstly reported in 1965 by M. Birnbaum during examining the surface damage on semiconductors (germanium, silicon, gallium arsenide, gallium antimonide, and indium antimonide) by a focused Ruby laser \cite{birnbaum1965semiconductor}. The parallel straight lines (or called grooves) were observed at laser powers lower than the formation of crack patterns. Later, this effect was also observed for pulsed lasers and it turned out that the laser wavelength, fluence, polarization, and number of irradiated pulses greatly affected the LIPSS generation\cite{van1982laser,bonse2012femtosecond,bonse2016laser}.  Almost in the same time, a number of theoretical studies were focused on the understanding of the LIPSS formation \cite{emmony1973laser,keilmann1982periodic,sipe1983laser,dufft2009femtosecond,bonse2009role,rudenko2016random,rudenko2019amplification}. The applications of these structures were widely examined by both scientific and engineering communities. One of the direct results was that the laser generated grooves served as diffractive optical elements manipulating the optical spectra. These nano-gratings were controlled by laser polarization, wavelength, fluence, and scan speed, which was useful for color marking, encryption, and optical data storage \cite{dusser2010controlled,vorobyev2008colorizing,ahsan2011colorizing,ou2014colorizing,yao2012selective}. Moreover, the contact angle was increased after the formation of LIPSS, so that superhydrophobic surfaces were produced \cite{barberoglou2009bio,kietzig2009patterned,kietzig2011laser}. In addition, the LIPSS were capable to increase the tribological performances \cite{yu1999laser,yasumaru2011frictional,wang2015reduction}.

Apart from the laser processing of solid or bulk materials, their porous counterparts have attracted more and more interest in the past decades. Comparing to the bulk glasses, such as fused silica and BK7, porous glasses (typical pore sizes range from 2 to 20 nm \cite{toquer2011effect,veiko2016femtosecond,burchianti2006reversible}) were easier to be densified by lasers and impregnated of nanoparticles or dyes, so that they presented promising applications in the field of optical filters \cite{alexeev1993porous,sendova2012plasmonic}, microfluidic channels \cite{liao2010three,ju2011fabrication,sugioka2012femtosecond}, and optical waveguides \cite{antropova2012peculiarities,simmons1979optical}. For example, the refractive index changes in a bulk glass were normally in the range of ($\Delta n \approx 10^{-3}$ to $10^{-2}$ \cite{eaton2011high}), which were realized because of the local mass-density or material changes after ultra-short laser induced heating or shock-waves. The solidification and decompaction in bulk materials were restricted owing to their intrinsic physical and chemical limitations; in contrast, the pores and their random connected networks inside porous glasses allowed larger changes in local mass density and in refractive index ($\Delta n \approx 0.12$) \cite{veiko2016femtosecond,simmons1979optical}. To date, most microfluidic devices were two-dimensional that possessed limited applications for integrated devices \cite{liao2014femtosecond}. A recent developed technique has enabled the fabrication of 3D microfluidic structures in porous glass with arbitrary sizes and geometries \cite{liao2010three,ju2011fabrication,liao2012rapid,ju2013direct}. The process was based on an ultra-fast laser writing of porous glass immersed in water followed by a post-annealing to collapse nanopores. Based on this technique, a square wave-shape channel\cite{liao2010three}, a large volume microchamber\cite{ju2011fabrication}, and a 3D microfluidic mixer inside a glass chip\cite{liao2012rapid} were successfully fabricated.  Furthermore, V. P. Veiko et al. used ultrafast lasers to process porous glass and  examined densification, decompatction, void formation regimes as a function of pulse energy and scanning speed \cite{veiko2016femtosecond}. These processes accompanied by heat diffusion and multi-pulse accumulation, as well as nonlinear light propagation, absorption and plasma scattering. Thus, it was hard to explain the obtained results without a detailed modeling.

In addition,  mesoporous (typical pore size ranges between 2 and 50 nm) glasses loaded with nanoparticles served as emerging platforms for quantum-dots applications \cite{litvin2013pbs,letant2006semiconductor}, optical recording \cite{royon2010silver},nano-wire synthesizing \cite{huang2000ag}, Q-switched laser \cite{zhang2019method}, light-controlled atomic dispensers \cite{burchianti2004light}, and drug delivering \cite{li2012mesoporous,slowing2007mesoporous} as their strong capabilities in absorbing nanocomposites. The kinetics of the embedded nanoparticles enabled additional possibilities in laser processing of porous glasses and, particularly, in a better control over their properties. It was shown by L. Marmugi and E. Mariotti et al. \cite{marmugi2014laser} that the adsorption/desorption and nanoparticle formation in nanoporous glasses were strongly affected by the laser irradiation. A properly chosen laser wavelength and illumination time was able to control the concentration of the formed NPs. In those materials loaded with nanopartilces, the heat transfer at nanometric scales differed from macroscopic scales was of great interest. Using the time-resolved optical pump-probe spectroscopy, N. D. Fatti et al. studied the thermal interface resistances of metal nanoparticles embedded in a homogeneous matrix \cite{gandolfi2018ultrafast, banfi2012temperature, stoll2015time}. Furthermore, the presence of silver nanoparticles in soda-lime glasses allowed the possibility of generating self-arranged periodic Ag nanoparticles by an intense UV laser\cite{tite2017ag,goutaland2013laser}. In this case, the LIPSS were formed under laser exposure before any surface damages appeared. Recently, a two-beam interference of a picosecond laser shining on the surface of a silver-nanocomposite porous glass was shown to cause grating formation and its transformation to grooves at higher laser fluences  \cite{andreeva2020picosecond}. At lower fluences, the structures were parallel to the interference pattern, while they were converted to the grooves orientated parallel to laser polarization as soon as distribution were generated at high fluences.

Despite the laser processing of solid-state materials, laser ablation in liquids (LAL) and LAL-based nanoparticle synthesis in solutions were extensively studied during the past decades \cite{geohegan1998time,kabashin2003synthesis,amendola2009laser,hashimoto2012studies}.
To better control nanoparticle size distribution, laser-induced nanofabrication in liquids with and without following laser-induced colloidal fragmentation was proposed. For example, in 2003, A. V. Kabashin et al. \cite{kabashin2003synthesis} reported synthesis of colloidal nanoparticles during femtosecond laser ablation of gold in water. It was found that the nanoparticle size was small (3-10 nm) at low laser fluences (<400$J/cm^2$) while large nanoparticles with a broad size 
 as soon as the fluence exceeded the identified threshold.

Unlike porous glasses that only acted as the serving platforms, porous semiconductors interacted with the encapsulated nanocomposites providing additional controlling degrees of freedom in transparent conductors \cite{granqvist2007transparent,han2008improved,liu2010transparent}, photoconduction \cite{hoyer1995photoconduction,zaban1998photosensitization,kamat2010beyond}, and photocatalytics \cite{chen2011new,wang2012plasmonic,awazu2008plasmonic,chen2010plasmonic,zheng2007ag,georgekutty2008highly}. In those hosting media, the most studied semiconductors were TiO$_2$ and ZnO as they presented significant photoactive behaviors, and were naturally abundant and chemically stable. The ability to generate conducting electrons and holes after absorbing photons could activate the oxidation and reduction of several chemical compounds. However, the bulk semiconductors were inefficient in the photocatalytic process as plenty of conducting electrons or holes were reduced by recombination during their migrations to the material surfaces before interacting with chemicals in the vicinity \cite{hashimoto2005tio2,zhang2013plasmonic,wang2012plasmonic,awazu2008plasmonic,chen2010plasmonic}. In contrast, the porous state of materials having larger contacting surfaces with the environment and smaller recombination probability could boost the photoactivated process. The dramatic boosting was observed by a recently proposed idea of the so-called "plasmonic photocatalysis" \cite{awazu2008plasmonic,kubacka2011advanced,chandrasekharan2000improving,kamat2002photophysical,hirakawa2005charge,zhao1996sol}, which encapsulated nanoparticles of noble metals such as Au and Ag into these semiconductors. The photocatalysis benefited from the light absorption in visible and the nearfield enhancement due to the surface plasmon resonances of Au and Ag nanoparticles. Typically, TiO$_2$ and ZnO had relatively large band gaps ($E_g = 3.2 eV$), so that they only absorbed the UV light. The plasmonic absorption by metallic nanoparticles occurs across the visible to infrared and the induced free electrons were found effectively injected into the nearby semiconductors of suitable band gaps \cite{mongin2012ultrafast,tian2005mechanisms,hirakawa2005charge, yu2006size,takahashi2011solid}. In addition, the contacting of a noble metal with a semiconductor formed the Schottky junction that efficiently separated the electrons and holes to avoid the recombination by taking advantages of the internal electric field in a certain region. After the process, silver nanoparticles were positively charged and tended to dissolve into Ag$^+$ ions to keep the neutral state. On the other hand, electrons diffused from TiO$_2$ films and accepted by the Ag$^+$ ions thanks to the photocatalytic process when illuminated by the UV light. The nucleation increased the quantity of monomers in the matrix (hosting medium), which resulted into the growth of nanoparticles by Ostwald ripening or coalescence. Because of the size-dependent optical responses of Ag nanoparticles, the spectra or colors of these samples were reversible. This effect was recognized as "photochromism" \cite{ohko2003multicolour,naoi2004tio2,naoi2005switchable,diop2015magnetron}, which has numerous  applications in optical data storage \cite{wang2011optical,rakuljic1995optical}, and, particularly, in the re-writable colored images \cite{yokoyama2000fulgides}.

In the past decades, metamaterials composed of artificially designed structures had attracted tremendous interests, due to their abilities of engineering the optical amplitude, phase, and polarization at will that not found in nature. The applications were wide and not limited to electromagnetic/optical cloak \cite{schurig2006metamaterial,cai2007optical,valentine2009optical,ergin2010three}, aberration-free lens \cite{aieta2013aberrations,loo2012broadband,ma2010broadband}, ultra-thin lens \cite{chen2012dual,pors2013broadband}, metasurface based holography \cite{yu2011light,huang2015aluminum,zheng2015metasurface,ni2013metasurface,huang2013three}, polarization imaging \cite{arbabi2018full,li2015circularly}, and chirality-dependent pulse modulation \cite{yan2018chirality,yan2019electromagnetically}. However, these structures were normally in sub-wavelength scale and required high-precision technologies such as electron beam lithography, focused ion beam, and nanoimprinting that were either expensive or unsuitable for large-scale productions. In recent years, the "NANOPARTICULE" team of Laboratoire Hubert Curien developed a cost-effective way of producing nanoparticles and nano-gratings by using the continuous-wave and pulsed-lasers to direct write mesoporous TiO$_2$ films loaded with Ag nanoparticles \cite{crespo2010reversible,crespo2012one,destouches2014self,baraldi2016polarization,destouches2013photo}. The applications in color printing \cite{diop2017spectral} and image multiplexing \cite{sharma2019tailoring,sharmalaserdriven2019} were demonstrated.

The laser irradiation of the semiconductor loaded with Ag nanoparticles was identified as a complex process consisting of light absorption at plasmonic resonances, photo-oxidation, reduction, and nanoparticles growth. Based on a simplified model without heat diffusion, the writing-speed threshold of dramatic growth of Ag nanoparticles was understood \cite{liu2015understanding,liu2016selfthesis}. Below the certain writing-speed, only small nanoparticles were observed. Nevertheless, the formation of Ag nanoparticles and nanogratings as a function of laser writing-speed was not well understood.

In addition to numerous experimental studies, a series of theoretical and numerical models were developed to better understand the physical and chemical process involved. First part of the proposed models considered lasers just as heat sources, which can be valid for rather long laser pulses (nanosecond and longer) and for a continuous-wave laser irradiation \cite{tian2004temperature,tao2009thermal}. Several studies also examined ultra-short laser interactions \cite{couairon2002light,sprangle2002propagation,miyamoto2011evaluation,sun2013role,couairon2007femtosecond, rudenko2016random,gulley2012interaction,wu2005femtosecond, chekalin2015light}. In particular, T. Brabec and F. Krausz (1997) \cite{brabec1997nonlinear}, A. Couairon (2002) \cite{couairon2002light}, P. Sprangle (2002) \cite{sprangle2002propagation} and A. L. Gaeta (2000) et al. \cite{gaeta2000catastrophic} developed an envelope equation modeling the propagation of ultra-short pulses by assuming that the pules envelope is slowly changing along the direction of light transportation. For multiple pulses (repetition rate of 100 kHz), I. Miyamoto et al. proposed a simplified thermal conduction model and simulated the picosecond laser modifications of bulk glasses. By comparing with experiments at different laser energy and repetition rate, the nonlinear absorptivity was obtained and used for the thermal diffusion model. It was concluded that the laser energy was absorbed through avalanche ionization seeded by thermally excited free-electrons \cite{miyamoto2011evaluation}. In addition, for inertial confinement fusion \cite{duchateau2009simple}, G. Duchateau et al. proposed a model considering the solid-to-plasma transition to better simulate the laser-plasma coupling. It turned out that around 100 picoseconds were required to induce a full plasma state, suggesting that the process of solid-to-plasma transition could not be ignored for inertial confinement fusion applications.

To date, very limited numerical studies have been performed in understanding the laser interactions with porous materials \cite{liao2015high,antropova2012peculiarities,li2004comparison, rudenko2016random}.
In which, most studies \cite{antropova2012peculiarities,li2004comparison} were concentrated on a continuous wave laser irradiation.
For ultra-fast lasers, Y. Liao et al. analyzed the morphological features of nanogratings in porous glass and concluded that the formed nanovoids were a manifestation of spherical nanoplasma produced by multiphoton ionization \cite{liao2015high}. By assuming the nanoplasma with a size of tens of nanometers, they performed a finite difference time domain simulation to explain the formation of subwavelength-width cracks. In 2016, A. Rudenko et al. proposed a more rigours model  of ultrafast laser-plasma coupling and simulated the evolution from small spherical nanometric inhomogeneities to periodical nanoplanes \cite{rudenko2016random}.
 
Furthermore, laser-induced nanoparticle formation was modeled in several studies \cite{volkov2007numerical,boyer2012modeling,delfour2015mechanisms,liu2015understanding, liu2017three}. 
Many of the performed investigations considered that nanoparticles were generated in gases or in liquid solutions. In these studies, thermal effects, nucleation and growth of nanopartilces, laser absorption and hydrodynamics were discussed. There is, however, still a lack of coupled multi-physical models and simulations suitable for prediction of the nanoparticle size or their formation regimes.

Recently, experiments were performed by shining a scanning continues wave laser on porous TiO$_2$ films containing Ag ions and nanoclusters. Interestingly, the resulting nanoparticles are found to be well-organized into  nanograting-like periodic structures enabling unique optical properties and several promising applications, namely for imaging and security. To explain nanoparticle formation, a numerical approach was also proposed explaining only a part of the observed effects \cite{destouches2014self,baraldi2016polarization}. In 2015, Z. Liu et al. \cite{liu2015understanding} developed a model considering size shrinkage by photo-oxidation, growth via Ostwald ripening, and Ag$^+$ reduction into free Ag$^0$ to explain the laser-writing speed related growth. This model was based on Kumar's nanopaticle kinetics model \cite{kumar2008efficient} and was inspired by the nanoclusters formation modeling in glass performed by Kaganovskii et al. \cite{kaganovskii2007formation} In these previous models, however, temperature field was not well calculated and coupled to nanoparticle growth taking place in time and space. For instance, such important process as heat diffusion  was completely disregarded and only single nanoparticles were taken into account for in the laser-induced heating. 

Despite the limitations, previously studies clearly confirmed that laser interactions with materials containing nano-clusters was a set of complicated process involving many electromagnetic, chemical and physical mechanisms taking place at very different time and space scales. In particular, light absorption and thermal diffusion were important in the process of nanocluster evolution. Therefore, much more rigours models were required for well understanding the involved processes and providing guidelines for better controlling laser-induced micro- and nano-structures. There were many open questions, namely concerning the role of heat diffusion, the correct description of nanoparticle sizes and the related absorption maps, the explanation of several anti-intuitive effects, such as temperature rise with increasing laser scanning speed, etc.

This thesis was performed in the "LASERMODE" team at Hubert Curien Laboratory, UMR CNRS 5516/ Univ. Lyon/ Univ. Jean Monnet in Saint-Etienne, France (LabHC) in a close cooperation with the experimental "NANOPARTICULE" group in the same Laboratory. A part of the experiments was also carried out in The Laser Micro- and nano- structuring Laboratory of the ITMO University (Saint-Petersburg, Russian Federation) in the frame of the PHC KOLMOGOROV Formalas project operated by Campus France. The first motivation was to better understand the laser micro-processing of porous glass by pulsed-lasers. Secondly, since the laser nano-processing was important and interesting as it acted as an emerging promising and cost-efficient technique for constructing nanoscale building blocks for light manipulation, models for laser processing of mesoporous TiO$_2$ films load with Ag nanoparticles were developed. Then, simulations were carried out to better understand the involved processes and, thus, better control laser treatment, and, thus, to get the desired sub-wavelength structures. 
The thesis was financed by the French Ministry of Science and Education and Jean Monnet University. Several missions were supported by the Laboratoire Hubert Curien and by the FormaLas project. Numerical simulations were performed on the local cluster at  Hubert Curien Laboratory and the CINES supercomputer in France.

After the introduction presented in this Chapter, the thesis is organized as follows:

Chapter 2 discusses the results of a femtosecond laser inscription of periodic void arrays in porous glass. The periodic voids and high aspect ratio densification were presented. The formation of the periodic void structures were shown to rely on laser parameters such as deposited energy and scanning speed. A thermodynamic analysis was performed to explain the linear relation of period and number of applied laser pulses per spot. 

Chapter 3 is focused on the continuous-wave laser writing of silver nanoparticles inside mesoporous TiO$_2$ thin films. A coupled model of nanoparticles growth, photo-oxidation, reduction, and heat diffusion was proposed to understand the writing-speed dependent phenomena. The spatial size distribution was revealed to be non-uniform that resulted into the transmission inhomogeneity during laser scanning the sample, which was shown in accordance with in-situ experiments. The performed study also depicted a novel view that Ag NPs grew ahead of the beam center due to an expanded temperature field. The reduction process was insignificant due to the low concentration of the reducing agents. Exhaust simulations were performed on the study of parameters such as activation energy of Ag diffusion, photo-ionization rate, and initial concentration of Ag$^+$ and Ag$^0$ in the mesoporous TiO$_2$. The phenomenon of temperature increasing with writing speed was discussed. Simulations showed the same trend with the analysis of the phase transition of TiO$_2$. The multi-scale model was simulated in three-dimensional and the results were compared with the in-situ transmission maps.

Chapter 4 considers femtosecond laser writing of Ag nanoparticles in mesoporous TiO$_2$ thin films. Compared to the continuous-wave cases, the ultra-fast lasers provided additional possibilities in generating surface grooves on TiO$_2$. The laser propagation was simulated based on the nonlinear Schr\"{o}dinger equation coupled with plasma equation. Due to the low pulsed-energy and loosely focusing conditions, the laser deposited energy was shown to be negligible for the studied laser setups. Then, the heat accumulation models were developed that were coupled with silver nanoparticles growth were simulated, which showed similar trends with the experiments.

Chapter 5 presents general conclusions and perspectives of this work.

\afterpage{\null \newpage}	

\chapter{Femtosecond laser inscription of periodic void structures and high aspect ratio densification in porous glass}

\hspace{0.5cm}  This chapter reports a thermodynamic analysis of ultra-fast laser induced periodic voids and high aspect ratio densification in porous glass. Previously, the femtosecond laser irradiation of porous glass, leading to stress-free densification, decompaction and void formation were demonstrated as a function of pulse energy and laser writing speed \cite{veiko2016femtosecond}. It lacked, however, well understanding of the multi-physical processes usually accompanied by nonlinear propagation of pulsed beams, light absorption, heat diffusion, and hydrodynamics. Furthermore, the formation of perfectly controlled periodic voids structure was shown to rely on laser energy per pulse and writing speed \cite{ma2017well}, which remained unclear. For this sake, a model based on the nonlinear Schr\"{o}dinger equation coupled with electron plasma generation is proposed to estimate the absorption of each pulse, which was then used for the multi-pulses heat accumulation. The estimated dimensions by temperature simulations are well correlated with the obtained shapes of the densified regions. The formations of the periodic voids can be understood by a superposition of two energy sources: the moving laser and the created static heat source. The model can by used by coupling hydrodynamics in the future to have a better understanding of the involved process.

\newpage

\section{Introduction}

The capacity of femtosecond laser systems to locally modify transparent materials has been used for many promising applications in different areas ranging from photonics to microfluidics \cite{gattass2008femtosecond}. The main advantage of the ultra-short laser pulses is in an extremely high precision of the energy deposition. Therefore, laser machining can be performed in volume of different glasses enabling three-dimensional inscription of numerous structures, such as photonics crystals, optical memories, waveguides, gratings, couplers, chemical and biological membranes and other devises \cite{graf2007pearl,bellouard2011femtosecond}. Previous studies have already revealed major mechanisms of femtosecond laser irradiation of typical glasses, such as fused silica, BK7 and calcium fluoride \cite{schaffer2001laser, stuart1995laser, rethfeld2006free, taylor2008applications, richter2013formation, bulgakova2015modification}. Such phenomena as photo-ionization and avalanche ionization leading to laser-induced breakdown  were in focus of both experimental and theoretical studies starting from the invention of high-power systems \cite{bloembergen1997brief,schaffer2001laser}. It was shown that very small regions could be efficiently treated in a well-determined way by using only a central part of the Gaussian radial distribution of the laser beam \cite{joglekar2004optics}. 

The results of laser machining can considerably differ if instead of a single laser pulse, trains of high repetition rate ultra-short laser pulses are used \cite{schaffer2001micromachining, miyamoto2011evaluation}. In fact, when time between laser pulses is shorter than that required for energy diffusion outside laser-treated volume, a considerable heat accumulation may lead to a pronounced enhancement in the size of the laser-affected zone. This effect was observed in the experiments with nanojoule laser energy at laser repetition rate as high as $25$MHz. For such multi-pulse laser irradiation regime  Miyamoto et al. \cite{miyamoto2011evaluation}  calculated thermal fields revealing stationary regimes for several optical materials and laser parameters. Moreover, it was shown that even upon a single ultra-short laser interaction with glass, the induced thermal effects could last for a significant time up to nanoseconds in dielectric materials \cite{veiko2014effective}. 

Additionally, femtosecond laser-induced formation of nanopores and voids was investigated for the initially non-porous optical materials \cite{watanabe2000optical, juodkazis2006laser, bellouard2011femtosecond, lancry2013ultrafast}. Such effects as strong explosion and shock wave formation were shown to play a role in a single void formation for tightly focusing conditions \cite{gamaly2006laser, hallo2007model, mezel2008formation, bulgakova2015modification, rapp2016high}, whereas cavitation and nanopores were found to play a role in volume leading to nanograting formation when laser focusing was not so strong \cite{lancry2013ultrafast, rudenko2016random, cerkauskaite2017ultrafast}. 

It was demonstrated, furthermore, that silica glasses with different densities including fused silica, porous glass, and aerogel could be also successfully processed \cite{veiko2016femtosecond, cerkauskaite2017ultrafast}. For porous glasses, several regimes were identified \cite{veiko2016femtosecond} with different corresponding types of material modifications ranging from densification and microvoid formation to wider channels appearance. In fact, laser-induced densification is expected in the regime of lower energy deposition, whereas the formation of channels was reported for larger laser absorbed energy. For intermediate absorbed energy, the so-called "decompaction" regime was identified accompanying by microscopic void formation during transverse laser scanning of porous glass in volume. We note that local control of porosity also allows the formation of waveguides and other embedded elements. Additionally, porous materials can be used for laser-assisted nanoparticle formation with promising photonic, plasmonic or photochromic properties. These investigations also open up a way toward an easier fabrication of optical sensors based on porous materials to monitor changes in the environment.

\begin{figure}[ht!]
  \centering
     \includegraphics[width=0.6\textwidth]{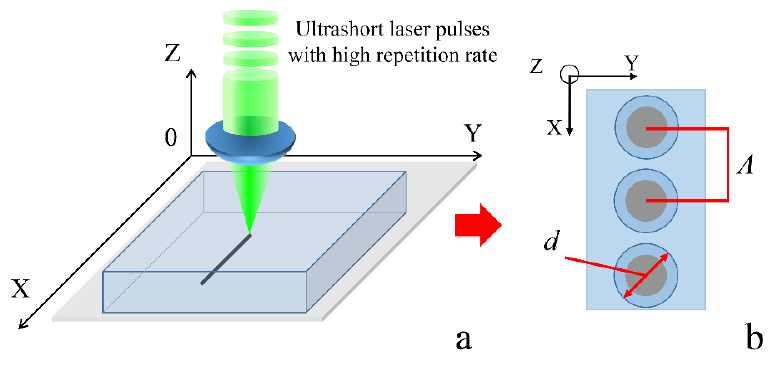}
     \caption{Illustration of transverse laser scanning geometry (a) and the final structure (b).}
  \label{fgr:a}
 \end{figure}

Previously, ultra-short lasers were shown to induce voids both in the direction of laser propagation and in the laser scanning direction. The former structures were attributed to such effects as self-focusing and filamentation \cite{kanehira2005periodic}, standing electron plasma wave \cite{sun2007standing} or spherical aberration \cite{song2008formation, wang2011fabrication, ahsan2014formation}. The transverse laser scanning of non-porous glass (Fig. \ref{fgr:a}) was performed only in a few studies and mostly for non-porous glasses \cite{bellouard2011femtosecond, richter2013formation, cvecek2014gas, matsuo2015spontaneous}. In particular, femtosecond laser pulses were shown to generate  self-organized bubble patterns \cite{bellouard2011femtosecond} in fused silica  at repetition rate as high as $9.4 MHz$ and scanning speed up to $31 mm/s$. A transition between chaotic and self-organized patterns at high scanning rate (above $10 mm/s$) was revealed and attributed to similarities with rather complicated non-linear dynamical systems. 
 
Thus, an efficient control over laser inscription is still puzzling, particularly for porous glasses. In fact, no systematized control possibilities over threshold, size or period have been proposed so far. That is why the present study is aimed at the identification of the major physical mechanisms involved the ultra-short laser-induced void structure formation in volume of a porous glass. Such parameters as structure periodicity, size and the required pulse energy will be explained.

\section{Results and discussion}

\subsection{Experimental study of void formation}

\begin{figure}[ht!]
  \centering
     \includegraphics[width=0.8\textwidth]{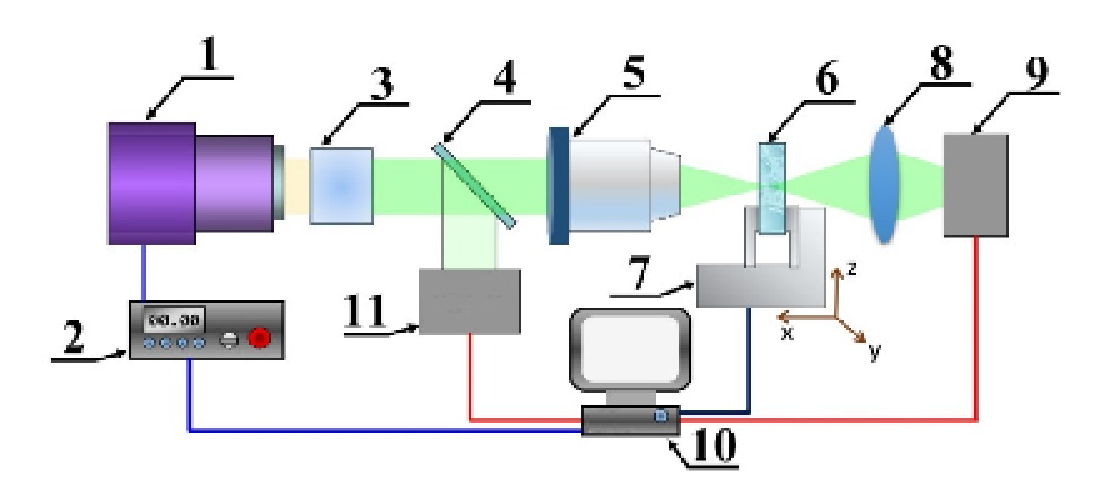}
     \caption{Illustration of the experimental setup, where tightly focused  
   (20$\times$, NA $= 0.4$) ultra-short laser pulses ($200 fs$, $500 kHz$, $515 nm$) 
  create decompaction region inside of the porous glass.}
  \label{fgr:b}
 \end{figure}

A series of experiments have been performed for porous silica glass plates. Their chemical composition is (mass fraction,\%): 94.73SiO$_2$-4.97B$_2$O$_3$-0.30Na$_2$O and the expected trace content ($\leqslant 0.1$ mass \%) of Al$_2$O$_3$. The average pore radius is $\approx 2 nm$, the porosity $ 0.26 cm^3 / cm^3$, and the specific surface area of the pores is $\approx 210 m^2/g$. The transmission of the used porous glass is high in the visible and near-IR range of wavelengths ($0.2-2.5 \mu m$). The experimental set-up is shown in Fig. \ref{fgr:b}. Here, Satsuma Yb-doped fiber laser with wavelength $\lambda = 515 nm$ at second-harmonic, temporal pulse width $\tau = 200 fs$ passes through a nonlinear optical crystal (3) and a plate (4) placed at $45^{o}$ angle to the laser beam direction (1). Material modifications are obtained by varying sample moving speed, $V_{s}$, from $0.0125 - 3.75 mm/s$ with respect to the focused laser beam and by changing laser pulse energy $E_p$ from 1.5 to 2.34 $\mu J$ at a constant repetition rate ($\nu = 500 kHz$). The experiments were performed by Dr. Roman Zakoldaev (ITMO University) and Dr. Maksim Sergeev (ITMO University) at Lebedev Physical Institute, Moscow. 

In this case, around 5 percents of laser energy is reflected from the plate (4) and arrives at the power measuring device (11). Then, the laser beam is focused by the objective (5) (LOMO, $20 \times$, $NA = 0.4$) in volume of the glass film/plate (6) at the plane of  the laser modified region formation (MR). When laser (1) is turned on, the optical table (7) (Standa 8SMC1-USBhF) starts moving along one of the axis X or Y. At the same time, a part of the transmitted laser radiation is registered by the power-meter (9) located behind the collective lens (8) placed right after the glass plate (6).  Another part of the laser energy that is reflected by the place (4) is used to control laser power. The control over the optical table (7) movement and its synchronization with laser power supply (2) is provided by the personal computer (10).

\begin{figure}[ht!]
 \centering
    \includegraphics[width=0.42\textwidth]{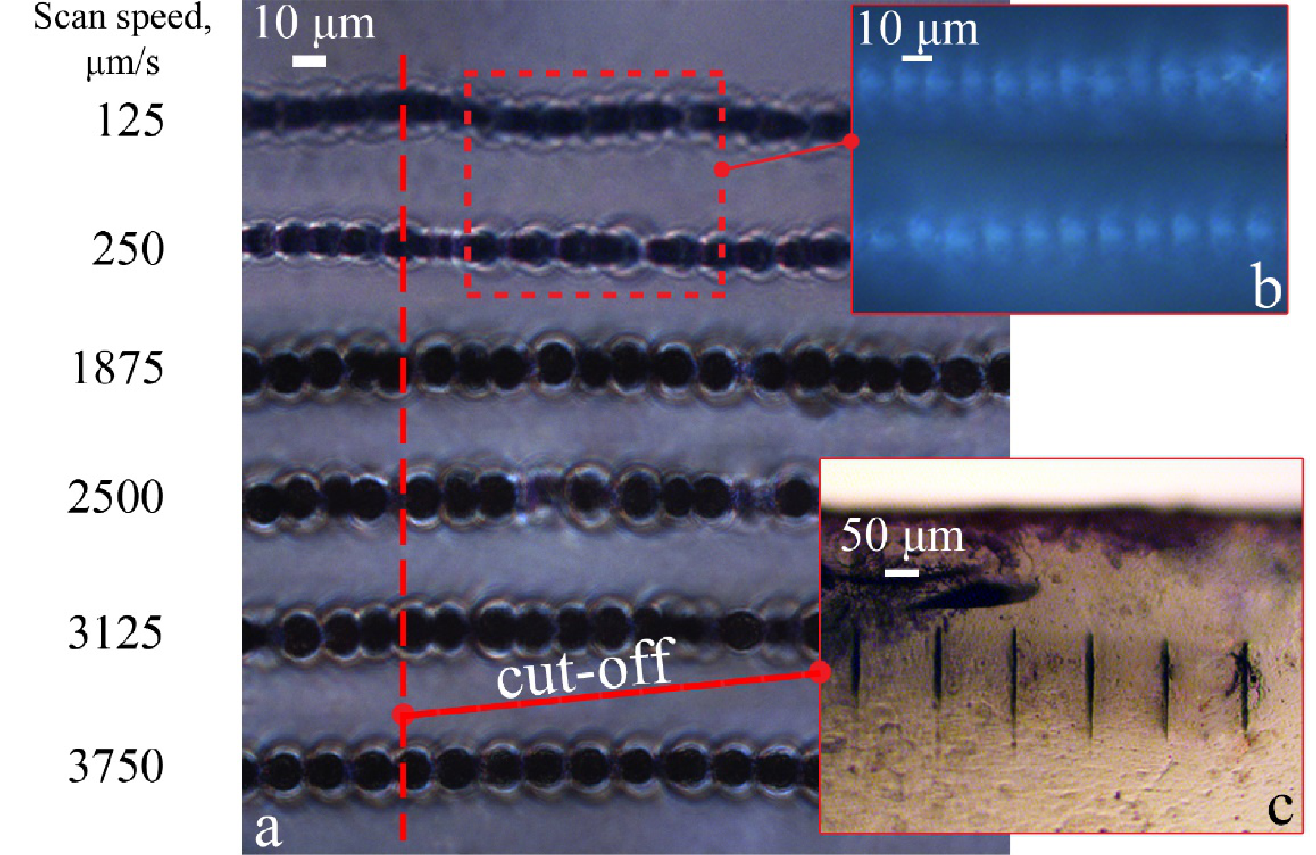}
        \includegraphics[width=0.35\textwidth]{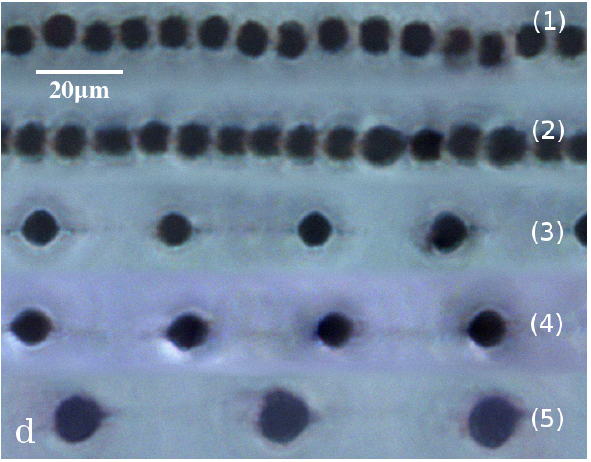}
    \caption{(a-c) Optical transmission microscopy of decompaction regions obtained for different writing speeds; (d) image of periodic lines of voids formed inside of the porous glass at a constant pulse energy ($Ep = 2.2\mu J$) and with different number of pulses per focusing spot: (1) 2600, (2) 3200, (3) 5330, (4) 32000 and (5) 80000. Experimental data from Ref. \cite{ma2017well}.}
\label{fgr:c}
\end{figure}

\begin{figure}[ht!]
 \centering
    \includegraphics[width=0.45\textwidth]{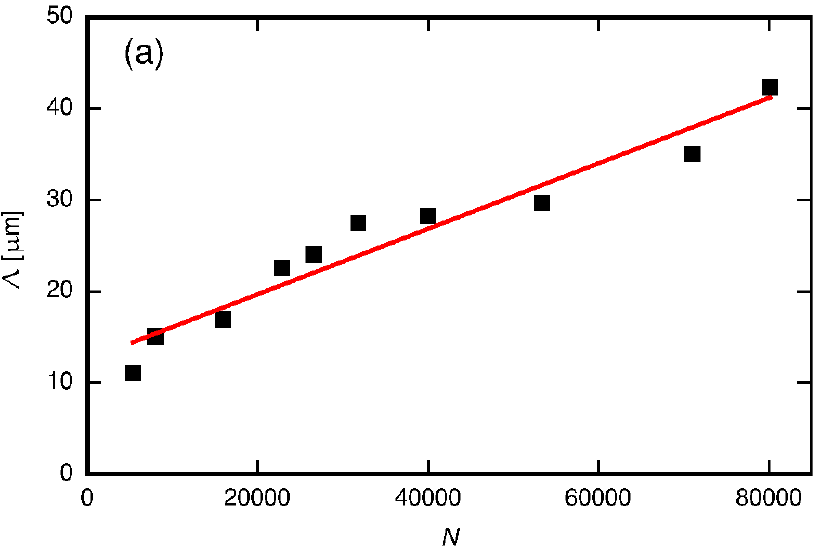}
\includegraphics[width=0.45\textwidth]{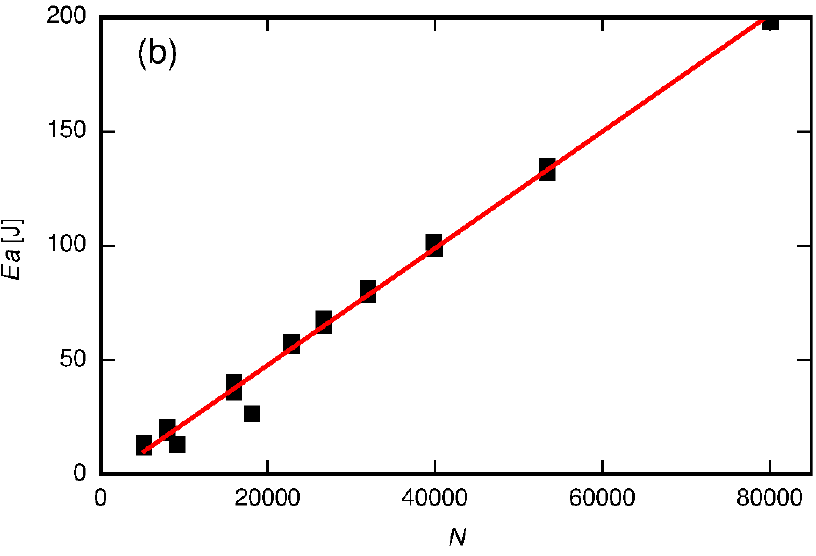}
    \caption{(a) The period of the decompaction regions as a function of the number of laser pulses per spot at a constant pulse energy ($E_p = 2.2 \mu J$); (b) the total threshold energy required to obtain the decompaction of PG using mentioned experimental setup as a function of the number of pulses per focusing spot. Experimental data from Ref. \cite{ma2017well}.}
\label{fgr:d}
\end{figure}

At rather low laser energy and the number of laser shots per focusing spot, a moderate densification, or, compaction is observed in the porous glass. In the intermediate range of laser energy per pulse and of the number of laser pulses per spot, the decompaction structure is observed (Fig. \ref{fgr:c}) and is typically characterized by a series of voids with a size up to ten micrometers, which changes only slightly with the number of laser pulses per focusing spot.  Importantly, the observed microvoids appear with a periodicity from a few microns to several tens of micrometers (Figs. \ref{fgr:c} and \ref{fgr:d}(a)) in the considered laser irradiation regime. The distance between the voids can be varied by the total laser energy delivered per focusing diameter, $E_p$,  or by the number of laser pulses per spot, $N$ , determined by the scanning speed. Here, a couple of interesting observations can be done.

Firstly, the period  $\Lambda \sim \xi N$, increases almost linearly with  $N$ (or, decreasing scanning speed), where $\xi$ is a constant length defined by the slope of the line in Fig. \ref{fgr:d}(a). Here,  $\xi \approx 0.62 nm$. Secondly, the performed experiments clearly demonstrate that the laser energy required for the formation of the decompaction structures (energy threshold) also linearly increases with the number of the applied laser pulses per laser irradiated spot (Fig. \ref{fgr:d}(b)). Here, $E_a = E_0 \times N$, where $E_0$ depends on the focusing conditions and is on the order of 2.5 $\mu$J.

\subsection{Experimental study of high aspect ratio densification}

\begin{figure}[htbp]
\centering\includegraphics[width=0.9\textwidth]{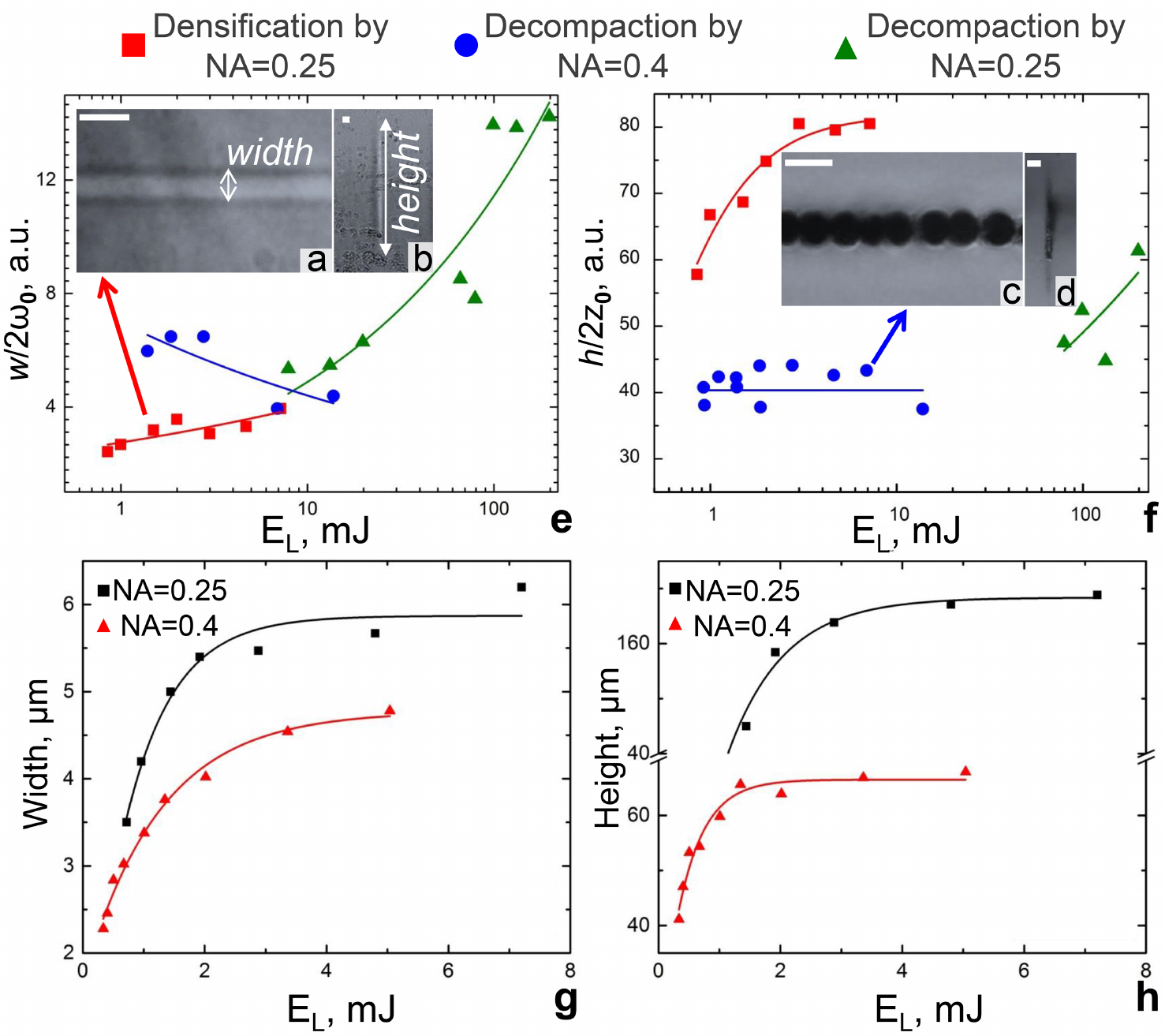}
\caption{Top view and cross-section microphotos of fabricated structures in PG: densification region (a,b) obtained by $E_L$ = 2 mJ and 10X, NA = 0.25 lens, and voids (c,d) by $E_L$ = 2 mJ and 20X, NA = 0.4 lens. Measured ratio of $w$ to the laser beam waist diameter (e) for two lenses and $h$ to the Rayleigh length ($2 z_0$) (f) as a function of $E_L$. Experimental dependencies obtained for different NA: width (g) and height (h) of the modified region as a function of $E_L$. The scale bar is 10 $\mu$m. Experimental data from Ref. \cite{itina2019ultra}.}
\label{fgr:Chapt4Exp2all}
\end{figure}

For high repetition rate multi-pulse laser irradiation, the obtained results depend not only on laser pulse energy, $E_p$, but also on the number of laser pulses per laser spot, $N$. In this case, it is convenient to use the total incident laser energy, which is defined as follows $E_L = E_p N$. Thus, in the present study a series of ultra-short laser PG scanning were performed for two different focusing conditions by varying $E_L$. When $E_L$ is small, porous glass densification is proven to take place \cite{veiko2016femtosecond,ma2017well, veiko2018direct} (Fig. \ref{fgr:Chapt4Exp2all}(a). In the intermediate range of $E_L$, a series of voids are typically formed with a size up to ten micrometers (decompaction) (Fig. \ref{fgr:Chapt4Exp2all}(b)). Further increase of $E_L$ leads to the channel formation.

Here, we focus attention at the possibilities to obtain the deepest (the highest) possible PG densification. The focusing position was the same for all the cases, $300 \mu$m below the surface. For this, we firstly  examine the shapes of the laser-modified zones, their heights ($h$) and widths ($w$), as presented as a function of $E_L$ (Fig. \ref{fgr:Chapt4Exp2all}). The obtained results show that when the lens with $\mathrm{NA} = 0.4$  is used, only decompaction regime is observed (Fig. \ref{fgr:Chapt4Exp2all}(a) and Fig. \ref{fgr:Chapt4Exp2all}(b)). This regime is characterized by a big void formation at all the considered laser energies \cite{veiko2016femtosecond, ma2017well}.  In this case, an increase in $E_L$ leads to a more spherical laser-affected zone.

On the contrary, PG densification without void formation is possible if the total incident laser energy is below 7.5 mJ at $\mathrm{NA} = 0.25$. Figs. \ref{fgr:Chapt4Exp2all}(c) and (d) demonstrate how the dimensions of the laser-densified volume evolve at $\mathrm{NA} = 0.25$ as a function of laser energy. We note a considerable difference in the shape and energy-dependent behavior of the laser-affected volume: while the corresponding ($w$) increases both for $\mathrm{NA} = 0.4$ and $\mathrm{NA} = 0.25$, an increase in $h$ is much more pronounced when $\mathrm{NA}$ is smaller. 

Based on the obtained results for $\mathrm{NA} = 0.25$, three different sub-regimes can be distinguished as follows:

(i) densification regime favorable for waveguide recording at $E_L$ smaller than 2 mJ;

(ii) densification regime with a considerably increase the depth-to-width aspect ratio at $E_L$ from 2 to 7.5 mJ;

(iii) decompaction regime at  $E_L$ higher than 7.5 mJ.

 It is noted that the second regime is more appropriate for barrier formation, where care should be taken to elongate the densified region in depth. The use of lenses with even smaller numerical aperture leads to an unacceptable broadening of the laser-affected zone.

\section{Modeling and discussion}

To explain the obtained experimental results, a thermodynamic analysis of the laser heating is performed. Every applied femtosecond laser pulses generates and heats free carries inside the glass. The subsequent electron-ion/matrix relaxation takes place on a picosecond time scale. In the considered multi-pulse regime, however, material temperature, or "base temperature" \cite{miyamoto2011evaluation,wu2005femtosecond} rises during longer time, $t_h$, corresponding to the focal spot irradiation time and creating an almost spherical region of the energy accumulation (Fig. \ref{fgr:e}(a)). This region can be then considered as a static heat source which appears consecutively in addition to the moving laser. Note, that the involved physical processes occur on very different time scales, where the energy relaxation stage is much longer than the laser absorption and heating ones.

The thermodynamic analysis is based on a modification of the procedure proposed by Miyamoto et al.   \cite{miyamoto2011evaluation,sun2007standing} for multi-pulse laser irradiation regime. In our model, both light propagation and energy absorption  per shot are calculated by using nonlinear Schr\"{o}dinger equation (NLSE) coupled with material ionization including photo-ionization, avalanche and electron trapping\cite{wu2005femtosecond}. The effective optical index of the studied porous glass \cite{veiko2016femtosecond}, $n_0$, is estimated  $\approx$1.342 according to the effective medium theory, e.g. Maxwell-Garnett theory. Moreover, the photo-ionization rate is calculated by using Keldysh theory \cite{keldysh1965diagram}. The NLSE equation used here is written as follows

\begin{equation}
    j \frac{\partial}{\partial z} \varphi = - \frac{1}{2k} {\nabla}^2_{\tau} \varphi + \frac{\upsilon_g}{2} \frac{{\partial}^2 \varphi}{\partial t^2} - k \Delta n \varphi
	\label{eqn:Eq01}
\end{equation}

\noindent where $\varphi(x,y,z,t)$, ${\nabla}_{\tau}$, $k$ and  $\upsilon_g \approx 361 {fs}^2/cm$ are the envelop function of the electric field, the Laplacian operator in the XY plane, the wave number and the group velocity dispersion coefficient, respectively. The term $\Delta n$ includes the optical index change and photo-ionization, 

\begin{equation}
    \Delta n = j \frac{W_{PI} U_g } {n \varepsilon_0 c_0 {|\varphi|}^2 k} + n_2 \frac{n \varepsilon_0 c_0 {|\varphi|}^2}{2} + j \frac{\sigma_B \rho}{2 k} - \frac{\sigma_B \omega \tau_s \rho}{2 k}
	\label{eqn:Eq02}
\end{equation}

\noindent where $W_{PI}$ is the photo-ionization rate and the related to material properties, such as band gap $U_g = 9 eV$, electromagnetic wave angular frequency $\omega$, effective mass of electron $m \approx 0.86 m_e$, and the light intensity  \cite{keldysh1965diagram}. $m_e \approx 9.1 \times 10^{-31} kg$ is the free electron mass. The temporal optical index during pulsed laser, $n$, is estimated by Drude model, and $n_2 \approx 3.5 \times 10^{-16} cm^2/W$, $\varepsilon_0 \approx 8.854 \times 10^{-12} F/m$, $c_0 \approx 2.9979 \times 10^8 m/s$, $\tau_s \approx 1.0 fs$ are the Kerr-effect coefficient, vacuum permittivity, light speed in vacuum and electron collision time, respectively. $\sigma_B$ is the cross section for inverse Bremsstrahlung absorption, $\sigma_B = (e^2\tau_s / m(\omega^2 \tau_s^2 + 1) \cdot (1/n c_0 \varepsilon_0))$, where $e \approx 1.6022 \times 10^{-19} C$ is the elementary charge. The free electron density is described by the following single rate equation during photo-ionization \cite{wu2005femtosecond}
\begin{equation}
    d \rho / d t = (W_{PI} + \eta I \rho) (1-\rho / \rho_m) - \rho / \tau_p
	\label{eqn:Eq03}
\end{equation}

\noindent which takes into account the multi-photon ionization, avalanche $\eta = \sigma / U_{eff}$ and electron relaxation time $\tau_p \approx 150 fs$. Here, $\omega_0$, $\lambda$, $\rho_m \approx 2.2 \times 10^{22} cm^{-3}$, $I = n c_0 \varepsilon_0 |\varphi|^2 / 2$ and $U_{eff} = (1+m/m_e) \cdot (U_g + e^2 |\varphi|^2 / 4m \omega^2)$ \cite{kaiser2000microscopic} are the waist radius, maximum plasma density, laser wavelength in vacuum, laser intensity and effective band gap, respectively.The solution of these equations results in the following distribution of the total absorbed energy
\begin{equation}
	q(z) =  \pi \omega_0^2 [1+(\frac{\lambda z}{\pi \omega_0^2 n_0})^2] \int_t \frac{4\pi \cdot imag(n)}{\lambda} I(z,t) dt
	\label{eqn:Eq04}
\end{equation}

\noindent Finally, thermal behavior at time $t$ is considered as a heating by consecutive laser sources separated along the scanning direction with distances of $Vs / \nu$. The analytical solution of the thermal equation is derived by Green's function method as follows \cite{miyamoto2011evaluation}
\begin{equation}
	\Delta T(x,y,z,t) = \frac{1}{\pi \rho_g c_g} \sum_{i=0}^{N} \frac{1}{\sqrt[]{\pi \alpha (t-i/\nu)}} Q_i
	\label{eqn:Eq05}
\end{equation}

\begin{equation}
    Q_i = \int \frac{q(z)}{\omega_z^2 + 8\alpha(t-i/\nu)} exp[-\frac{2\{[x+V_s(t-i/\nu)]^2 + y^2\} }{\omega_z^2+8\alpha(t-i/\nu)} - \frac{(z-z')^2}{4\alpha t}] dz'
    \label{eqn:Eq06}
\end{equation}

\noindent where $\omega_z = \omega_0 \sqrt[]{1+(\frac{\lambda z}{\pi \omega_0^2 n_0})^2}$ is the estimation of laser beam radius along $z$ axis. Thermal parameters such as mass density $\rho_g$, heat capacity $c_g$ and thermal diffusity $\alpha$ are shown in Table \ref{table:a}. With all these equations at hand, we estimate the temperature distribution and evolution during multi-pulse laser scanning. Fig. \ref{fgr:e} shows the 2D base temperature distribution and evolution at various scanning speed. 

\subsection{Periodical void formation}

The formation of the periodic patterns in multi-pulse regime is often attributed to the heat accumulation \cite{graf2007pearl, matsuo2015spontaneous}. In ultra-short laser processing, heat accumulation is typically considered as non-negligible when time between successive laser pulses, $\delta t = 1 / \nu$,  becomes shorter than the characteristic cooling time of the focal spot.  When these times are comparable as in the present study (laser repetition rate is $500 kHz$), heat can escape from the focal region avoiding strong thermo-mechanical effects. 

\begin{table}[ht!]
\centering
\caption{Parameters summary used in simulation}
\begin{tabular}{llll}
\hline
\hline
Parameter & Description & Value or definition &Reference\\ \hline
$ \lambda $ & Laser wavelength in vacuum & $515$ nm &\\
$n_0$ & Effective optical index of porous glass & 1.342 &\\ 
$\upsilon_g$ & Group velocity dispersion coefficient & $361$ {fs}$^2$/cm & Ref(\cite{wu2005femtosecond,gulley2012interaction})\\ 
$ m_e $ & Free electron mass & $9.1\times 10^{-31}$kg &\\
$ m $ & Effective electron mass & $0.86 m_e$ &Ref(\cite{wu2005femtosecond})\\
$ n_2 $ & Nonlinear refractive index & $3.5 \times 10^{-16}$ cm$^2$/W &Ref(\cite{wu2005femtosecond})\\
$ \varepsilon_0 $ & Vacuum permittivity & $ 9.954 \times 10 ^{-12}$ F/m &\\
$ c_0 $ & Light velocity & $2.9979\times 10^8$ m/s &\\
$ \tau_s $ & Electron collision time & $1.0$ fs &Ref(\cite{wu2005femtosecond})\\
$ \tau_p $ & Electron relaxation time & $150$ fs &Ref(\cite{wu2005femtosecond})\\
$ e $ & Elementary charge & $1.6022\times 10^{-19}$ C &\\
$ \rho_m $ & Maximum plasma density & $2.2\times 10^{22}$ cm$^{-3}$ &Ref(\cite{wu2005femtosecond})\\
$ U_g $ & Material band gap & $9$ eV &\\
$ \tau $ & Pulse duration & $200$ fs &\\
$ \nu $ & Repetition rate & $500$ kHz &\\
$ V_s $ & Scanning speed & $0.125 - 100$ mm/s &\\
$ \omega_0 $ & Laser beam waist radius & $2.45 - 2.6\mu $m &\\
$ E_p $ & Pulse energy & $1.5 -  2.3$ $\mu$J &\\
$ \rho_g $ & Mass density of porous glass & $2.1 -  2.2$ g/cm$^3$&\\
$ c_g $ & Heat capacity & $0.8 -  1.6$ J/g/K&\\
$ \alpha $ & Thermal diffusity & $2.7 \times 10^{-7}$ m$^2$/s&\\

\hline
\hline
    \label{table:a}
\end{tabular}
\end{table}

In our case, laser source moves continuously in the direction transverse to the propagation one. As a result, laser beam passes the focal volume with lateral diameter $d_f=2R_f$ during the heating time $t_h=d_f/V_{s}$ by applying $N=\nu\frac{2R}{V_{s}}$ laser pulses. Thus, the linear dependencies revealed in Fig. \ref{fgr:c}  can be presented as follows
\begin{equation}
     \Lambda (N) = \xi N = \xi N_0 + \xi (N - N_0) = d_v + \xi N_1
	\label{eqn:Eq1}
\end{equation}
\noindent for the period of the void structure, and

\begin{equation}
     E_a(N) = E_0 N = E_0 N_0 + E_0 (N - N_0) = E_v + E_0 N_1
\label{eqn:Eq2}
\end{equation}
\noindent for the threshold laser energy respectively, where $N_0$ is the  minimum number of laser pulses per focusing spot in the decompaction regime; $N_1 =( N - N_0)$; $d_v$ is the typical void diameter, and $E_v = E_0 N_0$ is the minimum laser energy per focusing spot required for decompaction. Thus, these dependencies indicate that only a small fraction of energy goes to the void formation. When the number of laser pulses overcomes $N_0$, then a large fraction of energy is lost, and this amount rises linearly with the number of laser pulses as well as structure period. 

\begin{figure*}[ht!]
  \centering
    \includegraphics[width=0.6\textwidth]{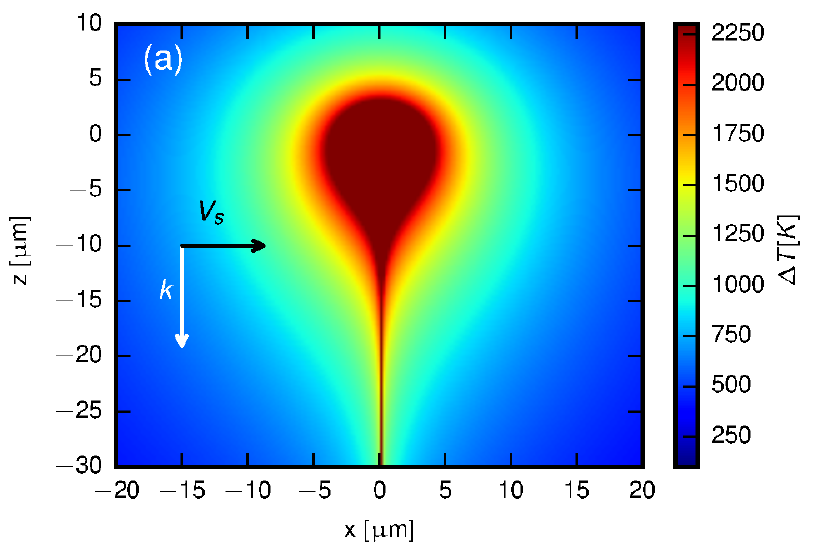}\\
        \includegraphics[width=0.4\textwidth]{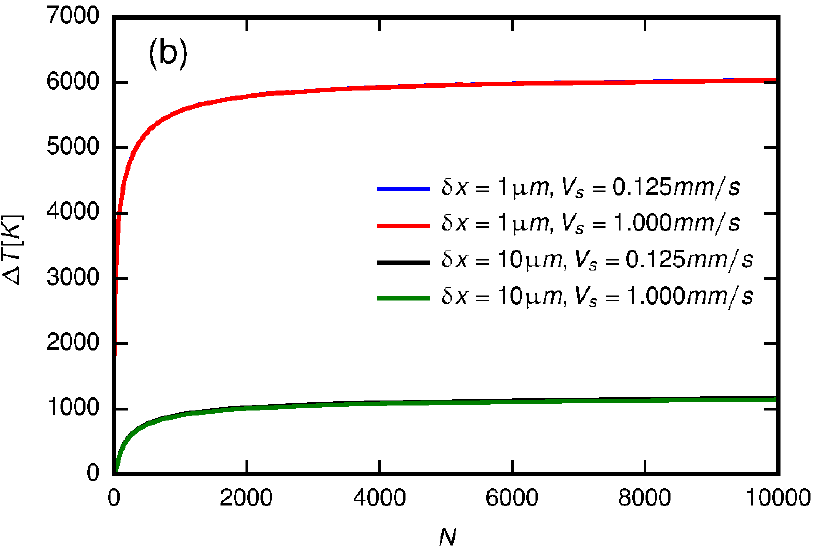}
            \includegraphics[width=0.4\textwidth]{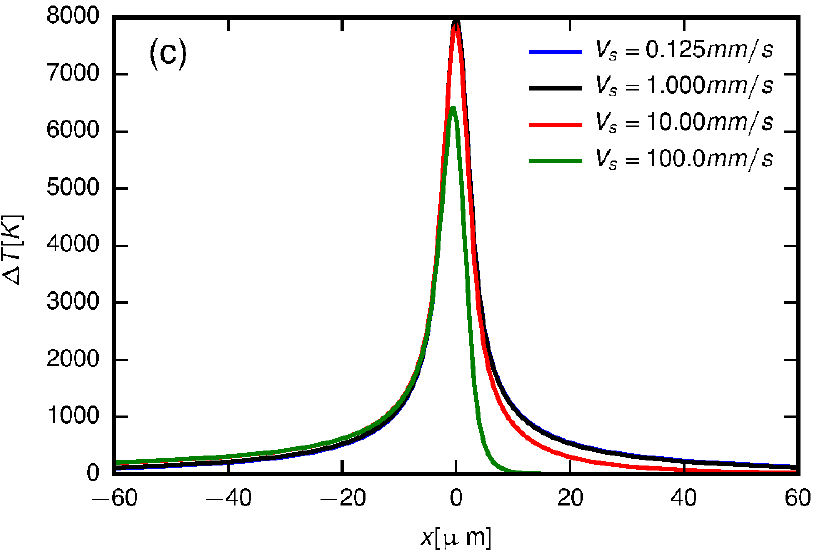}
    \caption{(a) XZ view of base temperature distribution ($E_p = 2.2 \mu J$, $V_s=1 mm/s$, $\nu=500 kHz$, $\omega_0=2.45 \mu m$) after 10,000 pulses irradiated on the porous glass; (b) temporal profiles of base temperature increase in positions of (1,0,0)$\mu m$ and (10,0,0)$\mu m$ at various laser scanning speed; (c) 1D temperature increase along X axis at various scanning speed after 10,000 pulses illuminated on the material, as one can see, at speed lower than $10$ mm/s the heat front propagates faster than the laser scan; Note, in this calculation, the coordinate is moving at speed of $V_s$; for simplicity, heat parameters are set as constants with $c_g = 1.6J/g/K$, $\rho_g = 2.2 g/cm^3$ and $\alpha = 2.7 \times 10^{-7} m^2/s$.}
 \label{fgr:e}
\end{figure*}

Material decompaction occurs when the absorbed energy is high enough to heat material in the focal region (at $d_f / 2$) up to a temperature exceeding the glass softening temperature and when the energy source supplies enough energy for the void formation. It was estimated that $\approx 2.5 \mu J$ per pulse is required for the tensile stress in the localized molten region of fused silica to exceed several MPa, which is enough for a void formation (see Ref. \cite{bulgakova2015modification}). As one can see in Fig. \ref{fgr:d}(b), in our case $\approx 2.2 \mu J$ is the laser energy per pulse, but at least $E_v$ is required to create a void structure and this energy rises with $N$. The energy losses include energy spent for light reflection, scattering, and transmission. Additionally, laser energy is required for such effects as material ionization (both field and avalanche ionization processes), plasma heating, phase transitions (here, softening, densification, plasma formation), and it is also dissipated via pressure waves, heat conductivity, convection and viscous flow motion, and even radiation.  As a result, only a very small fraction of the absorbed energy actually leads to the void formation. 

Apparently, energy losses increase linearly with $N$, so that the most effective regime for void structure formation is when $N$ is rather small, or at the highest laser writing speed required for the void structure formation (see Fig. \ref{fgr:d}(b)) . In this case, voids are closely packed and $\Lambda = d_v$. The typical void size is $\approx 10 \mu m$ corresponds to the radius of the molten and superheated region and exceeding the diameter of the focal region (around $4 \mu m$), where gas/plasma phase is formed.

To estimate void size, we define the temperature required for void formation for the corresponding laser irradiation conditions by applying the Grady's criterion for spall in liquid \cite{grady1988spall}. According to the viscoelastic energy conservation law, the elastic energy of the deformation is defined as $\approx Y\sigma{t^2}$, where $t$ is the deformation time limited by the time between two pulses $2 {\mu}s$ and $\sigma =\frac{\Delta{\rho}}{\rho t}$ is the laser-induced strain rate, should be sufficient to do the work against the dissipation forces defined by ${\eta}\sigma$. The viscosity of porous glass is lower than that of the pristine fused silica and its dependencies on the glass porosity have been previously reported \cite{sura1990viscosity, boccaccini1995viscosity}. For instance, for porosity of $26\%$, the viscosity is approximately ten times lower than it for the pristine silica. This way, it is possible to derive the maximum viscosity required for cavitation as follows $\eta_{max} = Y\sigma{t}^2 \approx 10^6 Pa \cdot s$. These viscosities correspond to the temperature $T_d$ of order $2500 K$, exceeding boiling temperature of the material. The results of the performed calculations presented in Fig. \ref{fgr:e} indicate that the thermo-affected zone with the temperatures exceeding $2500 K$ is around $10 {\mu}m$, which agrees well with the experimental results (Fig. \ref{fgr:c}).    

We note that for the solid material, another equation \cite{gamaly2006laser, cerkauskaite2017ultrafast} can be derived based on the energy conservation law as follows: assuming that the internal energy of the material with volume of $\frac{4}{3}\pi r^3$ is around the absorbed laser pulse energy, one gets $\frac{4}{3}\pi Y r_{0}^3 \approx E_{abs}$. The fact that voids grow only slightly with laser energy
means that upon a certain laser energy above the densification threshold, the absorbed fraction is almost constant, whereas energy losses should grow linearly. Then, for the diameter of a spherical void can be estimated as follows $d = C\sqrt[3]{\frac{6 \alpha E_p}{\pi Y}}$, where $\alpha$ is the fraction of laser energy $E_p$ absorbed by the material; $Y$ is Young modulus; $C = \sqrt[3]{\frac{\Delta{\rho}}{\rho}}$ is the compression ratio. For fused silica, Young modulus $Y = 74.5 GPa$, whereas for porous glass it can be smaller, for instance $Y = 17.5 GPa$ was reported by Cerkauskaite et al. \cite{cerkauskaite2017ultrafast}. For the absorbed energy on the order of $2 \mu J$, the estimation gives void diameter $\approx 10 \mu m$, which is more than twice larger than the focusing diameter $d_f$.

\begin{figure}[ht!]
  \centering
    \includegraphics[angle=0, width=0.5\textwidth]{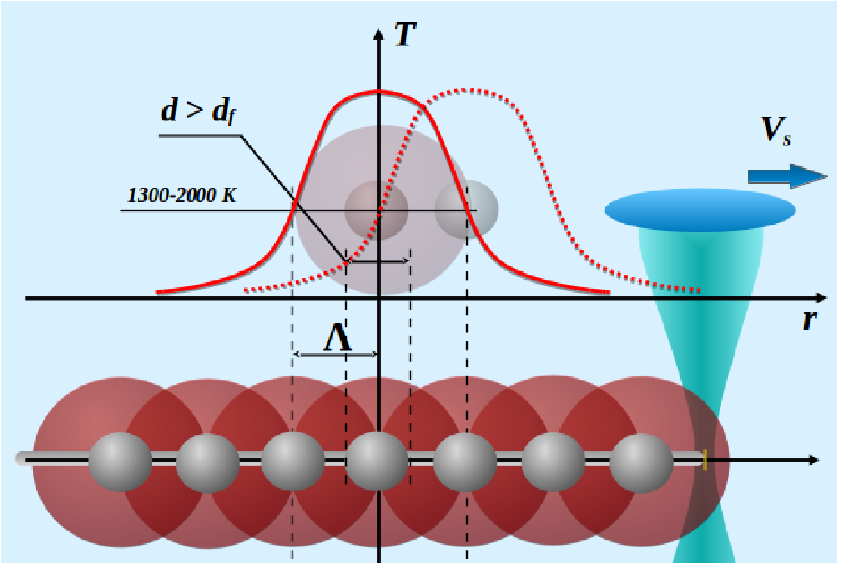}
    \caption{Schematics illustrating the void structure formation}
 \label{fgr:f}
\end{figure}

It remains to explain structure periodicity, and, namely, the observed almost linear period dependency on the number of laser pulses per focusing spot.  Considering that cooling time is rather long and a scanning heating source supplies energy all the time, a stationary regime is expected in the absence of the void formation. In this regime, once void is formed, the temperature field outside the void is induced by two heat sources: (i) a moving laser source; and (ii) a static spherical heat source formed at the location of the previously formed void (Fig. \ref{fgr:f}). The solution of a static thermodynamic problem is straightforward and gives  $T = Q / 4 \pi k r$  outside of the void, where $k$ is thermal conductivity and $Q$ can be considered to be proportional to $E_0 N$, indicated by the revealed linear dependency of the decompaction threshold. Heat diffusion induces densification and softening. The fronts of these phase transitions propagate longer distances when more energy is absorbed, therefore  $\Lambda=r_d = Q / 4 \pi k T_d$. Once laser overcomes the modified region, again enough energy can be spent for the next void. When $r_d$ is smaller than $r_f/2$, we get a closely packed void structure, otherwise the voids are spaced with a distance $\Lambda$ that rises linearly with $N$.

\subsection{Densification}

The observed results rely on the physical processes involved in the formation of laser-induced modifications. When femtosecond laser is used, non-linear ionization generates electron plasma (ELP). Both defect formation and ELP relaxation lead to the local modifications in the refractive index. As a result, optical properties, such as scattering and absorption, are modified. 

\begin{figure}[!ht]
\centering\includegraphics[width=0.9\textwidth]{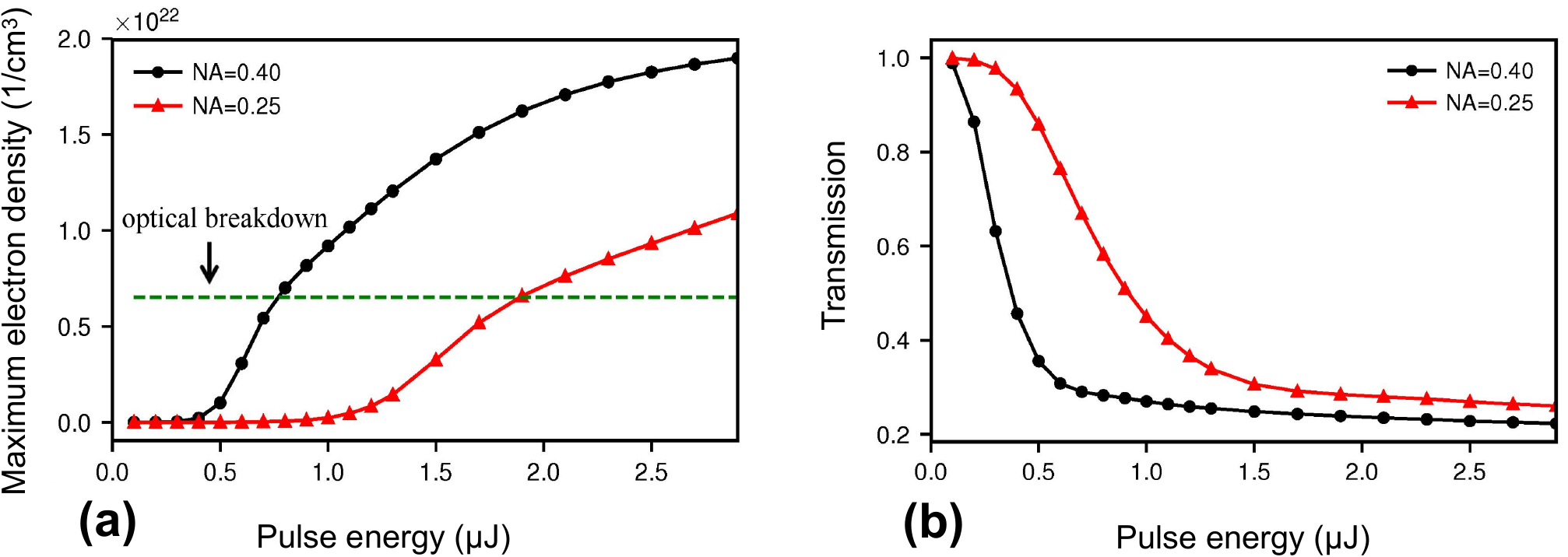}
\caption{ Calculated maximum electron density (a) and light transmission (b) for single ultra-short laser pulse. The model parameters are given in \cite{ma2017well}. }
\label{fgr:Chapter4_exp_4all}
\end{figure}

The generated ELP strongly affects laser propagation. Our calculations \cite{ma2017well} show that the maximum ELP density depends on both  laser focusing conditions and laser pulse energy (Fig. \ref{fgr:Chapter4_exp_4all}(a)). Importantly, optical breakdown takes place at laser energy of $~1.9 \mu$J for $\mathrm{NA} = 0.25$, whereas for $\mathrm{NA}=0.4$ it is expected at at a much smaller laser energy ($~0.7 \mu$ J). The transmission also drops at much smaller energy for $\mathrm{NA}=0.4$ than for $\mathrm{NA} = 0.25$ ((Fig. \ref{fgr:Chapter4_exp_4all}(b)). These results explain the difference in the material modifications regimes observed for these two different laser focusing conditions.

Importantly, if laser power is above the critical one $P_{cr} = 3.77\lambda ^2 / 8 \pi n_0 n_1$, where $n_0$ and $n_1$ stand for the linear (1.34) and nonlinear refractive ($\approx 0.4\times 10^{-19} m^2/W$) indexes of the material \cite{cimek2017experimental}, $P_{cr} = 7.4\times 10^5 W$), the laser propagation of a Gaussian laser pulse is accompanied by such effects as self-focusing and filamentation \cite{couairon2007femtosecond}. Further increase in laser power becomes inefficient (a so-called "intensity clamping effect" \cite{xu2008effect}). 

In all the experiments,  laser power exceeded the critical one.  In fact, $P_0 = E_p/\tau$, it ranges from $5.0\times 10^5 W$ to $1.2 \times 10^7 W$ for $E_p$ at $0.1\mu J$ and $2.3 \mu J$ respectively. Despite the fact that the critical power does not depend on the laser focusing, both the characteristic length of the filament appearance $z_f$ (1) and filament dimension strongly depends on the laser beam waist \cite{couairon2007femtosecond}, where

\begin{equation}
z_f = \frac{0.367 k \omega^2_0}{\sqrt{[(P_0 / P_{cr})^{1/2} - 0.852]^2 - 0.0219}}
\label{eqn:eq1}
\end{equation}

\noindent where, $k$ is a coefficient. Note that the equation would be invalid for incident power $P_0 > 100 P_{cr}$ \cite{couairon2007femtosecond}. Because $z_f$ is proportional to $\omega_0^2$, even rather small difference in $\omega_0$ affects the filamentation process. In our case, when $\mathrm{NA} = 0.25$ ($2 \omega _0 = 5.7 \mu m$), and if $\mathrm{NA}=0.4$ ($2 \omega_0 = 3.6 \mu m$), so that the ratio of $z_f$ parameter in our case is $\approx 2.5$. This means that for $\mathrm{NA} = 0.25$ laser propagates $\approx 2.5$ longer distance before the possible appearance of self-focusing than at $\mathrm{NA}=0.4$. During the propagation path, the absorption and heating of ELP typically lead to laser energy losses. In fact, denser plasma formation and heating lead to the decompaction (void formation) rather than to the formation of a densified region. Thus, at smaller values of $E_L$, mostly densification is observed for $\mathrm{NA} = 0.25$ (Fig. \ref{fgr:Chapt4Exp2all}). 

\begin{figure}[!ht]
\centering\includegraphics[width=0.8\textwidth]{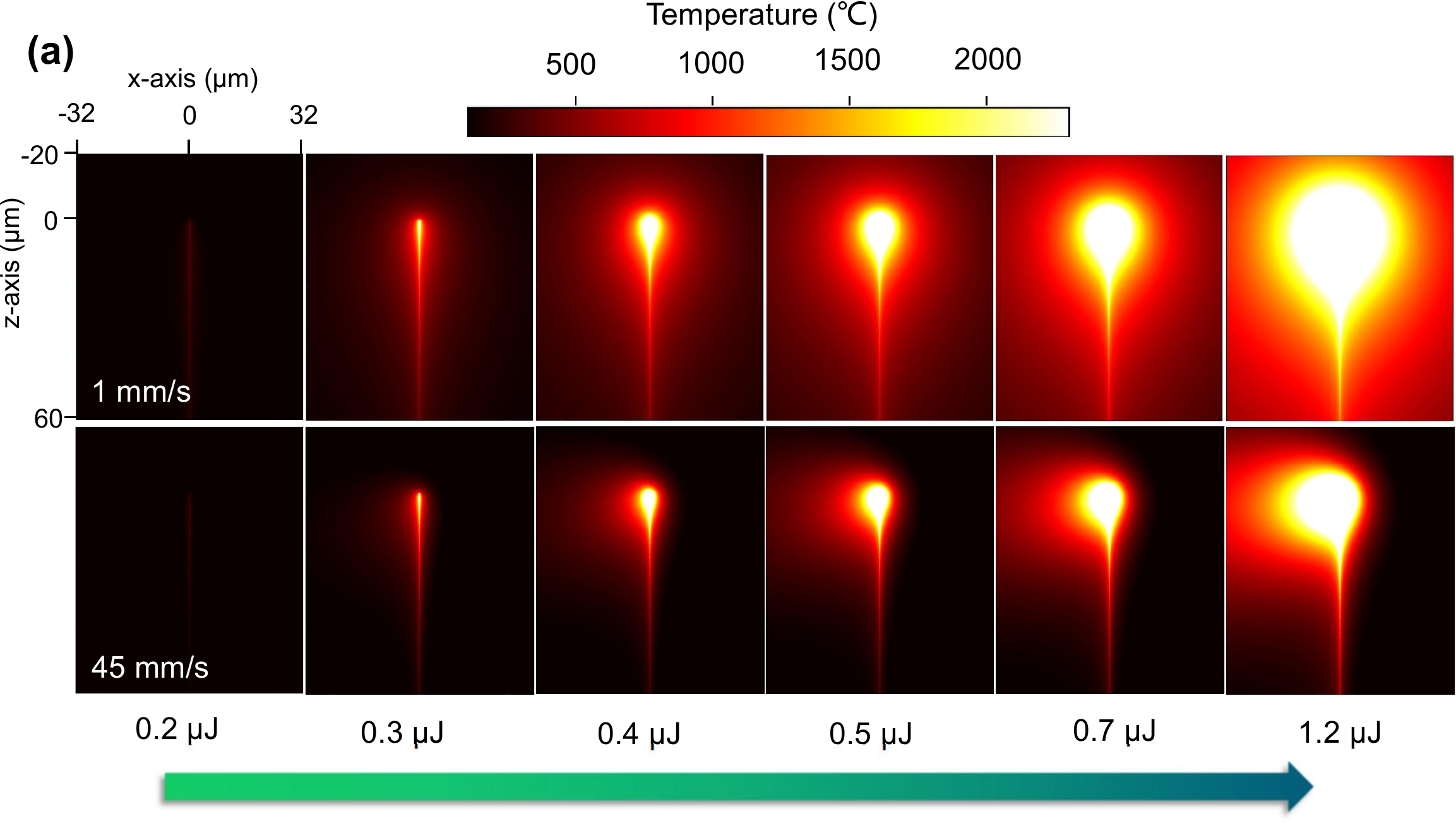}\\
\centering\includegraphics[width=0.8\textwidth]{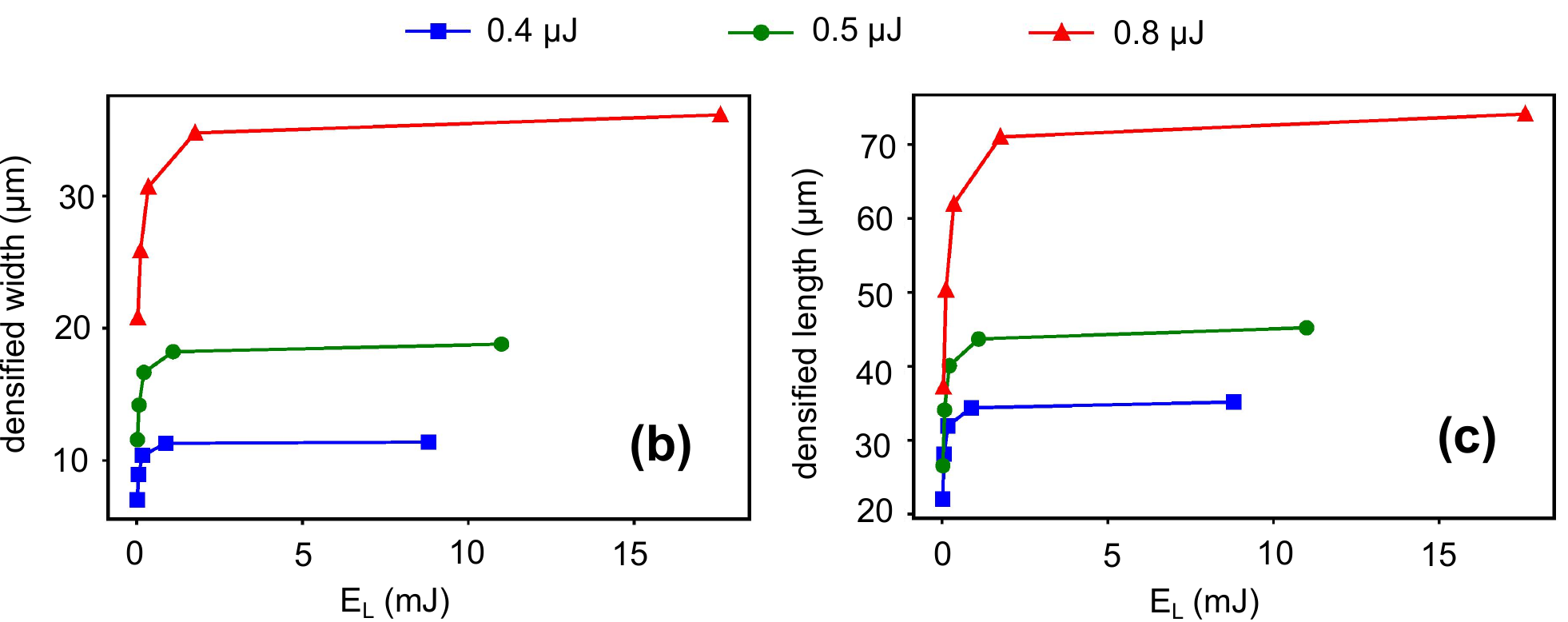}
\caption{Heat accumulation by an ultra-fast laser of high repetition-rate (500 kHz $\mathrm{NA} = 0.25$) at various pulse energies and scanning speeds (a).  Laser moves in the positive X direction (from left to right). Dimension of the densified area: width (b) and elongation (c).  Here, these dimensions are estimated based on the temperature of $1000^\circ\text{C}$ \cite{veiko2016femtosecond}. }
\label{fgr:Chapter4_exp_5all}
\end{figure}

As we have noted in the Introduction, additional effects, such as defect, electronic and/or thermal accumulation, play a role in a multi-pulse regime. Particularly, heat accumulation depends on laser pulse width, repetition rate, laser spot dimensions and scanning speed, as well as on the material properties.

For this case, a combined model was developed \cite{ma2017well} that carefully accounts for ultra-short laser light propagation,  material ionization, and heating. A series of calculations are performed for the case of $\mathrm{NA} = 0.25$ (Fig. \ref{fgr:Chapter4_exp_5all}). Based on the calculated temperature distributions (Fig. \ref{fgr:Chapter4_exp_5all}(a)), we  evaluate the sizes of the laser-modified region (Fig. \ref{fgr:Chapter4_exp_5all}(b) and (c)). In the presented experiments discussed above, the system is designed to control the total incident laser energy $E_L$. To better compare the multi-pulse simulations with experiments, the results are shown as a function of $E_L$, which is calculated as $E_L = \frac{2\omega_0 f_p E_p}{Vs}$, where $f_p = 500 kHz$ is the repetition rate, $Vs$ is the laser scanning speed. The performed simulations reveal how the corresponding temperature profiles change with the increase in laser pulse energy for two laser scan speeds (1 mm/s and 45 mm/s). At high speed, a tail appears in the temperature distribution if laser pulse energy is high enough. This tail is commonly observed behind powerful moving heat sources. The higher the speed of the heat source, the longer is the tail.  At lower laser energies, this tail is, however, less pronounced.  

Importantly, Figure \ref{fgr:Chapter4_exp_5all} confirms the experimentally observed rise and saturation in the dimensions of the laser-modified area. Here, they were estimated based on the dimension of the region where temperature exceeds softening one. The observed saturation in the dimensions of laser-densified region is mainly caused by the reflection and scattering effects of the ELP. 

We note, furthermore, that laser-induced modifications in local optical properties are known to cause local field enhancement\cite{pereira2008laser, rasedujjaman2018polarization} and pulse-to-pulse elongation of the  laser-affected region. Upon many pulse, depending on laser energy, repetition rate, induced refractive index change, and the size of the modified area, such accumulation effects either lead to a further increase in the aspect ratio of the laser-modified region\cite{rudenko2016random, mcleod2008subwavelength}, or to the appearance of a big void in its center if laser energy is high enough \cite{watanabe2000optical, veiko2016femtosecond, bulgakova2015modification}. It is estimated that about 2.5$ \mu J$ per pulse is required the void formation \cite{bulgakova2015modification, ma2017well}.

\section{Conclusion}

The periodic void structures produced in a well-controlled manner inside porous glasses by femtosecond laser pulses. The trains of ultrashort laser pulses with a properly chosen energy per pulse and writing speed not only provide the required energy deposition, but also allow an efficient inscription of the periodic arrays of voids in these materials. Contrary to powerful single laser pulses that are known to produce explosion-like effects, pulse trains can be used to considerably reduce non-desirable effects, such as material cracking. 

The performed thermodynamic analysis have shown that the possibility of an efficient control over laser micromachining in volume relies on a better understanding of the physical mechanisms involved. For the given pulse repetition rate, the main controlling parameters being the laser scanning speed and the mean energy per pulse, while the former defines both the structure period and the required laser energy per focusing spot. The period has been found to depend on the size of the densified region induced by the created almost static spherical heat source located at the void position. The final structure is induced by a superposition of two energy sources: the moving laser and the created static heat source. 

Additionally, the performed analysis has indicated that the void formation is more efficient when the minimum number of laser pulses per focusing spot corresponds to the writing speed as high as about $0.47mm/s$. Under these conditions, a compact void structure has been written in the porous glass, so that the structure period is equal to the void diameter. For lower scanning speed, both threshold and period increase linearly as a result of the rising energy losses. The lost energy goes mostly for material heating and densification thus leading to a reduced absorption at the passages between the voids.

The performed study thus provides a way of a well-controlled laser inscription of void arrays in the bulk of porous materials. The obtained results can be useful for advancing not only a wide range of photonics applications, but also for many applications in microfluidics, biology and other fields.

Furthermore, the ultra-short laser interaction with PG in a high repetition rate multi-pulse regime was investigated. Conditions favorable for the inscription of either highly symmetric wave-guides or high aspect-ratio membranes and barriers were determined, which was explained based on a series of numerical calculations. Laser focusing affects filament appearance distance. In the considered case, this effect is minor compared to the effect of the laser-induced ionization since the created ELP strongly affects thermal field as well as the shape of the laser-affected zone. The characteristic dimensions of the calculated temperature field are well correlated with the obtained shapes of the densified regions. The maximum density of the created electron plasma was found, furthermore, to strongly depend on the focusing conditions. 

Laser-based high aspect ratio densification is promising for numerous applications in microfluidics and for the fabrication of the integrated multi-purpose sensors. The identified laser irradiation conditions required for the efficient void-free material densification thus provide a way toward a better control over direct laser writing of membranes, barriers and various integrated devices in porous glasses.

\afterpage{\null \newpage}

\chapter{Continuous-Wave laser writing of silver NPs in mesoporous TiO$_2$ thin films}

\hspace{0.5cm} Comparing to porous dielectrics, the semiconducting counterparts such as TiO$_2$ and ZnO exhibit additional possibilities in photoactive applications. The photosensitive abilities were activated by the process of porous semiconductors interact with nanoscale dopants through transferring free electrons. In recent years, the interaction of semiconducting media with metallic nanoparticles demonstrated laser controlling behaviors, which was promising for optical data storage, color printing, and multiplexing \cite{destouches2014self,liu2015understanding,baraldi2016polarization,sharma2019laser}.

 In this chapter, we present numerical insights in continuous-wave laser writing silver nanoparticles (NPs) inside mesoporous TiO$_2$ thin films. Controlling the formation of NPs by direct laser writing ensures an efficient way of engineering the optical responses of nanocomposite materials through adjusting laser parameters such as power, focusing and writing speed. The process of a NP's growth, photo-oxidation, and reduction affect the light absorption by an ensemble of laser-activated NPs. However, those self-consistent process are not well understood. Previous studies \cite{liu2015understanding, liu2016laser, liu2016selfthesis} explained the speed threshold switching the growth or shrinkage of silver NPs by neglecting heat diffusion, it remained unclear, however, how the NP size can be controlled by the laser writing speed. In this chapter, based on the coupled calculations of size-variation, light absorption, and heat diffusion by an ensemble of NPs, we propose a two-dimensional model taking account of spatial information of size and temperature to have a further investigation into the process. The spatial size distribution is revealed to be non-uniform leading to the transmission inhomogeneity along the laser written lines. This fact is confirmed by the in-situ transmission experiments. The performed study also depicts a novel view in which NPs grow ahead of the laser beam center due to the heat diffusion. The nonlinear growth never stops until it exhausts the majority of the free Ag$^0$ in the matrix, while the amount of Ag$^0$ by reduction cannot compensates the consumption. After that, the photo-oxidation dominates the process and finally acts as the controlling role in NP size depending on the writing speed. The simulations also show that it is not the activation energy of Ag$^0$ diffusion but the ionization efficiency affect the final NP size, which helps understand in how to improve the laser processing of differently prepared samples. 
 
 Reproduced in part with permission from [H. Ma et al., J. Phys. Chem. C 123.42 (2019): 25898-25907.]  Copyright [2019] American Chemical Society.

\section{Introduction}

Plasmonic nanoparticles (NPs) and nanostructures have great potential in photovoltaics \cite{atwater2010plasmonics, nakayama2008plasmonic}, nano-imaging \cite{kawata2009plasmonics}, bio-sensing \cite{anker2008biosensing}, photoelectric sensors \cite{brongersma2015plasmon, clavero2014plasmon} and light engineering metamaterials \cite{yin2013photonic,luk2010fano, ergin2010three}. A lot of plasmonic metasurfaces have been created during the last decade to design new optical functionalities. Most of them resulted from perfectly arranged metallic oligomers and were produced by focused ion beam and electron beam lithography, two techniques that are still expensive, time consuming and not well suited to large surfaces. Nanoimprint lithography of metallic nanostructures overcomes these limitations and is expected to become a way to spread the use of plasmonic metasurfaces at the industrial level. This technique is, however, limited to reproduction of masks. Laser direct writing is another technology that provides more flexibility, is well suited for large scale processing and allows a control of plasmonic properties at sub-wavelength level.

Among the metasurfaces that have great promises for applications is titanium dioxide (TiO$_2$) loaded with metallic NPs. TiO$_2$ is one of the most studied and widely applied host-materials due to its unique characteristics, such as excellent optical transmittance in visible wavelength, high refractive index, and stable chemical properties \cite{matsumoto2001room, clavero2014plasmon}. Additionally, TiO$_2$ is a good n-type semiconductor material and it allows fast electron acceptance because of the high density of states in the conduction band\cite{clavero2014plasmon}. However, only ultraviolet light can activate the photovoltaic process due to the intrinsic wide-bandgap limitations of TiO$_2$. Embedding silver NPs in such a material forms a Schottky barrier, which boosts the photovoltaic efficiency in the visible region. The plasmonic resonance induces a high density of free electrons whose energies exceed the Schottky barrier and makes an increase in the quantum efficiency possible. This so-called "plasmonic hot-carrier" phenomenon has attracted wide interest in research communities during the past decade \cite{clavero2014plasmon, brongersma2015plasmon, tatsuma2017plasmon}. By injecting hot-electrons from silver NPs to the surrounding semiconductor matrix, NPs become unstable and tend to dissolve into Ag$^+$ ions, a mechanism also named photo-oxidation \cite{naoi2004tio2}. In the meantime, hot electrons also relax through electron-phonon coupling that heats the Ag NPs. This heating leads to other mechanisms, such as chemical reduction of silver ions and NP growth, mechanisms compete with the aforementioned photo-oxidation tending to shrink the NPs \cite{liu2015understanding, baraldi2016polarization, liu2017three}.

Using scanning laser writing for such TiO$_2$:Ag film provides an elegant way of controlling the NP size, organization and anisotropy, and thus control the optical properties of such plasmonic metasurfaces. Previously, the growth of Ag NPs in mesoporous films of amorphous TiO$_2$ was investigated as a function of scanning speed for different laser powers and focusing. The first developed model \cite{liu2015understanding} was able to simulate the experimentally observed phenomenon showing that NPs grow only above a speed threshold and tend to oxidize below this threshold. In this model, a set of differential equations was solved numerically to demonstrate the competition between shrinkage and growth during laser writing. Later, the model was improved to simulate the size distribution of the grown NPs as a function of time and speed \cite{liu2017three}. These previous numerical studies suffered, however, from several limitations. Namely, they modeled the evolution of the system in a point without considering the effect of heating by the NPs grown few micrometers apart and still under the laser beam. The temperature was then calculated by assuming a homogeneous spatial size distribution of NPs under the laser spot. This led to a good description of the final NP size distributions as a function of speed, but also to a rather questionable growth kinetics and temperature history.

Here, we propose a model, which accounts for both thermal and spatial size distribution during laser writing to better investigate the growth kinetics during the laser writing and examine the relative role of the underlying physico-chemical mechanisms involved in the laser-induced nanoparticle size evolution. The Galerkin method and adaptive time stepping techniques are used to solve the nonlinear diffusion-controlled growth model based on the open-source finite element method library (Deal.II 8.5.0) \cite{arndt2017deal}. A set of 2D and 3D simulations are carried out and compared to new experimental results obtained based on in-situ transmission measurements. The modeling part summarizes the major physico-chemical processes involved and presents how the thermal equation is coupled. The last part analyzes the obtained results and provides new insights into the dynamics of the system under continuous wave laser scanning.  In particular, we examine the  importance and interplay of the different mechanisms involved.

\section{Experimental}
\subsection{Sample preparation and in-situ characterization setup}
Mesoporous films of amorphous TiO$_2$ are prepared according to a previously published procedure based on a sol-gel process \cite{crespo2014changes}. The obtained films have pore sizes from 5 to 20 nm and a thickness estimated to  230$\pm$50nm by scanning electron microscopy (SEM - FEI Nova nanoSEM 200) on the cross-section. Silver is introduced in the pores by soaking the mesoporous sample into an ammoniacal silver nitrate solution and small silver nanoparticles are created in the film by exposing it to UV light (400 $\mathrm{\mu W cm^{-2}}$ at 254 nm wavelength). Ag NPs are observed after this process with an averaged size around 3nm \cite{liu2017haadf}. The detailed descriptions of the process were reported previously \cite{liu2015understanding,crespo2010reversible,crespo2012one,crespo2014changes, babonneau2018real}. The experiments were carried out by Dr. Said Bakhti at Lawrence Berkeley National Laboratory under the supersion of Nathalie Destouches (Laboratoire Hubert Curien) and Daniel S Slaughter (Lawrence Berkeley National Laboratory).

\begin{figure*}[ht!]
 \centering
    \includegraphics[width=0.8\textwidth]{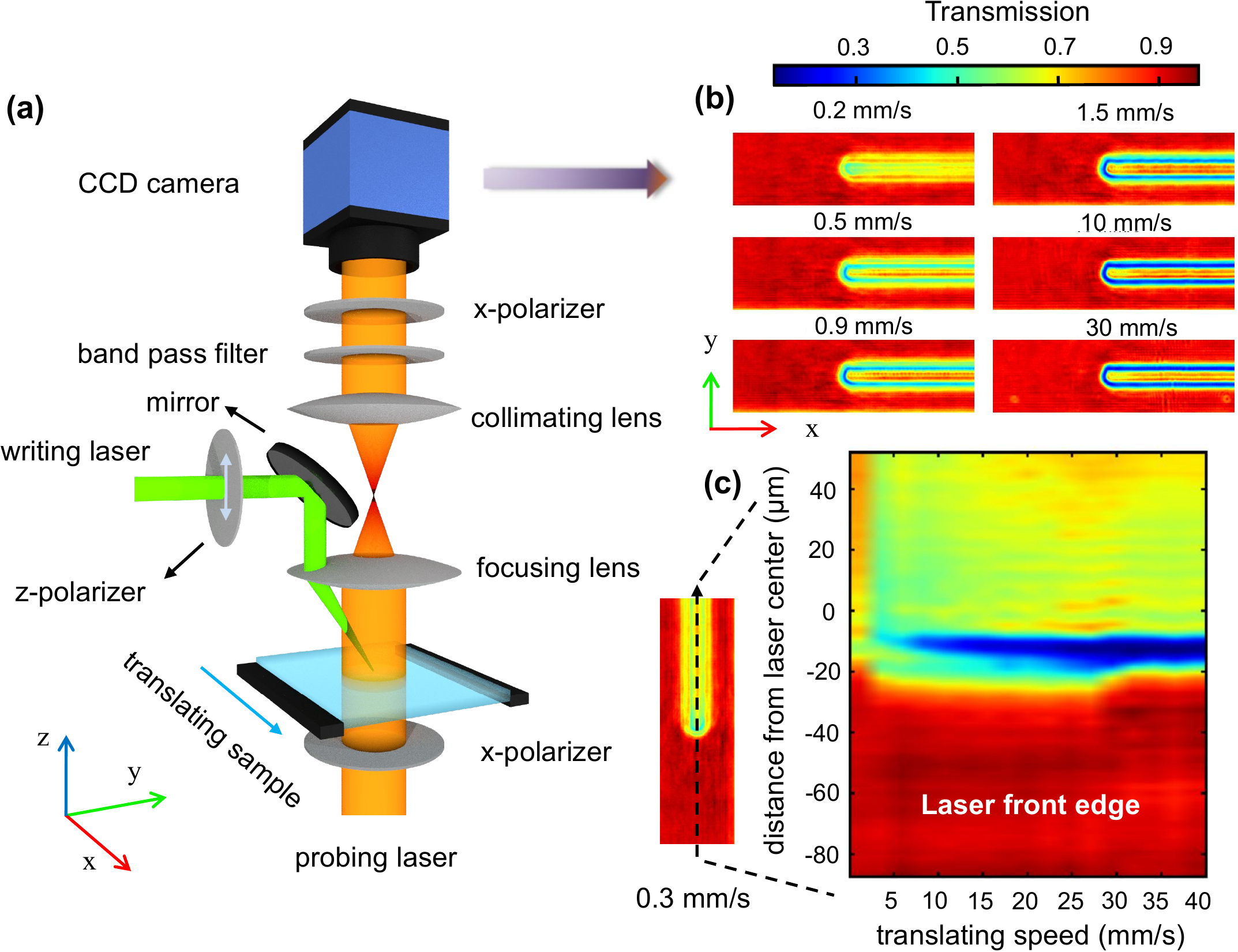}\\
    \includegraphics[width=0.8\textwidth]{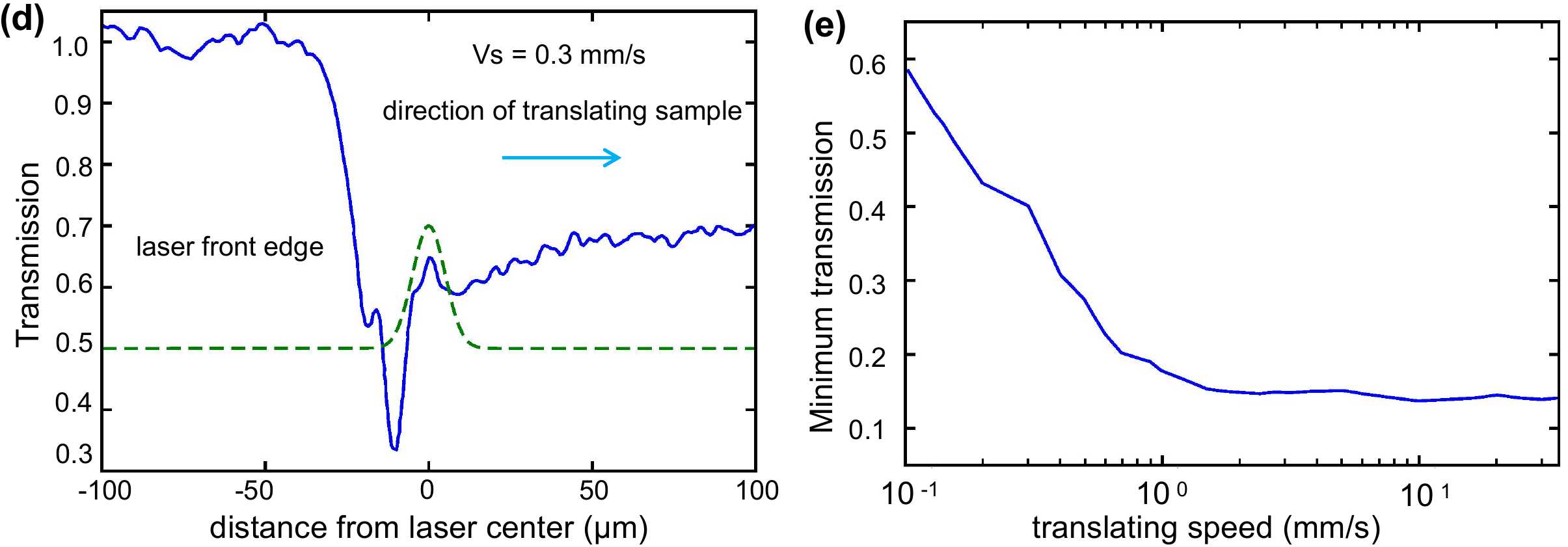}
    \caption{In-situ transmission experiments. Schematic of the laser setup (a),  transmission images of the sample under writing at various translating speeds (b), map showing the transmission along a line parallel to the scanning direction and passing through the beam center (dashed line in the left-handed image) (c), transmission profile along the dashed line of (c) at 0.3 mm/s (d), and the minimum value of transmission along the scanning direction versus scanning speed (e). The green dashed line stands for the writing laser of spot size $20 \mu m$ at $1 / e^{2}$ ($17 \mu m$ FWHM). The collimated beam before focusing is 1.7 mm full width at $1 / e^{2}$ (1 mm FWHM). Experimental results are taken from Ref. \cite{ma2019laser}.}
\label{fgr:1all}
\end{figure*}

Figure \ref{fgr:1all}(a) shows the setup used for in situ measurements of the sample transmission during laser writing. The writing laser is a 1W continuous-wave laser emitting at 532 nm and linearly polarized along the y-direction on the sample plane. It is focused on the sample surface by a 5 cm focal length lens. Imaging is performed thanks to a probe nanosecond laser emitting at 527 nm, a wavelength absorbed by growing Ag NPs. The latter is collimated on the sample and then imaged on a CCD camera through a couple of lenses, the first one being the same as the one used to focus the writing beam. Without writing beam, the sample transmittance is fully stable over time, meaning that the nanosecond laser does not modify the nanocomposite film. The imaging beam is linearly polarized (x-axis) with a polarization perpendicular to the writing beam. A linear polarizer just before the camera and a band-pass filter stop all photons from the writing beam before the camera. The writing beam is thus not seen directly on the images where only the transmission changes of the samples are observed. The writing beam position on the images is determined before carrying out in situ experiments. During the experiments, the sample is translated at a constant speed along the x-axis, in the positive direction (according to the sketch in Figure \ref{fgr:1all}(a)).

\subsection{Results of the in-situ transmission}
Transmission images are calculated by normalizing the raw images by the averaged intensity recorded in the laser front area. Images are recorded for various scanning speeds (0.1-40 mm/s) and a selection is shown in Figure \ref{fgr:1all}(b). In all cases, transmission varies in the laser written line and our attention is focused on the area located around the laser beam  corresponding to the distance "0" in Figure \ref{fgr:1all}(d). The profiles along the line center show the evolution of transmission from a position 100 $\mu m$ in front of the laser beam center to a position 100 $\mu m$ after the beam center. The measured transmission strongly decreases in the front (leading) edge of the laser beam before increasing back near the beam center and reaching a constant value in the back edge, as shown in Figure \ref{fgr:1all}(c) and (d). Unexpectedly, the lowest transmission is before the laser center. Moreover, its value decreases with increasing speed until saturation (Figure \ref{fgr:1all}(e)). The transmission inhomogeneity indicates the variation of the NPs' characteristics (such as size, density, shape and order) along the translation direction. To better understand the mechanisms involved in the NP growth and their kinetics during the laser scan, we perform numerical simulations of the measured behaviors accounting for photo-oxidation, thermal diffusion, diffusion-limited growth, reduction, and light scattering.

\section{Modeling}
\begin{figure}[!hb]
 \centering
    \includegraphics[width=0.8\textwidth]{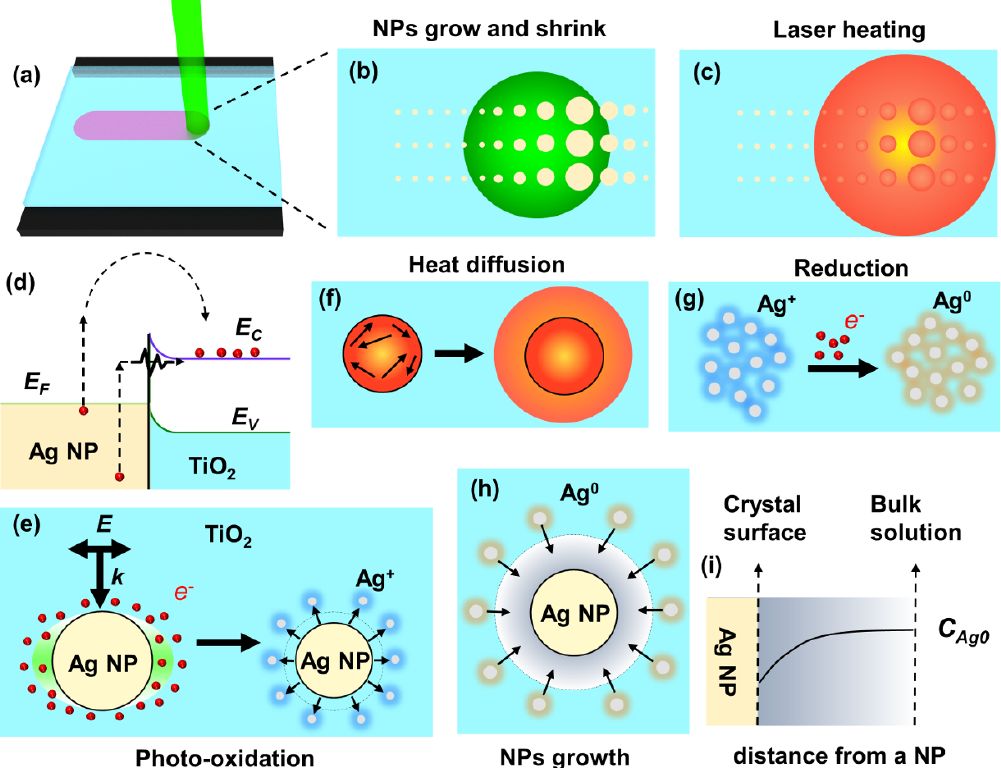}
    \caption{Illustration of the laser writing of a colorful line (a) and the mechanisms involved during laser excitation of Ag NPs inside the TiO$_2$. Ag NPs grow ahead of the laser beam and then shrink due to photo-oxidation (b). NPs inside the laser spot act as the heat sources and contribute to the temperature field due to the thermal diffusion (c). Injection and tunneling of plasmon-induced hot-electrons (d), process of photo-oxidation (e), heating and diffusion (f), reduction (g), and diffusion-limited growth (h) and (i).}
\label{fgr:2all}
\end{figure}

In this model, we are going to present how the isolated Ag NPs grow and shrinkage affecting the heat absorption inside the laser spot and vice versa. A complicated interplay during the dynamical process is revealed. Figure \ref{fgr:2all}(a) illustrates the laser writing of a colorful line by controlling writing speed. We also emphasize that Ag NPs start to grow ahead of the laser beam center as shown in Figure \ref{fgr:2all}(b). The average temperature indicating the heat absorption is a phenomenon relying on a group of NPs. The variation of a single NP size could barely explain the results shown in (Figure \ref{fgr:2all}(c)), which makes great difference with the results of the previous model \cite{liu2015understanding}.

\subsection{Photo-oxidation}
This mechanism is the one that leads to electron emission from NPs upon photon absorption. As sketched in Figure \ref{fgr:2all}(d), when Ag NPs are in contact with TiO$_2$, the electron energy required to allow electrons to leave the NPs is lowered to the value of the Schottky barrier that is created at the interface between the metal and the n-type semiconductor. Electrons excited through the localized surface plasmon resonance of NPs at the laser wavelength (2.4 eV) can cross the Schottky barrier by injection or tunneling. The band bending prevents the injected electrons from escaping back to the metallic crystals. As a result, the positively charged NPs are unstable and tend to dissolve into Ag$^+$ ions to keep the neutral state (Figure \ref{fgr:2all}(e)) \cite{liu2015understanding}. The number of silver atoms leaving a nanoparticle per unit time due to the above processes was calculated as \cite{liu2015understanding}:

\begin{equation}
    n_{oxi}(t) =\eta_0 \frac{ \sigma_{abs}(R,\lambda) I(x, y, z, t)} {h \nu}
	\label{eqn:oxidationRate}
\end{equation}

\noindent where $\eta_0$ is the ionization efficiency, $\sigma_{abs}$ the absorption cross-section of a nanoparticle of radius $R$ at laser wavelength $\lambda$, $I$ the laser intensity, $h$ the Planck constant and $\nu$ the light frequency. The ionization efficiency was estimated experimentally in reference \cite{liu2015understanding}. $\sigma_{abs}$ is calculated by using the classical Mie theory.

\subsection{Reduction}
Figure \ref{fgr:2all}(g) depicts the reduction process where Ag$^+$ ions can diffuse  and recombine with electrons from TiO$_2$ conduction band or from compounds absorbed in the film such as H$_2$, CO from the atmosphere or NO$^-_3$ from the initial silver nitrate solution \cite{crespo2011irradiance, kazuma2011nanoimaging, kazuma2012photoelectrochemical, clavero2014plasmon, kawahara2005electron}. Silver ions can come from the initial non-reduced silver species present in the film or from the photo-oxidation process. The reduction process leads to Ag$^0$ monomers that can contribute to the NP growth according to the next mechanism. We assume a homogeneous Ag$^+$ ion and Ag$^0$ monomer distribution for a small cuboid, in which matter conservation is presumed. The decrease of Ag$^+$ concentration due to the reduction per unit time has the following form \cite{liu2015understanding, kaganovskii2007formation}:

\begin{equation}
        \frac{d C_{Ag^+}}{d t} |_{red} = \exp \Big( -\frac{E_p}{N_A k_B T} \Big) \cdot D^{red}_{0} \exp \Big( -\frac{E_D}{N_A k_B T} \Big)
                                \cdot C_{red} C^{2/3}_{Ag^+}
	\label{eqn:reduction}
\end{equation}

\noindent where $E_p$ and $E_D$ are the activation energies, $D^{red}_{0}$ is the diffusion coefficient of the reducer at infinite temperature, $C_{red}$ the concentration of reducing agent, and $C_{Ag^+}$ the concentration of Ag$^+$. $N_A$ is the Avogadro constant, $k_B$ the Boltzman constant and $T$ the absolute temperature.
The first exponential function describes the probability of reduction, while the second term stands for the temperature dependent diffusion of the reducing agents.

\subsection{Diffusion-controlled growth}
Based on previous experiments \cite{liu2015understanding,crespo2010reversible,crespo2012one,crespo2014changes, babonneau2018real} we suppose that primary tiny NPs, nanoclusters, or "monomers" are formed in the porous matrix due to ultraviolet illumination. A uniform distribution of small and monodisperse NPs is considered in the initial state. Thermal diffusion is considered to be the main initiator of the Ag NP growth in the TiO$_2$ matrix under laser irradiation (Figure \ref{fgr:2all}(h) and (i)), as was shown by Liu et al. \cite{liu2015understanding,liu2016laser}.

To determine the total number of monomers absorbed by an Ag NP of radius $R$ per unit time, the Fick's first law is applied:

\begin{equation}
    n_{abs}(t) = 4 \pi R^2 D_{Ag^0}(T) \frac{d C_{Ag^0}}{d r}|_{r=R}
	\label{eqn:fick}
\end{equation}

\noindent where the diffusion coefficient $D_{Ag^0}(T)$ is described by the Arrhenius law as follows
\cite{mcbrayer1986diffusion, butrymowicz1974diffusion}:
\begin{equation}
    D_{Ag^0}(T) =  D_{0Ag^0}\exp \Big( -\frac{E_{Ag^0}}{N_A k_B T} \Big)
    \label{eqn:DAg0defi}
\end{equation}

The homogeneous distribution $C_{Ag^0}(r)$ is calculated by solving the steady state diffusion equation \cite{landau1981course}:

\begin{equation}
    \frac{1}{r} \frac{\partial ^2}{\partial r^2}rC_{Ag^0}(r) = 0
	\label{eqn:densityMonomer}
\end{equation}

\noindent with boundary conditions $C_{Ag^0}(r\to \infty) = C_{Ag^0}(\infty)$ and $C_{Ag^0}(r=R) = C_{Ag^0}(R)$. The analytic solution is obvious:

\begin{equation}
    C_{Ag^0}(r) = C_{Ag^0}(\infty) - [C_{Ag^0}(\infty) - C_{Ag^0}(R)] \frac{R}{r}
	\label{eqn:densityMonomerSolution}
\end{equation}

\noindent The equilibrium concentration at the surface of the NP is defined by the following thermodynamic equation \cite{liu2015understanding, kaganovskii2007formation, landau1981course, thanh2014mechanisms,kwon2011formation}:

\begin{equation}
    C_{Ag^0}(R) = S_{Ag^0} ( 1 + \frac{2 \gamma \omega}{R k_B T} )
	\label{eqn:densityMonomerSurface}
\end{equation}

\noindent where, $S_{Ag^0}$ is the Ag$^0$ solubility in mesoporous TiO$_2$ film, $\gamma$ the nanoparticle-matrix interfacial tension, and $\omega$ the atomic volume of Ag in crystal.

Substituting Eq. \ref{eqn:densityMonomerSolution} and Eq. \ref{eqn:densityMonomerSurface} into Eq. \ref{eqn:fick}, the total amount of monomers absorbed by the Ag NP per unit time is given by:

\begin{equation}
    n_{abs}(t) = 4 \pi R D_{Ag^0}(T) [C_{Ag^0}^{BS}- S_{Ag^0} ( 1 + \frac{2 \gamma \omega}{R k_B T} )]
	\label{eqn:fick2}
\end{equation}

\noindent where $E_{Ag^0}$ the activation energy of diffusion, and $C_{Ag^0}^{BS} \equiv C_{Ag^0}(\infty)$ is the concentration of neutral silver monomers in the matrix.

\subsection{Light absorption by Ag NPs and thermal diffusion}
In addition to re-radiation and injection, the rest of the absorbed laser energy decays via the electron-phonon and phonon-phonon interactions that heat the irradiated region (Figure \ref{fgr:2all}(f,c)). The electron relaxations and electron-phonon coupling occur on the time scale of few picoseconds, while phonon-phonon interactions take up to a few nanoseconds \cite{delfour2015mechanisms, metwally2015fluence, pustovalov2016light, plech2004laser}. Since Ag NPs grow more slowly than the electron-phonon interaction speed, the classical thermal diffusion equation is introduced:

\begin{equation}
    C_m \frac{\partial T}{\partial t} =\nabla \cdot ( k_{m} \nabla  T) + \alpha_{abs} \cdot I(x,y,z,t) 
	\label{eqn:thermal}
\end{equation}

\noindent where $k_m$ is the heat conductivity, and $C_m$ the volumetric heat capacity. 

Although the absorbed laser energy by a Ag NP is partially relaxed via injection or tunneling, the majority of the energy is finally converted into heat \cite{brongersma2015plasmon, clavero2014plasmon,kawahara2005electron}. It is reasonable to estimate the heating source by taking into account the total absorbed energy. Normally, a laser beam attenuates exponentially in a composite having the mixture of nano-inclusions. Many efforts \cite{malasi2014mie, doyle1989optical, rysselberghe1932remarks} have been devoted to effective medium theories (EMTs) aiming to bridge the Fresnel medium together. Their key common ideas are that, as the concentration of inclusions goes to zero, the effective dielectric function should behave asymptotically as the host medium, and follow the Kramers-Kronig relation \cite{malasi2014mie}. In such media, the Beer-Lambert law describes both light absorption and  transmission. Herein, the laser intensity at any position is given by:

\begin{equation}
    I(x,y,z,t) = [1 - \Lambda(x,y) ] I_0(x, y, t) \exp (-z \alpha_{abs})
	\label{eqn:intensity02}
\end{equation}

\noindent where $\alpha_{abs} =  \sigma_{abs} \cdot C_{NP}$ is the absorption coefficient, $\Lambda(x,y)$ the surface reflection, and $I_0(x,y,t) = \frac{P_0}{\pi w_0^2 / 2} \cdot \exp\{-\frac{-2[(x-V_s t)^2+y^2]}{w_0^2}\}$ the Gaussian shape laser intensity at time $t$ and point $(x, y)$, where $x$ is the coordinate along the translation direction on the top surface of the TiO$_2$ film. Here, $C_{NP}$ is the NP concentration, $w_0$ the beam radius at $1/e^2$, $P_0$ the laser power, and $V_s$ the laser scanning speed.

For simplicity, we use the Maxwell-Garnett (MG) theory to estimate the effective index of the thin film. The surface reflection $\Lambda(x,y)$ is then estimated using the Fresnel equation.

\subsection{The coupled equations}
A link between NP concentration $C_{NP}$ and radius $R$ was previously revealed \cite{liu2015understanding} by analysis of \textit{post mortem} HAADF-STEM micrographs of samples produced at different speeds. The dependency is obtained by fitting the experiments according to an empirical law. It is given as follows \cite{liu2015understanding}: 
\begin{equation}
    C_{NP} = \frac{1}{(a_1 R + a_2)^3}
	\label{eqn:CNP}
\end{equation}
\noindent where $a_1 \approx 2.6$ and $a_2 \approx 8.7 \mathrm{nm}$.
The amount of crystallized silver changes in two ways: the variation of size and of the total number of NPs in the volume. Therefore, the differential of concentration is:

\begin{equation}
    \frac{d C_{Ag_{cst}} }{d t} = N_{Ag_{cst}} \frac{d C_{NP}}{d t} + C_{NP} [n_{abs}(t) - n_{oxi}(t)]
	\label{eqn:ConcenAgcrystal}
\end{equation}

\noindent where $N_{Ag_{cst}} = 4 \pi R^3 / 3 \omega$ is the number of crystallized Ag atoms in a nanoparticle. For each nanoparticle, the variation of total number of atoms is due to growth and oxidation. Therefore, the radius of a nanoparticle changes as:

\begin{equation}
    \frac{d R}{d t} = [n_{abs}(t) - n_{oxi}(t)] \cdot  \frac{\omega}{4 \pi R^2}
	\label{eqn:dRnp}
\end{equation}

\noindent Thus, the amount of crystallized Ag atoms in the studied unit cell varies due to the changes of NP concentration as \cite{liu2015understanding}:

\begin{equation}
    \begin{aligned}
        N_{Ag_{cst}} \frac{d C_{NP}}{d t} &=  N_{Ag_{cst}} \frac{d C_{NP}}{d R} \frac{d R}{d t}  \\
                    & = \frac{R}{3} [n_{abs}(t) - n_{oxi}(t)] \frac{d C_{NP}}{d R} 
	\label{eqn:dNcst}
    \end{aligned}
\end{equation}

\noindent where, $C_{NP}$ is the concentration of NPs in the unit cell. The right part of Eq. \ref{eqn:dNcst} implies its contribution to oxidation. The mass conservation of Ag in an unit cell leads to the following equations \cite{liu2015understanding}:

\begin{equation}
        \frac{d C_{Ag^+}}{d t} =n_{oxi}(t) [ C_{NP}(t) + \frac{R(t)}{3} \frac{d C_{NP}} {d R} ]
                                 - \frac{d C_{Ag^+}}{d t} |_{red}
	\label{eqn:ConcenAg1}
\end{equation}

\begin{equation}
        \frac{d C_{Ag^0}^{BS}}{d t} = - n_{abs}(t) [ C_{NP}(t) + \frac{R(t)}{3} \frac{d C_{NP}} {d R} ]
                                  + \frac{d C_{Ag^+}}{d t} |_{red}
	\label{eqn:ConcenAg0}
\end{equation}

\subsection{The extended 2D model}
Eq. \ref{eqn:dRnp} to Eq. \ref{eqn:ConcenAg0} describe the full coupled optical, thermal, photo-oxidizing and  growing field. The Arrhenius description of diffusion coefficients implies a nonlinear system that makes the 3D calculation extremely time-consuming. 

The extended 2D model provides an appropriate approach with less computational efforts to simulate the multi-scale problem and access the spatial information of NP size and temperature in a sample. We simulate the process in cross-section, which consists of two coupled steps. In the first step, the temperature field is computed by FEM. With the knowledge of the heat distribution, the growth, oxidation, and reduction are calculated adaptively by using the Bulirsch-Stoer algorithm \cite{press2007section}. The approach provides second-order accuracy with comparatively few calculation steps. The forecasting time step at each grid is obtained by the Bulirsch-Stoer algorithm. The minimum time step is used for the calculation of temperature and growth in the next step. In this way, few computational efforts are required and the accuracy is guaranteed. The studied structure consists of a TiO$_2$ film 200nm thick on top of a glass (Figure \ref{fgr:3all}(a)). The effective refractive index of the mesoporous titania film is estimated to $n = 1.7$ and the initial diameter of Ag NPs homogeneously distributed in the film is assumed to be 3 nm \cite{liu2015understanding}. The width of the structure is set to $200 \mu m$ and the laser scans the sample from $x = -50 \mu m$ to $x = 50 \mu m$ to avoid the boundary absorption of heat (Figure \ref{fgr:3all}(b)). Figure \ref{fgr:3all} shows the simulated results on the top surface (as illustrated in Figure \ref{fgr:3all} (c)) of the TiO$_2$ film at a laser writing speed of 50 $\mathrm{\mu m/s}$. Table \ref{table:Chapter5a} summarizes the parameters used in simulations.


  \begin{table*}[ht!]
  \centering
  \caption{A summary of the parameters in simulation}
  \begin{tabular}{llll}
  \hline
  \hline
      Parameter & Description & Value or definition &Reference\\
  \hline
  $n$ & optical index of TiO$_2$ film & 1.7 & Ref. \cite{liu2015understanding}\\ 
  $P_0$ & laser power & 180mW & \\ 
  $w_0$ & waist radius & $\mathrm{10 \mu m}$&\\
  $h_f $ & thickness of TiO$_2$ film & 200 nm & Ref. \cite{baraldi2016polarization}\\
  $\omega$ & atomic volume of crystallized Ag & 10.27 $\mathrm{cm^3/mol}$ &Ref. \cite{singman1984atomic}\\
  $E_{Ag^0} $ & activation energy of Ag$^0$ diffusion & 0.95eV &Ref. \cite{mcbrayer1986diffusion, butrymowicz1974diffusion, liu2015understanding}\\
  $D_{0Ag^0}(500K) $ & Ag$^0$ diffusion coefficient at 500K &0.6$\times 10^{-14} \mathrm{m^2/s}$ &Ref. \cite{mcbrayer1986diffusion, butrymowicz1974diffusion, liu2015understanding}\\
  $\eta _0$ & ionization efficiency & 2.3 $\times 10^{-5}$ &Ref. \cite{liu2015understanding, crespo2010reversible}\\
  $E_p$ & activation energy of reduction & 0.037eV &Ref. \cite{liu2015understanding, kaganovskii2007formation}\\
  $C_{red} $ & concentration of reducing agent (RA) & 2.258 $\mathrm{\times 10^{-4} mol/m^3}$ &\\
  $E_D $ & activation energy of RA diffusion & 0.4eV &Ref. \cite{haran1998determination, torresi1987hydrogen}\\
  $D_0^{red}(500K) $ & RA diffusion coefficient at 500K  &  2.5$\times 10^{-14} \mathrm{m^2/s}$ &Ref. \cite{haran1998determination, torresi1987hydrogen}\\
  $S_{Ag^0} $ & solubility of Ag$^0$  &  $1.66 \mathrm{mol/m^3}$ &Ref. \cite{liu2015understanding, kaganovskii2007formation, prudenziati2004dissolution}\\
  $\gamma$ &Ag NP - TiO$_2$ interfacial tension  &  $0.506 \mathrm{J/m^2}$ &Ref. \cite{liu2015understanding, kaganovskii2007formation}\\

  $C_{Ag^0}^{BS}(t=0 s)$ & initial Ag$^0$ concentration in matrix &  $1.495 \mathrm{\times 10^{4} mol/m^3}$ &Ref. \cite{liu2015understanding, liu2016laser}\\
  $C_{Ag^+}(t=0 s)$ & initial Ag$^+$ concentration in matrix &  $0.996 \mathrm{\times 10^{4} mol/m^3}$ &Ref. \cite{liu2015understanding, liu2016laser}\\
  $T(t=0s)$ & initial temperature  &  300K &\\
  $R(t=0s)$ & initial radius  & 1.5nm &Ref. \cite{liu2015understanding}\\
  
  \hline
  \hline
      \label{table:Chapter5a}
  \end{tabular}
  \end{table*}

\section{Simulation results}
\begin{figure*}[!ht]
 \centering
    \includegraphics[width=0.9\textwidth]{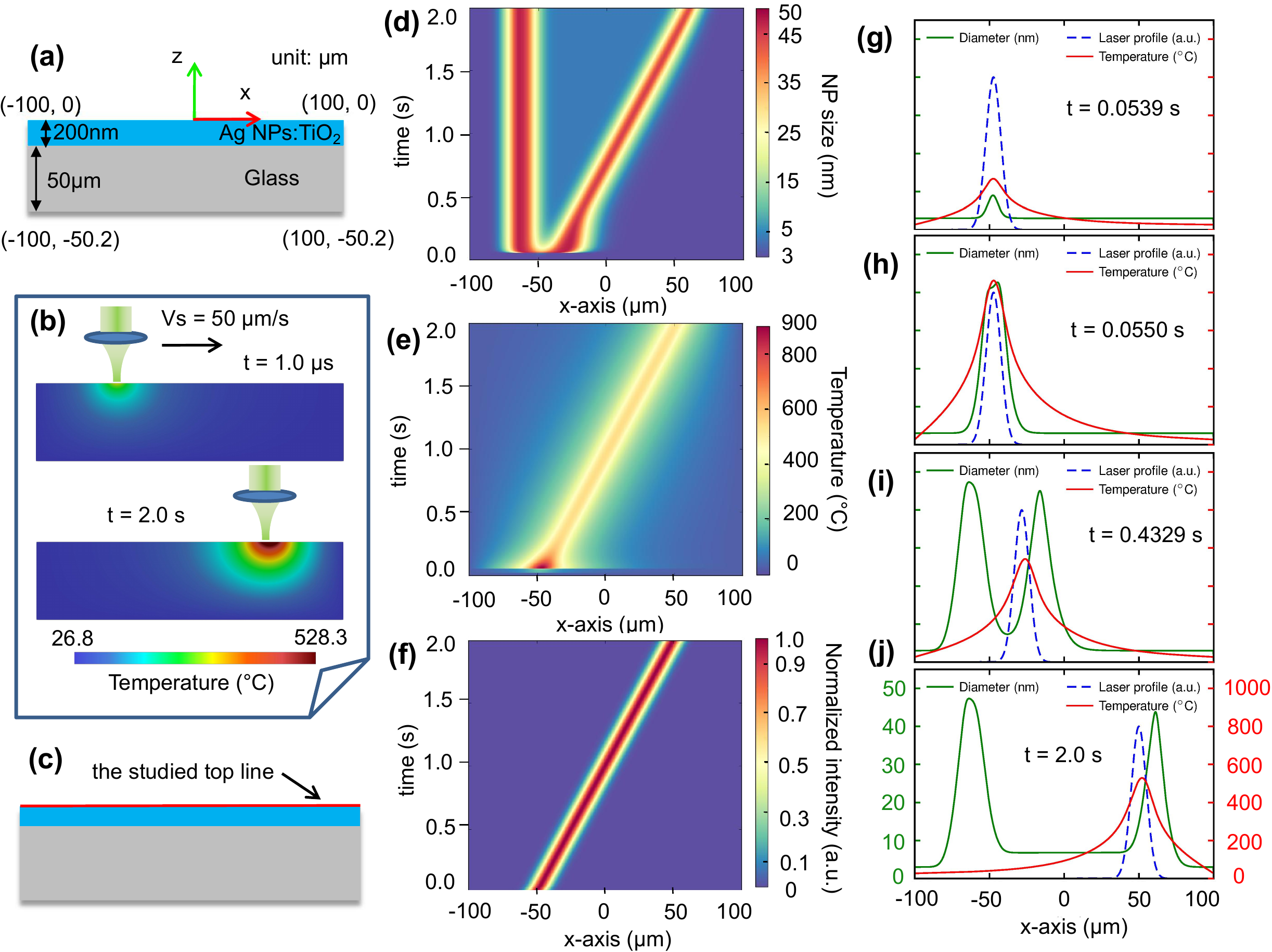}
    \caption{Simulations of the extended 2D model. The structure used in calculations (a), and an example of the temperature distributions at different times (b). Time variation maps of NPs size (d), temperature (e), normalized light intensity (f) along the translation of the laser beam at the top surface (c) of the film. (g), (h), (i) and (j) stand for the different states of the sample at four different times. All simulations are carried out at a scan speed of 50 $\mu m/s$.}
\label{fgr:3all}
\end{figure*}

The 2D model clearly demonstrates the spatial inhomogeneity during the kinetic process (Figure \ref{fgr:3all}). For instance, as one can see in Figure \ref{fgr:3all}(j), the NP's spatial-size-distribution (SSD) varies from uniform (3 nm in diameter) at the initial stage to non-uniform upon laser writing with big NPs formed at the beginning of the line (43 nm) and under the laser beam (20 nm), whereas the final NP size at the middle of the laser line is 6.8 nm after laser irradiation. Simultaneously, the temperature profile (Figure \ref{fgr:3all}(e)) changes with its maximum value rapidly increasing to 870 $\mathrm{^{\circ} C}$ at the beginning and more slowly decreasing to a steady one at 528 $\mathrm{^{\circ} C}$ after translating the sample by about 25 $\mu m$. Figure \ref{fgr:3all}(f) shows the laser during the writing.

To clarify the whole process from the beginning, the NP's SSD (solid green line), temperature (solid red line) and laser profiles (dashed blue line) are plotted in the same figures at various times (Figure \ref{fgr:3all}(g) to (j)). The curves at different times clearly show the main stages during laser writing. At the first step, NPs inside the laser spot grow rapidly due to the positive feedback (Figure \ref{fgr:3all}(g) and (h)). Accordingly, the temperature increases dramatically to 870 $\mathrm{^{\circ} C}$ as illustrated in Figure \ref{fgr:3all}(d) and (h)). After then, the shrinkage of NPs by oxidation dominates the process. The positive correlation between NPs size and absorption leads to the temperature decrease. As a matter of fact, the decreasing of temperature slows down the reaction speed, which in turn decreases the maximum size in the laser front edge (Figure \ref{fgr:3all}(i)). The negative feedback continues until a steady state sets in. Afterward, the stationary stage persists till the end as plotted in Figure \ref{fgr:3all}(i) to (j). Indeed, the above processes are valid for all scanning speeds that are lower than the maximum speed for NP growth. It is obvious that for high writing speeds that are comparable with the reaction rate, there is less time for NPs to grow. In this case, the laser-induced NP growth is not expected. The discussion will be presented in the next section.

\begin{figure*}[!htb]
 \centering
    \includegraphics[width=0.95\textwidth]{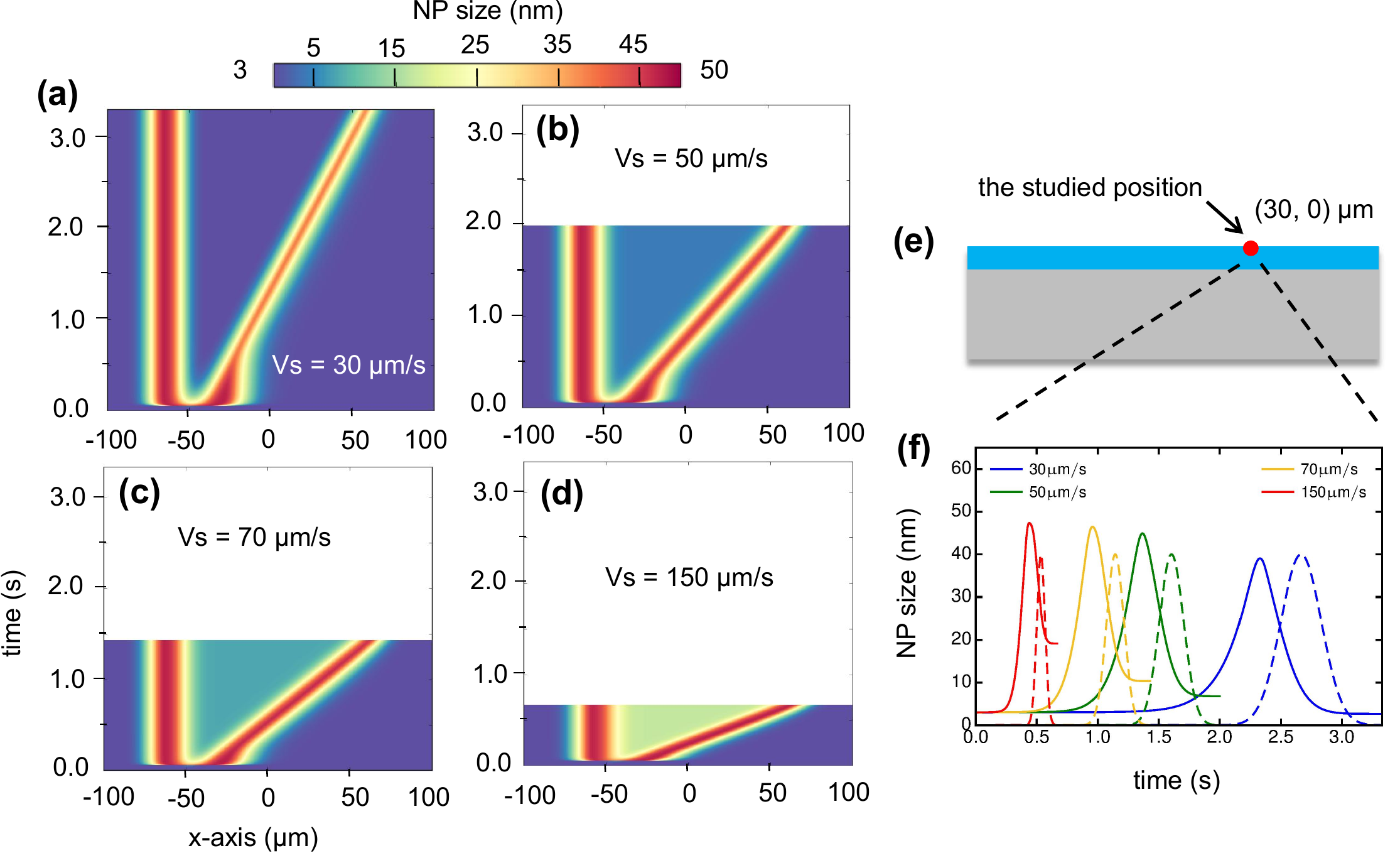}
    \caption{Time variation of the NP size along the translation direction at various writing speeds (a) to (d). To compare results at various speeds, the time variation of NP size (f) at the studied position (e) are plotted. Dashed lines show when the laser beam reaches the studied location.}
\label{fgr:4all}
\end{figure*}

Transforming the coupled equations into a moving coordinate system (MCS) \cite{jensen1980solution}($x=x'-\int_{0}^{t} Vs(t')dt'$ and $t=t'$) implies the steady-state equations (see Eq. \ref{eqn:Chapter5supeq1}). The stationary solutions depend on moving boundary conditions that are computationally difficult by FEM. The transient results of the extended 2D model reveal the existence of these kinds of solutions for various writing speeds (as shown in Figure \ref{fgr:3all} and Figure \ref{fgr:4all}). 

\begin{equation}
    \left\{
    \begin{aligned}
         \boldsymbol{V_s} \cdot \nabla R  +  \frac{(n_{abs} - n_{oxi})\omega}{4 \pi R^2}  &= 0      \\
         \boldsymbol{V_s} \cdot \nabla C_{Ag^0}^{BS} - n_{abs}[C_{NP} + \frac{1}{3} R \frac{d C_{NP}} {d R}] + \frac{d C_{Ag^+}}{d t} |_{red} &= 0  \\
         \boldsymbol{V_s} \cdot \nabla C_{Ag^+} + n_{oxi} [C_{NP} + \frac{1}{3} R \frac{d C_{NP}} {d R}] - \frac{d C_{Ag^+}}{d t} |_{red} &=0    \\
         C_m \boldsymbol{V_s} \cdot \nabla T+ \nabla \cdot ( k_{m} \nabla  T) + \alpha_{abs} \cdot I(x, y, z) &= 0    
	\label{eqn:Chapter5supeq1}
    \end{aligned}
    \right.
\end{equation}
\noindent where $\boldsymbol{V_s}$ is the moving speed of the laser at laboratory coordinate.

The photo-oxidation process is particularly important in the understanding of the speed-dependent final size. For all writing speeds, as discussed above, the NPs size profiles are nearly the same in the first stage owing to their rapid growth. In fact, the positive feedback stops as soon as the free Ag$^0$ atoms in matrix are exhausted as implied by Eq. \ref{eqn:fick2}. After size-shrinkage and temperature decrease, the stationary states are set with the total amount of photo-oxidized Ag$^0$ depending on the integration time of oxidation. In other words, there is more time at low scanning speeds to oxidize the NPs, so that lower temperatures are reached. The SSD width is broader at the beginning and thinner at stationary state as shown in Figure \ref{fgr:4all} (a) to (d). Note that reduction has little influence on the growth due to the low concentration of reducing agent, as reported by Liu et al.\cite{liu2015understanding}.

To get rid of the rapid growth at the beginning of the laser writing and to consider only the steady state behavior of the sample, the studied position is set at $x = 30 \mu m$ (shown in Figure \ref{fgr:4all}(e)) with its distance 80$\mu m$ away from the laser center at $ t = 0 s$. Figure \ref{fgr:4all}(f) plots the time variations of the NP size in this location for various writing speeds. The growth process lasts from a few tenths of a second to seconds, and the growth rate depends on the writing speed. According to Eq. \ref{eqn:fick2}, the temperature strongly affects the growth rate. This means that the sample experiences different temperatures at different speeds, which is clearly shown in Figure \ref{fgr:4all}(g). In all cases, the Ag NP growth is comparatively much slower than what was shown in reference \cite{liu2015understanding} that ignored the collective behavior of NPs and the thermal diffusion.

Heat diffusion is strongly contributes to the inhomogeneous SSD of NPs. According to Figure \ref{fgr:4all}(f), for all writing speeds, NPs start to grow even before the laser arrives and the size increase is stopped before the maximum intensity is reached. It is clear that the heat diffusion due to phonon-phonon interaction is faster than the growth by atoms diffusion. Thus, the temperature has broader profiles than NP's SSD. That is why the NPs in the laser vicinity are pre-heated and once the temperature exceeds the activation threshold, NPs start growing. Nevertheless, the temperature-dependent growth rate is faster than the studied writing speeds. Otherwise, the end of the growth process would be expected after reaching the maximum of temperature, or laser intensity.

\begin{figure*}[ht!]
 \centering
    \includegraphics[width=0.9\textwidth]{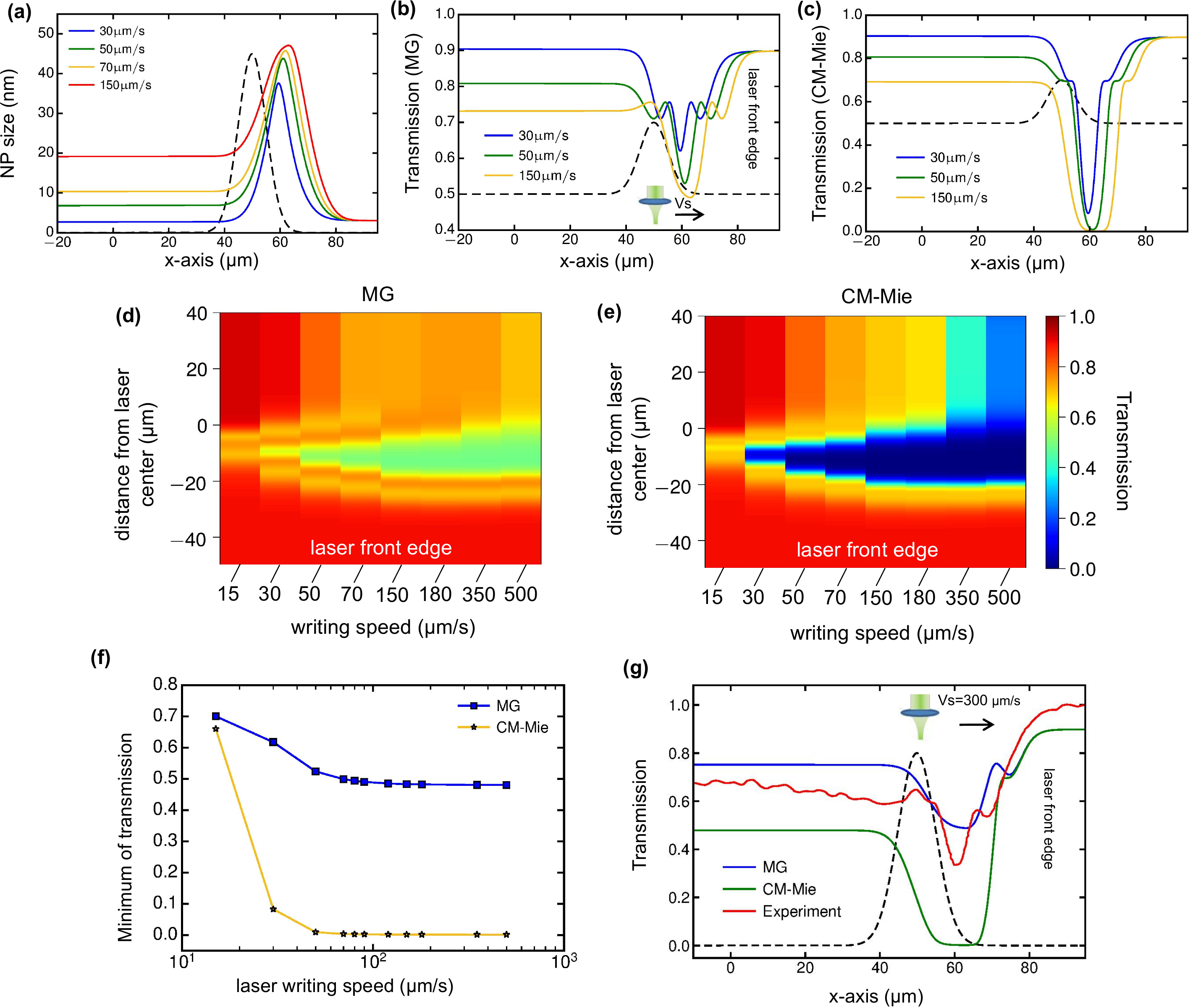}
    \caption{Simulated NP size and transmission along the writing line at the end of the calculation for various writing speeds. NP size (a). Sample transmission calculated by Maxwell-Garnett (b) or Mie coupled with Clausius-Mossotti model (c). Calculated transmission maps by MG (d) and CM-Mie (e) model. Minimum transmission versus writing speed for the two models (f). Comparison of transmission by simulation (MG and CM-Mie) and in-situ experiment (g). The dashed black lines stand for the laser profiles.}
\label{fgr:5all}
\end{figure*}

\subsection{Comparison with experiments}
Experimentally, the transmissions are measured in the steady state far from the starting positions. Herein, the discussion is then limited to the comparison with simulated results at steady-state. For each writing speed, we obtain the NPs' size profile after the laser moves $100 \mu m$. The NP's SSDs are shown in Figure \ref{fgr:5all}(a). Here, laser moves toward the positive x-axis with its intensity profile shown by the dashed black line. As discussed above, the maximum sizes appear before the laser-center arrives at all the writing speeds. To compare with the \textit{in-situ} experiments, transmission profiles are retrieved based on two different effective-medium models. The Maxwell-Garnett approximation is commonly used in literature \cite{pedrueza2011novel}. However, it is only valid for the NPs whose sizes are much smaller than laser wavelength \cite{malasi2014mie}. For larger NPs, compared with MG, the Mie theory associated with Clausius-Mossotti equation \cite{rysselberghe1932remarks} provides a size-dependent correction by accounting for the multipole contributions (CM-Mie) \cite{malasi2014mie, doyle1989optical}. The transmission profiles at wavelength 527 nm are calculated according to the Fresnel equation under normal incidence:

\begin{equation}
    \begin{aligned}
        &\mathrm{Trans} = \frac{4 Z_{air} Re(Z_{sub}) }{|Z_{air} B + C|^2} \\
        \mathrm{where} \\
        \left[
            \begin{array}{c}
                B\\
                C
            \end{array} 
        \right]
    &= 
        \left[
            \begin{array}{cc}
                cos \delta _m & j sin \delta _m / Z_m \\
                j Z_m sin \delta _m & cos \delta _m
            \end{array} 
        \right]
        \left[
            \begin{array}{c}
                1\\
                Z_{sub}
            \end{array} 
        \right] \\
    \delta _m &= 2 \pi n_m h_f / \lambda 
	\label{eqn:fresnel}
    \end{aligned}
\end{equation}

\noindent where $Z_i$ is the media impedance (i = air, effective medium or glass substrate), $h_f = 200 nm$ the film thickness, and $n_m$ the effective index.

Figure \ref{fgr:5all}(b) and (c) show the retrieved transmission profiles at various writing speeds based on MG and CM-Mie respectively. Though the value of transmission by MG and CM-Mie differ, the trends are quite similar: the transmission decreases in the laser front edge before growing back inside the irradiated area. No matter which EMT model is used, the position at transmission minimum is always located at the laser front edge for every writing speed. The calculations show that the transmission minimum is close to the location of the largest NPs in the laser front edge.
Furthermore, the transmission maps calculated using MG (Figure \ref{fgr:5all}(d)) and CM-Mie model (Figure \ref{fgr:5all}(e)) have similar trends with the results of the in-situ experiment shown by Figure \ref{fgr:1all}(c). Though the writing speed in our experiments and simulations can differ, the transmission dip (shown by green or blue color in the maps) is becoming broader and moves farther toward the laser front edge when the writing speed is increased.
In addition, the transmission minimum decreases and saturates while increasing the writing speed (Figure \ref{fgr:5all}(f)). These results are consistent with the in-situ experiments as shown in Figure \ref{fgr:5all}(g). The location-dependent transmissions are due to the NPs formation along the laser scanning direction. Nevertheless, we note that the value of the transmission in the simulation and experiment are not exactly the same. The differences are probably due to several modeling assumptions. For example, a single-size model was used, while the size of Ag NPs is dispersed. Secondly, the effective medium theory ignores the NP-NP interactions, which can also affect the electromagnetic response if nanoparticle density is getting high. Interestingly, the transmission obtained by using the CM-Mie model is underestimated. In principle, CM-Mie is an improvement that should work better than the MG model by taking into account the contributions of multipoles, such as quadrupole, octupole and etc. These differences are shown in Figure \ref{fgr:5all}(d), (e), and (g) indicating that the effective medium theory could only approximately describe the true composition of the porous thin films with a few layers of NPs inside.

\begin{figure}[ht]
 \centering
    \includegraphics[width=0.4\textwidth]{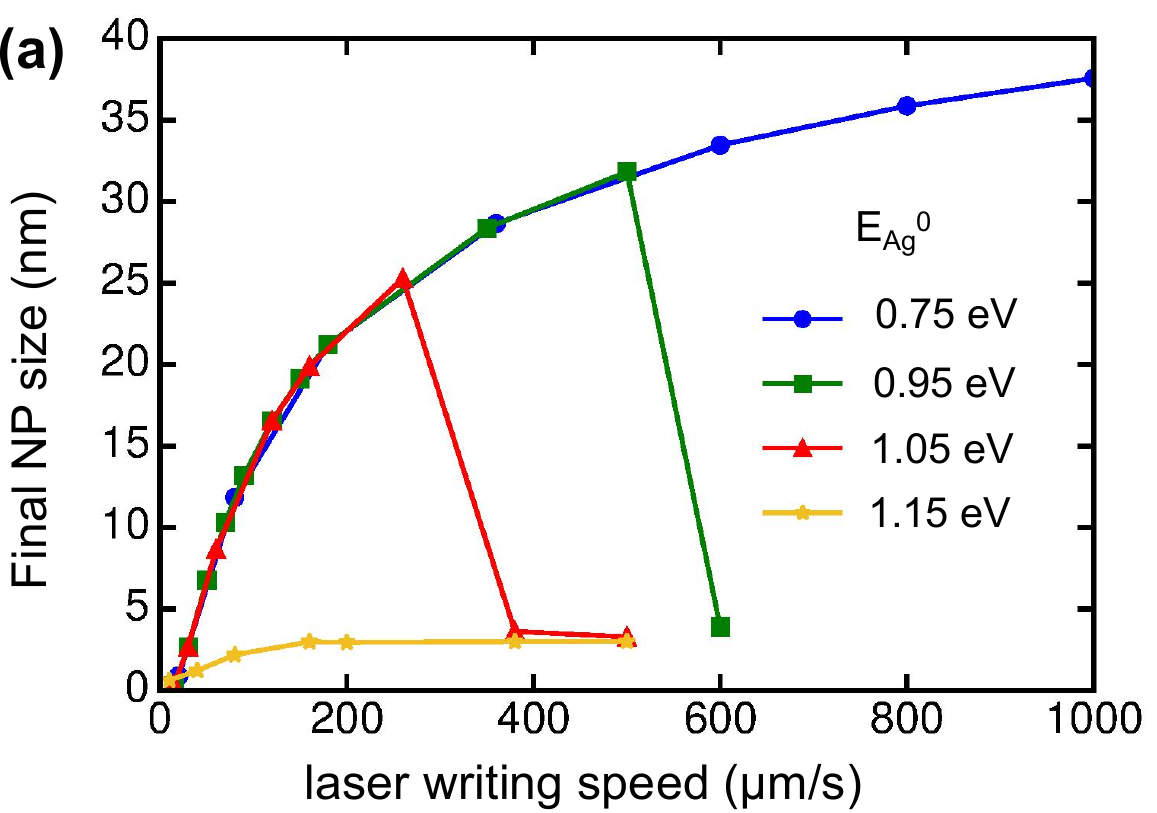}
    \includegraphics[width=0.4\textwidth]{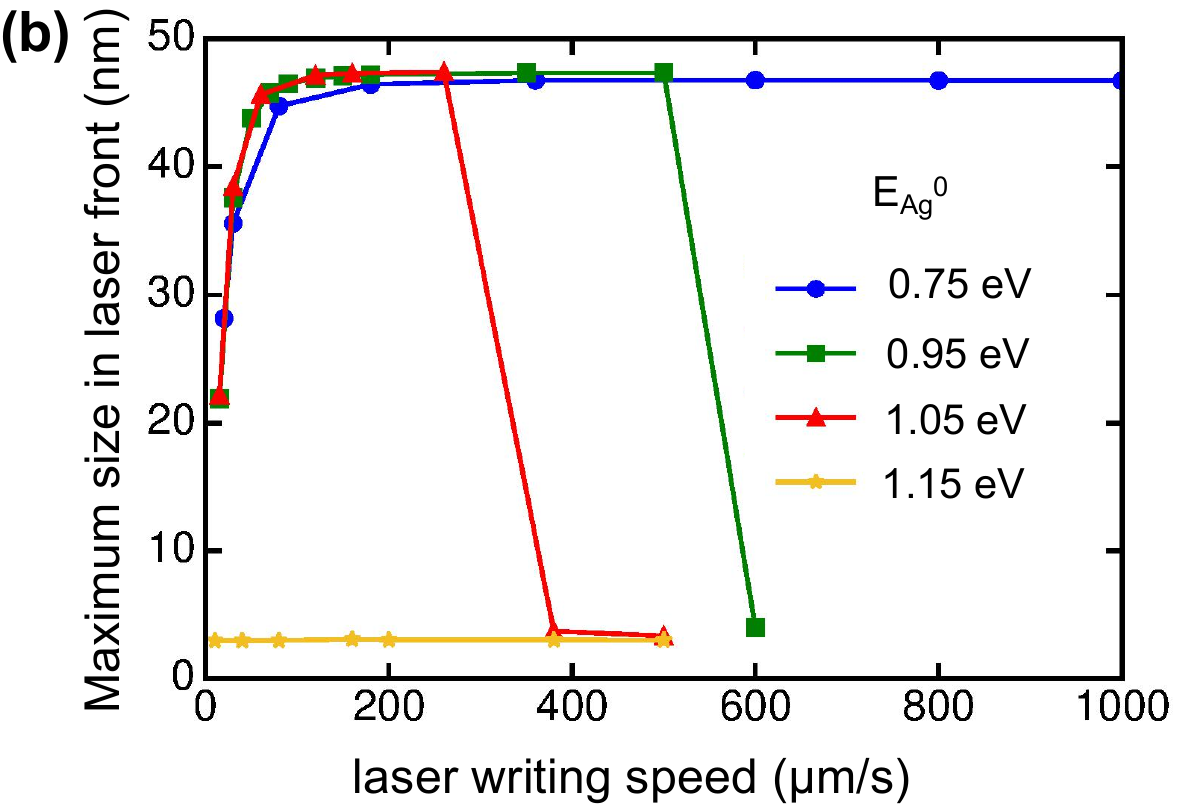}\\
    \includegraphics[width=0.4\textwidth]{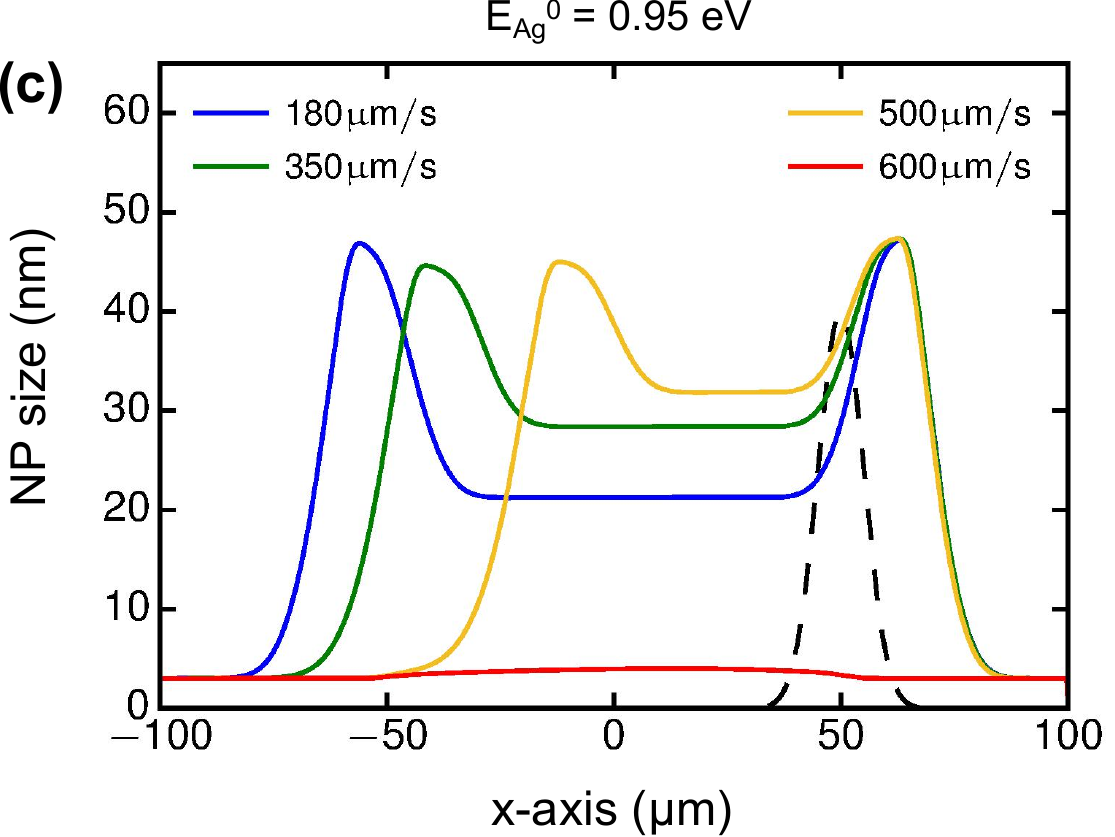}
    \caption{Influences of the activation energy of free Ag$^0$ diffusion ($\mathrm{\eta _0 = 2.3 \times 10^{-5}}$). The final NP size after laser scanning (a) and the maximum size in the laser front (b) for various activation energy, and (c) the spatial distribution of NP size with $E_{Ag^0} = \mathrm{0.95eV}$.}
\label{fgr:6all}
\end{figure}

\subsection{Activation energy of diffusion}
The activation energy of metal ion diffusion in the studied system affects NP's growth threshold. The exact activation energy in the mesoporous TiO$_2$ film is not known to the best of our knowledge. The activation energy is estimated from the Ag diffusion in silicon dioxide and copper, which gives the value around 1.0 eV \cite{liu2015understanding,mcbrayer1986diffusion, butrymowicz1974diffusion}. Thus, a reasonable value for Ag in mesoporous TiO$_2$ ranges from 0.5eV to 1.5 eV \cite{liu2015understanding}.

To understand how the activation energy influences the NP growth, the NP final size and maximum size in the laser front edge are calculated at various scanning speeds with different activation energies. They are shown in Figure \ref{fgr:6all}(a) and (b). Interestingly, the maximum size never exceeds 4 nm for activation energy $E_{Ag^0} =\mathrm{1.15eV}$. In this case, the activation energy is too high to allow the rapid positive feedback to appear. The size increase and saturation occur for lower energies. Moreover, the activation energy never alters the final size and maximum size profiles except for the high writing speeds. It means that the activation energy has a little impact on the saturation process. Instead, an increase of the activation energy decreases the growing speed that has a great influence on the up-threshold of writing speed. To understand what happens at high speed, NP size profiles are plotted in Figure \ref{fgr:6all}(c) at various scanning speeds along x-axis. At high speed the growth never starts since there is not enough time for NPs in a rapid moving laser spot (e.g. $\mathrm{600 \mu m/s}$) to grow large enough to activate the feedback promptly.

\subsection{Oxidation rates}
\begin{figure}[ht!]
 \centering
    \includegraphics[width=0.43\textwidth]{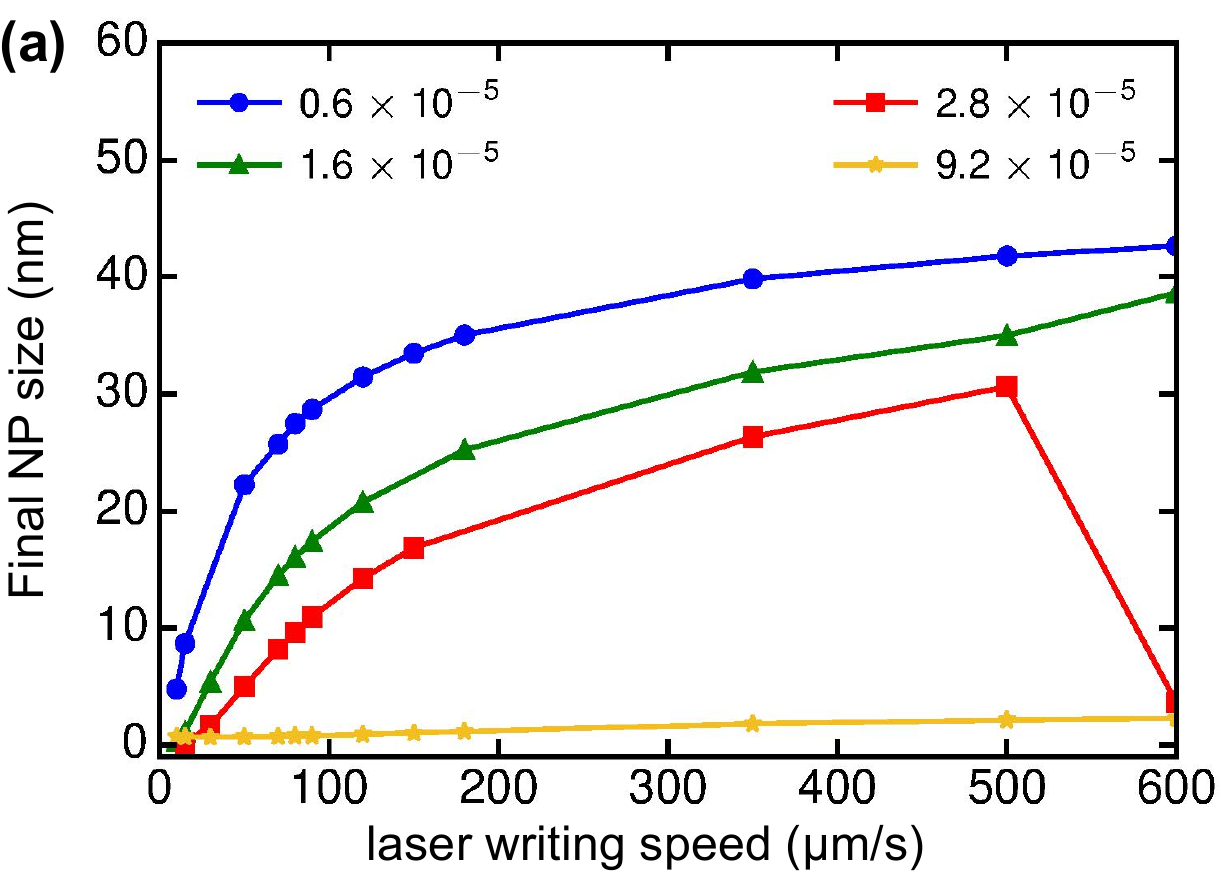}
    \includegraphics[width=0.43\textwidth]{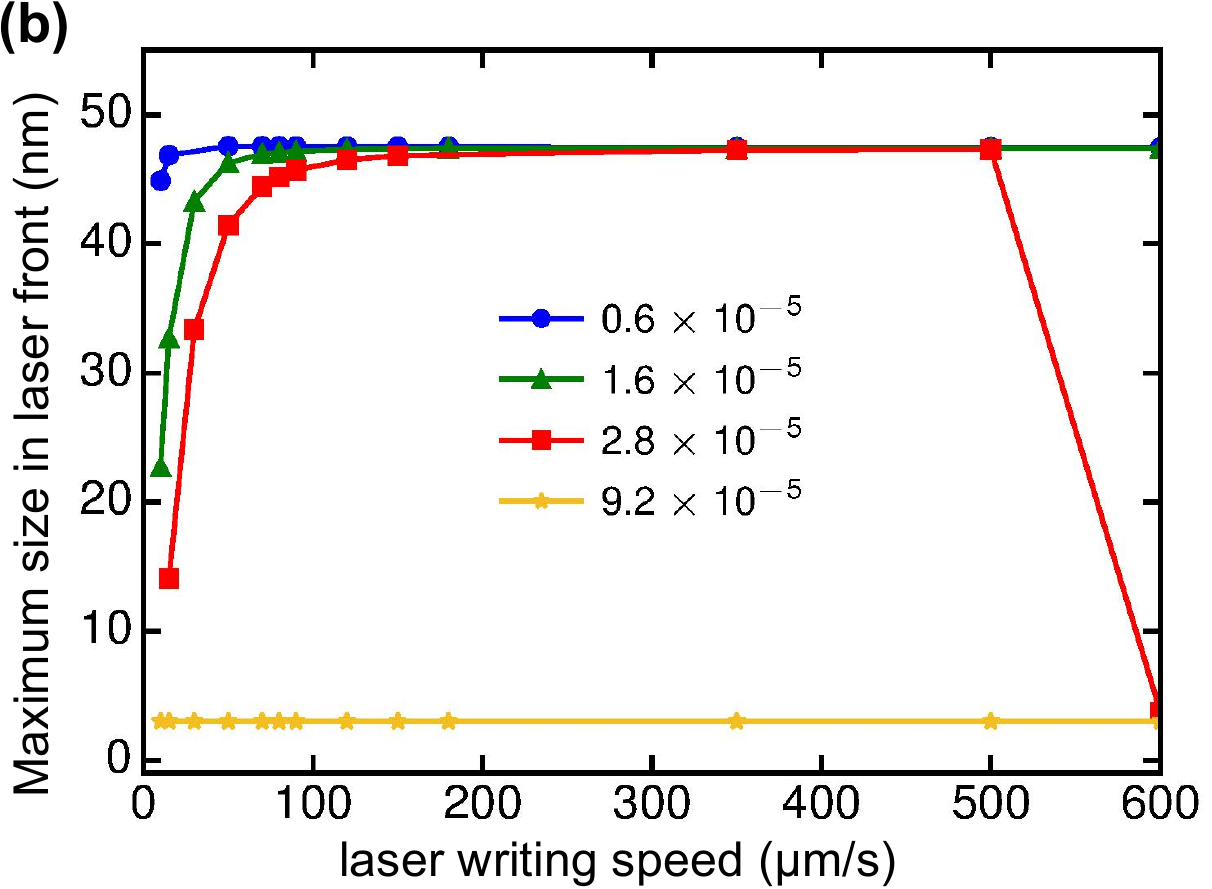} 
    \caption{Influence of ionization efficiency ($\mathrm{E_{Ag^0} = 0.95 eV}$). Final NP size after laser writing (a) and Maximum size in the laser front edge (b).}
\label{fgr:7all}
\end{figure}

The photo-oxidation process strongly affects the steady-state parameters. Figure \ref{fgr:7all} shows the dependence of the NP final size and maximum size on the writing speed at various oxidation rates. The two size-profiles saturate at lower speeds while decreasing the oxidation rate. The result again confirms the fact that the oxidation process contributes greatly to the scanning speed dependent phenomenon. Once the rapid feedback is activated, the fast growth of NPs never stops until the concentration of Ag$^0$ decreases to the value close to Ag$^0$ solubility. This is also the reason why the activation energy ($E_{Ag^0}$) never alters the saturation profile. On the contrary, the oxidation leads to the size-shrinkage with an amount inversely proportional to the scanning speed. The suggested NP size increases while decreasing the oxidation rate for the same scanning speed. Owing to the positive correlation between absorption and size, lower oxidation rates saturate at lower speeds. Furthermore, oxidation affects the positive feedback. For a given activation energy, an increase of the oxidation rate  means more time is required for a NP to grow above a threshold size. This is the case for oxidation rate $\eta_0 = 1.6 \times 10^{-5}$ and $\eta_0 = 2.8 \times 10^{-5}$ at $\mathrm{600\mu m/s}$. For the high oxidation rate $\eta_0 = 9.2 \times 10^{-5}$ as shown in Figure \ref{fgr:7all}, the rapid positive feedback never appears.

\begin{figure}[ht!]
 \centering
    \includegraphics[width=0.55\textwidth]{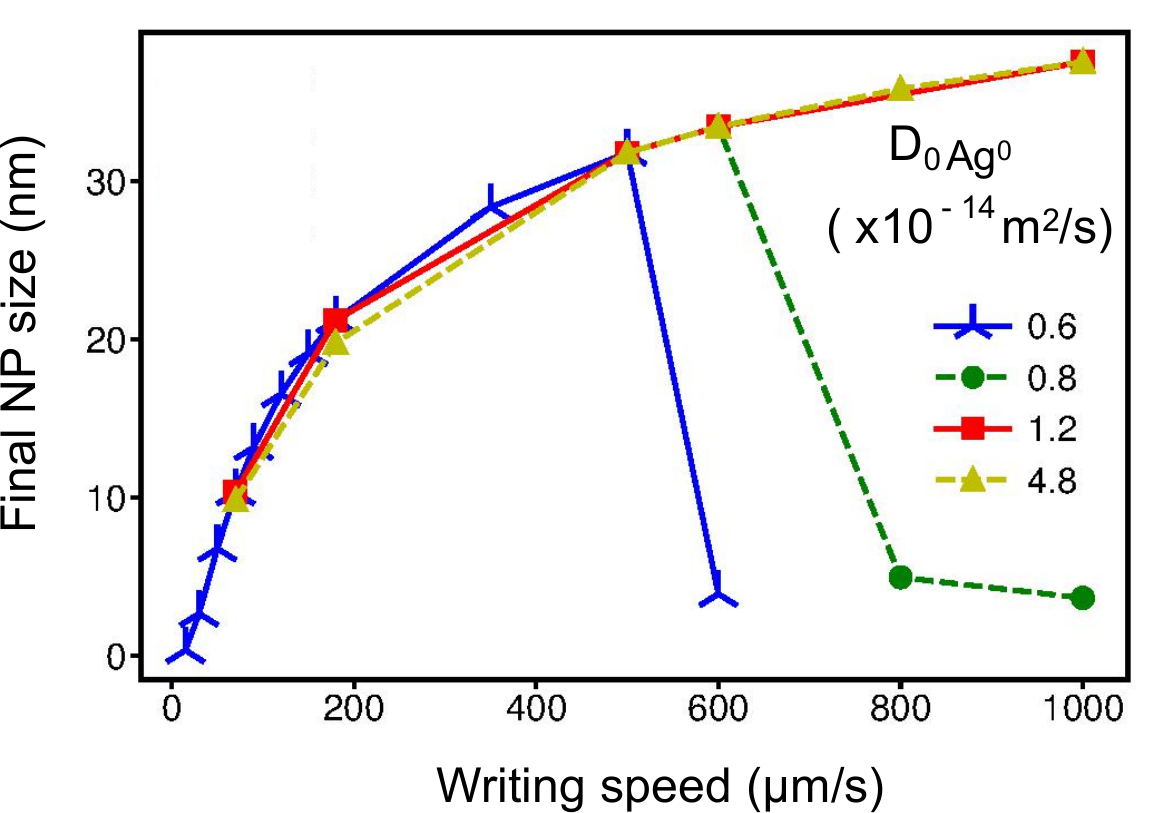}
    \caption{Influence of Ag$^0$ diffusion coefficient ($\mathrm{E_{Ag^0} = 0.95 eV}$, and $\mathrm{\eta_0 = 2.3 \times 10 ^{-5}}$). }
\label{fgr:chapter5_DEAg0_p1}
\end{figure}

\subsection{Diffusion coefficient at 500K}
The Ag $^0$ diffusion coefficient is important to silver NP's growth speed. In the previous discussions we have seen the abrupt descending of the NP size (as shown in Figures \ref{fgr:6all} and \ref{fgr:7all}) at fast writing speeds. The mechanism has been attributed to the low growth speed. In porous material, especially in mesoporous TiO$_2$ thin films, the diffusion coefficient of silver can be larger than it is inside the bulk material, such as the silicon, and fused silica. Based on this assumption, we perform simulations by increasing the diffusion coefficient of silver at 500K. Figure \ref{fgr:chapter5_DEAg0_p1} shows the final NP size over different writing speeds. The abrupt descending tends to a right move to the high-speed end while increasing the diffusion coefficient, which disappears for $\mathrm{D_{0Ag^0}(500K) = 1.2 \times 10 ^{-14} m^2/s}$ and $\mathrm{D_{0Ag^0}(500K) = 4.8 \times 10 ^{-14} m^2/s}$ even at the highest writing speed studied (namely, $\mathrm{1000\mu m /s}$). Moreover, the curve of the final NP size with speed does not depend on the silver diffusion coefficient. The mechanism lies in the fact as we discussed previously that the photo-oxidation is the key role in laser writing-speed controlled phenomena. Due to the plasmonic size-dependent absorption, once the NPs inside the laser spot start to grow, the increment in size leads to a higher power absorption. The increased temperature further speed up the growth process. As a result, the growth is quite non-linear and it happens before the oxidation be a role. The abrupt decrements in the Figure \ref{fgr:chapter5_DEAg0_p1} are because of the insufficient time for the size-accumulation to activate the rapid growth as shown by Figure \ref{fgr:6all} (c).

\begin{figure}[ht!]
 \centering
    \includegraphics[width=\textwidth]{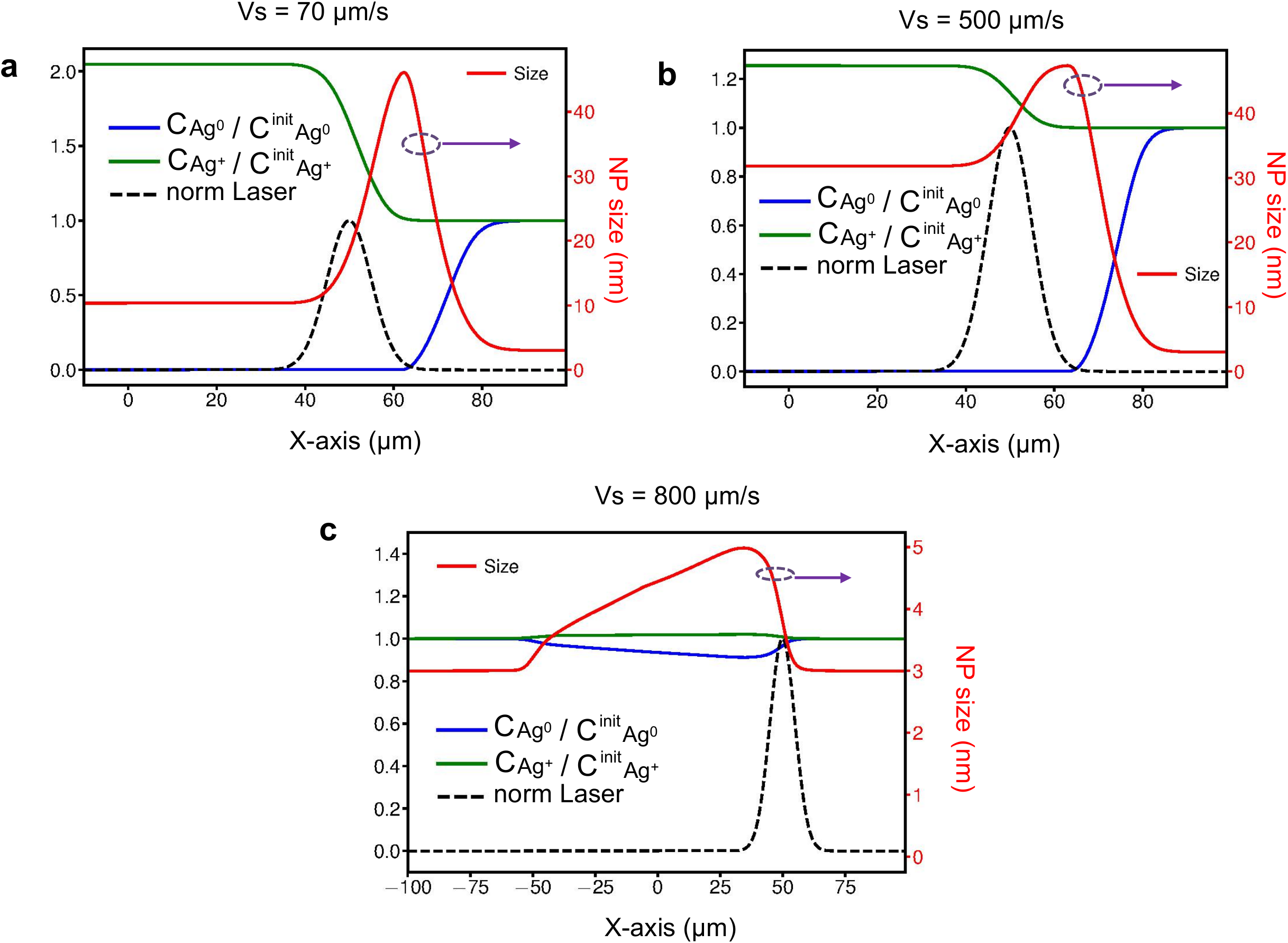}
    \caption{Concentrations of silver in the forms of ions and free atoms in matrix ($\mathrm{E_{Ag^0} = 0.95 eV}$, $\mathrm{\eta_0 = 2.3 \times 10 ^{-5}}$, and $\mathrm{D_{0Ag^0}(500K) = 0.8 \times 10 ^{-14} m^2/s}$). Laser scans from left to the right. The initial concentrations of silver ions and atoms are shown in Table \ref{table:Chapter5a}.}
\label{fgr:chapter5_CAg0Ag1}
\end{figure}

\subsection{Concentrations of free Ag$^0$ and Ag$^+$}
The previous discussions have shown that silver NPs starts to grow even before the laser center arrives and it never stops until the majority of the Ag$^0$ atoms in the matrix are consumed. To better illustrate the underlying process, here we plot spatial distributions of Ag$^+$ ions, Ag$^0$ atoms and NP size alongside with laser profile in Figure \ref{fgr:chapter5_CAg0Ag1}. In the regime of laser-activated rapid growth of NPs (namely, Figure \ref{fgr:chapter5_CAg0Ag1} (a) and (b)), the thermal diffusion allowed NPs' growth appears at positions more than 30$\mu m$ ahead of the laser beam center. This, on the other hand, can easily be seen from the drop-down positions of the Ag$^0$ concentration curves. In the region from $x = 60 \mu m$ to $x = 80 \mu m$, no changes are observed for the Ag$^+$ concentration. Thus, this region only exists the process of thermal-activated growth of silver NPs. The size of Ag NPs reaches the maximum value until it exhausts the majority number of the Ag$^0$ atoms due to the fast growth speed. Thereafter, the growth is nearly stopped when the laser arrives and  the photo-oxidation tends to dissolve the NPs into Ag$^+$ ions. The green curves represent the ratio of the real time quantity and the initial number of Ag$^+$ ions. The maximum sizes of the Ag NP ahead of the laser beam are close for the two scanning speeds ($\mathrm{V_S} = 70 \mu m/s, 500 \mu m/s$). After passing-through of the laser beam, the quantity of Ag$^+$ is increased by a factor of 1.26 and 2.05 for $\mathrm{V_S} = 500 \mu m/s$ and $\mathrm{V_S} = 70 \mu m/s$, respectively. The slower writing speed has longer time for the oxidation to shrink the NPs, therefore, larger amount of ions are observed. in addition, no abrupt transformations of Ag$^+$ into Ag$^0$ are presented in the simulations. This is because of the low efficiency of the reduction process studied here (Table \ref{table:Chapter5a}, Eq. \ref{eqn:reduction}). Therefore, the process is not the traditional photochromism \cite{naoi2004tio2,crespo2010reversible}. The samples are changed and are not reversible unless efficient methods are introduced to pump-up the Ag$^0$ quantity in the matrix after the laser processing.

Figure \ref{fgr:chapter5_CAg0Ag1} (c) explains the abrupt decrements of the NP size curve (green dashed line in Figure \ref{fgr:chapter5_DEAg0_p1}) at $\mathrm{V_S} = 800 \mu m/s$. The process is not stable and can not be discussed using the stationary-state descriptions used above. However, it is clear that the size of Ag NP is increased when the laser scans from left to the right. The decreasing concentration of Ag$^0$ along the X-axis shows the accelerating of the growth process, which is thermal activated as we discussed previously. According to Mie absorption, for the studied Ag size range (3 nm to 50 nm in diameter) at 532 nm wavelength, the larger the size the higher the absorption is. The higher temperature results to the faster diffusion of Ag$^0$ to be absorbed by a NP. Due to the limited computational resources, the simulated distances for the laser scanning is set to 100 $\mu m$. The distances required for the laser to scan before activating the rapid growth at $\mathrm{V_S} = 800 \mu m/s$ is unknown.

\begin{figure}[ht!]
 \centering
    \includegraphics[width=\textwidth]{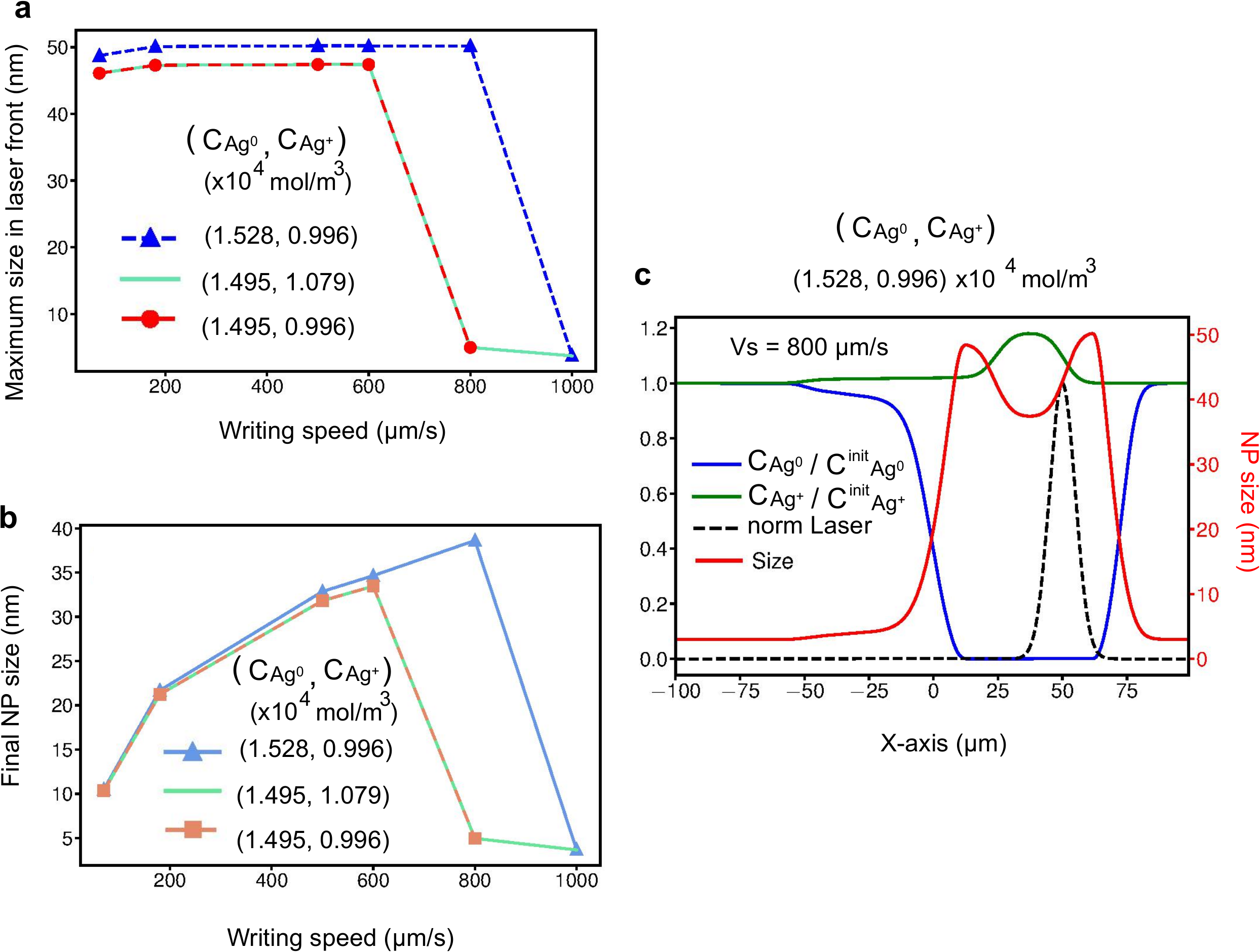}
    \caption{Simulated NP size at maximum in the laser front (a) and at final stage after laser scans (b) for different initial concentrations of Ag$^+$ and Ag$^0$ in TiO$_2$ thin films. (c) the influence of initial $C_{Ag^0}$ by increasing the value from $1.495 \times 10^4 mol/m^3$ to $1.528 \times 10^4 mol/m^3$, as a comparison to Figure \ref{fgr:chapter5_CAg0Ag1}(c). (Note that $\mathrm{E_{Ag^0} = 0.95 eV}$, $\mathrm{\eta_0 = 2.3 \times 10 ^{-5}}$, and $\mathrm{D_{0Ag^0}(500K) = 0.8 \times 10 ^{-14} m^2/s}$).}
\label{fgr:chapter5_CAg0Ag1_p2}
\end{figure}

The influences by initial concentrations of Ag$^+$ and Ag$^0$ are shown in Figure \ref{fgr:chapter5_CAg0Ag1_p2}. From the previous discussions we find that once the process is activated, the nonlinear growth of NPs never stops until the initial amount of Ag$^0$ are dramatically consumed. This means the maximum size in the laser front should be larger while increasing the concentration of Ag$^0$, which is confirmed by the simulation as shown in Figure \ref{fgr:chapter5_CAg0Ag1_p2}(a). Moreover, since the size-shrinkage is monotonically decreasing with writing speed of the laser, the final NP is also increased (Figure \ref{fgr:chapter5_CAg0Ag1_p2}(b)). The initial Ag$^+$ ion concentration has negligible impacts on the studied model (Figure \ref{fgr:chapter5_CAg0Ag1_p2}(a-b)). This illustrates the inefficient reduction to covert the Ag$^+$ to Ag$^0$. In addition, the increments of Ag$^0$ concentration accelerates the growth speed in average. Comparing the dashed blue curve in Figure \ref{fgr:chapter5_CAg0Ag1_p2}(a) to the dashed green curve in Figure \ref{fgr:chapter5_DEAg0_p1} at point $V_S = 800 \mu m/s$, it is obvious that the maximum size in the laser front increases from 5.0 nm ($C_{Ag^0} = 1.495 \times 10^4 mol/m^3$) to 50.2 nm ($C_{Ag^0} = 1.528 \times 10^4 mol/m^3$). The spatial distribution of Ag$^+$ and Ag$^0$ concentration, and Ag NP size that are plotted in Figure \ref{fgr:chapter5_CAg0Ag1_p2} (c) clearly demonstrate the nonlinear activation of the growth process. Because the positive feedback of size increase with light absorption, it results to the rising of the maximum size in laser front during laser scanning. At first, the size increases very slowly as shown both in Figure \ref{fgr:chapter5_CAg0Ag1} (c) and Figure \ref{fgr:chapter5_CAg0Ag1_p2} (c). However, the averaged growth speed is larger for $C_{Ag^0} = 1.528 \times 10^4 mol/m^3$ than the case of $C_{Ag^0} = 1.495 \times 10^4 mol/m^3$, so that the Ag size in the laser front can be larger and once it exceeds the specific threshold size, the rapid growth is observed as shown in Figure \ref{fgr:chapter5_CAg0Ag1_p2} (c).

\section{The relation of temperature-rise and writing speed}
The first empirical relation of the temperature-rise and the laser writing speed \cite{liu2016selfthesis} comes from the observations of TiO$_2$ phase changes. As reported \cite{kholmanov2003influence, ocana1992low, yuan1995characterization, mosaddeq2000refractive}, a thermal annealing of initial amorphous TiO$_2$ prepared by processing of sol-gel \cite{yuan1995characterization, bersani1998raman, vinogradov2014low}, physical and chemical vapour deposition\cite{goossens1998gas}, and colloidal chemistry methods \cite{ma1998investigation} results to the phase transition of amorphous-to-anatase at 400 - 600 $^{\circ} C$ and anatase-to-rutile at 800 - 1000$^{\circ} C$. Indeed, the temperature at phase transitions differs from different processes and conditions \cite{bersani1998raman, kholmanov2003influence, sharma2019crystal, yanagisawa1999crystallization}. The phase transition from anatase to rutile normally appears at higher temperature than it from amorphous to anatase.

\subsection{Raman spectrum acts as the signature of temperature history}

\begin{figure}[ht!]
 \centering
    \includegraphics[width=0.85\textwidth]{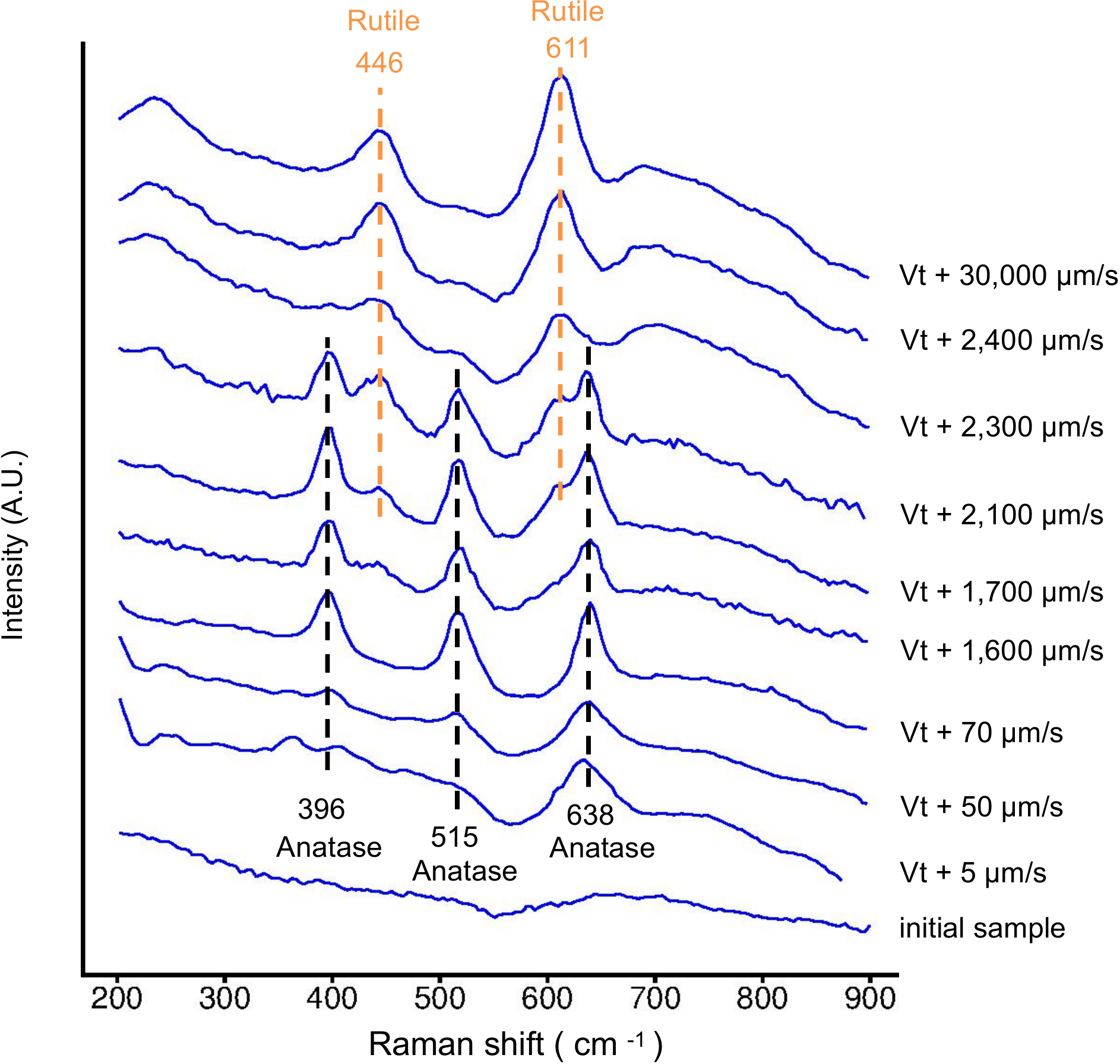}
    \caption{Raman spectra of samples after different laser writing speeds. The excitation wavelength for Raman characterization is 633 nm. The experimental data are taken from Ref. \cite{liu2016selfthesis}. }
\label{fgr:chapter5_tempRise_raman}
\end{figure}

In the studied laser writing Ag embedded TiO$_2$ thin films, the phase of TiO$_2$ can be estimated by the Raman spectroscopy. Figure \ref{fgr:chapter5_tempRise_raman} shows the Raman spectra of the laser irradiated samples over different writing speeds. The experimental results are from Ref. \cite{liu2016selfthesis}, which were carried out by Zeming Liu at Laboratoire Hubert Curien under the supervision of Nathalie Destouches. The characteristic peaks of anatase phase are 144 , 197 and 639 cm$^{-1}$ at E$_g$ modes, 399 and 519 cm$^{-1}$ at B$_{1g}$ modes, and 513 cm$^{-1}$ at A$_{1g}$ mode \cite{chen2007titanium, swamy2005finite, ohsaka1978raman}. The fundamental peaks of rutile phase are E$_g$ mode (447 cm$^{-1}$), B$_{1g}$ (143 cm$^{-1}$), A$_{1g}$ mode (612 cm$^{-1}$),  and B$_{2g}$ (826 cm$^{-1}$) \cite{porto1967raman, swamy2006size}. It is clear that the initial sample before laser treatment is amorphous. The appearance of the peaks at anatase phase starts from writing speed of Vt + 5 $\mu m/s$, where Vt stands for the threshold speed of NP growth \cite{liu2016selfthesis}. The peaks are progressively distinct while increasing the writing speed. The rutile phase appears from Vt + 1,700 $\mu m/s$. The anatase peaks are completely disappear at Vt + 2,300 $\mu m/s$. Based on the above analyses, the temperature-rise inside the Ag:TiO$_2$ film shows a positive correlation with the writing speed.

A fine illustration of the temperature-Vs relation shown in the range from Vt + 5 $\mu m/s$ to Vt + 2,100 $\mu m/s$ can be revealed by further exploitation of the size effect of anatase TiO$_2$ on Raman spectra \cite{chen2007titanium, zhang2000raman, li2005raman, bersani1998raman, ivanda1999effects}. The size influence on Raman lineshape comes from the phonon confinement that leads to the breakdown of the phonon momentum selection rule at $q \approx 0$\cite{zhang2000raman}. This is the case for nanocrystals due to the spatially confined phonons all over the Brillouin zone, which will contribute to the first-order Raman scattering \cite{zhang2000raman, chen2007titanium, li2005raman}. For spherical nanocrystals at first-order scattering, the Raman intensity is estimated as \cite{tiong1984effects, shen1984raman, zhang2000raman, li2005raman, bersani1998raman, ivanda1999effects}:

\begin{equation}
    I(\omega) \propto \int_{BZ}\frac{|C(0,\mathbf{q})|^2}{(\omega-\omega(\mathbf{q}))^2 + (\Gamma_0/2)^2} d^3 \mathbf{q}
	\label{eqn:phonon-confinement_p1}
\end{equation}

\noindent where $\omega$ is the Raman shift expressed in wavenumber, $\omega(\mathbf{q}) = \omega_0 + \Delta [1 - cos(qa)]$ is the phonon dispersion function, $a = 0.3768$ nm is the lattice parameter, $\Delta = 20 cm^{-1}$ is the width of the phonon dispersion curve, $\omega_0 = 144 cm^{-1}$ is the zone center phonon frequency, $|C(0,\mathbf{q})|^2 = \exp(-q^2 D^2 / 16 \pi^2)$ is the confinement function, $D$ is the size of the nanocrystals, $\Gamma_0 = 7 cm^{-1}$ is the Raman linewidth at room temperature (the constant value of 7 $cm^{-1}$ corresponds to bulk anatase), and $d^3 \mathbf{q} \propto q^2 dq$ for a three-dimensional confinement that is suitable for power, and quantum dots; $d^3 \mathbf{q} \propto q dq$ for quantum wires (two-dimensional confinement); and $d^3 \mathbf{q} \propto dq$ for quantum wells (one-dimensional confinement) \cite{li2005raman, chen2007titanium}. For nanocrystals, the three-dimensional confinement is the most convincing model \cite{li2005raman, chen2007titanium, lei2001fabrication, kelly1997raman, swamy2005finite, barborini2002engineering}.

\begin{figure}[ht!]
 \centering
    \includegraphics[width=0.9\textwidth]{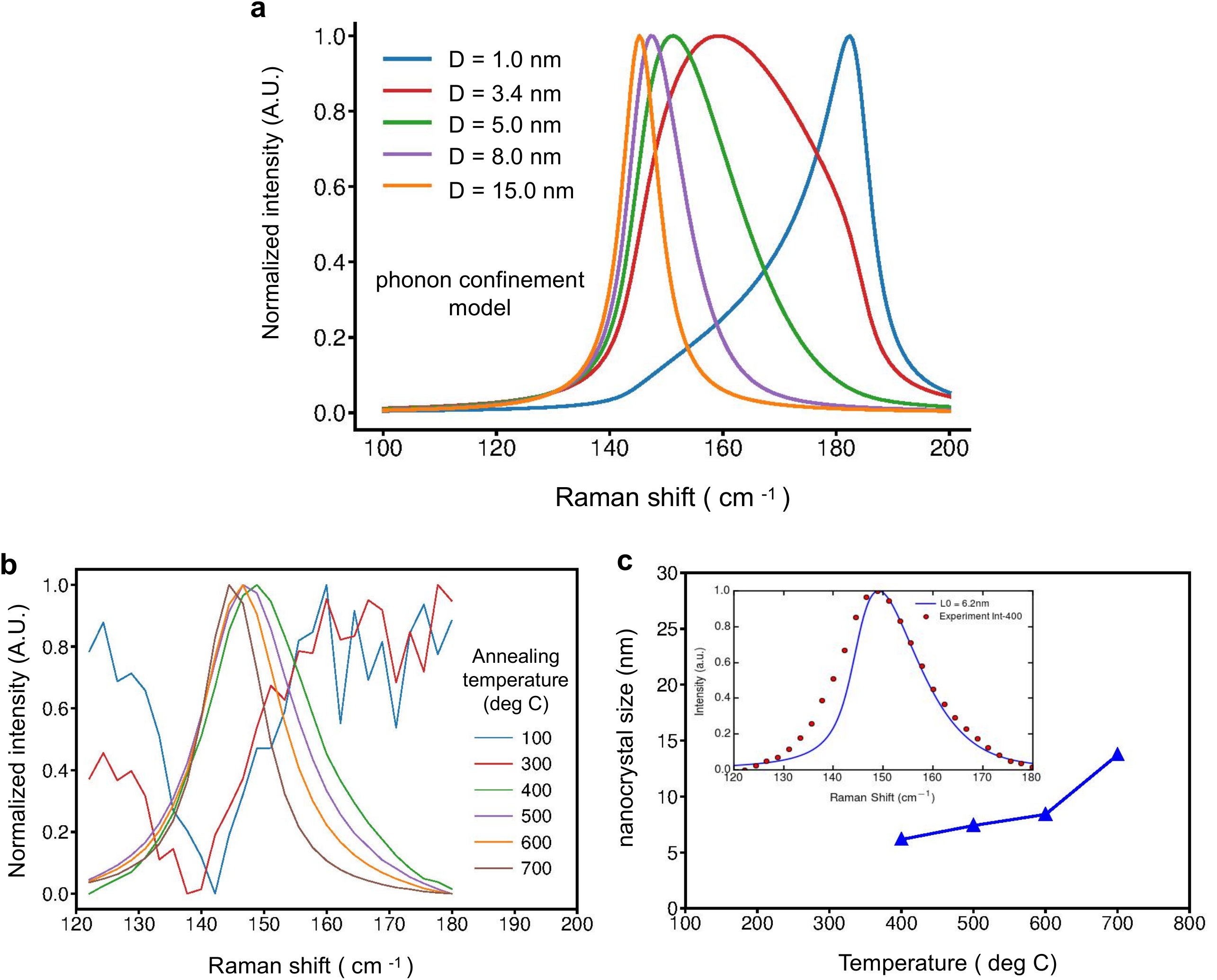}
    \caption{Raman spectra by phonon confinement model of various nanocrystal sizes (a). Experimental Raman spectra of amorphous TiO$_2$ samples after annealing at various temperature, the data are taken from Ref. \cite{sharma2019crystal}, and are normalized in the range of 120-180 cm$^{-1}$ as $I_{norm}(\omega) =  \frac{I(\omega) - min[I(\omega)]}{max\{I(\omega) - min[I(\omega)]\}}$ (b). The relation of TiO$_2$ nanocrystal size and annealing temperature (c). The crystallized size are obtained by fitting the experimental data from (b) to phonon confinement model. The insert-map of (c) shows the fitting curve (blue) along with the experimental data (red dot).}
\label{fgr:chapter5_tempRise_raman_p1}
\end{figure}

\begin{figure}[ht!]
 \centering
    \includegraphics[width=0.85\textwidth]{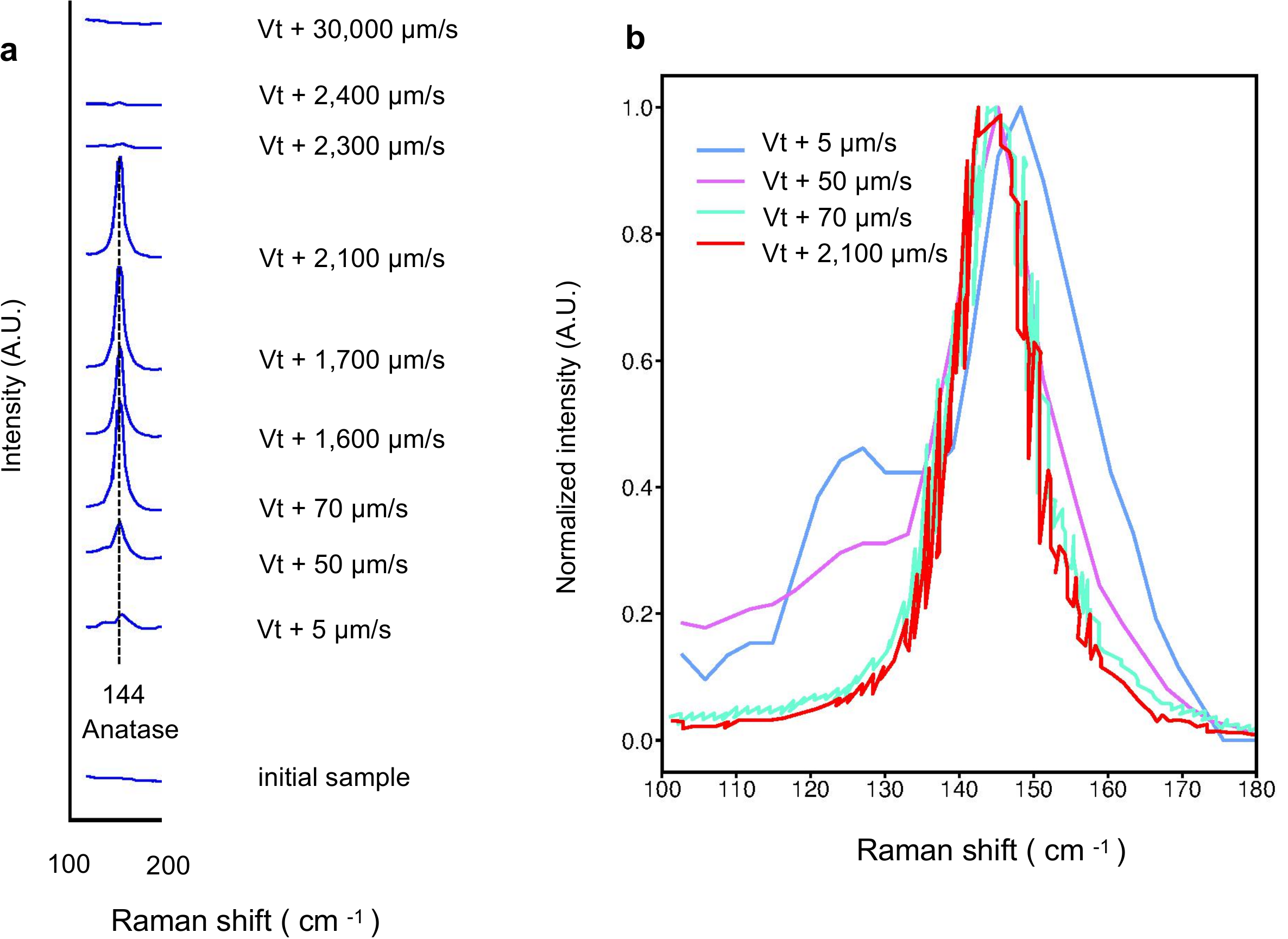}
    \caption{Raman spectra in the range of 100-200 cm$^{-1}$ of samples after different laser writing speeds (a). The excitation wavelength for Raman characterization is 633 nm. The experimental data are taken from Ref. \cite{liu2016selfthesis}, and are normalized in the range of 100-200 cm$^{-1}$ as $I_{norm}(\omega) =  \frac{I(\omega) - min[I(\omega)]}{max\{I(\omega) - min[I(\omega)]\}}$ (b).}
\label{fgr:chapter5_tempRise_raman_p2}
\end{figure}

The typical Raman spectra calculated by the three-dimensional phonon confinement model with different size of TiO$_2$ nanocrystals are shown as Figure \ref{fgr:chapter5_tempRise_raman_p1} (a). As the nanocrystal size decreases, the spectrum is broadened asymmetrically and shifted towards the higher frequency end. These features can be used for estimating the TiO$_2$ nanocrystal size by fitting the theoretical Raman intensity with the experiments \cite{sharma2019crystal} of annealing initial amorphous samples at various temperatures (see Figure \ref{fgr:chapter5_tempRise_raman_p1}(b-c)). The experiments were carried out by Nipun Sharma at Laboratoire Hubert Curien under the supervision of Nathalie Destouches \cite{sharma2019crystal}. The samples calcined at temperature between 100 and 300 $^{\circ} C$ are still amorphous. The distinct peaks of the Raman curves appear at temperature from 400 $^{\circ} C$. The lineshape tends to be more symmetric and narrowing as the temperature increases. The fitting between the phonon confinement and the experiments reveals a relation of nanocrystals' growth and annealing temperature. It is reasonable that at higher temperature the crystals are larger.

The temperature history of Ag:TiO$_2$ thin films under CW laser irradiation can then be estimated by the method discussed above. To show this, the Raman spectra in the range of 100 - 200 $cm^{-1}$ at different writing speeds are shown in Figure \ref{fgr:chapter5_tempRise_raman_p2}(a). The experimental data are taken from Ref. \cite{liu2016selfthesis} and are then normalized for the comparisons of line characteristics (see Figure \ref{fgr:chapter5_tempRise_raman_p2}(b)). To compare with the TiO$_2$ calcining experiments discussed above \cite{sharma2019crystal}, the data showing strong anatase peaks are selected (speed from Vt+50$\mu m/s$ to Vt+2,100$\mu m/s$). Though the resolution of the curves are limited, it is clear that the Raman spectrum are becoming more symmetric and narrowing while the writing speed is increased. The TiO$_2$ are more crystallized into larger nanocrystals for higher writing speed. The conclusion is that the maximum temperature inside Ag:TiO$_2$ films increases with laser writing speed in the range of Vt+50$\mu m/s$ to Vt+2,100$\mu m/s$).

In short summary of this subsection, the annealing experiments of TiO$_2$ that are reported in many researches \cite{chen2007titanium, swamy2006size, swamy2005finite} showing the TiO$_2$ phase-transitions, which roughly provides the temperature history. Moreover, the phonon confinement model in accordance with the calcined mesoporous experiments performed by N. Sharma et al. \cite{sharma2019crystal} provides fine information of the temperature. Based on such analyses, the post mortem Raman spectra \cite{liu2016selfthesis} of Ag:TiO$_2$ reveal a progressively increasing nature of the temperature with the writing speed of a CW laser. However, this phenomenon looks counter-intuitive since it seems less deposited laser energy leads to higher temperature rise. Is there something missing while mentioning this behavior? In the next subsection, we will go into the depth of the temperature discussion by the assistance of the multi-physical model.

\subsection{Multi-physical simulation of the temperature during laser writing}

In the previous sections, we have discussed the importance of temperature in the growth of Ag NPs inside TiO$_2$, but rarely mentioned the relation of temperature and laser writing speed shown by the model. To clarify the temperature behavior, we are going to discuss the transient maximum-temperature inside the sample during the laser scanning and then concentrates on the steady-state. The reason why we can split the process into a few stages including rapid growth and steady-state actually arises from, for one thing, variations of the spatial size-distribution of Ag NPs, for another, mostly the temperature evolutions.

\begin{figure}[ht!]
 \centering
    \includegraphics[width=0.85\textwidth]{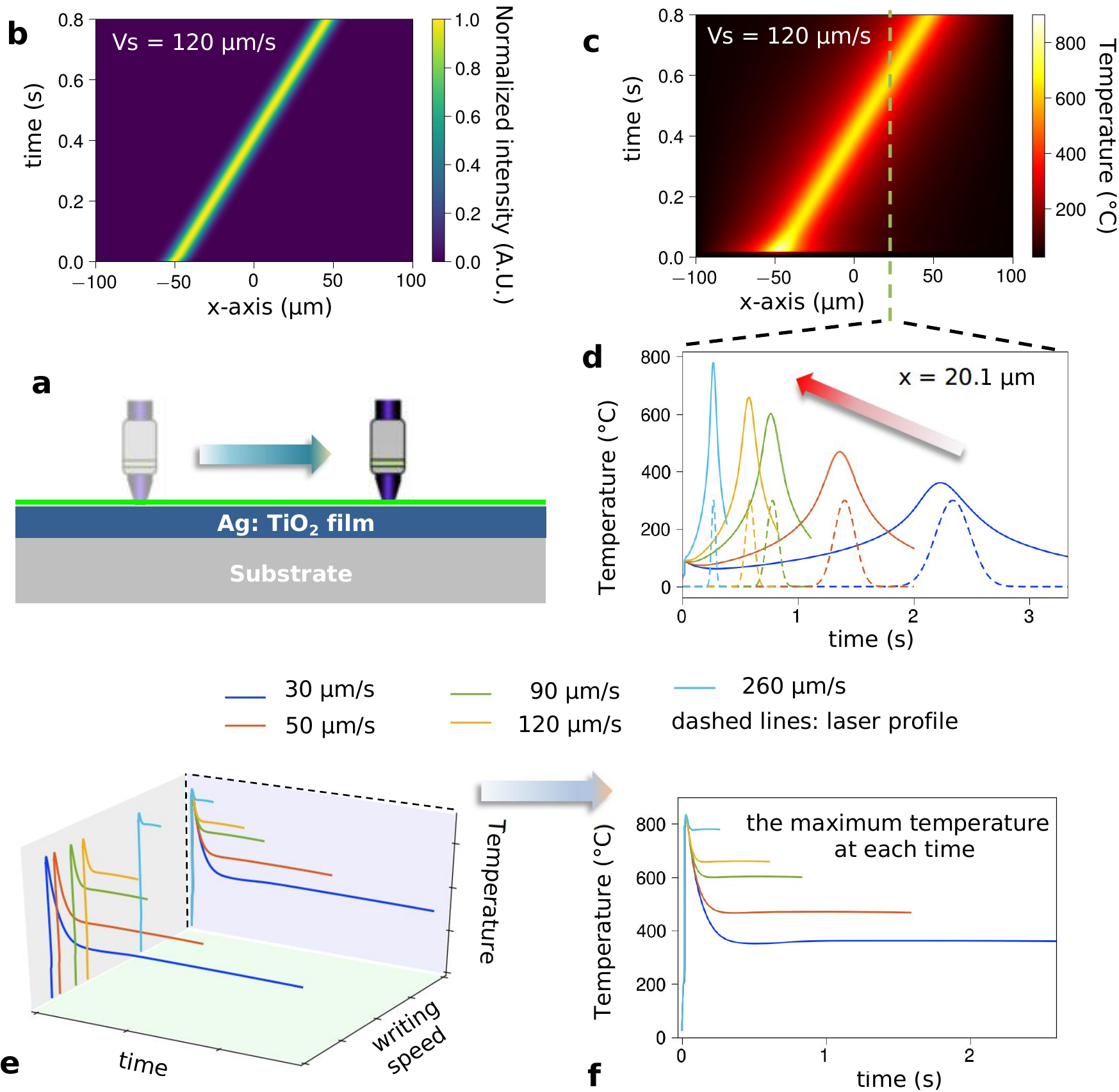}
    \caption{Results of the two-dimensional model. Illustration of the laser scanning direction (a). The top green line represents the locations for retrieving laser profile (b) and temperature distribution (c) from the model. The dynamical temperature history is taken at position $x = 20.1 \mu m$ (c). The maximum temperature in the TiO$_2$ as a function of time during laser writing (e-f). The activation energy is $\mathrm{E_{Ag^0} = 0.95 eV}$, and the ionization rate is $\mathrm{\eta_0 = 2.3 \times 10 ^{-5}}$. Other parameters are shown in Table \ref{table:Chapter5a}.}
\label{fgr:chapter5_tempRise_p1}
\end{figure}

The whole history of the temperature distribution (on the top surface of TiO$_2$) is simulated and plotted in Figure \ref{fgr:chapter5_tempRise_p1}. To better compare the temperature field, the maximum temperature of the structure versus the time is plotted in Figure \ref{fgr:chapter5_tempRise_p1}(e-f). As it is previously shown but rarely discussed in Figure \ref{fgr:3all}(e), the temperature increases from room temperature to a rather high value dramatically before decaying slowly till a steady distribution. This trend is continuously repeated for all the performed simulations at different writing speed. For curves standing for writing speed from 30 $\mu m/s$ to 260 $\mu m/s$, there are time ranges where the maximum temperature inside the structure is constant. Since NPs' growth and reduction process are greatly affected by the temperature, the maximum temperature provides the information of the highest reaction rate inside the TiO$_2$ thin films. Thus, the reactions during the time ranges with constant temperature are steady. Due to this reason, these states can be regraded as steady states.

\begin{figure}[ht!]
 \centering
    \includegraphics[width=0.9\textwidth]{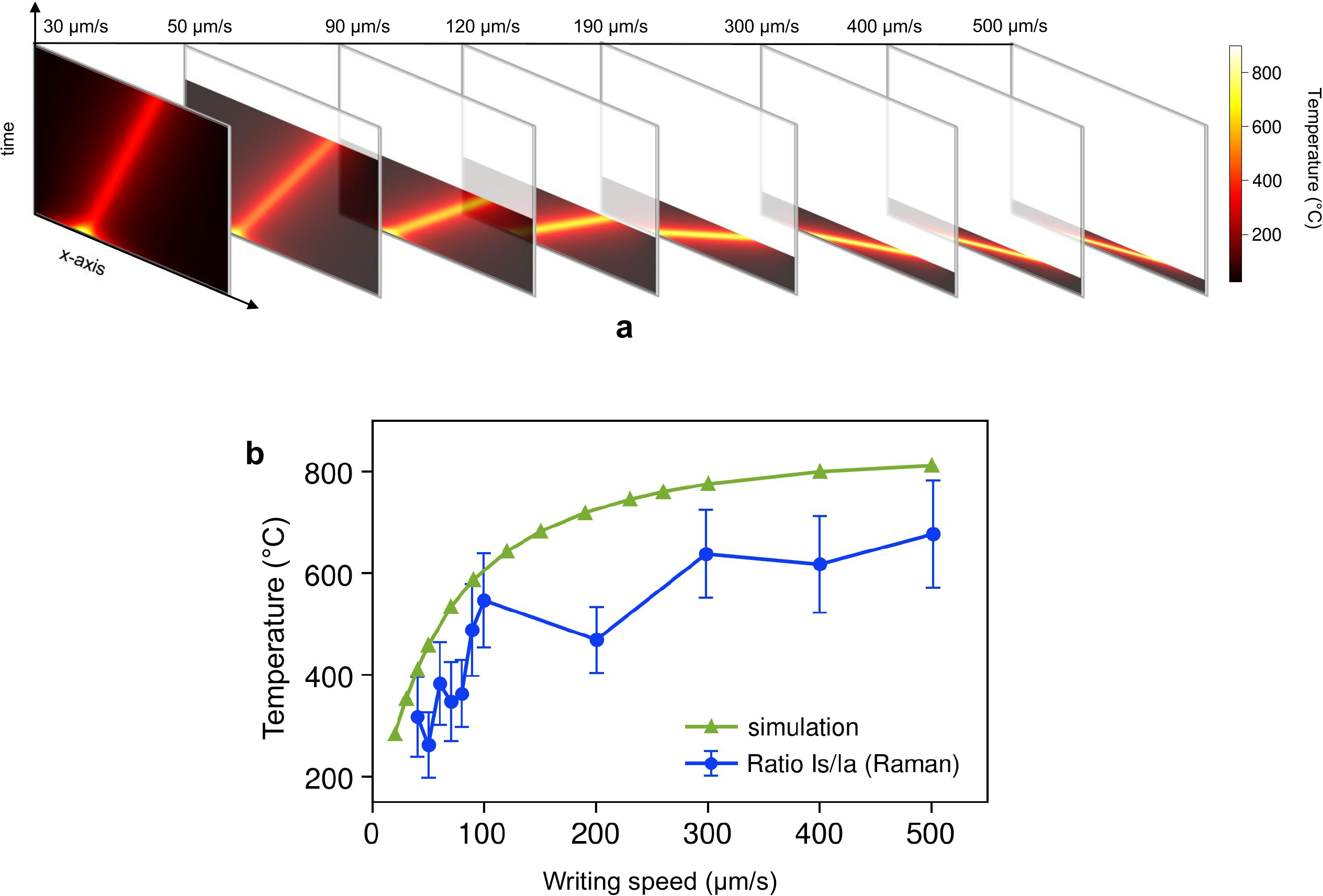}
    \caption{Simulation and comparisons with experiments. The x-time map of the temperature (a). The maximum temperature at steady state versus laser writing speed (b). The experimental data are taken from Ref. \cite{liu2016selfthesis}. The activation energy is $\mathrm{E_{Ag^0} = 0.95 eV}$, and the ionization rate is $\mathrm{\eta_0 = 2.3 \times 10 ^{-5}}$. Other parameters are shown in Table \ref{table:Chapter5a}.}
\label{fgr:chapter5_tempRise_p2}
\end{figure}

The temperature at steady state increases with the writing speed (Figure \ref{fgr:chapter5_tempRise_p1}(e) and (f)). This is also shown in Figure \ref{fgr:chapter5_tempRise_p1} (d) by studying the dynamical temperature at position $x = 20.1 \mu m$. The position can be anywhere locating at the steady state shown by Figure \ref{fgr:chapter5_tempRise_p1}(e) and (f). Expect the maximum value has a positive correlation with the writing speed, the temperature evolves differently. The faster the laser writing speed, the faster the temperature changes. This can be useful for calcined or biological applications where the speed of the annealing and the temperature are required to be well controlled. To investigate the relation of maximum temperature with the laser writing speed, simulations at writing speed from 30 $\mu m/s$ to 500 $\mu m/s$ are performed and shown in Figure \ref{fgr:chapter5_tempRise_p2}. The two-dimensional maps in Figure \ref{fgr:chapter5_tempRise_p2}(a) show the time-evolution of the  spatial-distribution of temperature on top of the TiO$_2$ thin film. The tilted the angle is, the higher the temperature it has in the TiO$_2$. The maximum temperature is found to increase at first following by a saturation as the writing speed increases (Figure \ref{fgr:chapter5_tempRise_p2}(b)).

The comparisons of temperature by experiments can be performed by the in-situ Raman spectroscopy. In fact, the Raman scattering or inelastic scattering has two possible outcomes: the photons having lower energy than the incident (known as the Stokes shift); and the photons is of higher energy than the incident (known as the anti-Stokes shift). Though the peak-locations of the Stokes and anti-Stokes in the spectra are symmetric around $\omega = 0$ $cm^{-1}$, the intensities differs because of the different populations of initial states. The initial states of Stokes modes has larger population than that of the ant-Stokes modes for a thermodynamic equilibrium system \cite{maher2008raman,cui1998noncontact,nemanich1984raman,santoro2004situ}. The anti-Stokes modes obeys the Boltzmann distribution $\exp[-\hbar \omega_{\nu} / (k_B T)]$, which as a result gives the ratio of the anti-Stokes and Stokes intensities \cite{maher2008raman,cui1998noncontact,nemanich1984raman,santoro2004situ}:

\begin{equation}
    \frac{I_{AS} (\nu)}{I_S (\nu)} = B_0 \frac{(\omega_L + \omega_{\nu})^4}{(\omega_L - \omega_{\nu})^4} \exp[-\hbar \omega_{\nu} / (k_B T)]
    \label{eqn:Raman_temp_p2}
\end{equation}

\noindent where $\hbar$ is the reduced Planck's constant, $\omega_{\nu}$ and $\omega_{L}$ are circular frequencies of the vibration modes and the probing laser, $k_B$ is the Boltzmann constant, $T$ is the absolute temperature, and $B_0$ is a constant. The temperature can be retrieved from in-situ Raman experiments based on the Eq. \ref{eqn:Raman_temp_p2}. Here, the experimental data are taken from Ref. \cite{liu2016selfthesis}. The writing speed by the experiments was limited to 500 $\mu m/s$ in order to get a sufficient acquisition time ($\lambda = 533 nm$). The experiments show an increasing tendency in the temperature as the writing speed increases, which confirms the simulations. Noting that only anatase nanocrystals can be observed for the cases at writing speed below 500 $\mu m/s$ \cite{liu2016selfthesis}, the temperature by simulations saturates at speeds of 200-300 $\mu m/s$ is thus inconsistent. Before going further to discuss how the simulations can be fitted with the experiments, the following discussions will be how to understand the temperature behavior.

\begin{figure}[ht!]
 \centering
    \includegraphics[width=0.85\textwidth]{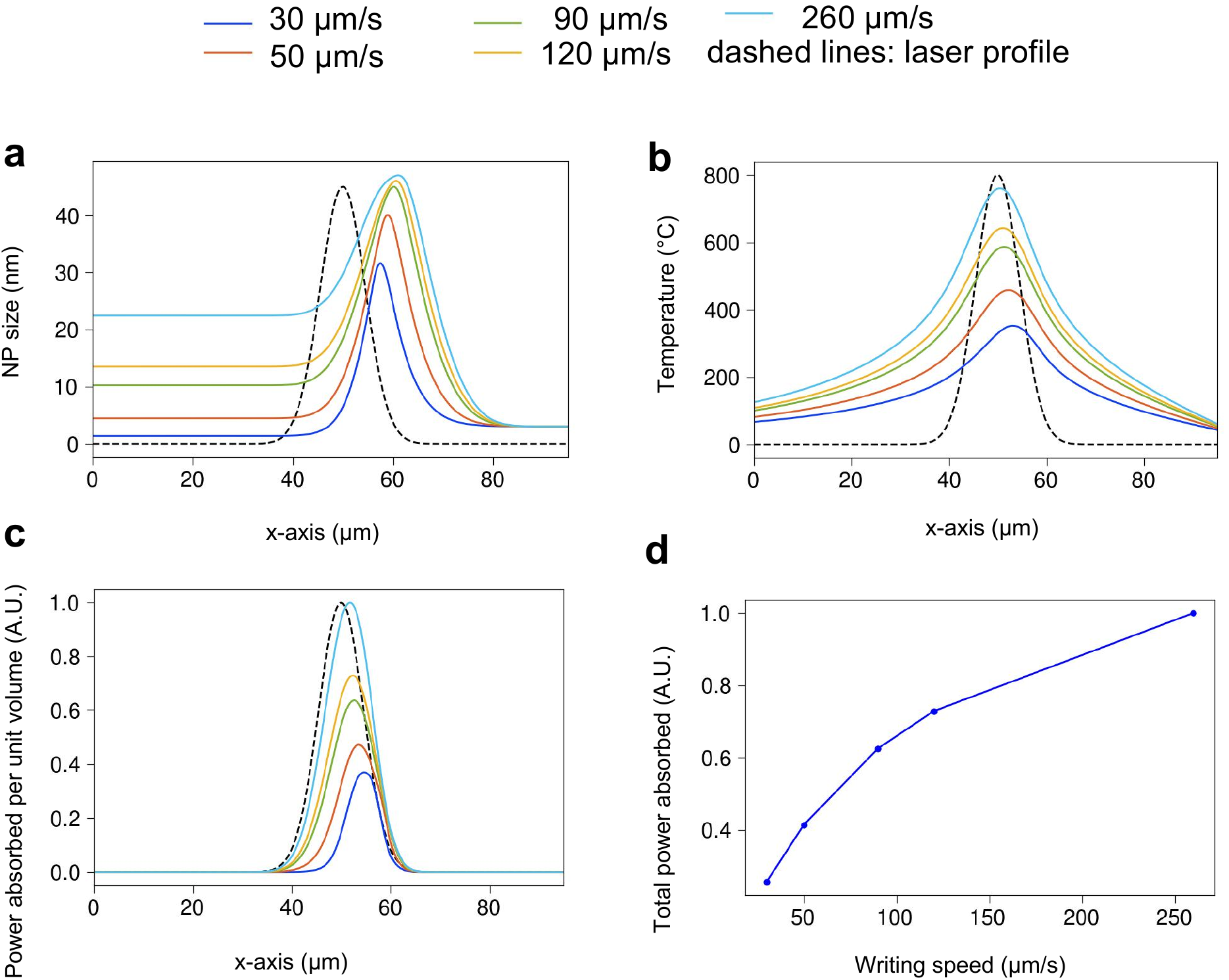}
    \caption{The instant light absorption during laser scanning. The Ag NP size (a), temperature (b) are plotted as the laser-center locates at $x=50 \mu m$. The absorbed laser power per unit volume (c) and the total laser power absorbed (d) by the film are calculated. ($\mathrm{E_{Ag^0} = 0.95 eV}$, $\mathrm{\eta_0 = 2.3 \times 10 ^{-5}}$).}
\label{fgr:chapter5_tempRise_p3}
\end{figure}

In order to understand why the less deposited-energy (higher writing speed) leads to higher temperature rise in the studied system, the instant absorption along x-axis are calculated and shown in Figure \ref{fgr:chapter5_tempRise_p3}. It is noted that the laser-center locations at $x=50 \mu m$ representing the steady state. The starting position of the laser locates at $x= - 50 \mu m$. The second term on the right-hand side of Eq. \ref{eqn:thermal} is calculated for different writing speeds and shown in Figure \ref{fgr:chapter5_tempRise_p3}(c). The corresponding NP size and temperature distributions are shown in Figure \ref{fgr:chapter5_tempRise_p3}(a) and (b). It is obvious that the area below the absorbing curves increases with the writing speed. The total power absorbed by the Ag:TiO$_2$ thin film are calculated by integrating the heat source term of Eq. \ref{eqn:thermal} over the whole thin film (Figure \ref{fgr:chapter5_tempRise_p3}(d)). Clearly, the absorption at-peak and in-average at steady state are both increased at higher writing speed. Consequently, as shown by Figure \ref{fgr:chapter5_tempRise_p3}(b), the maximum temperature has a positive correlation with the writing speed.

\begin{figure}[ht!]
 \centering
    \includegraphics[width=\textwidth]{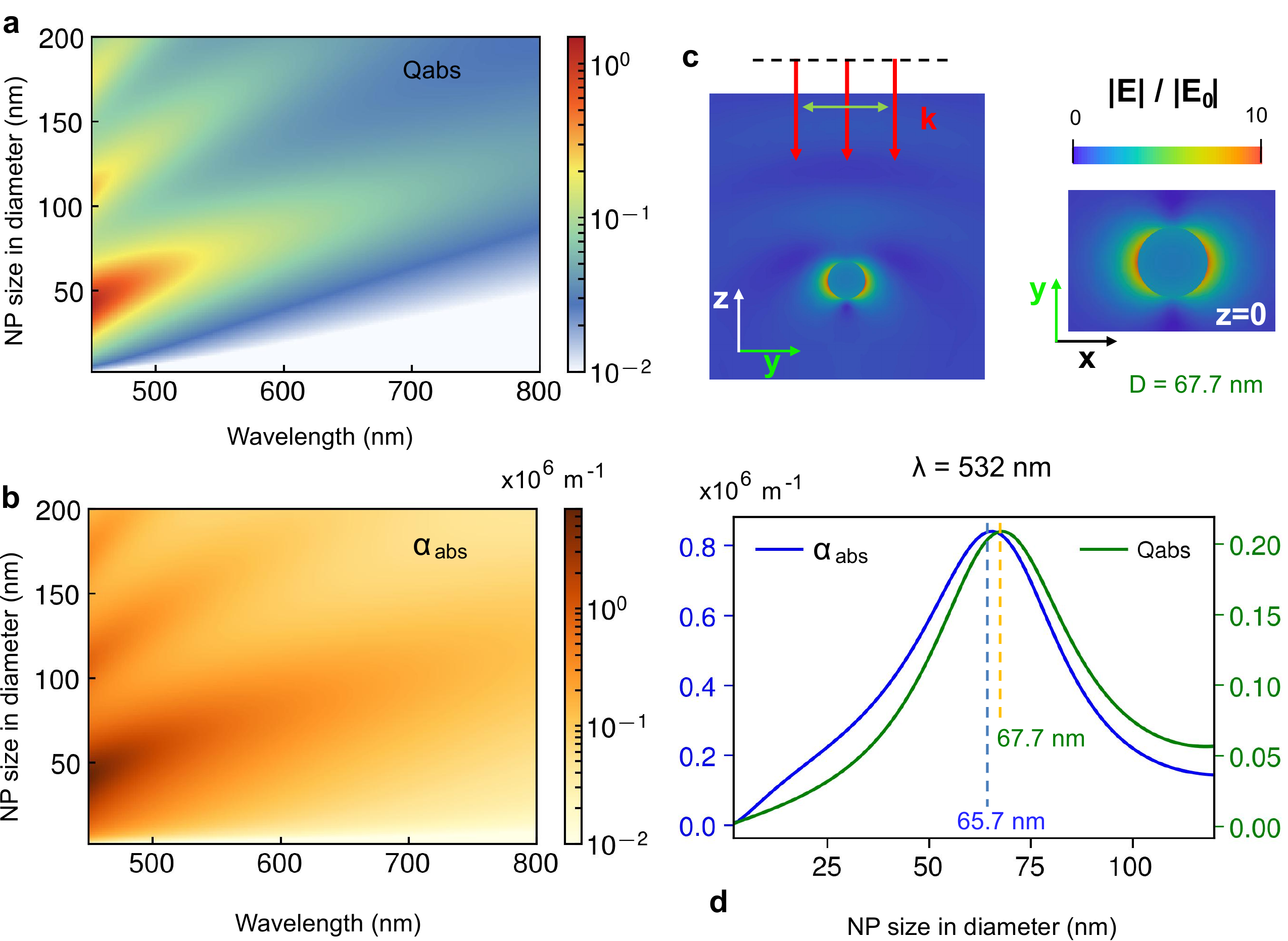}
    \caption{Maps of absorption efficiency factor (a) and absorption coefficient (b). Electromagnetic simulations (COMSOL) of plane wave ($\lambda = 532 nm$) scattering by a Ag NP (c). The absorption efficiency factor ($Q_{abs}$) and absorption coefficient ($\alpha_{abs}$) as functions of Ag NP size at wavelength 532nm (d).}
\label{fgr:chapter5_mieCal}
\end{figure}

Further investigations of the heat equation (Eq. \ref{eqn:thermal}) and the NP size distribution (Figure \ref{fgr:chapter5_tempRise_p3}(a)) reveal that the Ag NP's intrinsic plasmonic absorption is responsible for the boosted power absorption. In the modeling, we used the Mie theory to calculate the NP's absorption by assuming that NPs are randomly distributed and the interactions between NP-NP can be ignored. Figure \ref{fgr:chapter5_mieCal}(a) shows the map of the absorption efficiency factor ($Q_{abs} = \sigma_{abs} / \pi R^2$), where $\sigma_{abs}$ is the cross-section of absorption and $R$ is the radius of Ag NP. In heat equation (Eq. \ref{eqn:thermal}), recall that $\alpha_{abs} = C_{NP} \times \sigma_{abs}$, the NP's abundance or concentration ($C_{NP}$) determines the averaged absorption at any positions. The absorption coefficient $\alpha_{abs}$ is plotted in Figure \ref{fgr:chapter5_mieCal}(b). At wavelength $\lambda = 532 nm$, the absorption efficiency factor ($Q_{abs}$) and the absorption coefficient ($\alpha_{abs}$) peaks at 67.6 nm and 65.7 nm in diameter, respectively (see Figure \ref{fgr:chapter5_mieCal}(d)). At plasmonic resonance, the near-field enhancement of electric field is shown in Figure \ref{fgr:chapter5_mieCal}(b). The maximum NP size by the multiphysical modeling is around 46 nm, which is below the peak-absorption size. For the possible sizes of Ag NPs studied by the model are in the left range where the absorption increases while the NP is larger. We have known that, during the laser irradiation, the process forms large NP ahead of laser beam center following by the shrinkage due to photo-oxidation. It results to comparatively larger size at higher writing speed at any positions behind the locations of the maximum size (Figure \ref{fgr:chapter5_tempRise_p3}(a)). Thus, according to the size-dependent absorption shown in Figure \ref{fgr:chapter5_mieCal}, both the maximum and averaged temperature-rise in the laser spot increases as the writing speed is faster. The limitation of the modeling is obvious that the single-size may give negative response of temperature as the NP size exceeds the peak-absorption size (65.7 nm or 67.7 nm). In fact, the NP size is dispersed according to experiments \cite{liu2017haadf, liu2016selfthesis, liu2015understanding}, and the temperature never decreases as the writing speed increases to the upper limit (around $30 mm/s$) of the setups \cite{liu2016selfthesis, destouches2014self, liu2016laser}.

\begin{figure}[ht!]
 \centering
    \includegraphics[width=0.9\textwidth]{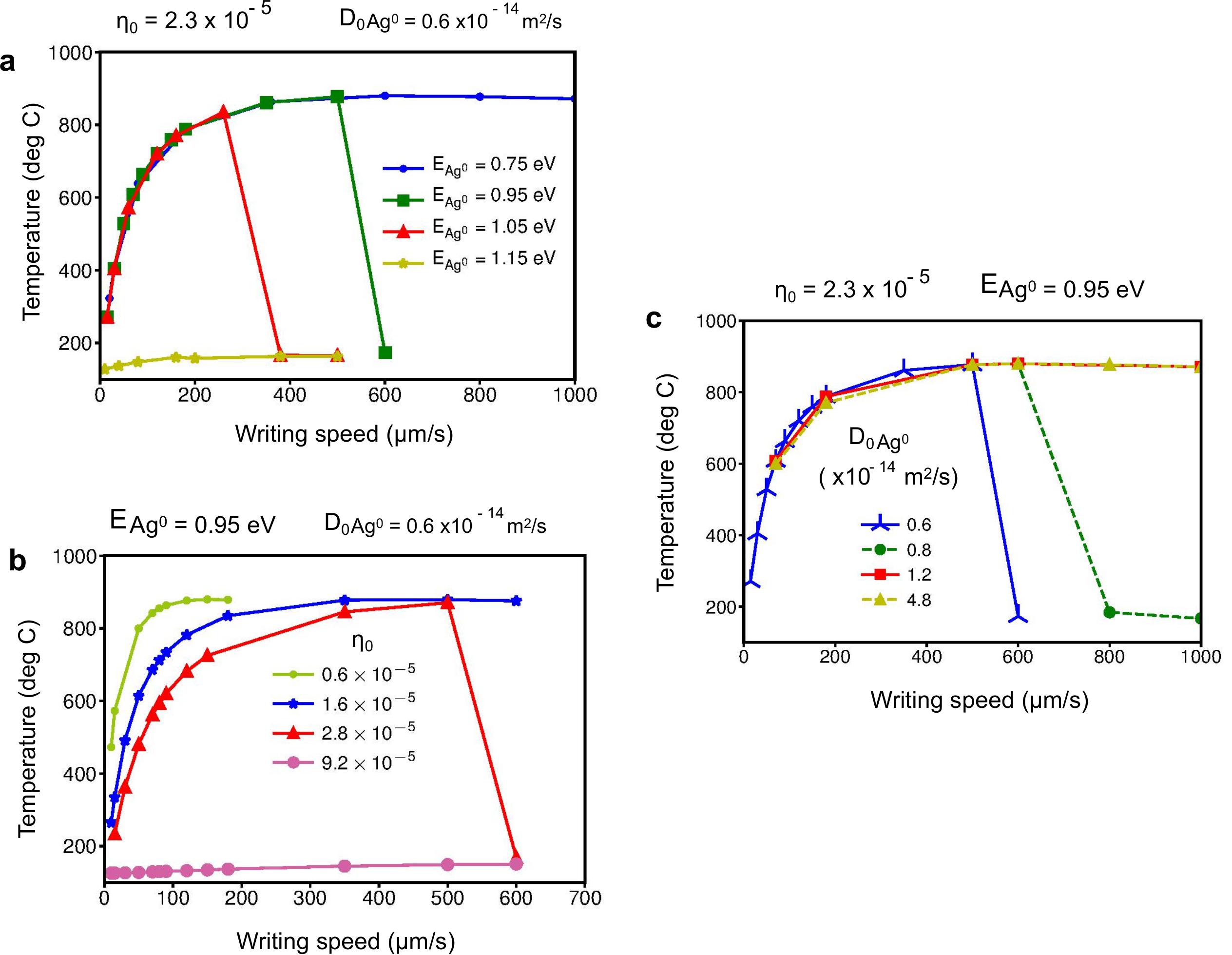}
    \caption{The maximum temperature at steady state as a function of the writing speed influenced by activation energy of Ag$^0$ (a), ionization efficiency (b), and the diffusion ability $\mathrm{D_{0Ag^0}(500K)}$ (c).}
\label{fgr:chapter5_DEAg0_maxT_p1}
\end{figure}

Due to lack of the direct experimental proof by in-situ TEM characterization, the in-direct diagnostics by in-situ transmission and or in-situ temperature can be used for interpolating the model parameters. To do so, one needs to know how do Ag diffusion and ionization rate affect the measurable quantities and what are the trends by these coefficients. For this purpose, here, this discussion will focus on the temperature-rise. We have seen from the previous sections that the Ag diffusion activation energy $E_{Ag^0}$, the diffusion ability $D_{0Ag^0}(500K)$, and the ionization rate $\eta_0$ have great influence on the NPs' final size after laser treatment, and the maximum size in the laser front. The resulted temperature-rise is also affected. Figure \ref{fgr:chapter5_DEAg0_maxT_p1} shows the influences of $E_{Ag^0}$, $D_{0Ag^0}(500K)$, and $\eta_0$ to the maximum temperature-rise at steady states. Hereafter in this subsection, for simplicity, the mentioned maximum temperature-rise stands for the steady states.

\noindent Firstly, an increase in activation energy ($E_{Ag^0}$) slow-down the NP's growth, which can be seen from Figure \ref{fgr:chapter5_DEAg0_maxT_p1}(a) and Figure \ref{fgr:6all}(c) with $E_{Ag^0}$ ranges from 0.75 ev to 1.15 eV. The break firstly appears at $Vs = 600 \mu m/s$ for $E_{Ag^0} = 0.95 eV$, which decreases to $V_S = 300 \mu m/s$ for $E_{Ag^0} = 1.05 eV$. Secondly, the diffusion ability ($D_{0Ag^0}(500K)$) has positive consequences to the NP's growth. The influence, for instance, as shown in Figure \ref{fgr:chapter5_DEAg0_maxT_p1}(c) speed-up the NP's growth. The speed at break position increases from $600 \mu m/s$ to $800 \mu m/s$ before completely disappearing as increasing the diffusion ability ($D_{0Ag^0}(500K)$) from 0.6 to 1.2 $\times 10^{-14} m^2/s$. A further increase of the diffusion ability does not change the curve. Moreover, the Ag diffusion does not affect the maximum temperature-rise (Figure \ref{fgr:chapter5_DEAg0_maxT_p1}(a,c)), on the contrary, the photo-oxidation rate ($\eta_0$) shift the saturating position of the curve (Figure \ref{fgr:chapter5_DEAg0_maxT_p1}(b)). The essential lies in the growth nature of Ag NPs in the studied system which the Ag NPs' growth is ahead of the laser beam due to the heat-diffusion; the growth stops by consuming the majority of the $Ag^0$ in the matrix while the quantity of $Ag^0$ by reduction is too small to feed back into the growth, so that the photo-oxidation dominates the process. As consequences, the oxidation rate affect the spatial-distribution of NP size along the x-axis, which according to the absorption relation shown by Figure \ref{fgr:chapter5_mieCal} affect the maximum temperature-rise. It is obvious that increasing the ionization rate $\eta_0$ shall decrease the NP's size and the temperature. As what is found previously, too high ionization rate shall affect the first rapid-growth (e.g. Figure \ref{fgr:6all}(c)), as a result the temperature at $\eta_0 = 9.2 \times 10^{-5}$ never exceeds $200 ^{\circ}C$. 

\begin{figure}[ht!]
 \centering
    \includegraphics[width=0.95\textwidth]{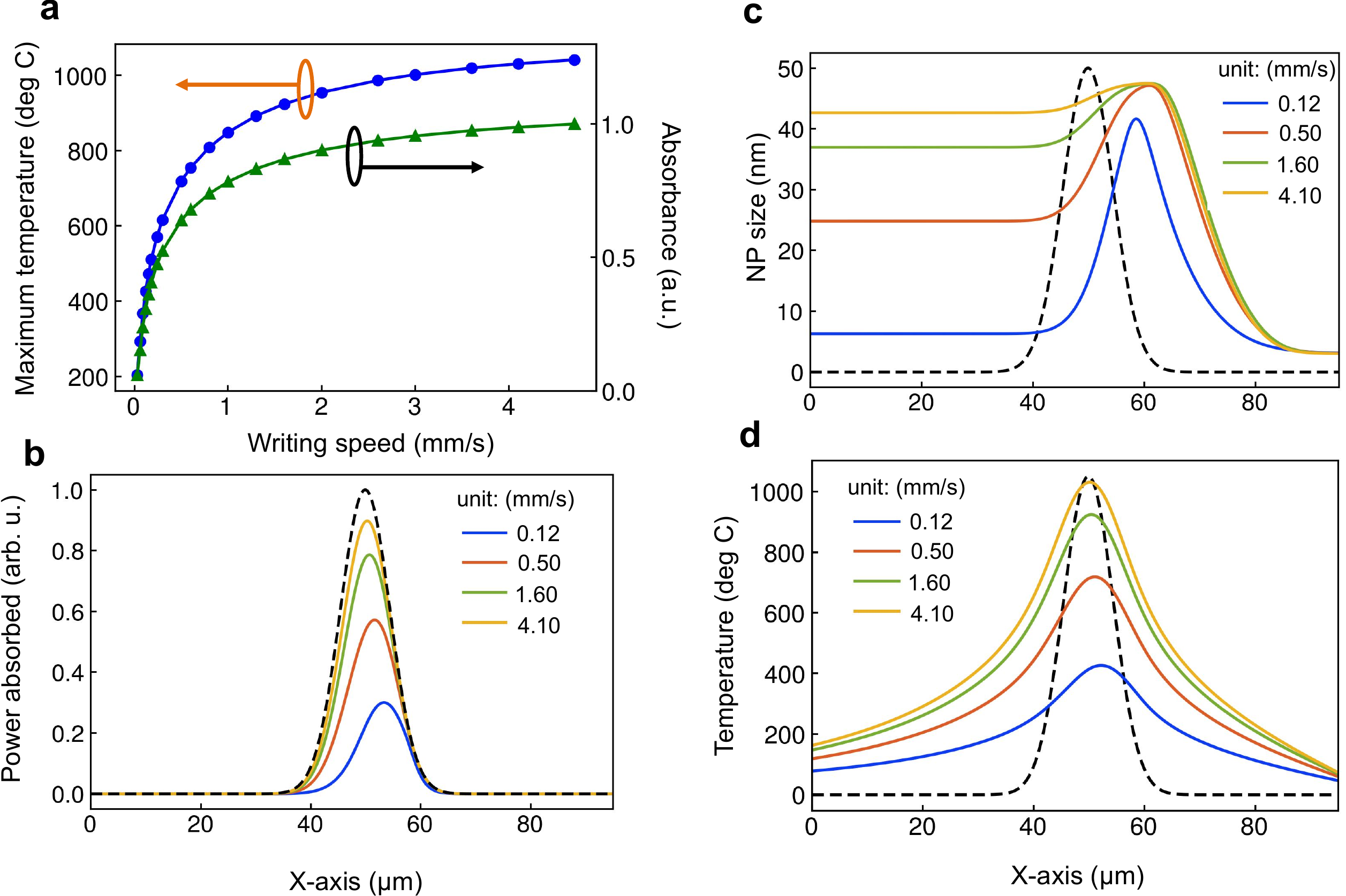}
    \caption{The temperature as a function of the laser writing speed and the corresponding absorbance. ($\mathrm{E_{Ag^0} = 0.75 eV}$, $\mathrm{\eta_0 = 6.9 \times 10 ^{-5}}$, and $\mathrm{D_{0Ag^0}(500K) = 4.0 \times 10 ^{-14} m^2/s}$). Laser intensity profiles are shown in black dashed lines.}
\label{fgr:chapter5_tempRise_p4}
\end{figure}

Indeed, the temperature by experiments increases from speed at a few tens of $\mu m/s$ to 30 mm/s (Figure \ref{fgr:chapter5_tempRise_raman}). The simulated temperature till now seems to saturate and even breaks at speed much lower than the experiments. Now, the question is whether the model can be interpolated close to that experimental findings by adjusting the diffusion parameters and the ionization rate. Before coming back the question, a referring to the parameter discussion would help to understand why one can do this. First of all, the ionization rate is the only one alter the curve shape and the saturation; it, however, suffers from temperature break by simply increasing the ionization rate. Furthermore, a decrease of the activation energy $E_{Ag^0}$ removes the break points. The last but not the least, the increase of diffusion ability ($D_{0Ag^0}(500K)$) accelerates the NP's growth and thus removes the break points in the temperature curve. Figure \ref{fgr:chapter5_tempRise_p4}(a) shows the temperature vs laser writing speed at $E_{Ag^0} = 0.75 eV$, $\eta_0 = 6.9 \times 10 ^{-5}$, and $D_{0Ag^0}(500K) = 4.0 \times 10 ^{-14} m^/2$. The break in the temperature curve disappears at this choice of Ag diffusion and ionization parameters. The absorbance in Figure \ref{fgr:chapter5_tempRise_p4}(a) is calculated as follows:

\begin{equation}
    \mathrm{Absorbance} =\int_{x = -100 \mu m}^{x = 100 \mu m} \int_z \alpha_{abs} \cdot I(x,z) dxdz
	\label{eqn:thermal_absorbance}
\end{equation}

\noindent where, the term $\alpha_{abs} \cdot I(x,y,z)$ stand is the second term of the heat equation \ref{eqn:thermal}, the distributions on top surface of the structure are shown in Figure \ref{fgr:chapter5_tempRise_p4}(b). The saturation of the temperature-rise vs writing speed is due to the small size change by photo-oxidation in the laser beam (Figure \ref{fgr:chapter5_tempRise_p4}(c)). The temperature profiles at studied writing speed along with the laser intensity profile (black dashed line) are shown in Figure \ref{fgr:chapter5_tempRise_p4}(d). Obviously, both the maximum temperature and the averaged temperature look increase with the writing speed.

\section{3D simulation}

From the previous calculations, we find that solving the coupled equations is non-trivial due to the intrinsic multi-scale problem. Firstly, the growth speed of NPs strongly depends on the temperature, thus making time-steps range from 0.1 ns to 1 $\mu s$ to keep second-order accuracy according to the maximum temperature it experiences. At the same time, the scanning speed of laser is comparatively small that varies from 100 $\mu m/s$ to 50 mm/s. For instance, to study the laser scans 100 $\mu m$ on the sample, in the
case of a typical value to blind computing, the time step is set to 1 ns, it requires more than 2,000,000 steps to finish the calculation. Moreover, the calculation of heat diffusion based on finite element method (FEM) is time consuming.

\begin{figure}[ht!]
 \centering
    \includegraphics[width=0.75\textwidth]{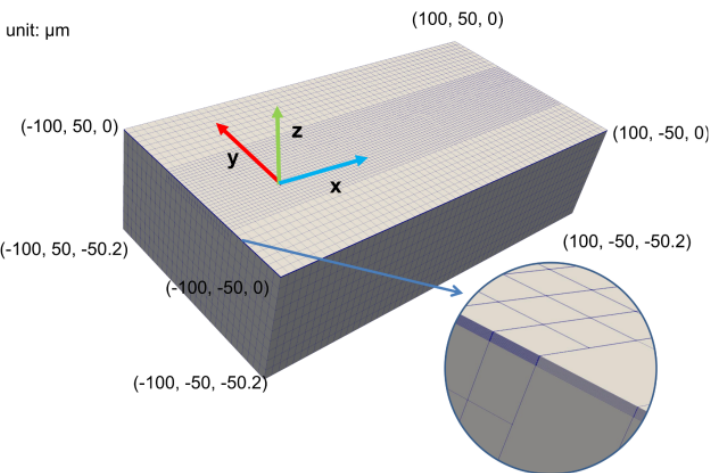}
    \caption{3D view of the meshes used in FEM calculations.}
\label{fgr:chapter5_3D_p1}
\end{figure}

Comparing to the 2D model, 3D simulations provide a much more complete information. Albeit calculating the temperature and NP size is more difficult in 3D due to the matrix assembling in FEM algorithm, a good design of the data structure in C/C++ enables a scalable and parallel computing on a supercomputer. To
accelerate the calculation, the code is developed based on the distributed-memory parallel programming algorithm. The computational domain consists of 200 nm thick TiO$_2$ film on top of 50 $\mu m$ glass substrate, with width in X and Y of 100 $\mu m$ and 200 $\mu m$, respectively. Figure \ref{fgr:chapter5_3D_p1} shows the meshes used in the FEM calculation. The number of degrees of freedom is 96,865. Thank to the CINES project, we used 32-cores of Intel(R) Xeon(R) CPU-E5-2690 v3 @ 2.6GHz for three weeks to perform the 3D calculations. 
Table \ref{table:Chapter5a_2} summarizes the parameters used in 3D simulations.

\begin{table*}[ht!]
  \centering
  \caption{A summary of the parameters in 3D simulation}
  \begin{tabular}{llll}
  \hline
  \hline
      Parameter & Description & Value\\
  \hline
  $P_0$ & laser power& 400mW \\ 
  $\lambda$ & laser wavelength& 532nm \\ 
  $w_0$ & waist radius & $\mathrm{9 \mu m}$\\

  $E_{Ag^0} $ & activation energy of Ag$^0$ diffusion & 0.75eV \\
  $D_{0Ag^0}(500K) $ & Ag$^0$ diffusion coefficient at 500K &$16 \times 10^{-14} \mathrm{m^2/s}$\\
  $\eta _0$ & ionization efficiency & 6.9 $\times 10^{-5}$ \\

  $C_{Ag^0}^{BS}(t=0 s)$ & initial Ag$^0$ concentration in matrix &  $1.495 \mathrm{\times 10^{4} mol/m^3}$\\
  $C_{Ag^+}(t=0 s)$ & initial Ag$^+$ concentration in matrix &  $0.996 \mathrm{\times 10^{4} mol/m^3}$ \\

  \hline
  \hline
      \label{table:Chapter5a_2}
  \end{tabular}
\end{table*}

\begin{figure}[ht!]
 \centering
    \includegraphics[width=0.82\textwidth]{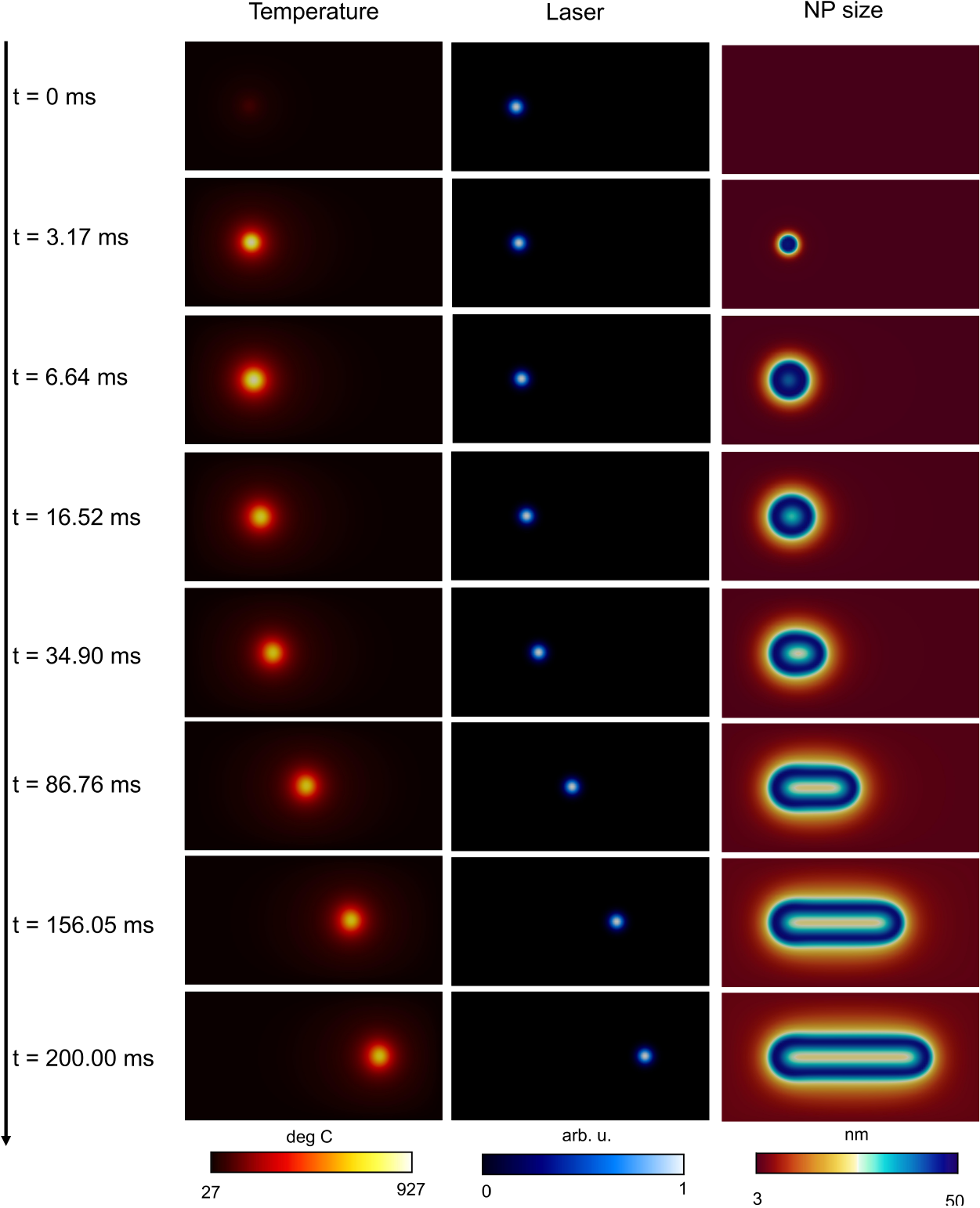}
    \caption{Top views of temperature, laser intensity and Ag NP size distributions at different time. The laser writing speed is $500 \mu m/s$.}
\label{fgr:chapter5_3D_p2}
\end{figure}

Figure \ref{fgr:chapter5_3D_p2} shows top views of laser intensity, temperature, and NP size at different time. In the presented results in Figure \ref{fgr:chapter5_3D_p2}, the laser writing speed is 500 $\mu m/s$; the beam waist size is $2w_0 = 18 \mu m$, at $e^{-2}$ of the peak intensity; the laser power is 400 mW; and the laser wavelength is 532 nm. In the calculation, the laser center goes from (-50,0,0) to (50,0,0) $\mu m$ in +x direction. The results present a general behavior for all scanning speeds. In particular, one can observe the fast growth of NPs at the stating position followed by the size shrinkage and by a transition to the steady state after laser leaves the starting position. At the starting position, due to the positive feedback of plasmonic absorption and NP size, a fast or explosive growth is observed accompanying by the dramatic temperature increase. The following photo-ionization shrinks the NP size in the laser beam, making asymmetry size from laser front to the tail. This process continues until the steady-state is set. The asymmetric NP size profile leads to the light absorption efficiency decrease from the laser front to the tail, which shifts the temperature distribution with its maximum value ahead of laser beam center. In addition, because of heat diffusion, NPs are found to grow even before the center of the laser beam arrives.

\begin{figure}[ht!]
 \centering
    \includegraphics[width=\textwidth]{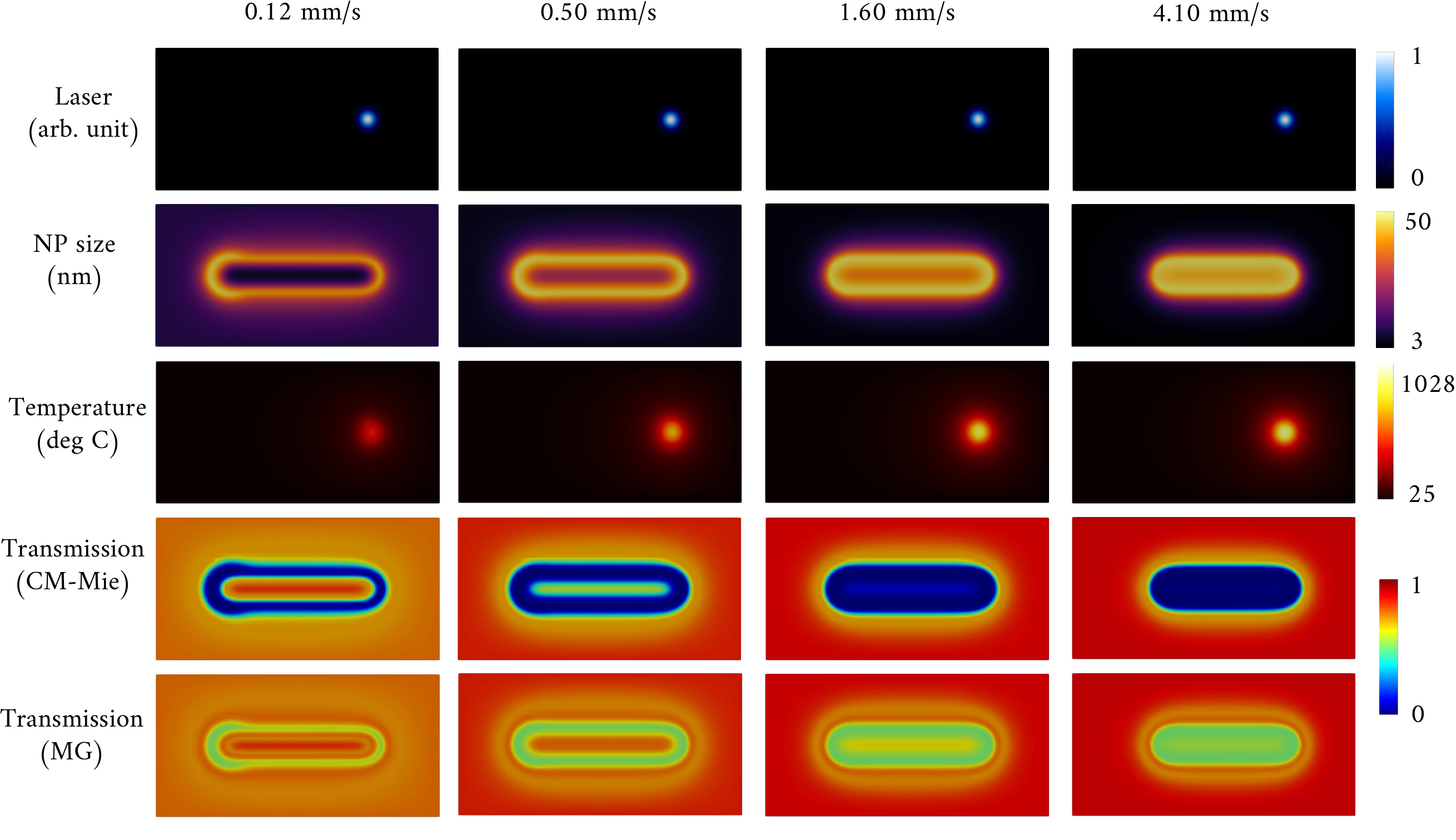}
    \caption{Top views of laser intensity, NP size, Temperature and the corresponding transmission maps for different writing speeds. The results are from the final step in each simulations.}
\label{fgr:chapter5_3D_p3}
\end{figure}

To compare the results of our 3D simulations with the in-situ experiments of transmission during laser writing, maps of laser intensity, NP size, temperature and the corresponding transmission by MG and CM-Mie models are presented in Figure \ref{fgr:chapter5_3D_p3}. The presented results correspond to the final step (the laser center stops at $x = 50 \mu m$) for each writing speed. The Figures show that, in contrast to the 2D model, here we can analyze NPs' size-distribution along the Y direction, which is perpendicular to the writing direction. The NP's size inside the laser spot is smaller than that at the periphery. It can be understood by the growth and shrinkage process. As indicated by Eq. \ref{eqn:fick2} and Eq. \ref{eqn:DAg0defi}, the NPs start the nonlinear growth as soon as the temperature exceed a specific value. The consequence is that, in all the studied writing speeds, the NP size in the laser front forms "\rotatebox{90}{U}" shapes. The photo-oxidation only affects the region inside the laser spot, which results to the small size of Ag NP inside the spot while remaining the outside region to be unaffected. The trend of size increase with writing speed revealed by the 2D model is valid inside the spot region. Since the larger NPs outside the laser spot have little contributions to the absorption and the heat conduction are assumed to be constant in this study, the relation of maximum temperature increase with writing speed is not affected.

Better comparisons with in-situ transmission experiments are done by directly calculating the transmission maps. The shape of the transmission maps by MG and CM-Mie are close to the in-situ experiments shown in Figure \ref{fgr:1all}(b). For both MG and CM-Mie mode, the transmission inside the laser spot affected line are higher than the vicinity, which is similar to the experiments. Moreover, the higher the writing speed is the lower the transmission in both the center of edge of the written line. Based on the NP size analyses above by the 3D simulations, one can deduce that the size of Ag NP inside the written lines are larger as the writing speed increasing in experiments. In addition, as the temperature is resulted from absorptions by the ensemble of NPs inside the laser spot and diffusion around the vicinity, getting the relation of maximum temperature verses single NP size is impossible. However, the in-situ experiments show better transparency inside the written lines than the model, which may result from particles movement, thickness reduction of the thin film under high temperature, imperfect considerations of size-variations (or size dispersion), a changing in heat conductance, and other changes inside the mesoporous material in reality.

\section{Conclusions}

\hspace{0.5cm}  We have proposed an extended and self-consistent model of laser-induced Ag NP growth in TiO$_2$ film. The multi-scale problem has been theoretically investigated by introducing the Bulirsch-Stoer algorithm coupled with Finite Element Method, so that the extended 2D model was solved numerically in an affordable time. The extended 2D model takes into consideration spatial size-distribution of NPs and provides new insights into the understanding of the writing speed-dependent phenomena. 

Firstly, the silver NPs have been shown to grow in the laser front edge due to pre-heating by thermal diffusion. Secondly, photo-oxidation has been found to be responsible for the writing speed-dependent size. The spatial size-distribution and temperature profile vary dynamically until a steady-state appears. The implied stationary solutions of the moving coordinate system are found to exist for various speeds by the transient analyses.  

Furthermore, the activation energy of Ag$^0$ diffusion has been shown to affect the growing speed, but to only slightly influence the size-saturation. On the contrary, the ionization efficiency is found to have a great impact on the saturation, which again proves the important role of oxidation in the speed controlled phenomena. In the meantime, high oxidation prohibits the rapid growth of NPs that levels up the threshold size of the positive feedback. 

The proposed study provides an overview of the influences of Ag$^0$ activation energy and ionization efficiency in mesoporous TiO$_2$ film on the NP size, which gives a new route for comparing the ionization efficiency in different samples.

We note, finally, that the proposed model can be also applied for many similar systems composed of metallic nanoparticles embedded in semiconductors such as, for instance, Ag:ZnO or Au:TiO$_2$.

\afterpage{\null\newpage}

\chapter{Multi-pulsed-laser writing of Ag NPs in TiO$_2$ thin films}

\hspace{0.5cm}  This chapter is focused on multi-physical modeling of ultrashort laser-induced Ag nanoparticle formation in TiO$_2$ films. Comparing with continuous-wave laser processing, pulsed-lasers having higher laser intensities and short pulses that are more likely in producing surface grooves (LIPSS), which provides additional possibilities in nano-structuring of porous semiconductors encapsulated with metallic NPs. It was recently shown that under specific fluence and writing speed, the surfaces grooves and Ag nano-gratings closed to the substrate were both generated \cite{liu2017three,liu2016selfthesis,sharmaTailoring2019,sharma2019laser}. The optical anisotropy was demonstrated as results of these subwavelength structures. It was convenient to control laser parameters for switching different structures. However, the multi-physical process consisting of Ag NPs growth, Ag$^+$ reduction, light absorption and heat accumulations were not well understood. To this end, this chapter will discuss the corresponding phenomena and propose simulation models to understand the underlying mechanisms. The performed simulations demonstrated an experimentally similar region of laser fluence and writing speed where Ag NPs grew dramatically. The proposed models paved the way for simulating Ag NPs and Ag nanograting formation by pulsed lasers in  future.

\newpage 

\section{Introduction}
Controlling nanoscale metallic structures is of great interest by research communities and in industrial applications. The laser based nano-structuring technique is emerging great potential as it is cost-efficient and able for large-scale production. Comparing with continuous-wave laser, the pulsed-laser has higher energy density and shorter time in processing that is more likely to generate nano-grooves on the material surface. For the hosting material, the presence of metallic nanoparticles (NPs) such as Ag, Al and Au NPs has two effects: on one hand, the optical response is tunable by the size-dependent plasmonic resonance; on the other, the plasmon-induced near-field enhancement decreases the laser fluence for damaging the material surface. As consequences, the NP shape is altered anisotropically by high intensity lasers \cite{stalmashonak2009intensity,voss2019situ,catone2018plasmon}. On the other hand, self-organizations of periodic grooves are more promising during laser processing. Based on this idea, several researches were done on the formation of regular gratings \cite{eurenius2008grating,yadavali2014dc}.

Recently, it was demonstrated that two kinds of gratings could be formed by the femtosecond laser irradiation of mesoporous TiO$_2$ films loaded with Ag NPs \cite{liu2017three,sharmalaserdriven2019, sharmaTailoring2019}. The surface grooves of period 490 - 500 nm (perpendicular to the laser polarization) closing to the laser wavelength (515 nm) and silver nanogratings of period 310 nm (parallel to the laser polarization) inside the TiO$_2$ film were formed at certain fluences and writing speeds. Table \ref{table:6b} summarized the experimental results. The fluence and speed dependent phenomena allowed potential applications in optical data storage and multiplexing \cite{liu2017three,sharmalaserdriven2019}.

\begin{table*}[ht!]
  \centering
  \caption{Femtosecond laser writing of TiO$_2$ doped with Ag NPs. Experimental data are taken from Ref. \cite{liu2016selfthesis,liu2017three}.}
  \begin{tabular}{lllll}
  \hline
  \hline
      Name&Fluence ($mJ/cm^{2}$)& Speed ($mm/s$) & Surface grooves & Silver Nanogratings \\
  \hline
  G1&42 - 62 & 50 -120 & 490 nm, $\perp \mathbf{E}$ & NPs within LIPSS valleys \\
  G2&46 - 54 & 5 - 40 & 500 nm, $\perp \mathbf{E}$ & 310 nm, $\parallel\mathbf{E}$ \\
  G3&37 - 46 & 5 - 30 & 500 nm, $\perp \mathbf{E}$ & homogeneous NPs (50 nm) \\
  G4&29 - 37 & 5 - 30 & None & Anisotropic NPs (<15 nm) \\
  \hline

  \hline
  \hline
      \label{table:6b}
  \end{tabular}
  \end{table*}

However, it lacks well understanding of the multi-pulsed laser processing the material, which is accompanied with light absorption, thermal diffusion, heat accumulation, Ag NP growth, and Ag$^+$ reduction. To this end, this chapter focuses on the modeling of the involved processes. Following the discussion of nonlinear propagation and absorption by the glass substrate, two models of heat accumulation are proposed and coupled with the Ag NP growth and Ag$^+$ reduction. Results are discussed and compared in the following section.

\section{Pulsed-laser propagation}
Substrate absorption and heating strongly depend on laser parameters. Particularly, a set of model modifications are required  if a pulsed laser is used instead of a CW laser. In this Chapter, attention is focused on multi-pulsed laser processing of mesoporous TiO$_2$ films with embedded Ag NPs. 

We first consider the experimental conditions reported in Ref. \cite{liu2016selfthesis,liu2017three,sharmaTailoring2019}. The fluence of the pulsed laser ranges from 20 to 62 $mJ/cm^{2}$, the spot size (at $1/e^2$ intensity criterion) is around 35 $\mu m$, the repetition rate is 500 kHz, the pulse duration is 300 fs, the wavelength in vacuum is 515 nm, and the writing speed ranges from 5 to 120 mm/s. The corresponding pulse energy varies from 0.1 to 0.3 $\mu J$ by their studies. To check whether the substrate absorption should be considered, 2D+1 (time) simulations based on the nonlinear Schr\"{o}dinger equation (Eq. \ref{eqn:Eq01}) coupled with the plasma equation (Eq. \ref{eqn:Eq03}) are performed. Different pulse energies and focusing conditions are studied. 

\begin{figure}[ht!]
 \centering
    \includegraphics[width=\textwidth]{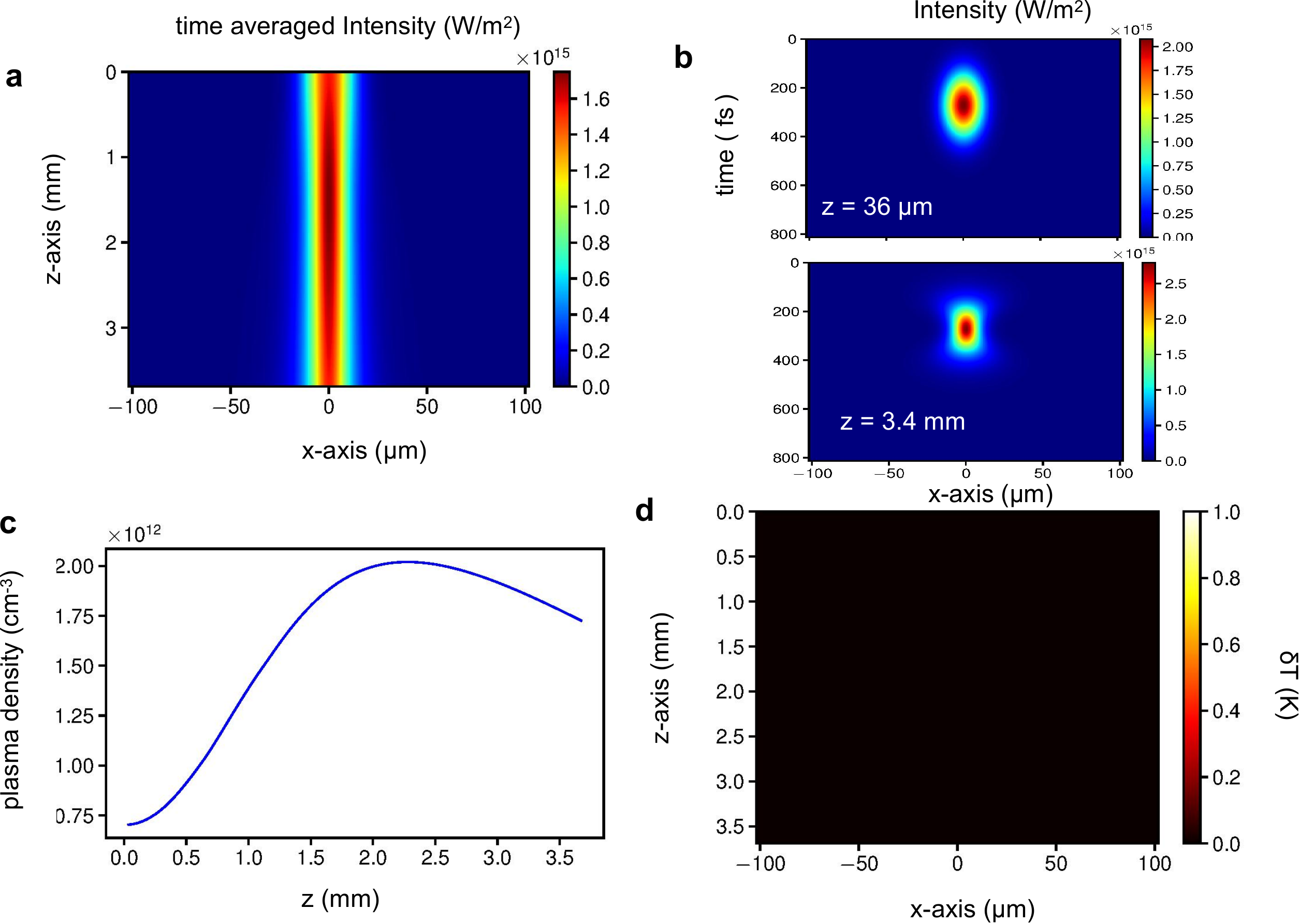}
    \caption{Simulated propagation of the time-averaged laser intensity (a), time-space distribution of laser intensity at $z=12 \mu m$ and $z = 3.4 mm$, maximum plasma density along z-axis (c), and laser induced temperature-rise (d). The incident pulse energy is 0.3 $\mu J$; and the spot size is $35 \mu m$.}
\label{fgr:chapter6_nlse_fig1}
\end{figure}

Figure \ref{fgr:chapter6_nlse_fig1} shows the evolution of the beam shape at different positions. The laser incidence is in the positive z-direction. For the conditions reported in Ref. \cite{liu2016selfthesis,liu2017three}, the beam shape changes very slowly as shown in Figure \ref{fgr:chapter6_nlse_fig1} (a-b) from z = 36 $\mu m$ to z = 3.4 $mm$. This result is in agreement with the self-focusing and filamentation conditions \cite{couairon2007femtosecond} $P_{cr} = 3.77 \lambda ^2 / 8\pi n_0 n_1$, where $n_0 = 1.52$ is the linear refractive index, and $n_1 = 3.5 \times 10^{-16} cm^2/W$ is the nonlinear refractive index. The calculated laser power approaches the critical power $P_{cr} \approx 7.5 \times 10 ^{5} W$.

The calculated time-averaged intensity is shown in Fig. \ref{fgr:chapter6_nlse_fig1} (a), which proves the self-focusing effect. The corresponding maximum plasma density increases from $0.75 \times 10^{12} cm^{-3}$ at the incident depth to around $2.0 \times 10^{12} cm^{-3}$ at z = 2.1 mm along the propagation direction before decreasing due to the self-focusing (Fig. \ref{fgr:chapter6_nlse_fig1} (c)). Nevertheless, the absorption in the glass is small in this case as shown in Fig. \ref{fgr:chapter6_nlse_fig1} (d). 


\begin{figure}[ht!]
 \centering
    \includegraphics[width=\textwidth]{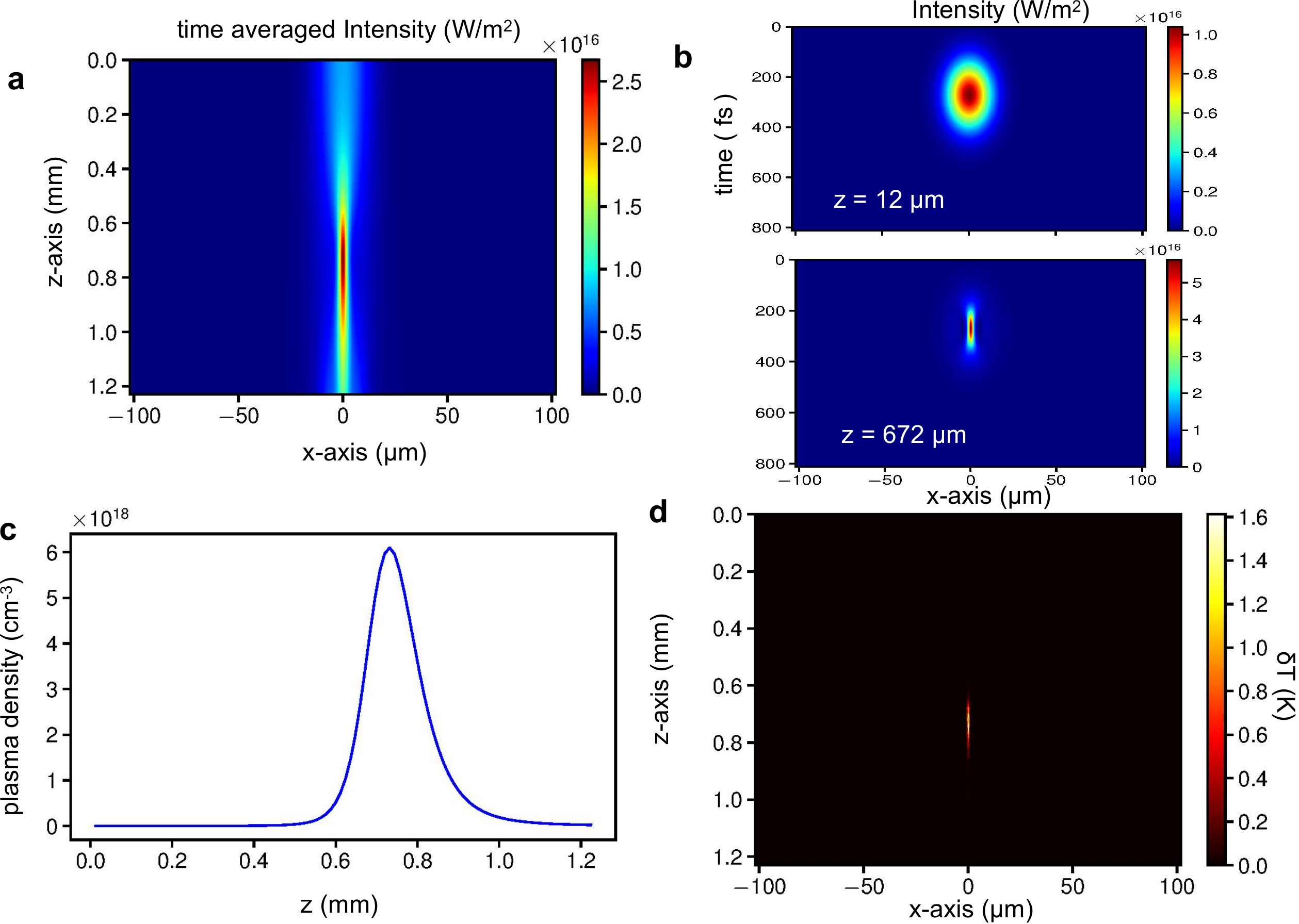}
    \caption{Simulated propagation of the time-averaged laser intensity (a), time-space distribution of laser intensity at $z = 36 \mu m$ and $z = 672 \mu m$, maximum plasma density along z-axis (c), and laser induced temperature-rise (d). The incident pulse energy is 1.5 $\mu J$; and the spot size is $35 \mu m$.}
\label{fgr:chapter6_nlse_fig2}
\end{figure}

To calculate sample temperature, two-temperature model is commonly used for ultra-short laser interaction with metals and semiconductors. Such approach is required to account for the absence of equilibrium electron-phonon/lattice/matrix during the laser pulse. Here, we are interested by much longer time scales. We consider only matrix temperature and the laser induced temperature rises in the calculations as follows:

\begin{equation}
\begin{aligned}
	\delta T &= \frac{1}{C_V} \int \sigma \mathbf{E} \cdot \mathbf{E} \\
	    & \approx \frac{1}{C_V} \int_t \frac{4\pi \cdot imag(n(x,z,t))}{\lambda} I(x,z,t) dt
	\end{aligned}
	\label{eqn:chapter6_eq_glass_abs_1}
\end{equation}

\noindent where $C_V$ is the volumetric thermal capacity; $I(z,t)$ is the laser intensity; and $n(x,z,t)$ is the refractive index. 

\begin{figure}[ht!]
 \centering
    \includegraphics[width=\textwidth]{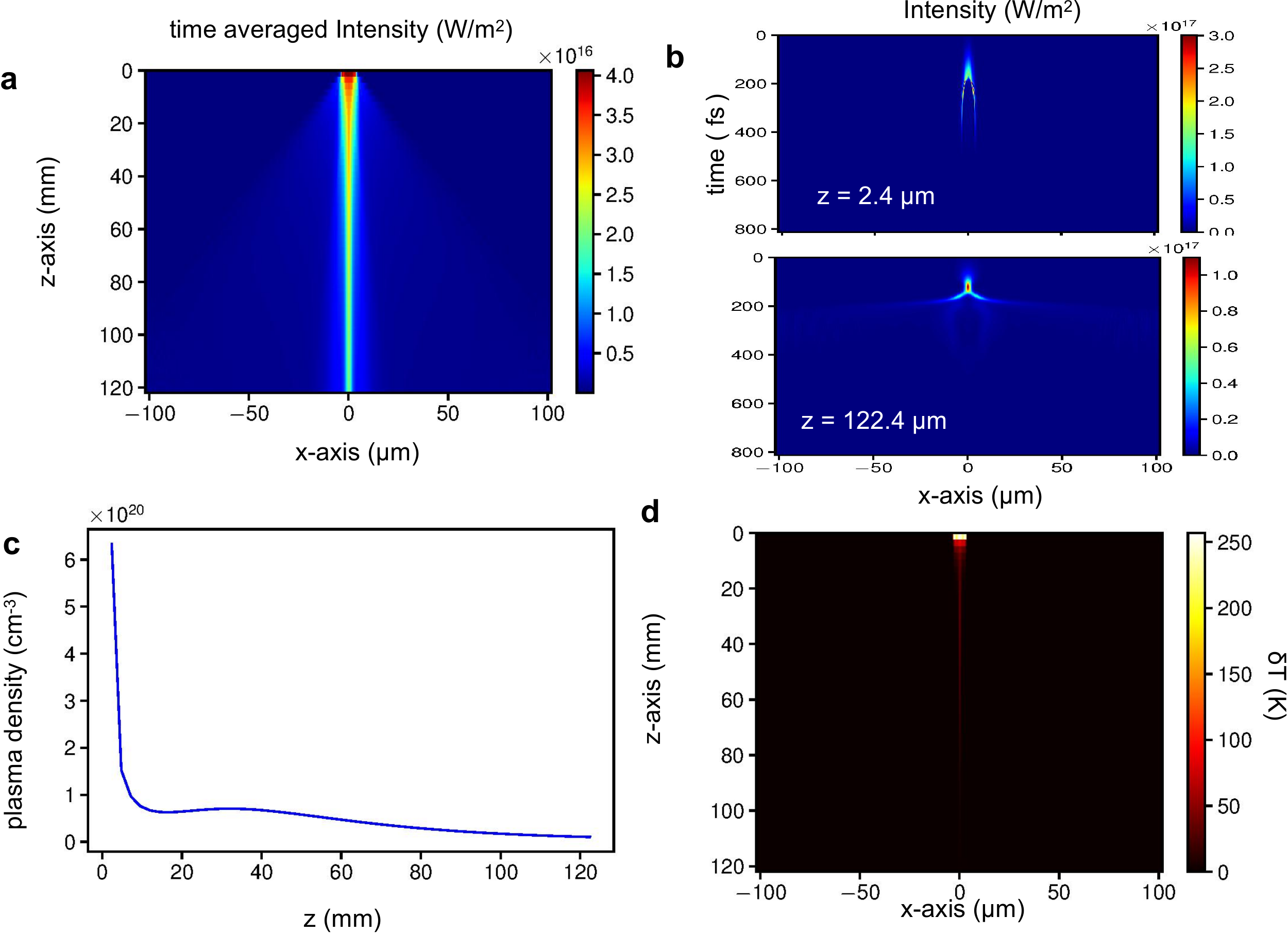}
    \caption{Simulated propagation of the time-averaged laser intensity (a), time-space distribution of laser intensity at $z = 2.4 \mu m$ and $z = 122.4 \mu m$, maximum plasma density along z-axis (c), and laser induced temperature-rise (d). The incident pulse energy is 1.5 $\mu J$; and the spot size is $6 \mu m$.}
\label{fgr:chapter6_nlse_fig3}
\end{figure}

To study the propagation of more powerful and tightly focused laser pulses inside the glass substrate, laser pulse energy of 1.5 $\mu J$, and spot size of 35 $\mu m$ are set in the simulations. The time-averaged intensity presented in Fig.\ref{fgr:chapter6_nlse_fig2} (a) shows a more tightly self-focused beam whose peak position is at around 0.76 mm after the incidence. Figure \ref{fgr:chapter6_nlse_fig2}(b) shows the temporal variations of intensities at z = $12 \mu m$ and $672 \mu m$. In this case, the beam shapes change greatly along z-axis. As a result, the maximum plasma density (Fig. \ref{fgr:chapter6_nlse_fig2} (c)) reaches $6\times 10^{18} cm^{-3}$. The laser induced temperature rise is smaller than 1.6 K as shown in Fig. \ref{fgr:chapter6_nlse_fig2} (d). 

Further focusing the beam into the spot size of 6 $\mu m$ (pulse energy of 1.5 $\mu J$) increases the calculated temperature rise. In this case, the plasma absorption and scattering are prominent, so that the beam shapes changes dramatically as shown in Fig. \ref{fgr:chapter6_nlse_fig3} (a-b). The time-averaged intensity clearly shows the scatterings in Fig. \ref{fgr:chapter6_nlse_fig3} (a). The maximum plasma promptly decreases from $6.2 \times 10 ^{20} cm^{-3}$ to $0.6 \times 10 ^{20} cm^{-3}$ in the first 20 $\mu m$ range (Fig. \ref{fgr:chapter6_nlse_fig3} (c)). The maximum temperature-rise reaches 250 K on the surface and decreases along the z-axis with a line shape whose radius is smaller than the beam radius (Fig. \ref{fgr:chapter6_nlse_fig3} (d)).

Because relatively low fluences are used in the discussed experiments, light absorption takes place inside the TiO$_2$ films, so that the absorption by the glass substrate can be ignored. Laser propagation and nonlinear effects are neglected in this study due to the lack of information about  nonlinear effects in TiO$_2$ and since rather low laser intensity is favorable for the considered applications. In addition, the near-field enhancement by the localized plasmonic resonances is also disregarded here. In the previous chapter, we have treated the film as an effective medium when considering the absorption; likewise, here, we introduce the same idea and try to simulate the growth of Ag NPs pulse-by-pulse.

\section{Heat accumulation}

The direct attempt to find the accumulated laser energy is to solve the general heat equation using the Green's function method as following:

\begin{equation}
\left\{
\begin{aligned}
    C_V \frac{\partial T}{\partial t} - \nabla \cdot k_T \nabla T &= 0\\
    T(x,y,z, t=0) &= \delta (x, y, z)
    \label{eqn:heatEQChap6_eq1}
\end{aligned}
\right.
\end{equation}

\noindent where $C_V$ is the specific heat capacity, $k_T$ is the thermal conductivity, and $\delta(x,y,z)$ is the Dirac delta function. The Green's function is found in the form \cite{beck1992heat}:
\begin{equation}
\begin{aligned}
    G_T(x,y,z,t) = \big(\frac{1}{\sqrt{4 \pi \beta_T t}}\big)^3 \exp{\big(-\frac{x^2 + y^2 + z^2}{4 \beta_T t}\big)}
    \label{eqn:heatEQChap6_eq1_2}
\end{aligned}
\end{equation}

\noindent Thus, the general solution of the heat transfer equation is given by:
\begin{equation}
\begin{aligned}
    T(x,y,z,t) = \int_{t_s} \int_{x_s,y_s,z_s} G_T(x-x_s,y-y_s,z-z_s,t-t_s) f(x_s,y_s,z_s,t_s) dt_s dx_s dy_s dz_s
    \label{eqn:heatEQChap6_eq1_3}
\end{aligned}
\end{equation}

\noindent where $T(x,y,z,t)$ is the temperature difference induced by the heat source, $\beta_T = k_T / C_V$ is the thermal diffusivity and $f(x_s,y_s,z_s,t_s)$ is the heat source. For pulsed-laser studied by Z. Liu et al. \cite{liu2017three, liu2016selfthesis}, the absorption inside the glass substrate can be ignored as discussed in the previous section. Thus, only thin film absorption is considered in this case. We remind that the laser pulse width is around 300 fs, while the growth of Ag NPs and thermal diffusion take place on nanosecond and longer time scales. For integration, the Green's function is close to the Dirac delta function. Therefore, the time integration term is written as:
\begin{equation}
\begin{aligned}
    \int_{t_s} G_T(x,y,z,t-t_s) f (x,y,z,t_s) dt_s = G_T(x,y,z,t-t_s^i) T_0(x,y,z,t=t_s^i)
    \label{eqn:heatEQChap6_eq1_4}
\end{aligned}
\end{equation}

\noindent where $t_s^i$ is time corresponding to the peak intensity of the pulse.

In addition, the initial temperature after absorbing a single pulse energy is calculated by neglecting  heat diffusion during such short time and is written as:

\begin{equation}
\begin{aligned}
    T_0(x,y,z,t=t_s) &= \frac{1}{C_V} \int_{\tau_p}\alpha_{abs}(x,y,z,t) I(x,y,z,t) dt\\
    &= \frac{1}{C_V} \int_{\tau_p} C_{NP} Q_{abs}(\omega) \pi R_{NP}^2 I(x,y,z,t) dt
    \label{eqn:heatEQChap6_eq1_4_2}
\end{aligned}
\end{equation}

\noindent where $I(x,y,z,t)$ is the laser intensity, $\tau_p$ is the pulse duration, $\alpha_{abs} = C_{NP} \pi R_{NP}^2 Q_{abs}$ is the absorption coefficient inside the film, $C_{NP}$ is the concentration of the Ag NP, $Q_{abs}(\omega)$ is the absorption efficiency factor, $\omega$ is the circular frequency of light, and $R_{NP}$ is the radius of an Ag NP.  Therefore, the Eq. \ref{eqn:heatEQChap6_eq1_3} is written as:
\begin{equation}
\begin{aligned}
    T(x,y,z,t) &= \int_{x_s,y_s,z_s} G_T(x-x_s,y-y_s,z-z_s,t-t_s^i) \\
    & \times \frac{1}{C_V}\int_{\tau_p} \alpha_{abs} I(x_s,y_s,z_s,t) dt dx_s dy_s dz_s, \\
    &(x,\: y,\: z,\: t \in \textbf{R}, \text{and} \: t > t_s^i)
    \label{eqn:heatEQChap6_eq1_5}
\end{aligned}
\end{equation}

\noindent where $t_s^i$ is the time of the incident pulse. For simplicity, laser intensity inside the TiO$_2$ thin film $I(x,y,z,t)$ can be written as:
\begin{equation}
\begin{aligned}
    I(x,y,z,t) = \frac{E_p}{\tau_p \pi w_0^2 /2} \exp{\big(-\frac{2(x^2+y^2)}{w_0^2}\big)} \exp{\big(-\frac{8(t-t_s^i)^2}{\tau_p ^2}\big)} \exp{\big(-\alpha_{abs} z \big)} 
    \label{eqn:heatEQChap6_eq1_6}
\end{aligned}
\end{equation}

\noindent where $E_p$ is the pulse energy, and $w_0$ is the beam waist (at $e^{-2}$ of the peak intensity). Recalling that $\int_{-\infty}^{\infty} \exp{-\frac{(x-x_0)^2}{2 a^2}}dx = a \sqrt{2 \pi}$, the one variable integration is as follows:
\begin{equation}
\begin{aligned}
    &\int_{x_s} G_T(x-x_s, t-t_s^i) \times T_0(x_s, t_s) dx_s\\
    &= \frac{1}{C_V} \int_{x_s} \frac{1}{\sqrt{4\pi \beta_T(t-t_s^i)}} \exp{\big(-\frac{(x-x_s)^2}{4 \beta_T (t-t_s^i)} \big)} \exp{\big(-\frac{2x_s^2}{w_0^2} \big)} dx_s\\
    &\times \alpha_{abs} I_0 \frac{\tau_p \sqrt{2 \pi}}{4}  \exp{\big(-\alpha_{abs} z \big)} \\
    &=\frac{1}{C_V} \sqrt{\frac{w_0^2}{w_0^2+8\beta_T(t-t_s^i)}} \exp{\big(- \frac{2x^2}{w_0^2+8\beta_T(t-t_s^i)}\big)} \\
    &\times \alpha_{abs} I_0 \frac{\tau_p \sqrt{2 \pi}}{4}  \exp{\big(-\alpha_{abs} z \big)}
    \label{eqn:heatEQChap6_eq1_7}
\end{aligned}
\end{equation}

\noindent where $ I_0 = \frac{E_p}{\tau_p \pi w_0^2 /2}$ is the peak intensity. It should be noted that the Eq. \ref{eqn:heatEQChap6_eq1_7} is only valid as the absorption coefficient $\alpha_{abs}$ is independent of $x$. 

\subsection{Homogeneous absorption in x and y}
If the absorption coefficient $\alpha_{abs}(x,y,z,t)$ is homogeneous in x and y, then $\alpha_{abs}(x,y,z,t) = \alpha_{abs}(z,t)$. The temperature difference induced is expressed as:
\begin{equation}
\begin{aligned}
    T^i(x,y,z,t;t_s^i) &=\frac{w_0^2}{C_V[w_0^2+8\beta_T(t-t_s^i)]} \exp{\big(- \frac{2(x^2+y^2)}{w_0^2+8\beta_T(t-t_s^i)}\big)} I_0 \frac{\tau_p \sqrt{2 \pi}}{4} \\
    & \times \int_{0}^{h_0} \frac{1}{\sqrt{4\pi \beta_T(t-t_s^i)}} \exp{\big(-\frac{(z-z_s)^2}{4 \beta_T (t-t_s^i)} \big)} \alpha_{abs}  \exp{\big(-\alpha_{abs} z_s \big)} dz_s
    \label{eqn:heatEQChap6_eq1_8}
\end{aligned}
\end{equation}

\noindent where $h_0$ is the thickness of the TiO$_2$ film, the rest of the integration along z-axis is zero because of the non-absorption in the glass substrate at the studied laser energies and focusing conditions. It should be noted that the variables $x$, $y$, and $z$ are relative to the pulse; in a global coordinate system, the transformation is $x_g = x + Vs \cdot t_s^i$, $y_g = y$, $z_g = z$. This is because the Green's function is written in such a coordinate system (see Eq. \ref{eqn:heatEQChap6_eq1} and Eq. \ref{eqn:heatEQChap6_eq1_2}). Note that this equation is valid when the substrate does not absorb laser energy. 

In the case of multi-pulse process, the temperature difference $T(x,y,z,t)$ is written as:
\begin{equation}
\begin{aligned}
    T(x,y,z,t) = \sum_i^N T^i(x - Vs \cdot t_s^i,y,z,t;t_s^i), \text{for} \: t \geq t_s^N
    \label{eqn:heatEQChap6_eq1_9}
\end{aligned}
\end{equation}

\noindent where $N$ is the total number of pulses irradiated on the sample, and the coordinate is in the global coordinate system. 

For a scanning laser beam, the laser-affected region is in the vicinity of the laser spot. We assume that laser scans in the positive x-axis direction at the speed of $V_s$. Then, we transform the Eq. \ref{eqn:heatEQChap6_eq1_8} into a moving coordinate system ($x'=x_g + V_s t, y'=y_g, z'=z_g, t'=t$). The temperature difference induced by a single pulse can be written as:
\begin{equation}
\begin{aligned}
    T_m^i(x,y,z,t;t_s^i) &=\frac{w_0^2}{C_V[w_0^2+8\beta_T(t-t_s^i)]} \exp{\big(- \frac{2[(x+V_s (t-t_s^i))^2+y^2]}{w_0^2+8\beta_T(t-t_s^i)}\big)} I_0 \frac{\tau_p \sqrt{2 \pi}}{4} \\
    & \times \int_{0}^{h_0}  \frac{1}{\sqrt{4\pi \beta_T(t-t_s^i)}} \exp{\big(-\frac{(z-z_s)^2}{4 \beta_T (t-t_s^i)} \big)} \alpha_{abs}  \exp{\big(-\alpha_{abs} z_s \big)} dz_s
    \label{eqn:heatEQChap6_eq1_10}
\end{aligned}
\end{equation}

\noindent whereas for the multi-pulse regime, one obtains the following expression
\begin{equation}
\begin{aligned}
    T_m(x,y,z,t) = \sum_i^N T_m^i(x,y,z,t;t_s^i), \text{for} \: t \geq t_s^N
    \label{eqn:heatEQChap6_eq1_11}
\end{aligned}
\end{equation}

In general,  Eq. \ref{eqn:heatEQChap6_eq1_8}, Eq. \ref{eqn:heatEQChap6_eq1_9}, Eq. \ref{eqn:heatEQChap6_eq1_10}, and Eq. \ref{eqn:heatEQChap6_eq1_11} have no analytical solutions because of the asymmetric function of $\alpha_{abs}(x,y,z,t)$, which dynamically changes due to the growth of Ag NPs while heating by laser pulses. It is noted that the deduced Eq. \ref{eqn:heatEQChap6_eq1_8}, Eq. \ref{eqn:heatEQChap6_eq1_9}, Eq. \ref{eqn:heatEQChap6_eq1_10} and Eq. \ref{eqn:heatEQChap6_eq1_11} are only valid for lasers whose pulse duration is much shorter than both the growth of Ag NPs and the heat diffusion time (normally in nanosecond scale). Otherwise, the dynamic changes of $\alpha_{abs}$ should be considered due to the size variation of Ag NPs inside the time integration term (Eq. \ref{eqn:heatEQChap6_eq1_4}) . Moreover, if the NP size changes, the obtained formulas from Eq. \ref{eqn:heatEQChap6_eq1_7} to Eq. \ref{eqn:heatEQChap6_eq1_11} are invalid. In this case, the convolution in Eq. \ref{eqn:heatEQChap6_eq1_7} has no semi-analytic solutions. Nevertheless, these equations can be used for the estimation of base temperature before growth of Ag NPs. This is correct, since the threshold temperature required for the growth of NPs is high comparing to the room temperature. As a result, multiple pulses to accumulate the base temperature so as to exceed the threshold for a dramatic growth of Ag NPs.


\subsection{Homogeneous absorption in y and z}

As in the Chapter focused on CW laser writing of Ag NPs inside the TiO$_2$ film, here the NP size changes dramatically along the laser scanning direction $x$. Furthermore, if the temperature variation along z-axis (direction of laser incidence) inside the film can be ignored, the single-pulse induced temperature difference $T^i(x,y,z,t)$ only has one integration of $x$. In this way, the resulted semi-anatylic equation can be largely simplified accelerating the temperature calculation. In this case, $\alpha_{abs}(x,y,z,t) = \alpha_{abs}(x,t)$ and the temperature difference is written as follows:

\begin{equation}
\begin{aligned}
    &T^i(x,y,z,t;t_s^i) =\\
    &\frac{1}{C_V}\sqrt{\frac{w_0^2}{[w_0^2+8\beta_T(t-t_s^i)]}} \exp{\big(- \frac{2(y^2)}{w_0^2+8\beta_T(t-t_s^i)}\big)} I_0 \frac{\tau_p \sqrt{2 \pi}}{4} \\
    & \times \int_{x_s} \alpha_{abs}(x_s,t_s^i) \frac{1}{\sqrt{4\pi \beta_T(t-t_s^i)}} \exp{\big(-\frac{(x-x_s)^2}{4 \beta_T (t-t_s^i)} \big)} \exp{\big(-\frac{2x_s^2}{w_0^2} \big)} \\
    & \times \int_{0}^{h_0} \frac{1}{\sqrt{4\pi \beta_T(t-t_s^i)}} \exp{\big(-\frac{(z-z_s)^2}{4 \beta_T (t-t_s^i)} \big)}  \exp{\big(-\alpha_{abs}(x_s,t_s^i) \cdot z_s \big)} dz_s dx_s
    \label{eqn:heatEQChap6_eq1_13}
\end{aligned}
\end{equation}

\noindent Since $\int_{x_0}^{x_1} \exp{-\frac{(x+c)^2}{2a^2}}dx = \sqrt{\frac{\pi}{2}} a \big[ \erf{\frac{x_1+c}{a\sqrt{2}} - \erf{\frac{x_0+c}{a\sqrt{2}}}} \big]$, the Eq. \ref{eqn:heatEQChap6_eq1_13} is expressed as follows:

\begin{equation}
\begin{aligned}
    &T^i(x,y,z,t;t_s^i) =\\
    &\frac{1}{C_V}\sqrt{\frac{w_0^2}{[w_0^2+8\beta_T(t-t_s^i)]}} \exp{\big(- \frac{2(y^2)}{w_0^2+8\beta_T(t-t_s^i)}\big)} I_0 \frac{\tau_p \sqrt{2 \pi}}{4} \\
    & \times \int_{x_s} \alpha_{abs}(x_s,t_s^i) \frac{1}{\sqrt{4\pi \beta_T(t-t_s^i)}} \exp{\big(-\frac{(x-x_s)^2}{4 \beta_T (t-t_s^i)} \big)} \exp{\big(-\frac{2x_s^2}{w_0^2} \big)} \\
    & \times \frac{1}{2} \bigg( \erf{\frac{h0+2\beta_T(t-t_s^i)\alpha_{abs}(x_s,t_s^i) - z}{\sqrt{4\beta_T(t-t_s^i)}}} - \erf{\frac{h_1|_{h_1 \to 0^-} + 2\beta_T(t-t_s^i)\alpha_{abs}(x_s,t_s^i) - z}{\sqrt{4\beta_T(t-t_s^i)}}} \bigg) \\
    & \times \exp{\bigg(-\alpha_{abs}(x_s,t_s^i) \big(z - \beta_T (t-t_s^i) \alpha_{abs}(x_s,t_s^i) \big) \bigg) } dx_s 
    \label{eqn:heatEQChap6_eq1_14}
\end{aligned}
\end{equation}

\noindent In most cases, we consider only the top surface $z = 0$ and the line along the laser scanning direction $y = 0$. Thus, the Eq. \ref{eqn:heatEQChap6_eq1_13} can be simplified as follows:
\begin{equation}
\begin{aligned}
    &T^i(x,t;t_s^i)|_{y=0,z=0} =\\ &\frac{1}{C_V}\sqrt{\frac{w_0^2}{[w_0^2+8\beta_T(t-t_s^i)]}}  I_0 \frac{\tau_p \sqrt{2 \pi}}{4} \\
    & \times \int_{x_s} \alpha_{abs}(x_s,t_s^i) \frac{1}{\sqrt{4\pi \beta_T(t-t_s^i)}} \exp{\big(-\frac{(x-x_s)^2}{4 \beta_T (t-t_s^i)} \big)} \exp{\big(-\frac{2x_s^2}{w_0^2} \big)} \\
    & \times \frac{1}{2} \bigg( \erf{\frac{h0+2\beta_T(t-t_s^i)\alpha_{abs}(x_s,t_s^i)}{\sqrt{4\beta_T(t-t_s^i)}}} - \erf{\frac{h_1|_{h_1 \to 0^-} + 2\beta_T(t-t_s^i)\alpha_{abs}(x_s,t_s^i)}{\sqrt{4\beta_T(t-t_s^i)}}} \bigg) \\
    & \times \exp{\bigg( \beta_T (t-t_s^i) \big(\alpha_{abs}(x_s,t_s^i) \big)^2 \bigg) } dx_s 
    \label{eqn:heatEQChap6_eq1_15}
\end{aligned}
\end{equation}

\noindent The numerical introduction of the above equation is quite tricky and normally requires that the condition $2\beta_T(t-t_s^i)\alpha_{abs}(x_s,t_s^i) < 100 h_0$ is fulfilled. Otherwise, the last exponential term $\exp{\bigg( \beta_T (t-t_s^i) \big(\alpha_{abs}(x_s,t_s^i) \big)^2 \bigg)}$ is divergent, so that the numerical errors can grow, so that the calculations can fail. This condition can be very useful in accelerating simulations. In fact, because of heat diffusion, for all pulses whose irradiation time $t_s^i$ smaller than $100h_0/(2\beta_T \alpha_{abs}(x_s,t_s^i))$ are less than 10\%. In the case of the studied Ag NPs inside TiO$_2$ films, the considered time is around 10 $\mu s$. A numerical solution of this issue should be included and will be discussed in the next section. However, this approximations can lead to an underestimated temperature. Furthermore, the Green's function should be carefully considered. For this, numerical introduction of the Dirac delta function can be of the form $\delta_{\epsilon}(x) = \frac{1}{\epsilon} \varphi(x/\epsilon)$ \cite{zahedi2010delta,engquist2005discretization,smereka2006numerical,beale2008proof,towers2007two,towers2008convergence}, where one can choose $\varphi(x) = \frac{1}{2}(1+cos(\pi x))$ for $x \leqslant 1$ and $\varphi(x) = 0 $ for $x > 1$. 
In the multi-pulses regime, the accumulated energy in the form of temperature difference is written as follows:

\begin{equation}
\begin{aligned}
    T(x,t)|_{y=0,z=h_0} = \sum_i^N T^i(x - Vs \cdot t_s^i,t;t_s^i)|_{y=0,z=h_0}, \text{for} \: t \geq t_s^N
    \label{eqn:heatEQChap6_eq1_16}
\end{aligned}
\end{equation}

\noindent where the coordinates $x$, $y$, $z$ and $t$ are in the global coordinate system.

The deduced Eq. \ref{eqn:heatEQChap6_eq1_15} and Eq. \ref{eqn:heatEQChap6_eq1_16} take into considerations spatial (along x-axis) and kinetic variations of absorption, which are the general descriptions of the temperature field. Because there is only one integration left in the calculation, the simulations can be drastically accelerated by coupling Eq. \ref{eqn:heatEQChap6_eq1_15} and Eq. \ref{eqn:heatEQChap6_eq1_16} with the NPs growth. In contrast, the three-dimensional calculations by finite element method are inefficient and time-consuming, requiring huge computational resources. However, the finite element method  (FEM) is more accurate, since it considers diffusion inside thin films. In contrast, the Eq. \ref{eqn:heatEQChap6_eq1_15} and Eq. \ref{eqn:heatEQChap6_eq1_16} assume that the film is very thin and that the heated layer can be regarded as a boundary condition.

\section{The coupled kinetic equations}

The modeling of Ag NPs formation is discussed in the previous Chapter. The coupled processes of growth, photo-oxidation and reduction are described at any position as follows \cite{liu2015understanding,liu2016selfthesis}:

\begin{equation}
\left\{
\begin{aligned} 
    &\frac{d R_{NP}}{d t} = [n_{abs}(t) - n_{oxi}(t)] \cdot  \frac{\omega_{Ag}}{4 \pi R_{NP}^2}\\
    &\frac{d C_{Ag^+}}{d t} =n_{oxi}(t) [ C_{NP}(t) + \frac{R_{NP}(t)}{3} \frac{d C_{NP}} {d R_{NP}} ] - \frac{d C_{Ag^+}}{d t} |_{red}\\
    &\frac{d C_{Ag^0}^{BS}}{d t} = - n_{abs}(t) [ C_{NP}(t) + \frac{R_{NP}(t)}{3} \frac{d C_{NP}} {d R_{NP}} ] + \frac{d C_{Ag^+}}{d t} |_{red}\\
 \end{aligned}
\right.
	\label{eqn:heatChap6_sec_couple_eq1}
\end{equation}

\noindent where $n_{oxi}$ is the number of silver atoms leaving a nanoparticle per unit time due to photo-oxidation, $\frac{d C_{Ag^+}}{d t} |_{red}$ is the decrease of Ag$^+$ concentration due to the reduction per unit time, and $n_{abs}$ is the total amount of monomers absorbed by the Ag NP per unit time, which are described in chapter 5 and in Ref. \cite{liu2015understanding, liu2016selfthesis, liu2016laser}. These variables can be written as:

\begin{equation}
\left\{
  \begin{aligned} 
    &n_{abs}(t) = 4 \pi R_{NP} D_{0Ag^0}\exp \Big( -\frac{E_{Ag^0}}{N_A k_B T} \Big) [C_{Ag^0}^{BS}- S_{Ag^0} ( 1 + \frac{2 \gamma \omega_{Ag}}{R_{NP} k_B T} )]\\
    &\frac{d C_{Ag^+}}{d t} |_{red} = \exp \Big( -\frac{E_p}{N_A k_B T} \Big) \cdot D^{red}_{0} \exp \Big( -\frac{E_D}{N_A k_B T} \Big) \cdot C_{red} C^{2/3}_{Ag^+}\\
    &n_{oxi}(t) =\eta_0 \frac{ \sigma_{abs}(R,\lambda) I(x, y, z, t)} {h \nu}
  \end{aligned}
\right.
	\label{eqn:heatChap6_sec_couple_eq2}
\end{equation}

The above coupled equations are in the simplified forms since the processes of Ag$^+$, Ag$^0$, and Ag NPs diffusion are neglected. In addition, the diversity in NP's size is not considered. The photo-oxidation process is regarded to be temperature independent \cite{liu2015understanding,liu2016selfthesis}. The parameters of the coupled equations are summarised in Table \ref{table:6a}.

\newpage

\begin{table*}[ht!]
  \centering
  \caption{The summary of parameters}
  \begin{tabular}{ll}
  \hline
  \hline
      Parameter & Description \\
  \hline
  & \\
  Ag NP growth, oxidation, and Ag$^+$ reduction&\\
  \hline
  $\eta _0$ & ionization efficiency\\
  $\omega_{Ag}$ & atomic volume of crystallized Ag\\
  $E_{Ag^0} $ & activation energy of Ag$^0$ diffusion\\
  $D_{0Ag^0}(500K)=D_{0Ag^0}\exp \Big( -\frac{E_{Ag^0}}{N_A k_B 500K} \Big)$ & Ag$^0$ diffusion coefficient at 500K\\
  $E_p$ & activation energy of reduction \\
  $C_{red} $ & concentration of reducing agent (RA) \\
  $E_D $ & activation energy of RA diffusion\\
  $D_0^{red}(500K) = D^{red}_{0} \exp \Big( -\frac{E_D}{N_A k_B 500K} \Big)$ & RA diffusion coefficient at 500K \\
  $S_{Ag^0} $ & solubility of Ag$^0$ \\
  $\gamma$ &Ag NP - TiO$_2$ interfacial tension \\
  $k_B$ & the Boltzmann constant\\
  $N_A$ & the Avogadro constant\\
  $T$ & temperature  \\
  $R_{NP}$ & radius of Ag NP \\
  
  $C_{Ag^0}^{BS}(t=0 s)$ & initial Ag$^0$ concentration in matrix \\
  $C_{Ag^+}(t=0 s)$ & initial Ag$^+$ concentration in matrix \\

  & \\
  Laser pulse and thermal diffusion&\\
  \hline
  $E_p$ & pulse energy \\
  $\tau_p$ & pulse duration \\
  $k_r$ & repetition rate\\
  $w_0$ & waist radius (at $e^{-2}$)\\
  $C_V$ & volumetric heat capacity \\
  $k_T$ & thermal conductivity\\
  $\beta_T$ & thermal diffusivity\\
  $t_s^i$ & time at peak intensity of the $i^{\text{th}}$ pulse\\
  $\alpha_{abs}(x_s,y_s,z_s,t_s^i)$ & absorption coefficient\\
  $x_s,y_s,z_s$ & the coordinates relative to the $i^{\text{th}}$ pulse\\
  $h_0 $ & thickness of TiO$_2$ film\\
  
  &\\
  
  \hline
  \hline
      \label{table:6a}
  \end{tabular}
  \end{table*}

\newpage
\subsection{Coupling algorithm for the semi-analytic function}
\begin{figure}[ht!]
 \centering
    \includegraphics[width=0.72\textwidth]{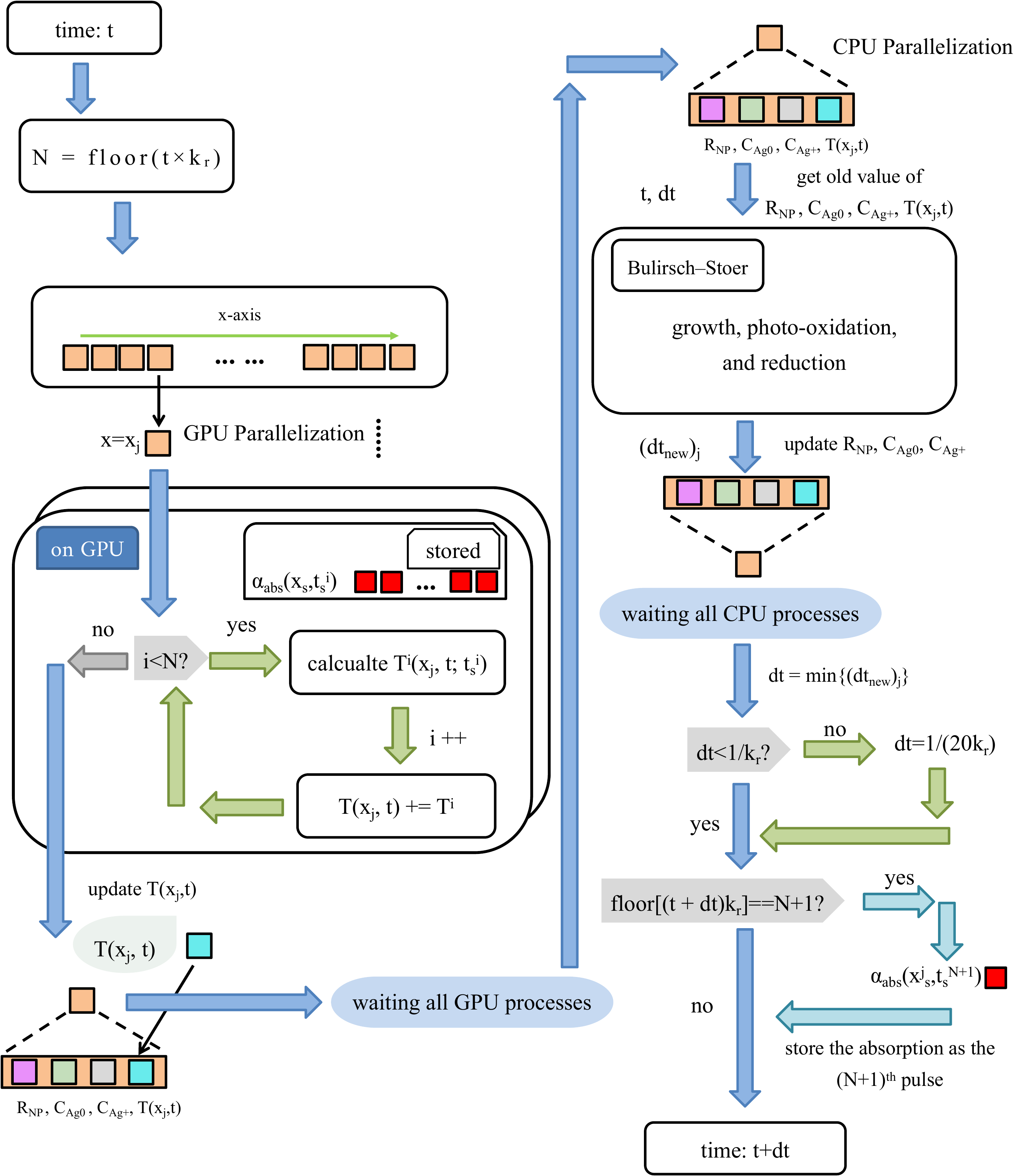}
    \caption{Schematics of the developed semi-analytic model. The temperature are calculated on the GPU and by pulse-by-pulse accumulations, which is then coupled with the Ag NP's growth, photo-oxidation, and reduction.}
\label{fgr:chapter6_coupledAna_1}
\end{figure}

Because of the simplified form of Eqs. (\ref{eqn:heatEQChap6_eq1_15} - \ref{eqn:heatEQChap6_eq1_16}), the calculation of temperature field can be paralleled and become scalable for GPU. Nowadays, GPU  has more than a few hundreds of cores, and is cheaper than CPU on a personal computer. To take advantages of such parallel schemes for accelerating the multi-scale problem, the developed model is shown in Fig. \ref{fgr:chapter6_coupledAna_1}. The simulations of Ag NPs growth, photo-oxidation, reduction, and heating are performed in 1D, which stands for the top surface of the TiO$_2$ thin film. The glass thermal diffusivity $\beta_T \approx 3.4 \times 10 ^{-7} m^{2} s^{-1}$ is considered in Eqs. (\ref{eqn:heatEQChap6_eq1_15} - \ref{eqn:heatEQChap6_eq1_16}). The Bulirsch-Stoer algorithm \cite{press2007section} is used for simulating the variations of Ag$^0$ atoms, Ag$^+$ ions, and Ag NPs inside the TiO$_2$ film. The discretization is of the second accuracy and can largely accelerate the simulation by estimating the time steps. 

\subsection{Coupling algorithm for the finite element method}

\begin{figure}[ht!]
 \centering
    \includegraphics[width=0.8\textwidth]{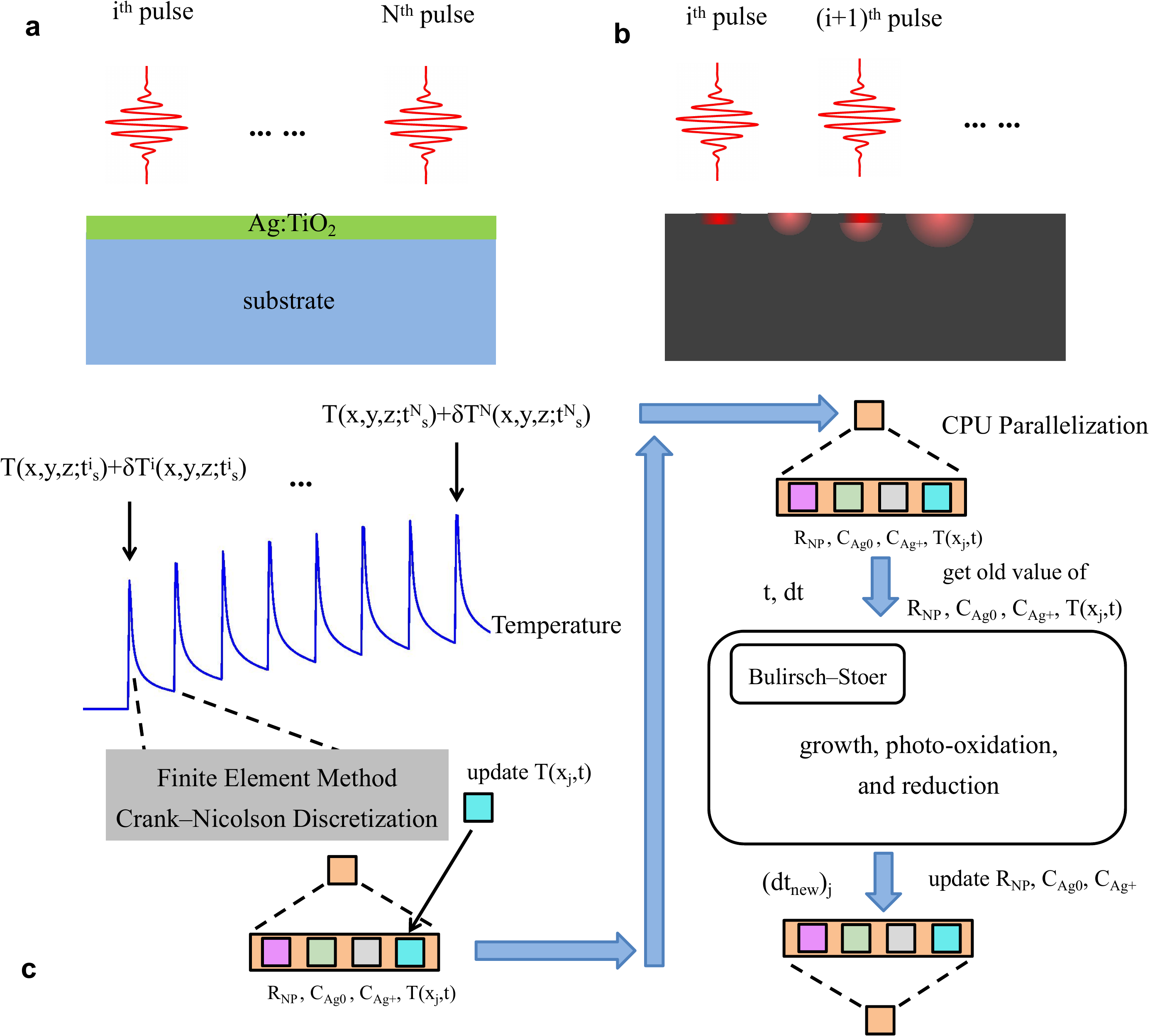}
    \caption{Schematics of the developed FEM-based model. }
\label{fgr:chap6_coupledFEM_1}
\end{figure}

The the implementation of coupling is direct in a finite element method (FEM). For an ultrafast pulse whose duration is much shorter the heat diffusion, the laser induced temperature difference, as given by Eq. \ref{eqn:heatEQChap6_eq1_4_2}. The heating by each pulse and the cooling between two pulses can be described as follows:

\begin{equation}
\left\{
\begin{aligned}
    & C_V \frac{\partial T}{\partial t} - \nabla \cdot k_T \nabla T = 0\\
    & \delta T^i(x,y,z;t_s^i) = \frac{1}{C_V} \int_{\tau_p} \alpha_{abs}(x,y,z,t_s^i) I(x,y,z,t) dt, \: i \in [1, N]
    \label{eqn:heatEQChap6_femCoupled_eq1}
\end{aligned}
\right.
\end{equation}

\noindent where, $N$ is the number of incident pulses. The Crank-Nicolson discreatization \cite{kays1967convective} is used for the heat transfer equations. The developed FEM model is depicted in Fig. \ref{fgr:chap6_coupledFEM_1} (a-c). In contrast to the semi-analytic model, the thermal properties of 200 nm TiO$_2$ film are considered in the FEM.

\noindent For simplicity, the specific heat capacity of anatase TiO$_2$ is taken from Ref. \cite{saeedian2012specific}, thermal conductivity is from Ref. \cite{batmunkh2014thermal}, the effective volumetric heat capacity and effective thermal conductivity of the mesoporous TiO$_2$ films are calculated by $C_{V_{eff}} = C_{V_{TiO_2}} (1-v_p) + C_{V_{air}} v_p $, and $k_{T_{eff}} = k_{T_{TiO_2}} (1-v_p) + k_{T_{air}} v_p$ \cite{smith-01115971}, where $v_p$ is the volume fraction of pores. The porosity is considered from 45\% to 60\% as reported in Ref. \cite{sharma2019crystal}. The calculated effective specific heat capacity is $C_{V_{eff}} \approx 1.008 \times 10 ^6 J/m^3/K$, and  effective thermal conductivity is $k_{T_{eff}} \approx 0.2 W/m/K$. For the glass, we consider a bulk material with specific heat of $C_V \approx 2.352 \times 10 ^6 J/m^3/K$, and thermal conductivity from $k_T \approx 0.5 W/m/K$ to $k_T \approx 1.8 W/m/K$.

\section{Results and Discussions}
This section deals with the results of the proposed models. Then, comparisons with the experiments are also discussed. In the first part, we present the base temperature by assuming that the Ag NPs do not grow during few first pulses. The results are compared with more rigorous calculations performed by using the finite element method model. In the second subsection, we report the semi-analytic results after laser scans 100 $\mu m$ length in the x-direction. Furthermore, the results of the FEM model are shown in the following subsection. The differences between the two models are compared and explained. Finally, the comparisons of the results obtained in simulations and experiments are presented.

\subsection{Base temperature before NP's growth}
In the previous Chapter, we have discussed the continuous-wave laser interactions of Ag NPs inside TiO$_2$ thin films. It was found that the Ag NPs grow very slowly in the beginning before the so-called "dramatic growth" takes place. This effect arises as the NP's size exceeds a certain value. In the region before the dramatic growth, one can estimate the temperature rise by using a homogeneous size assumption. In the case of pulsed laser writing of Ag NPs inside TiO$_2$ films, this can be done by looking at the base temperature and checking whether the Ag NPs can grow dramatically or not under the particular laser-irradiation conditions. In this case, the semi-analytic formula (Eq. \ref{eqn:heatEQChap6_eq1_10}) has an analytic solution as follows:

\begin{equation}
\begin{aligned}
    &T_m^i(x,y,z,t;t_s^i) =\\
    &\frac{w_0^2}{C_V[w_0^2+8\beta_T(t-t_s^i)]} \exp{\big(- \frac{2[(x+V_s (t-t_s^i))^2+y^2]}{w_0^2+8\beta_T(t-t_s^i)}\big)} \alpha_{abs} I_0 \frac{\tau_p \sqrt{2 \pi}}{4} \\
    & \times \frac{1}{2} \bigg( \erf{\frac{h0+2\beta_T(t-t_s^i)\alpha_{abs}}{\sqrt{4\beta_T(t-t_s^i)}}} - \erf{\frac{h_1|_{h_1 \to 0^-} + 2\beta_T(t-t_s^i)\alpha_{abs}}{\sqrt{4\beta_T(t-t_s^i)}}} \bigg) \\
    & \times \exp{\bigg( \beta_T (t-t_s^i) \big(\alpha_{abs} \big)^2 \bigg) }
    \label{eqn:heatEQChap6_eq2_21}
\end{aligned}
\end{equation}

\begin{figure}[ht!]
 \centering
    \includegraphics[width=0.9\textwidth]{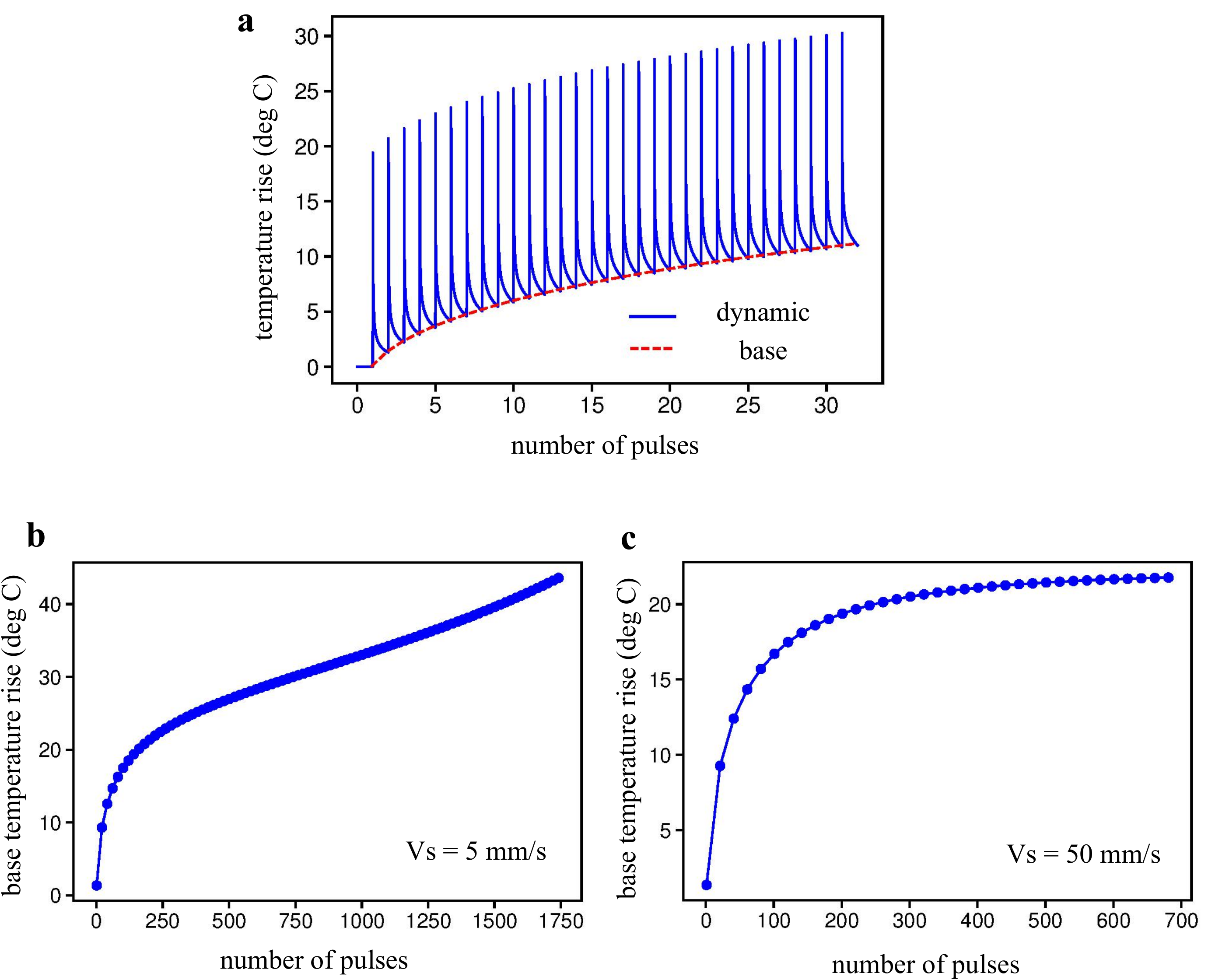}
    \caption {Temperature-rise pulse-by-pulse (a). The dashed red line stands for the base-temperature. The writing speed has little influence on the temperature profiles (for the first 30 pulses) in the range of 5 mm/s to 50 mm/s. Base temperature rise vs number of pulses for $V_s = 5$ mm/s (b) and $V_s = 50$ mm/s. The position is taken as x=0, y=0, and z=0.}
\label{fgr:chap6_alpha_homo_fig1}
\end{figure}

\noindent where $\alpha_{abs} = 5.0 \times 10^4 m^{-1}$ as we take the radius $R_{NP} = 1.5 nm$ for the initial Ag NPs \cite{liu2016selfthesis,liu2015understanding, liu2017haadf}. The cross-section of Ag absorption is calculated by the Mie theory at wavelength 515 nm, which corresponds to the experimental results obtained by using a femtosecond laser system with fundamental wavelength of 1030 nm, whose frequency is then doubled by a BBO crystal \cite{liu2017three,liu2016selfthesis, sharmaTailoring2019,sharmalaserdriven2019}. The laser spot size is $2w_0 = 35 \mu m$, the pulse duration is $\tau_p = 300 fs$, the laser energy is $E_p = 0.3 \mu J$, and the repetition rate $k_r$ is 500 kHz.

The analytical equation (Eq. \ref{eqn:heatEQChap6_eq2_21}) clearly demonstrates the pulse-by-pulse accumulation (see Fig. \ref{fgr:chap6_alpha_homo_fig1} (a)). For each pulse, the temperature-rise at laser center is around $19.3 ^{\circ} C$ and that decades quickly before the next pulse comes in. Because of the high repetition rate, the accumulation is observed looking at the dashed red line in Fig. \ref{fgr:chap6_alpha_homo_fig1} (a). To find out the influences of the laser writing speed, the base-temperature rises at $Vs = 5$ mm/s (Fig. \ref{fgr:chap6_alpha_homo_fig1} (b)) and $Vs = 50$ mm/s (Fig. \ref{fgr:chap6_alpha_homo_fig1} (c)) are calculated. In the case of $Vs = 5$ mm/s, the base-temperature rise increases quickly to 20 $^{\circ} C$ by the first 200 pulses and gradually to 40 $^{\circ} C$ after 1600 pulses. In comparison, for the writing speed of $Vs = 50$ mm/s, the temperature tends to saturate after 300 pulses. This result can be understood by the effective number of pulses ($N_{eff} = w_0 k_r / Vs$ ) inside the laser spot. The corresponding effective number of pulses are 1750 and 175 for $Vs = 5$ mm/s and $Vs = 50$ mm/s, respectively. For the writing speed of $Vs = 50$ mm/s, it is then clear that pulses whose number larger than 175 has gradually less impact on the accumulation because of the increasing distances for the heat to diffuse. As for $Vs = 5$ mm/s, this effect is relatively insignificant. Intuitively, if the writing speed is large enough that the diffusion can not catch up with, the contributions in accumulation by the effective number of pulses should decrease quickly. Here, set a limit at the speed of around 50 mm/s, which are the typical experimental values \cite{liu2017three, liu2016selfthesis, sharmaTailoring2019, sharmalaserdriven2019}.

\begin{figure}[ht!]
 \centering
    \includegraphics[width=\textwidth]{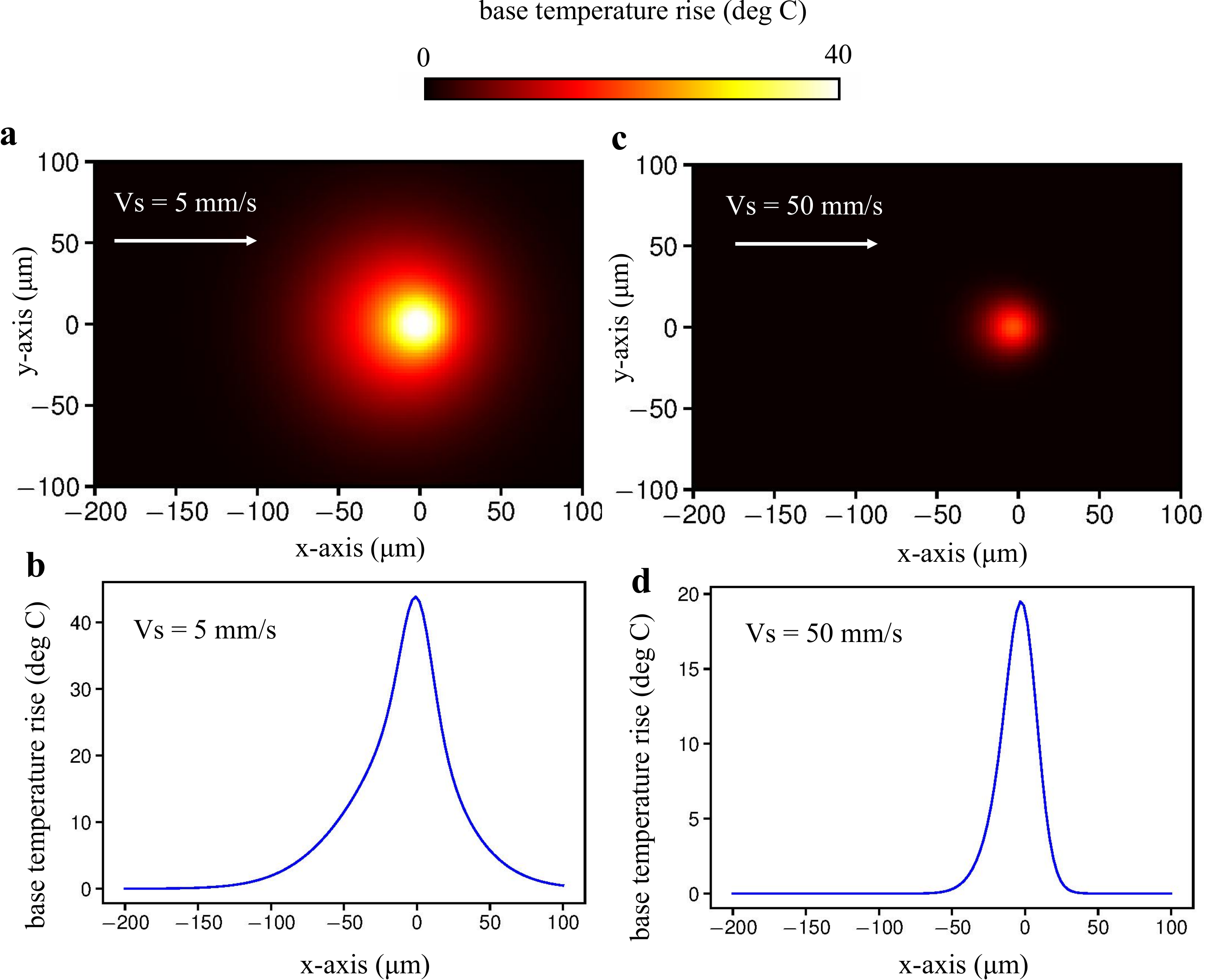}
    \caption{Base-temperature rise obtained by using the effective number of pulses $N_{eff} = 1750$ (a) and (b), and $N_{eff} = 175$ (c) and (d).  2D maps (a) and (c) stand for the top surface of TiO$_2$ films. Fig. (b) and (d) are temperature profiles along the line at y=0 on the top surface.}
\label{fgr:chap6_alpha_homo_fig2}
\end{figure}

The spatial distributions of base-temperature rise at two different writing speeds are shown in Fig. \ref{fgr:chap6_alpha_homo_fig2}. Because the Eq. \ref{eqn:heatEQChap6_eq2_21} is deduced from the 3D heat equation, the solution is accurate without ignoring the diffusion in the y-axis (as we did in the 2D finite-element-model). To show this, the 2D maps of base-temperature rise at $Vs = 5$ mm/s and $Vs = 50$ mm/s are calculated and plotted in Fig. \ref{fgr:chap6_alpha_homo_fig2} (a) and (c), respectively. The circle-shapes by the two figures clearly show the y-axis diffusion. The maximum temperature is the same as Fig. \ref{fgr:chap6_alpha_homo_fig1} due to the same equation. The temperature shapes become comma-alike as the writhing speed increases, which are the results of diffusion effects. The diffusion lags can be seen from the Fig. \ref{fgr:chap6_alpha_homo_fig2} (b). In the case of Fig. \ref{fgr:chap6_alpha_homo_fig2} (d), the temperature difference is small and only small number of pulses are calculated, so that the lag is not clearly seen. However, this can be revealed by performing the simulation for more number of pulses.

\begin{figure}[ht!]
 \centering
    \includegraphics[width=0.82\textwidth]{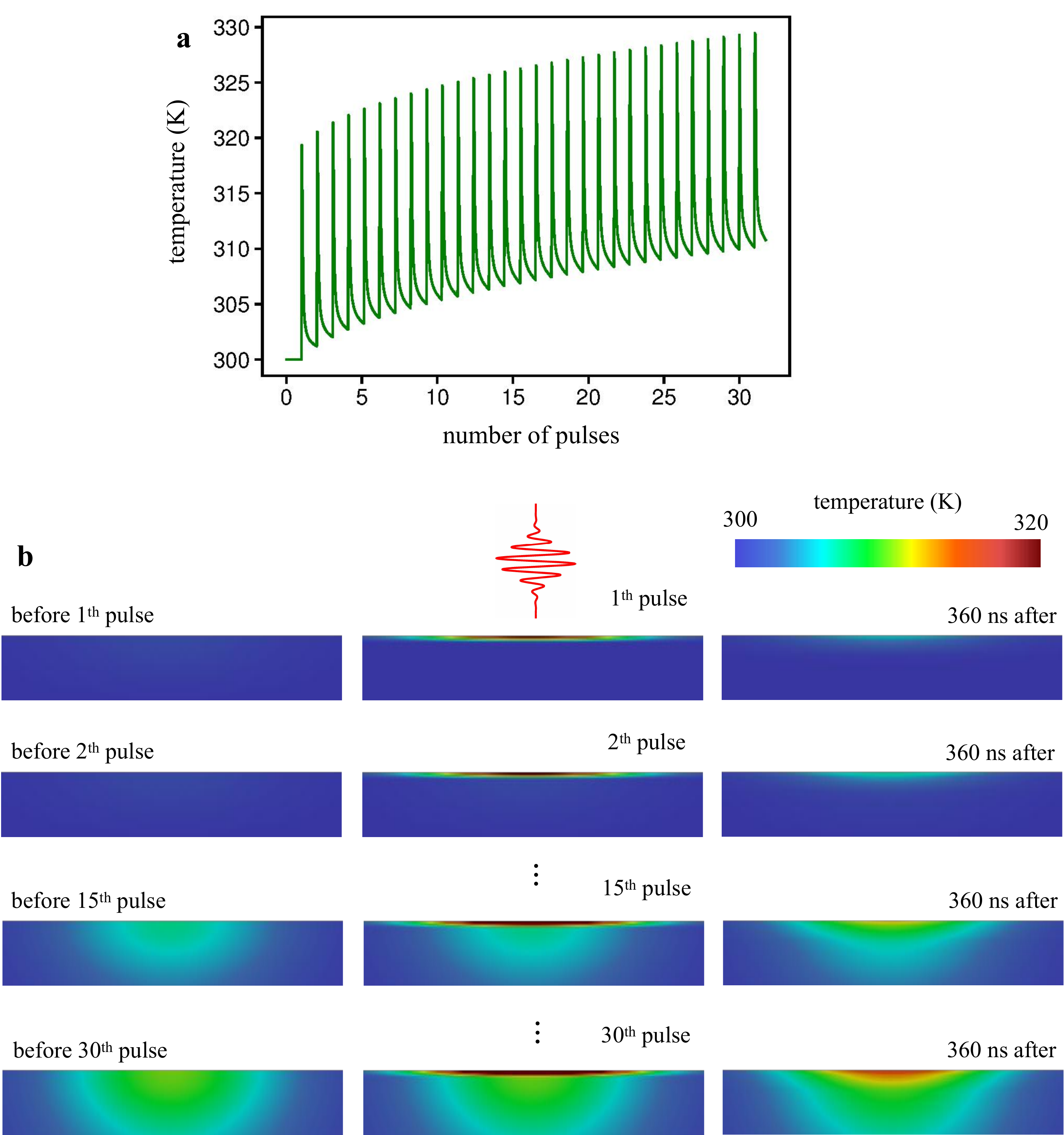}
    \caption{Results of the finite element method model for the first 30 pulses. The temporal temperature at laser center (a). The spatial diffusion shown in pulse-by-pulse (b).}
\label{fgr:chap6_fem_fewPulses_fig1}
\end{figure}

To check the validity of the homogeneous assumption, simulations are performed based on the FEM model. The results by the first 30 pulses ($Vs = 5$ mm/s) are shown in Fig. \ref{fgr:chap6_fem_fewPulses_fig1}. The temporal temperature shown in Fig. \ref{fgr:chap6_fem_fewPulses_fig1} (a) is identical to the Fig. \ref{fgr:chap6_alpha_homo_fig1} (a). Figure \ref{fgr:chap6_fem_fewPulses_fig1} (b) shows the process of pulse absorption in the TiO$_2$ film and heat diffusion in x- and z-axis. The accumulation effects are clearly depicted. After each pulse, the absorbed laser energy is confined inside the TiO$_2$ layer, which is, however, quickly damped by diffusing the energy into the media in the vicinity. Since the thermal diffusivity of glass is larger than the mesoporous TiO$_2$, the heat diffusion (into the glass substrate) is faster along the laser incident direction than the horizontal writing directions (see Fig. \ref{fgr:chap6_fem_fewPulses_fig1} (b)). This is neglected in the semi-analytic model. Within the first 30 pulses, the laser only moves 0.3 $\mu m$ for $Vs = 5$ mm/s. Though the temperature damped dramatically before the next pulse comes in, the residual temperature is accumulated pulse-by-pulse that can in some how greatly affect the growing speed of Ag NPs as soon as the base exceeds a certain value. However, this is not observed in the first 30 pulses. It is noted that the critical temperature depends on the diffusion properties of Ag$^0$ as shown in the previous chapter. For the case after hundreds of pulses irradiation, it will be shown in the next discussions that the homogeneous absorption is inaccurate. Therefore, the coupled models are required to understand the writing-speed influences on the pulsed-laser controlling Ag NPs growth inside TiO$_2$ films.

\subsection{Results of coupled semi-analytic equations}
The diffusion in z-axis (last erf and exponential terms in Eq. \ref{eqn:heatEQChap6_eq1_15}) is significant in the beginning after each laser pulse. The z-diffusion term is written as:
\begin{equation}
\begin{aligned}
&\varphi_z(t,t_s^i) = \\
&\frac{1}{2} \bigg( \erf{\frac{h0+2\beta_T(t-t_s^i)\alpha_{abs}(x_s,t_s^i)}{\sqrt{4\beta_T(t-t_s^i)}}} - \erf{\frac{h_1|_{h_1 \to 0^-} + 2\beta_T(t-t_s^i)\alpha_{abs}(x_s,t_s^i)}{\sqrt{4\beta_T(t-t_s^i)}}} \bigg) \\
    & \times \exp{\bigg( \beta_T (t-t_s^i) \big(\alpha_{abs}(x_s,t_s^i) \big)^2 \bigg) } 
    \label{eqn:heatEQChap6_eq1_22}
\end{aligned}
\end{equation}

\begin{figure}[ht!]
 \centering
    \includegraphics[width=0.6\textwidth]{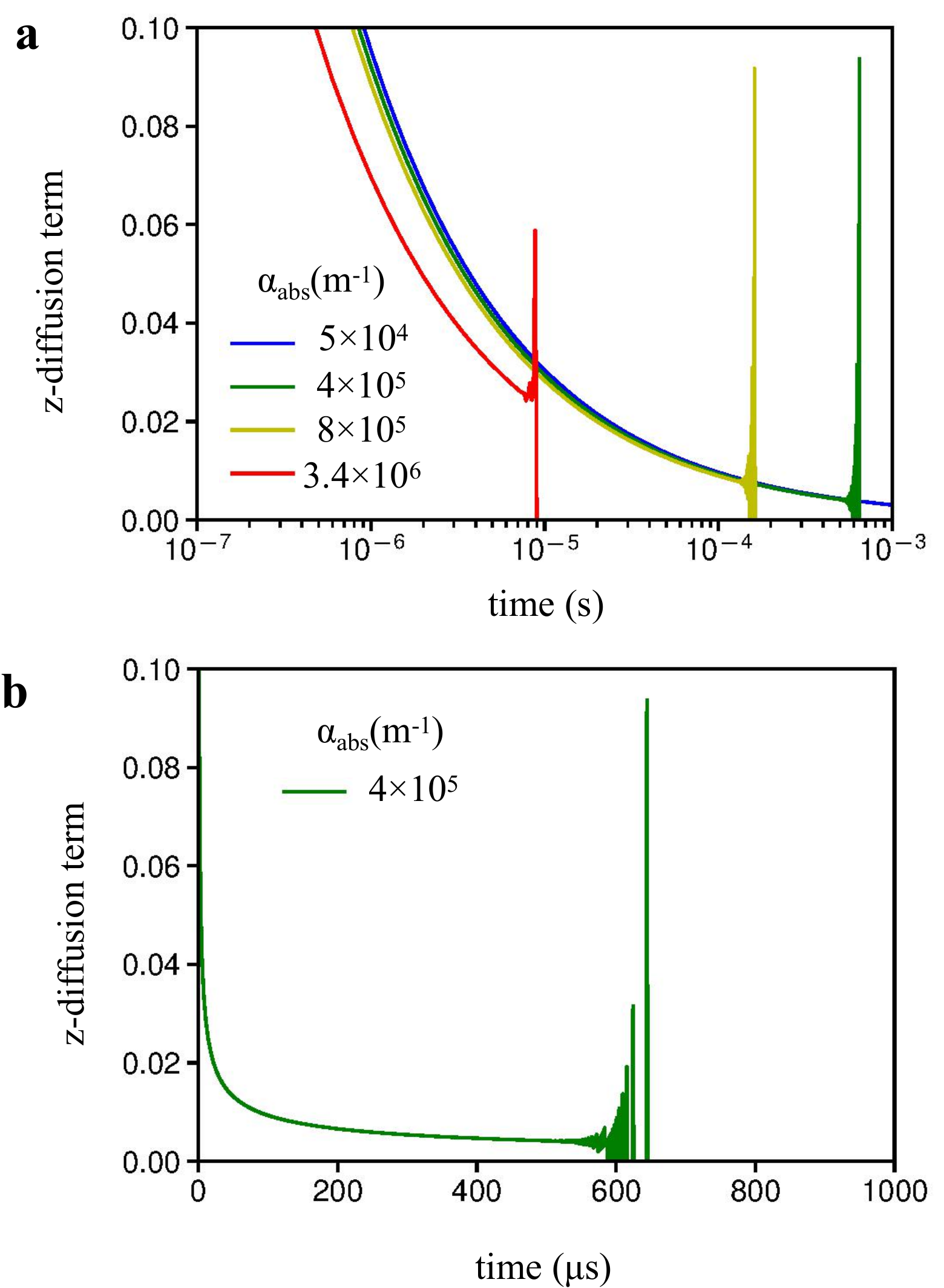}
    \caption{Illustrations of numerical errors of the z-diffusion part (Eq. \ref{eqn:heatEQChap6_eq1_22}).}
\label{fgr:chap6_coupledFit_fig1}
\end{figure}

In simulations of the heat accumulation by multiple pulses, however,  the z-diffusion term is inefficient at large number of pulses. Because heat diffusion becomes much slower after a certain time, we try to fit the diffusion function by using a very simple one. Another reason why we introduce this approximation is that the numerical errors become significant after a certain time, as shown in Fig. \ref{fgr:chap6_coupledFit_fig1}(a) and (b). It is found that an abrupt changes in the z-diffusion curves depends on the absorption coefficient. In addition, we found that the breakdown does not disappear even if one uses smaller time steps. This turns out be due to the numerical errors in multiplication of a divergent function with a infinitesimal. To accelerate the parallel calculation on a GPU, the for-loop should be avoided. In addition, we found that if the $t_s^i$ is smaller than $100h_0/(2\beta_T \alpha_{abs}(x_s,t_s^i)) \approx 10 \mu s$, the z diffusion part tends to be very small. Here, we use $\beta_T = 3.4 \times 10 ^{-7} m^{2} s^{-1}$ for a glass substrate, the maximum absorption coefficient $\alpha_{abs} \approx 3.4 \times 10 ^{6} m^{-1}$, and the film thickness is 200 nm. Figure \ref{fgr:chap6_coupledFit_fig1} shows that the diffusion parts become identical and asymptotically approaching to zero, which are less than 4\% after 20 $\mu s$ for all the studied absorption cases. However, the small portion can become significant in the accumulation regime, so that it cannot be neglected in the multi-pulses regime. In order to reduce the calculation of exponentials on a GPU (which is quite inefficient), we fit the z-diffusion part after $20 \mu s$ for the studied system. The resulted equation can be written as follows:

\begin{equation}
\varphi_z(t) = m/(t)^n
    \label{eqn:heatEQChap6_eq1_17_fit}
\end{equation}

\noindent where $m = 96.3\times 10^{-6} (s^{1/2})$, $n = 1/2$, and $t = t_g - t_s^i$, $t_g$ is the global time.

\begin{figure}[ht!]
 \centering
    \includegraphics[width=0.5\textwidth]{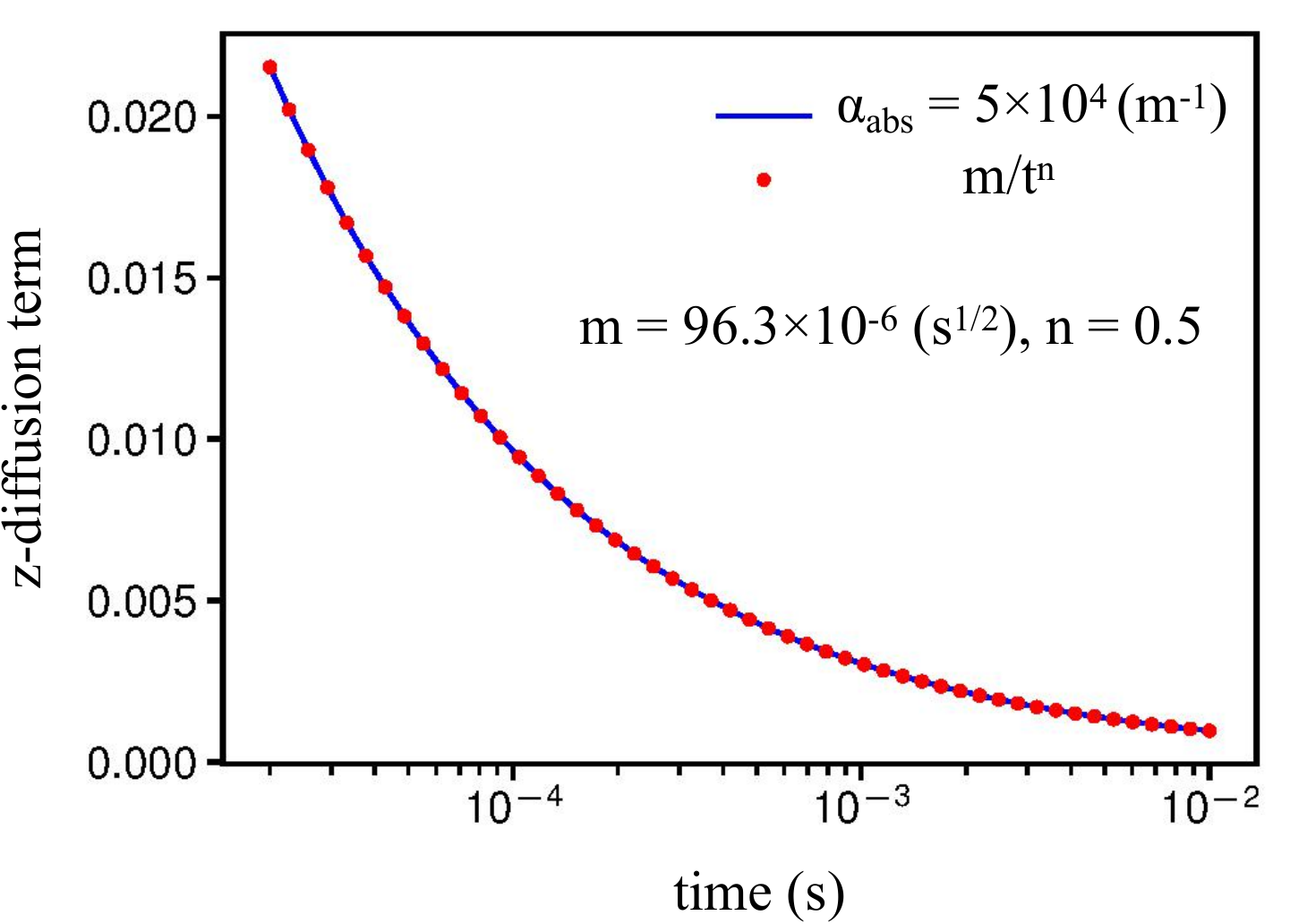}
    \caption{Calculated fitting function of the z-diffusion part.}
\label{fgr:chap6_coupledFit_fig2}
\end{figure}

\noindent Figure \ref{fgr:chap6_coupledFit_fig2} shows the comparisons of the fitting to the Eq. \ref{eqn:heatEQChap6_eq1_22}. The good fitting is in the range from 20 $\mu s$ to 10 $ms$ indicating that we can accelerate the simulation by replacing exponential calculations by such a simple function. Furthermore, the numerical error is avoided in multiplying infinitesimal with a divergent quantity.

To further decrease the simulation time for this multi-scale problem, unlike the procedure used in the step marching method, long loops should be simplified in Eq. \ref{eqn:heatEQChap6_eq1_16}. To do so, the accounted number of pulses is defined as $N_c$, which is the nearest to $N_c$ pulses at time $t$. In this case, the temperature contribution by the pulses before the time $t - Nc/k_r$ is ignored. The z-diffusion part is, as illustrated in Fig. \ref{fgr:chap6_coupledFit_fig1} (a) and Fig. \ref{fgr:chap6_coupledFit_fig2}, less than 4\% after $3 \mu s$ and 1\% after $100 \mu s$. Thus, a typical choice of the accounted number of pulses can be $N_c = 100$ and the corresponding time of the heat transportation is $200 \mu s$. The temperature error is estimated to be less than 1\% of each pulses. The numerical solution of the coupled semi-analytic equations is described in this way. The photo-oxidation is not considered because of the femtosecond time scale. 

\begin{figure}[ht!]
 \centering
    \includegraphics[width=0.85\textwidth]{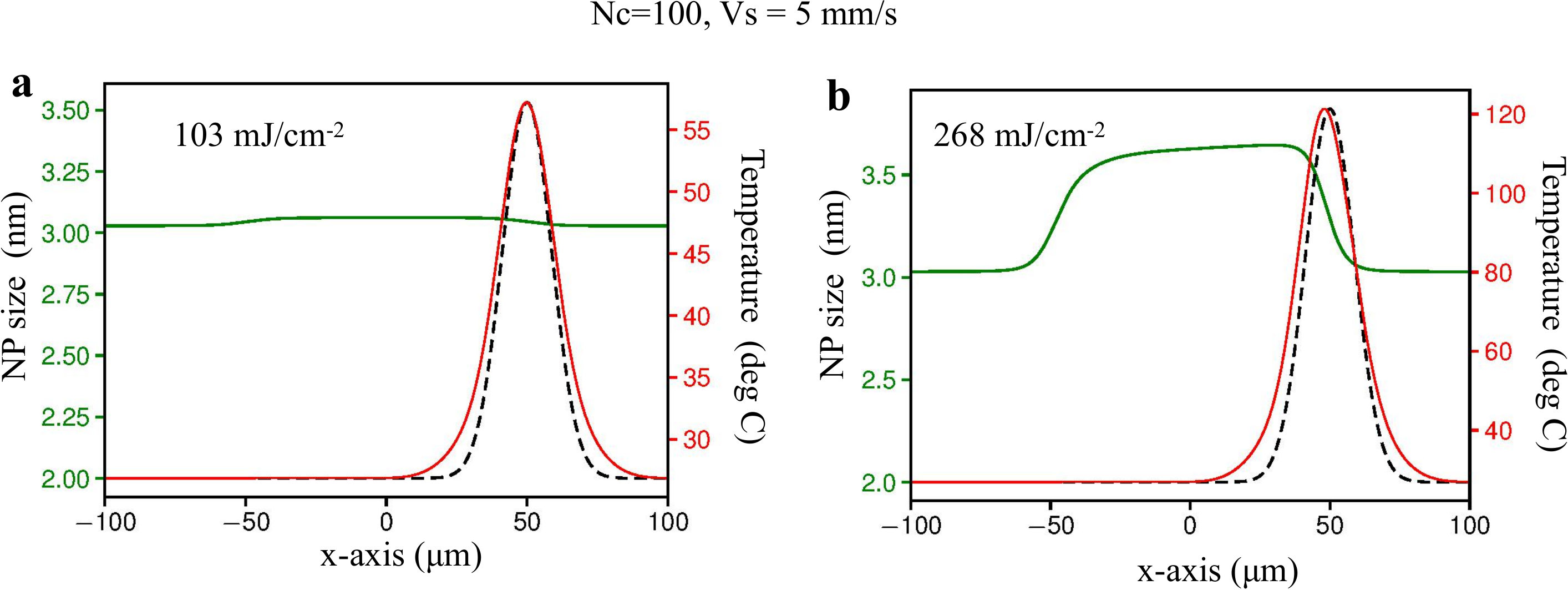}
    \caption{Calculated profiles of Ag NP size, temperature and laser intensity. The laser scans from $x=-50 \mu m$ to $ 50 \mu m$. Here, laser fluences are 103 $m J / cm^{2}$ for (a) and 268 $m J /cm^{2}$ for (b). $N_c$ stands for the number of accounted laser pulses at time $t$. The green lines show the NP size, red lines cprrespond to temperature, and the dashed black lines demonstrate normalized laser intensities.}
\label{fgr:chap6_coupled_ana_fig1}
\end{figure}

The results of the carried out simulations are shown in Fig. \ref{fgr:chap6_coupled_ana_fig1} for laser fluences of 103 $m J / cm^{2}$ (a) and 268 $m J / cm^{2}$ (b). The size of NPs by higher fluence is bigger than the lower fluence one. Though the considered fluences are around 2 and 4 times larger than the experimental ones (62  $m J / cm^{2}$), a so-called 'fast-growth' due to a strong coupling of size-increment and light absorption is never observed in the two cases. The temperature is not high enough to activate the nonlinear growth of Ag NPs so as to make the size exceeding a certain value for a fast-growth. 

\begin{figure}[ht!]
 \centering
    \includegraphics[width=0.85\textwidth]{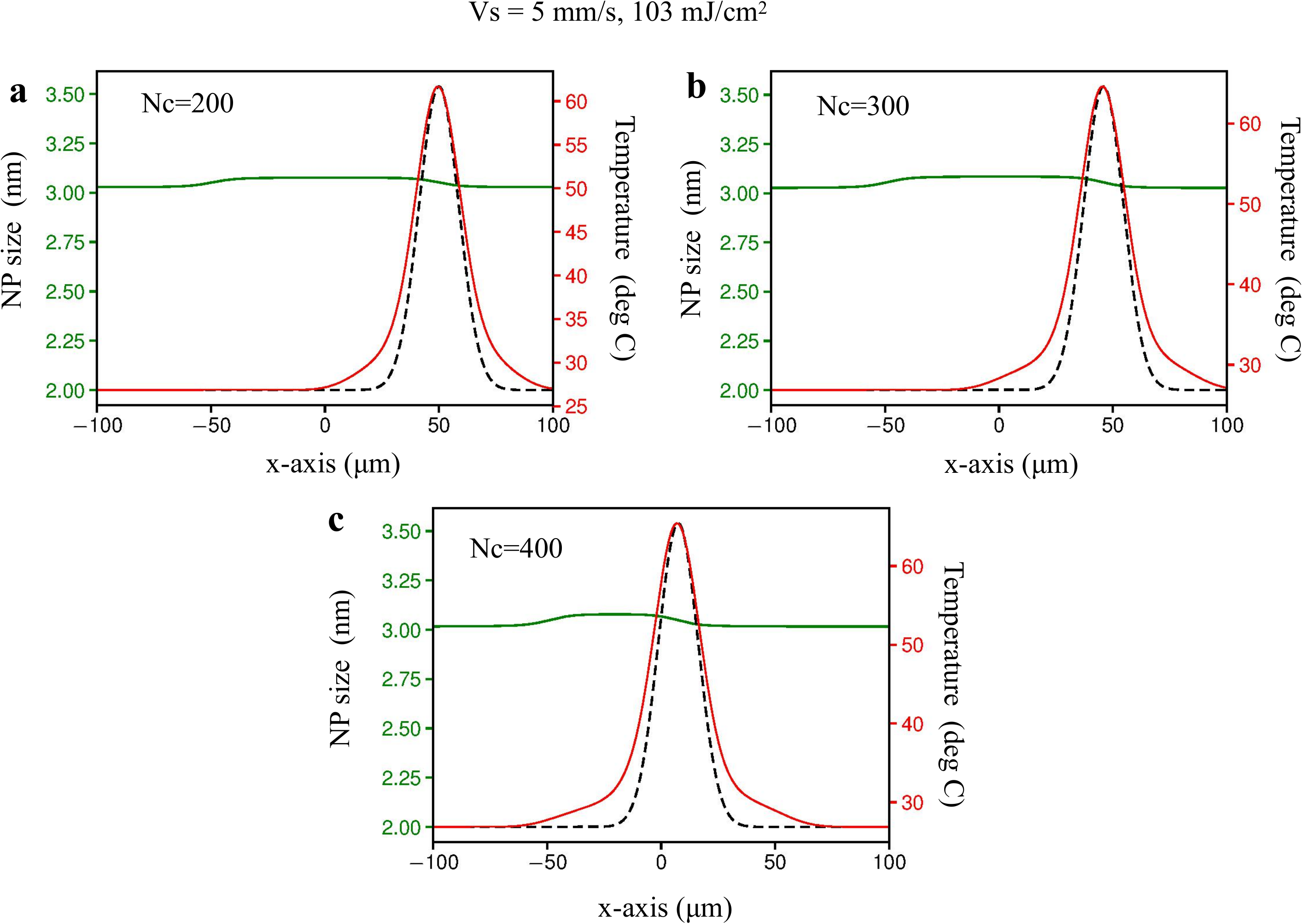}
    \caption{Simulation results by different number of accounted pulses: 200 (a), 300 (b), and 400 (c).}
\label{fgr:chap6_coupled_ana_fig2}
\end{figure}

To understand the reason why the temperature is underestimated, simulations are performed for different numbers of  laser pulses ($N_c$), as shown in Fig. \ref{fgr:chap6_coupled_ana_fig2}. Laser fluence is set to be 103 $m J / cm^{2}$, and the writing speed is $5 mm/s$. For the cases with the number of accounted pulses $N_c = 100$, 200, 300, and 400, the maximum temperature are 57.2, 61.7, 64.4 and 65.4 $^{\circ} C$, respectively. The maximum size of Ag NPs are the same and of 3.08 nm for the studied number of accounted pulses. Furthermore, the heat diffusion is overestimated since the calculations by Eq. \ref{eqn:heatEQChap6_eq2_21} only considers the TiO$_2$ as the boundary condition. As a result, the estimated threshold in laser fluence is overestimated.

\begin{figure}[ht!]
 \centering
    \includegraphics[width=\textwidth]{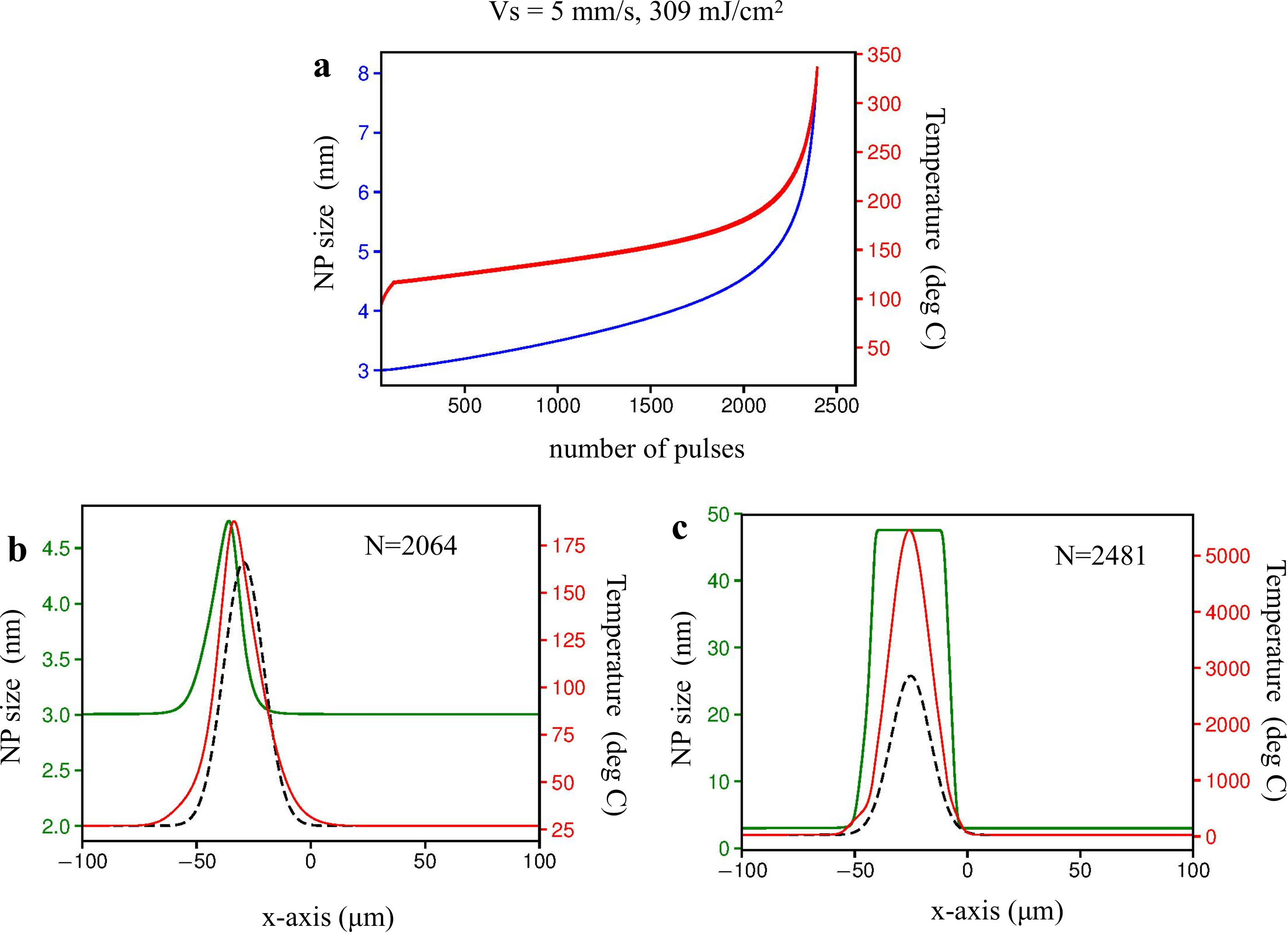}
    \caption{Simulations results of incident laser fluence at 309 $m J/cm^{2}$. The relation of the maximum NP size and temperature with number of pulses (a). The NP size, temperature and laser profiles at 2064 pulses (b), and 2481 pulses (c).}
\label{fgr:chap6_coupled_ana_fig3}
\end{figure}

As an an illustration of the fast growth, simulation are then performed for the laser fluence of 309 $m J/cm^{2}$ as shown in Fig. \ref{fgr:chap6_coupled_ana_fig3}. In this case, the reactions are much faster than in the case of 268 $m J/cm^{2}$. As a result, when the size of Ag NP exceeds 4.5 nm,  the so-called "fast-growth" is clearly observed (Fig. \ref{fgr:chap6_coupled_ana_fig3}(a)) due to the strong nonlinear relation of size-absorption at wavelength 515 nm. The spatial profiles of NP size, temperature, and laser intensity shown in Fig. \ref{fgr:chap6_coupled_ana_fig3}(b) depicts the state before the "fast-growth". Similar to the continuous-wave writing of Ag NPs in TiO$_2$ films, the peaks of temperature, NP size, and laser intensity are misaligned. However, the maximum temperature is observed behind the beam center because of the slow growth-speed before this time. It is anticipated that the temperature is higher after the fast-growth so that the growth of NPs stops before the laser center arrives. Figure \ref{fgr:chap6_coupled_ana_fig3}(c) shows the case after the fast-growth. It is found that the estimated temperature exceeds 5000 $^{\circ} C$ after the fast growth, which looks impossible by experiments \cite{liu2017three,sharmalaserdriven2019}. Previous experiments \cite{destouches2014self} have shown that the big NPs after growth tend to reside into a single layer toward the glass side. In this case, the effective medium assumption is unreasonable so that the laser absorption should be calculated based on more rigorous methods such as FDTD, FEM and coupled waves.

\subsection{Results of coupled equations by FEM}
In the previous section we have developed a semi-analytical model to simulate the pulsed laser writing of Ag NPs. Because of the large loops in numerical calculations, simplifications are made to accelerate the simulations by only considering the contributions of the nearest $N_c$ pulses to the temperature. Furthermore, the heat transportation inside TiO$_2$ is regarded as the one like glass. All these assumptions lead to the overestimation of the fluence threshold for the growth of Ag NPs. A more accurate method can be developed by directly solving the heat equation using the step-marching algorithm in the framework of finite element method. In this way, the heat diffusion inside the TiO$_2$ film is considered.

\begin{figure}[ht!]
 \centering
    \includegraphics[width=0.95\textwidth]{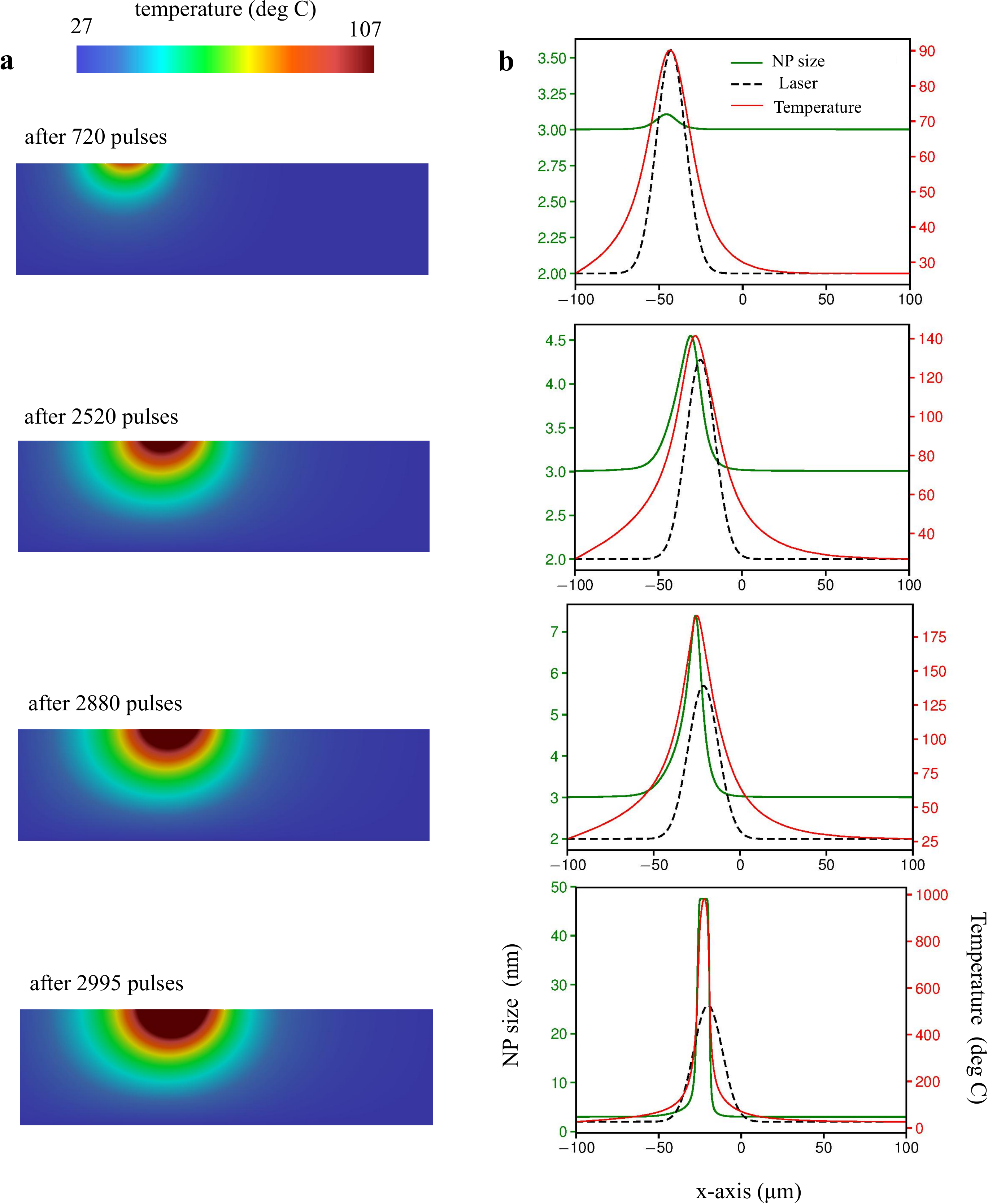}
    \caption{FEM simulations of Ag NPs growth under the pulsed-laser irradiation. The temperature distribution at different number of pulses (a), and the corresponding NPs, temperature and laser intensity on the top surface (b). The laser fluence is 103 $m J/cm^{2}$, the writing speed is 5 $mm/s$, and the spot size is 35 $\mu m$ at $1/e^2$.}
\label{fgr:chap6_fem_all_fig1}
\end{figure}

Figure \ref{fgr:chap6_fem_all_fig1} shows the finite element method simulations at different number of irradiated pulses. The writing speed is 5 mm/s, the fluence is 103 $m J/cm^{2}$, , the spot size 35 $\mu m$, and repetition rate is 500 kHz. The effective conductivity, capacity of porous TiO$_2$ are considered. The film thickness is set to 200 nm. The initial temperature is low so that the size of the Ag NP increases slowly before the laser moves a distance of 25 $\mu m$ in the positive x-axis direction (before 2520 pulses as illustrated in the Fig. \ref{fgr:chap6_fem_all_fig1} (a)). For the next 360 pulses, the growth speed is accelerated so that the maximum NP size increase from 4.5 nm to 7.1 nm. Since then, the nonlinear relation of size-absorption leads to the dramatic growth of Ag NPs in the next short 115 pulses. In contrast, this fast-growth is never observed by the semi-analytic model with the same laser conditions. The threshold, as discussed in the previous section, is around $309 m J/cm^{2}$ that corresponds to pulse energy of $1.5 \mu J$, which is 4-5 times larger than the experiments of $60 m J/cm^{2}$ \cite{liu2017three, liu2016selfthesis}. The finite element method model, however, gives the threshold around $103 mJ/cm^{2}$ is much closer.

\newpage

\newpage

\begin{figure}[ht!]
 \centering
    \includegraphics[width=0.5\textwidth]{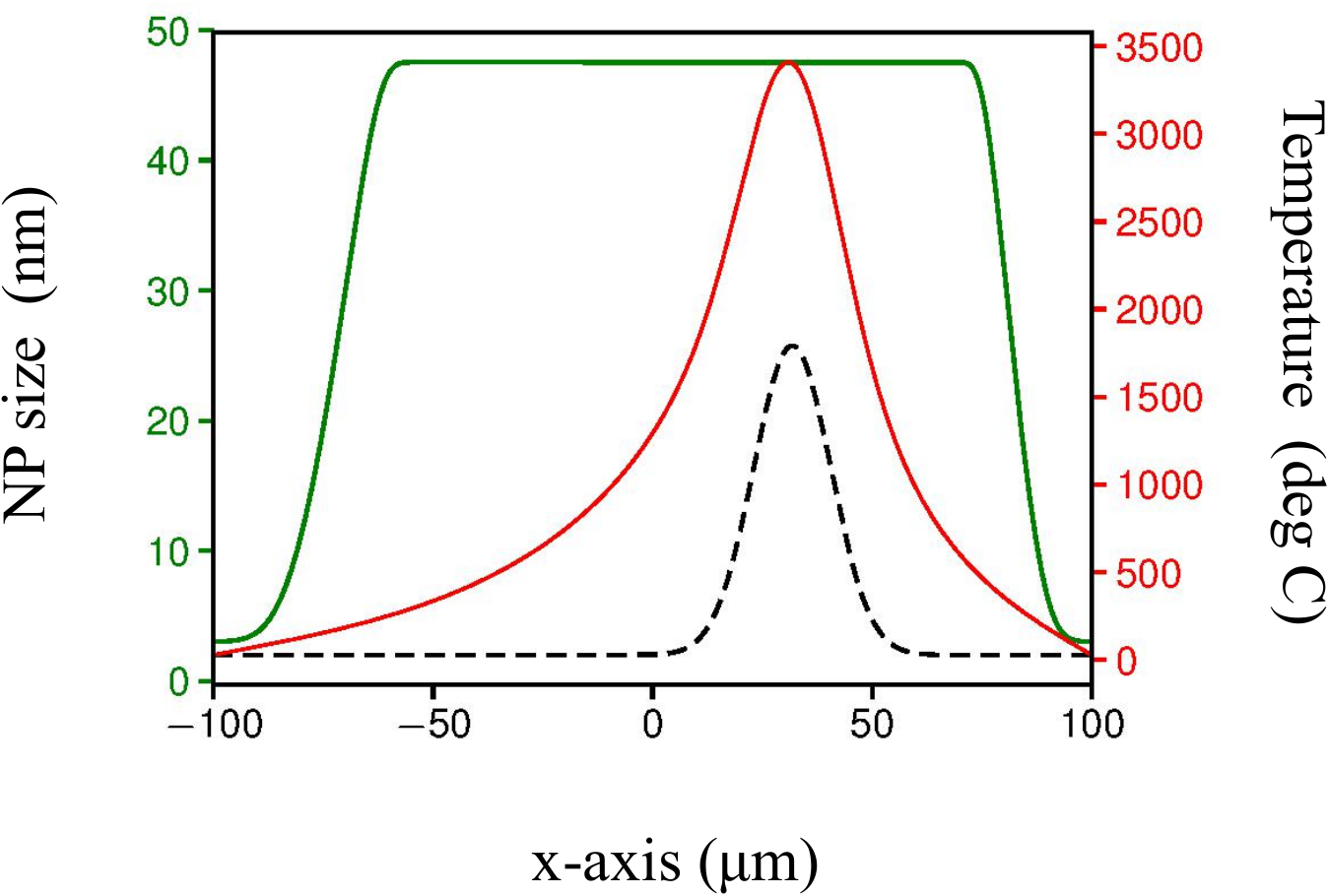}
    \caption{NP size, temperature, and laser profile after 8180 pulses irradiated on the sample. The dashed black line stands for the laser profile. Here, laser fluence is 103 $m J/cm^{2}$, the writing speed is 5 $mm/s$, and the spot size is 35 $\mu m$ at $1/e^2$.}
\label{fgr:chap6_fem_all_fig2}
\end{figure}

Figure \ref{fgr:chap6_fem_all_fig2} shows the size of NP, temperature, and laser profile after the laser moves a distance of 81.1 $\mu m$ from $x = -50 \mu m$, which corresponds to 8180 pulses. In reality, the maximum time-step is limited to less than 6 ns so as to precisely simulate the heat diffusion in TiO$_2$. The maximum size of Ag NP is around 48 nm, which is determined by the initial quantity of Ag$^0$ atoms inside the TiO$_2$ matrix. It it obvious that the NPs can grow rapidly due to the rather high temperature (around 3400 $^{\circ} C$). However, whether the temperature is overestimated is not known due to the lack of experiments. Intuitively, the glass is melt at this temperature. Recalling that the size of the formed NPs are comparable to the film thickness, the estimation of absorption obtained by using the classical effective medium approximation is no longer valid. Therefore, as indicated by the previous discussions, the model should be further modified to make a better prediction of the temperature.

\begin{figure}[ht!]
 \centering
    \includegraphics[width=\textwidth]{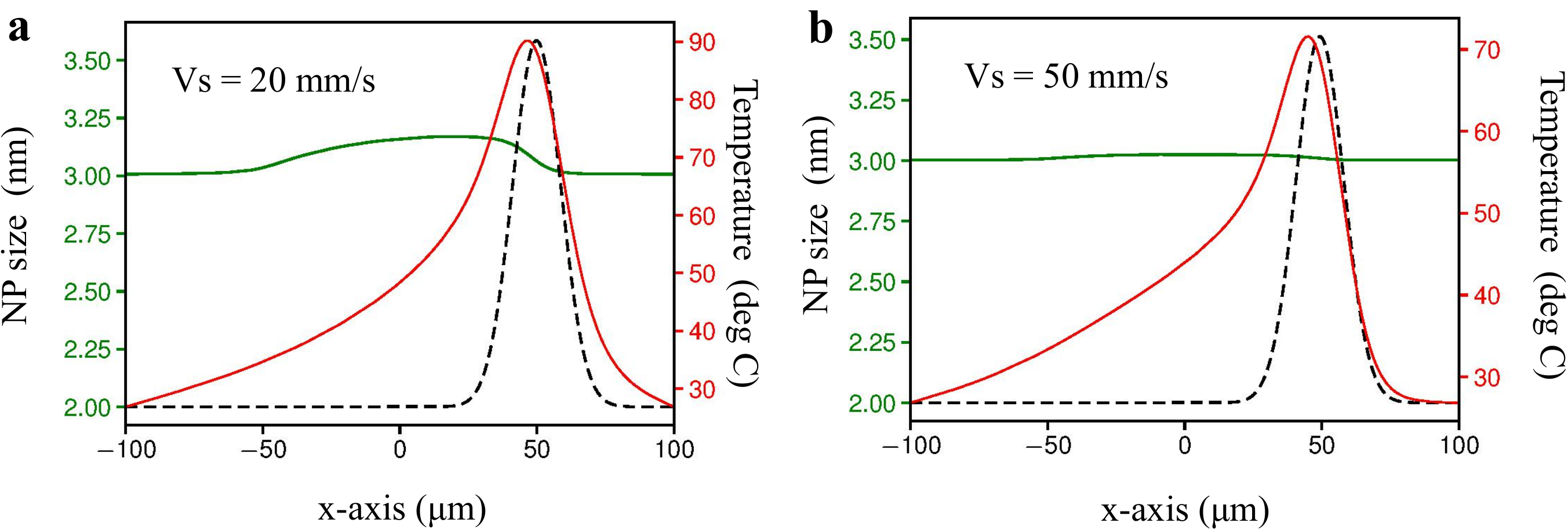}
    \caption{Calculated NP size, temperature, and laser profiles by laser writing speed of 20 $mm/s$ and $50 mm/s$. The dashed black line stands for the laser profile. The laser fluence is 103 $m J/cm^{2}$, and the spot size is 35 $\mu m$ at $1/e^2$.}
\label{fgr:chap6_fem_all_fig3}
\end{figure}

Most of the experiments \cite{liu2017three, liu2016selfthesis, sharmaTailoring2019, sharmalaserdriven2019} have shown that NPs are smaller than 10 or 20 nm for speed range from 50 mm/s to 120 mm/s. To understand the high speed writing of Ag NPs, simulations are performed at speed of 20 mm/s and 50 mm/s of laser fluence of $103 mJ/cm^{2}$ (Fig. \ref{fgr:chap6_fem_all_fig3}). For both cases, the fast-growth of NPs are not observed due to the lower accumulated temperature. The speed-threshold by experiments are actually higher than the simulation, which is due to the overestimation of the Ag$^0$ activation energy. As revealed in the study for continuous-wave laser, the model covers the spectrum range by experiments if model coefficients are properly chosen. To better fit the results, more efforts are required, particularly in the parameter optimization methodology. 

\begin{figure}[ht!]
 \centering
    \includegraphics[width=0.65\textwidth]{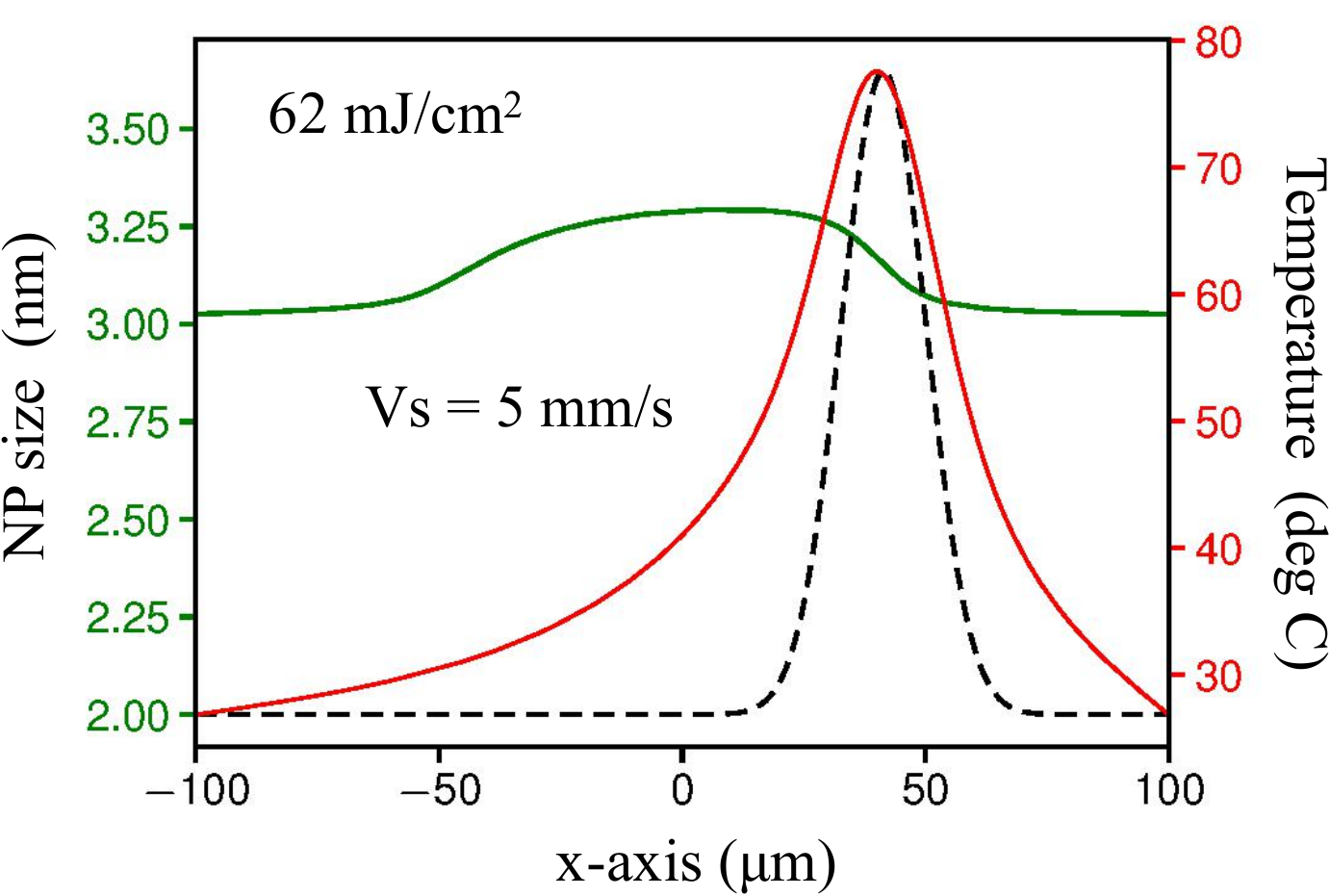}
    \caption{NP size, temperature, and laser profiles obtained for laser writing speed of 5 $mm/s$. The dashed black line stands for the laser profile. The laser fluence is 62 $m J/cm^{2}$, and the spot size is 35 $\mu m$ at $1/e^2$.}
\label{fgr:chap6_fem_all_fig4}
\end{figure}

Simulations performed for the same laser conditions ($62 mJ/cm^{2}$) with experiments \cite{liu2017three,liu2016selfthesis} are studied, which are shown in Fig. \ref{fgr:chap6_fem_all_fig4}. For the lowest speed that is around 5 mm/s, the size of Ag NPs never exceeds 3.5 nm, so that the fast-growth is not observed. Though the simulation is different from the experiments at the same condition, the results show that the existence of fluence threshold and the trend is in accordance with the experiments. 

\begin{figure}[ht!]
 \centering
    \includegraphics[width=0.8\textwidth]{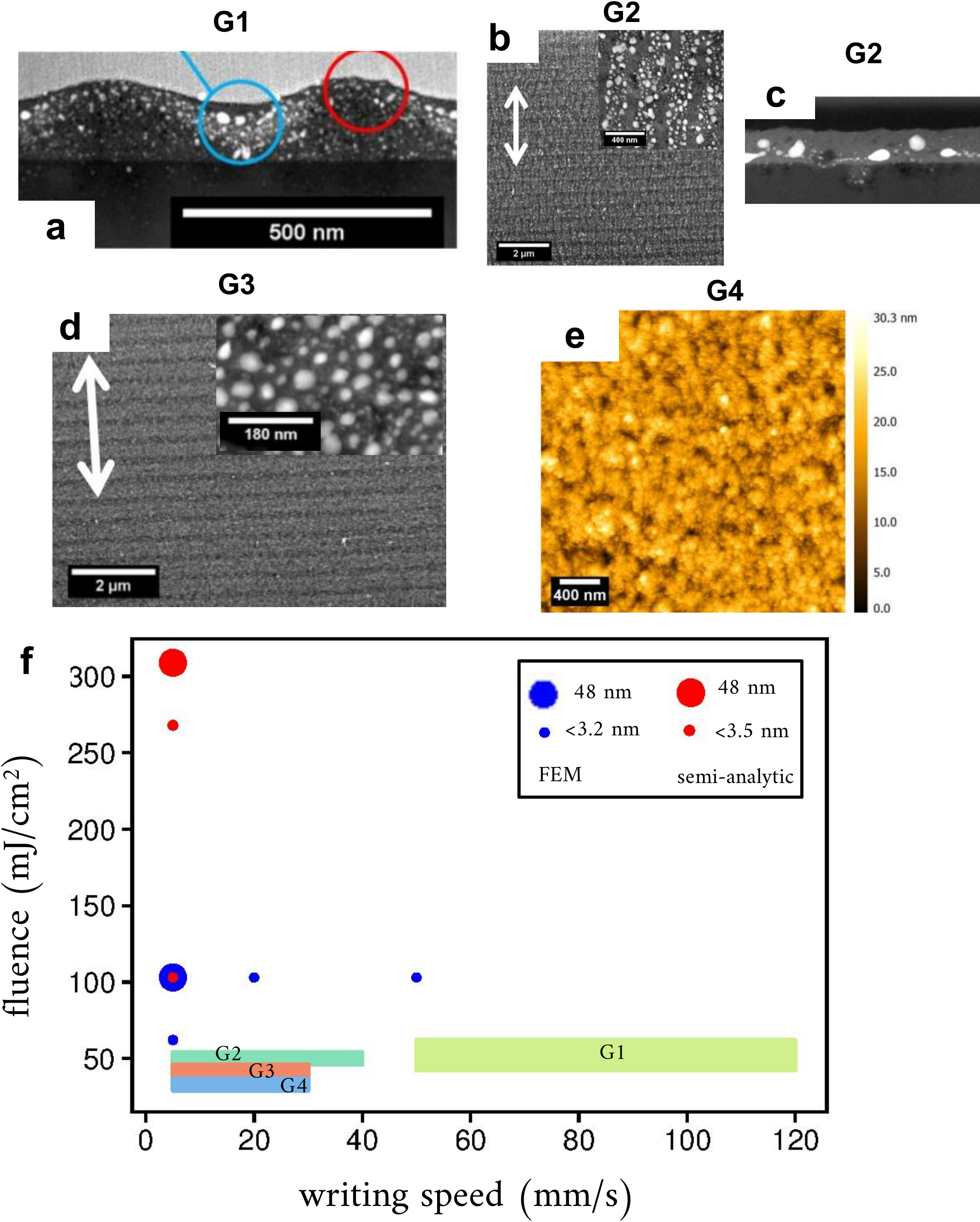}
    \caption{Comparisons of simulations with experiments by Ref. \cite{liu2016selfthesis}. The HAADF STEM image of G1 film cross-section (a); SEM images of G2 (b) and HAADF STEM top view of G2 (insert map), cross-section (c); EM images of G3 (d), the insert map is HAADF STEM image; AFM image of G4; the plot of Ag NP size by FEM and semi-analytic models ($Nc = 100$); the colored squred regions are from experiments by Ref. \cite{liu2016selfthesis}. White arrows represent the laser polarizations.}
\label{fgr:chapter6_exp_02}
\end{figure}

The simulated relation of Ag NP size with various fluences and writing speeds are summarized in Fig. \ref{fgr:chapter6_exp_02}. The experimental data are taken from Ref. \cite{liu2016selfthesis, liu2017three}. For G1 structure, LIPSS appeared on the TiO$_2$ surface which accompanied with Ag NPs (around 10-20 nm in diameter) within the grooves valleys and smaller NPs on the peak positions (Fig. \ref{fgr:chapter6_exp_02}(a)). As for G2, apart from the LIPSS on the material surface, Ag nano-gratings were formed closing to the glass substrate (Fig. \ref{fgr:chapter6_exp_02}(b-c)). 

For the case of G3, the NPs were formed homogeneously on the top surface (Fig. \ref{fgr:chapter6_exp_02}(d)). As soon as the fluence decreased, LIPSS and Ag nano-gratings were indistinguishable and NPs were small (5 - 15 nm) for G4. Though the proposed models do not consider the formation of LIPSS and Ag nano-gratings, the performed simulations show the same trend in terms of the Ag NP growth (Fig. \ref{fgr:chapter6_exp_02}(f)). For the semi-analytic model taking effective pulses as Nc = 100, the NP size do not increase until the fluence increases to $309 mJ/cm^2$. The simulations by the FEM model is much closer to the experiments where large NPs are observed. The differences of these two models come from two factors as explained in the previous section: firstly, not enough effective pulses are considered for the semi-analytic model; secondly, the diffusion in TiO$_2$ films are ignored by the semi-analytic model. Thus, the FEM results are more precise in this case. Nevertheless, the fluence of generating large NPs is $103 mJ/cm^2$ by simulation and $37 mJ/cm^2$ by experiments \cite{liu2016selfthesis, liu2017three}. Probably, the plasmonic induced near-field enhancement boosts the light absorption by nonlinear effects, which is not considered by the present models. Furthermore, the estimated activation energy and coefficient of Ag diffusion by simulations are far away from the practical cases. The presented threshold for NP growth only exists in a specific region of fluences and writing speed that is observed by the experiments. The corresponding temperature rises shown by simulations provide reasonable explanations of the temperature decrease as the writing speed increases or the fluence decreases.

\section{Conclusions}

We proposed models describing the ultra-fast laser processing of mesoporous TiO$_2$ films loaded with Ag NPs. The femtosecond laser propagation in glass substrate was simulated by the nonlinear Schr\"{o}dinger equation coupled with the plasma equation and concluded that the absorption by glass can be ignored under the studied laser fluences and focusing conditions. Based on this analysis, a semi-analytic heat accumulation equation by a multi-pulse laser was deduced and coupled with the growth of Ag NPs and Ag$^+$ reduction processes. To largely accelerate the simulation, a GPU based model was developed. Different effective number of pulses were compared. The threshold fluence of Ag growth was shown to be around $309 mJ/cm^2$. For comparisons, the heat accumulation was simulated by a finite element method and coupled with the Ag growth related equations. As a result, the threshold fluence of Ag growth was around $103 mJ/cm^2$, which was much closer to the experiments by Ref. \cite{liu2017three, liu2016selfthesis}. The differences by the two models were explained and it concluded that the finite element method model was more precise comparing to the semi-analytic model. In the finite element method model, the heat diffusion in the mesoporous TiO$_2$ was considered. The simulations showed a similar region of fluence and writing speed where Ag NPs grew dramatically; outside the region, only small NPs were obtained. Because of the complexity of the practical process, the formation of surface grooves (LIPSS) and NP gratings were not considered by the simulation. Nevertheless, the proposed models paved the way for the modeling of Ag NPs formation and nano-gratings by pulsed-lasers in the future.

\afterpage{\null\newpage}
\chapter{General conclusions and outlook}

\section{Conclusions}

\hspace{0.5cm}  This thesis provided the descriptions of several recently developed models and the results of numerical simulations of laser interaction with porous glass and mesoporous TiO$_2$ loaded with Ag nanoparticles. Part of the simulations were performed for porous materials without nanoclusters.  In the another part, porous films were impregnated with metallic ions, irradiated by a weak UV light forming primary nanoparticles affecting material absorption and enabling their further growth or shrinking by the considered following laser irradiation.

Firstly, the periodic micro-void structures produced inside porous glass by femtosecond laser pulses were explained by a thermodynamic analysis. The laser propagation model was based on the nonlinear Schr\"{o}dinger equation coupled with electron plasma equation. The simulations were conducted in 1D+1 (time) taking advantage of the GPU parallelization. The performed thermodynamic analysis showed the possibility to efficiently control laser micro-machining in volume. For the given pulse repetition rate, the laser writing speed had great influences on the size and period of the formed voids. Furthermore, the densifications at low pulse energies were studied by simulation at different focusing conditions. The characteristic dimensions of the calculated temperature field were well correlated with experiments. The performed study provided a way of understanding the laser inscription of void arrays and densifications in porous glasses, which are of interest for porous glass based applications in optical waveguides, filters, and other integrated devices for microfluidics.

Comparing to the porous glass, laser writing of mesoporous TiO$_2$ films loaded with Ag nanoparticles enabled additional possibilities in the generation of different NP sizes as a function of laser power, focusing or writing speed. To elucidate the optimum laser irradiation conditions, a model taking into considerations Ag nanoparticle shrinkage via photo-oxidation, Ag$^+$ ion reduction, growth by Ostwald ripening, and plasmonic (size-dependent) light absorption was developed. The performed simulations showed  dynamical changes in the variables such as Ag nanoparticles size, temperature field, and Ag$^+$ concentration. It was shown that the laser writing speed controlled the Ag nanoparticles size. As a result, the calculated transmission profile was inhomogeneous along the writing direction, which was demonstrated to be writing speed dependent. In order to accelerate the multi-scale simulation, the nanoparticle related processes such as growth, reduction, and photo-oxidation were solved using the Bulirsch-Stoer algorithm. The performed calculations depicted a novel view that Ag nanoparticles grow ahead of the laser spot center due to the heat diffusion. The mechanisms were attributed to the following processes

(\rom{1}) thermal activated nonlinear growth of nanoparticles that never stopped until the majority of free Ag$^0$ in the matrix were consumed; 

(\rom{2}) the amount of Ag$^0$ created by reduction failed to compensate the Ag$^0$ consumption process (\rom{1});

(\rom{3}) as the growth process stopped,  the photo-oxidation dominated the size-variation and finally played the controlling role; 

As a result, the maximum size of Ag nanoparticles appeared in the laser front edge. The final size of Ag nanoparticle depended on the laser writing speed due to the different time for photo-oxidation. The simulated transmission profiles showed similar trend with the results of the in-situ transmission experiments. The differences were attributed to the  remaining model limitations. The single size model, effective medium theory for thin film doped with Ag nanoparticles were considered as the origins of these differences.

Both spatial size distribution of nanoparticles and their plasmonic size-dependent absorption resulted in the temperature variations observed at different laser writing speeds. In particular, the temperature increase with writing speed was shown by the simulation, which seemed to be abnormal as less deposited energy led to higher temperature. This effect was explained by that fact that the Ag nanoparticle size was larger at a higher writing speed. Thus, it was the same for the absorption. In other words, the light absorption was more efficient at higher writing speeds. This result was confirmed by the TiO$_2$ phase-transition experiments. By studying the influences of the activation energy of Ag diffusion, diffusion coefficient of Ag, photo-oxidation rate, and initial concentration of Ag$^0$ and Ag$^+$, the model was shown to be robust in explaining the results of Raman experiments. Furthermore, a three-dimensional model was developed and simulated thanked to the high performances of the parallelization and scalable finite element method. The three-dimensional simulations reproduced the laser written lines in agreement with the experimental results.

In addition, the multi-physical modeling of Ag nanoparticle growth by a multi-pulsed laser was studied. This topic is interesting because the pulsed laser provided additional abilities in generating two kinds of nanostructures: the laser induced periodic surface grooves (LIPSS) and Ag nanogratings inside the TiO$_2$ film. To estimate the deposited laser energy,  femtosecond laser propagation inside the glass substrate was studied. It was concluded that the absorptions inside the glass volume could be ignored at the considered wavelength, fluences and focusing conditions. Based on these results, a semi-analytic model was developed to estimate the heat accumulation by multiple pulses, which was coupled with the Ag nanoparticle growth equations. The simulations were accelerated by using a GPU. The threshold fluence of Ag growth was shown to be around $309  mJ/cm^2$. For comparisons, a finite element method based model was proposed for simulating the heat diffusion coupled with Ag nanoparticle growth. The threshold fluence was estimated to be around $103    
 mJ/cm^2$, which was closer to the reported experimental values. The finite element method  was more precise comparing to the semi-analytic model since the heat diffusion inside the TiO$_2$ thin films was considered. The simulations  were carried out for different fluences and writing speeds  showing a region  where Ag nanoparticles grew dramatically, similar to the reported experiments. 

It should be noted that the developed models can be further extended to account for the formation of LIPSS and Ag nano-gratings in such media by coupling with nanoparticle migrations, surface melting and hydrodynamics. Thus, the range of the potential applications is rather wide. This work is, however, out of the scope of the present manuscript.

\section{Perspectives}

In laser processing of mesoporous TiO$_2$ films loaded with Ag NPs,  self-organized Ag nanogratings with period in subwavelength scale are particularly attractive for optical applications, such as optical data storage \cite{rakuljic1995optical}, color printing\cite{diop2017spectral}, and image multiplexing \cite{sharmalaserdriven2019,sharmaTailoring2019}. In such films, standing waves are known to be formed\cite{destouches2014self}. The number of supported modes depends on film thickness, refractive index and wavelength. The presence of the guiding modes allows optical modulations of spatial distribution of Ag NPs in a sub-wavelength scale. Previously, a model based on the coupled mode theory \cite{destouches2014self} was proposed to predict the period of such nanostructures. It was concluded that guided waves enhanced due to the growth of Ag NPs.  Among the involved mechanisms, several researches discussed possible contributions of the electromagnetic forces \cite{eurenius2008grating}, diffusion due to temperature gradient (Soret effect) \cite{braibanti2008does}, and diffusion due to density gradients \cite{smetanina2016modeling}. In future, the possible roles of ion and cluster diffusion should be better accounted for. Particularly, one of the possible effects is based on the so-called Soret-Dufour equation involving thermophoresis. 

\subsection{Gratings induced by photo-oxidation}

\begin{figure}[ht!]
 \centering
    \includegraphics[width=0.95\textwidth]{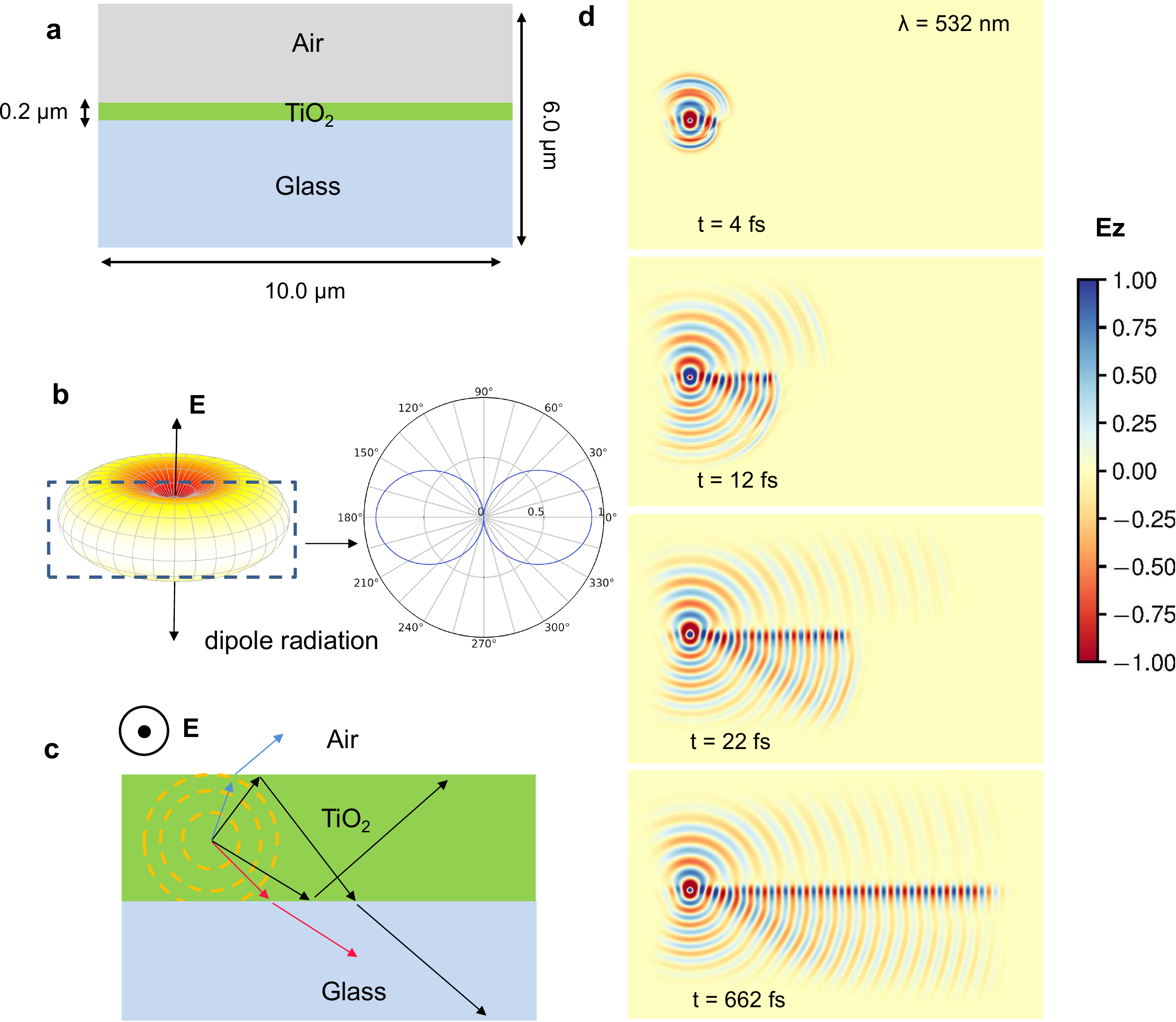}
    \caption{Thin film wave-guiding activated by a dipole source. The diagram of the structure (a); the dipole radiation pattern (b); illustrations of the propagation of waves at different angles (c); FDTD simulations (d).}
\label{fgr:chapter5_selfOrg_fig0}
\end{figure}

In Chapter 3, where CW laser writing of Ag NPs in TiO$_2$ films was considered, the roles of photo-oxidation was proportional to the number of the absorbed photons by each nanoparticle. In this case, mostly laser amplitude played a role. Nevertheless, the scattered photons have also a probability to flow to other nanoparticles to be scattered and partly absorbed in the considered nanocoposite films. Figure \ref{fgr:chapter5_selfOrg_fig0} (a) shows the studied structure. The thickness of the TiO$_2$ film is 200 nm and its effective refractive index equals to 1.7  \cite{liu2016selfthesis,liu2015understanding}. Under laser illumination, Ag NPs can be considered as absorbing and scattering centers. Under dipole approximation conditions, these NPs act, in fact, as dipole sources (Fig. \ref{fgr:chapter5_selfOrg_fig0} (b)). For the laser polarization perpendicular to the studied plane  (Fig. \ref{fgr:chapter5_selfOrg_fig0} (a,c)), waves with the incidence angle larger than the critical angle of the total reflection are reflected back into the film. As a result, the standing waves are formed, as the FDTD results have shown \cite{oskooi2010meep} in Fig. \ref{fgr:chapter5_selfOrg_fig0} (d). 

\begin{figure}[ht!]
 \centering
    \includegraphics[width=0.7\textwidth]{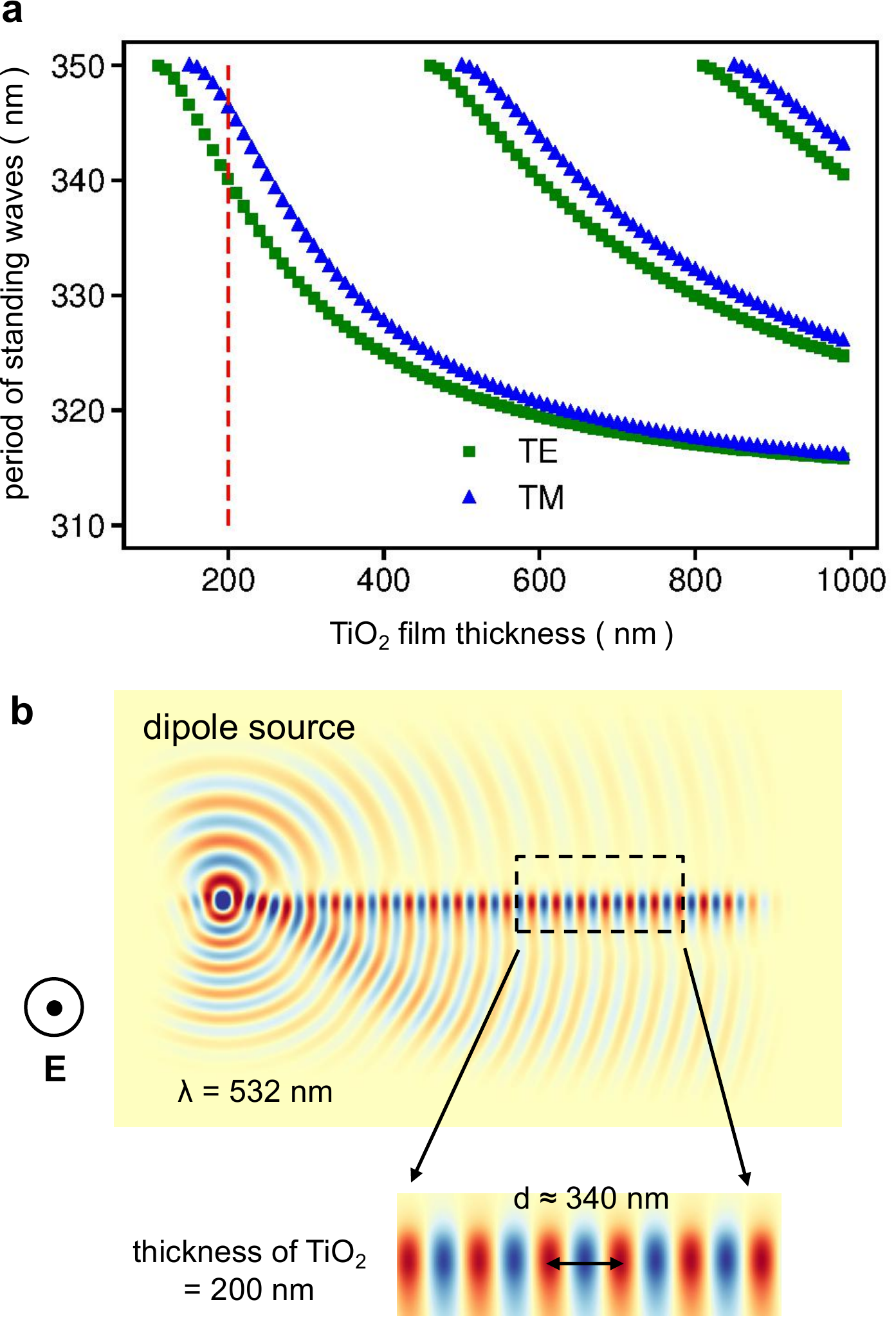}
    \caption{Standing wave period obtained by the mode expansion method for different thickness of TiO$_2$ films (a). Results of the FDTD simulation for a dipole-source activated standing wave.}
\label{fgr:chapter5_selfOrg_fig01}
\end{figure}

Only limited numbers of guiding modes are supported by the film, depending on the light wavelength, film thickness, and the refractive index. The guiding modes are obtained by the mode expansion method \cite{lohmeyer2002mode}. The period of standing waves are shown in Fig. \ref{fgr:chapter5_selfOrg_fig01} (a) for different TiO$_2$ thickness. For the laser wavelength of 532 nm, the 200 nm TiO$_2$ film supports only a single mode for TE and TM modes. 
In the case of Ag NPs, or dipole sources, scattered light polarization corresponds to the TE mode. The activated guiding waves are parallel to the laser polarization, which is the same orientation as for Ag nanogratings after laser processing \cite{liu2016selfthesis, baraldi2016polarization,destouches2014self}. 
The guiding-wave period obtained by the FDTD simulations is close to the mode expansion results as shown in Fig. \ref{fgr:chapter5_selfOrg_fig01} (b). As the thickness of the TiO$_2$ increases to around 420 nm and 810 nm, two and three modes appear for both TE and TM.

Considering the scattering centers (Ag NPs) as the TE oscillating sources, the standing waves contribute to the photo-oxidation. Therefore, the number of silver atoms leaving a nanoparticle per unit time due to the above-discussed processes is modified and can be written as:

\begin{equation}
    n_{oxi}(t) =\eta_0 \frac{I(x, y, z, t)} {h \nu} \bigg( \sigma_{abs}(R,\lambda) + \xi \sigma_{sca}(R,\lambda) sin^2(\frac{\pi x}{d}) \bigg)
	\label{eqn:conclu_eq1}
\end{equation}

\noindent where $I(x,y,z,t)$ is the laser intensity, $\eta_0$ is the ionization rate, $h \nu$ is the photon energy, $\sigma_{abs}(R,\lambda)$ is the absorption cross-section of the Ag NP with radius $R$, $\xi$ is the absorption ratio of scattered photons and ranges within (0, 1), $\sigma_{abs}(R,\lambda)$ is the cross-section of scattering, $d$ is the period of standing wave inside thin films, and $x$ is the coordinate. The squared sin function is a simplification of the guiding waves, which is invariant when laser moves.

\begin{figure}[ht!]
 \centering
    \includegraphics[width=0.95\textwidth]{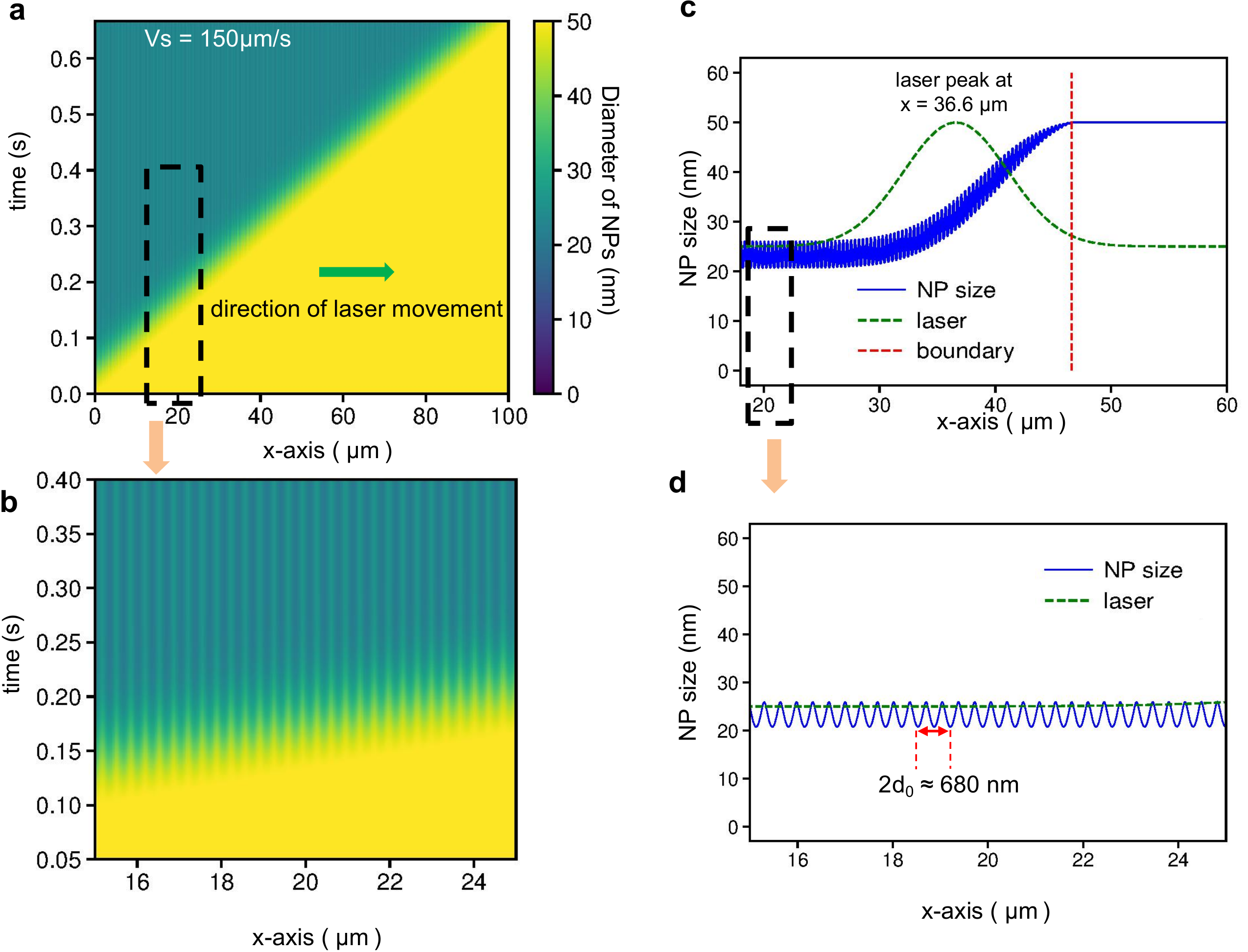}
    \caption{Ag NP size distribution obtained due to the photo-oxidation modulation at writing speed $Vs = 150 \mu m/s$. (a) the time-x map of Ag NP size; (b) the magnification of the black dashed region shown in (a); (c) the spatial distribution of Ag NP when the laser center locates at $x = 36.6\mu m$; (d) the magnified view of the black dashed region in (c). The read dashed line in (c) represents the location where NP has the maximum size.}
\label{fgr:chapter5_selfOrg_fig1}
\end{figure}

In the presented Thesis, it is found that Ag NPs start growing at the laser front edge due to heat diffusion. The nonlinear growth was fast and it never stopped until the initial Ag$^0$ atoms were exhausted. The maximum size was ahead of the laser beam at a distance typically in the range of a few ten micrometers. Based on this idea and to largely decrease the computation efforts, simulations are performed by supposing that the distance at the maximum size is $10 \mu m$ before the laser center; the maximum size is set to be 50 nm; the absorption ratio of scattered photons is taken as 0.3; the ionization rate is $\eta_0 = 6.9 \times 10^{-5}$; laser wavelength $\lambda = 532 nm$. Because the reduction of Ag$^+$ into Ag$^0$ only slightly affects the growth, the simulations can be regarded as the process after the fast growth. In this way, the Ag NP spatial size distribution is modulated by the photo-oxidation. To accelerate the simulation, the code is written in "CUDA C" taking advantages of the GPU parallelization. The studied distance is $100 \mu m$, the spatial discretization is $\delta x = 17 nm$, and the period of the standing wave is $d = 340 nm$. 

Figure \ref{fgr:chapter5_selfOrg_fig1}(a-d) shows the simulated Ag NP size at writing speed $Vs = 150 \mu m/s$. The laser scans from $x = 0\mu m$ to $x = 100\mu m$. Figure \ref{fgr:chapter5_selfOrg_fig1} (a-b) clearly shows the periodic profiles after photo-oxidation. While the laser moves from left side of $x = 0 \mu m$ to the right at $x = 100 \mu m$, the homogeneous spatial size distribution is modified and results into the formation of the nanoscale grooves. The further analysis as demonstrated in Fig. \ref{fgr:chapter5_selfOrg_fig1} (c-d) shows the period around 340 nm, which is exactly the period of the standing wave. The size at valley is around 20 nm and at peak is 25 nm.

\begin{figure}[ht!]
 \centering
    \includegraphics[width=0.95\textwidth]{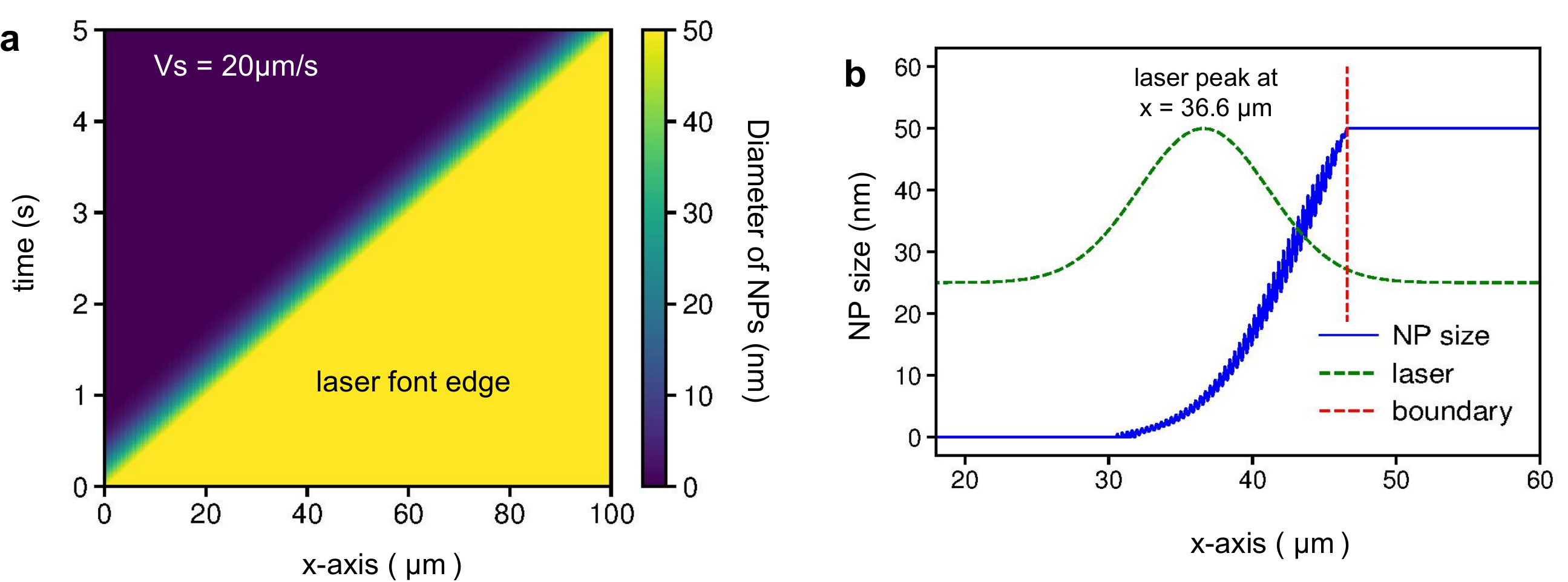}
    \caption{Ag NP size distribution by the photo-oxidation modulation at writing speed $Vs = 20 \mu m/s$. (a) the time-x map of Ag NP size; (b) the spatial distribution of Ag NP when the laser center locates at $x = 36.6\mu m$. The read dashed line in (b) represents the location where NP has the maximum size.}
\label{fgr:chapter5_selfOrg_fig2}
\end{figure}

For the laser writing speed of $Vs = 20 \mu m/s$ and smaller, the periodic grooves disappear. Figure \ref{fgr:chapter5_selfOrg_fig2} (a-b) shows the temporal variations of the Ag NP size along x-axis. It is found that at this writing speed, the grooves only exist inside the laser beam. Outside the region, the Ag NP size decreases to zero. In this case, the standing wave modulations are cancelled as the NP disappears by oxidation.

\begin{figure}[ht!]
 \centering
    \includegraphics[width=0.95\textwidth]{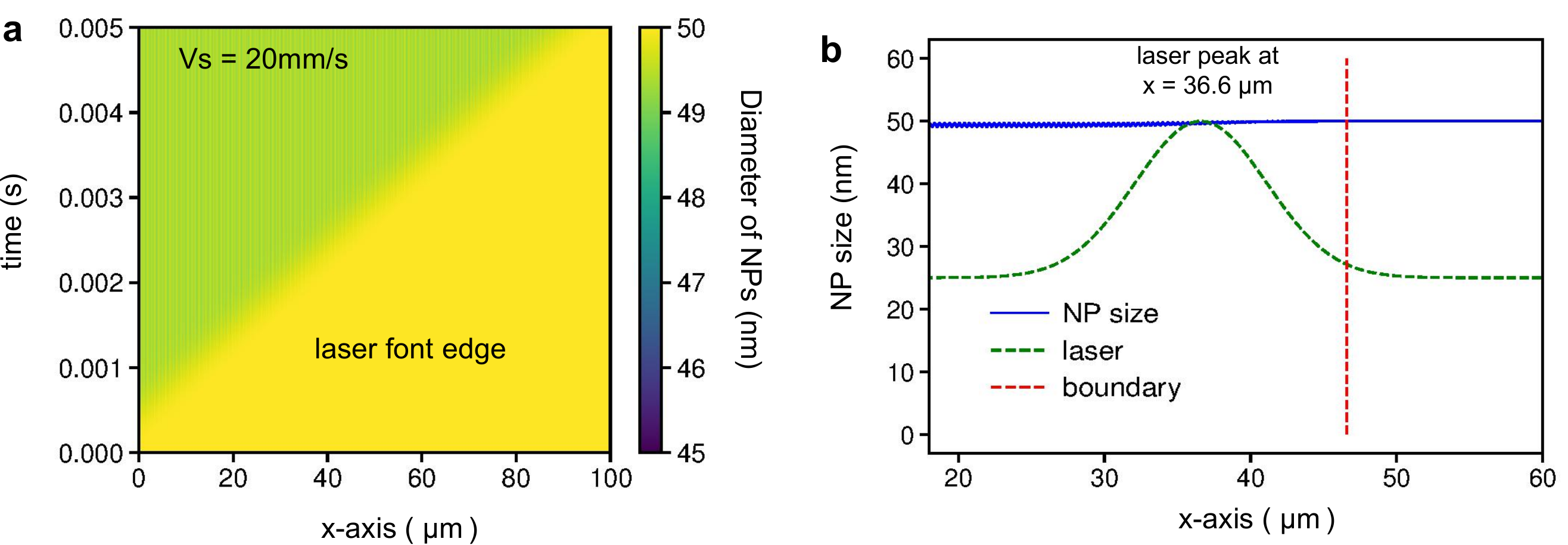}
    \caption{Ag NP size distribution by the photo-oxidation modulation at writing speed $Vs = 20 mm/s$. (a) the time-x map of Ag NP size; (b) the spatial distribution of Ag NP when the laser center locates at $x = 36.6\mu m$. The read dashed line in (b) represents the location where NP has the maximum size.}
\label{fgr:chapter5_selfOrg_fig3}
\end{figure}

In another opposite regime of the modulation at high writing speed, the resulted grooves are insignificant. Figure \ref{fgr:chapter5_selfOrg_fig3} (a) show the simulated Ag NP size along x-axis at different time. The difference in Ag NP size at peak and valley is small (less than 2 nm) as shown by Fig. \ref{fgr:chapter5_selfOrg_fig3}(b). In this case, due to the limited time for the photo-oxidation, the standing wave modulation is small.

\begin{figure}[ht!]
 \centering
    \includegraphics[width=0.9\textwidth]{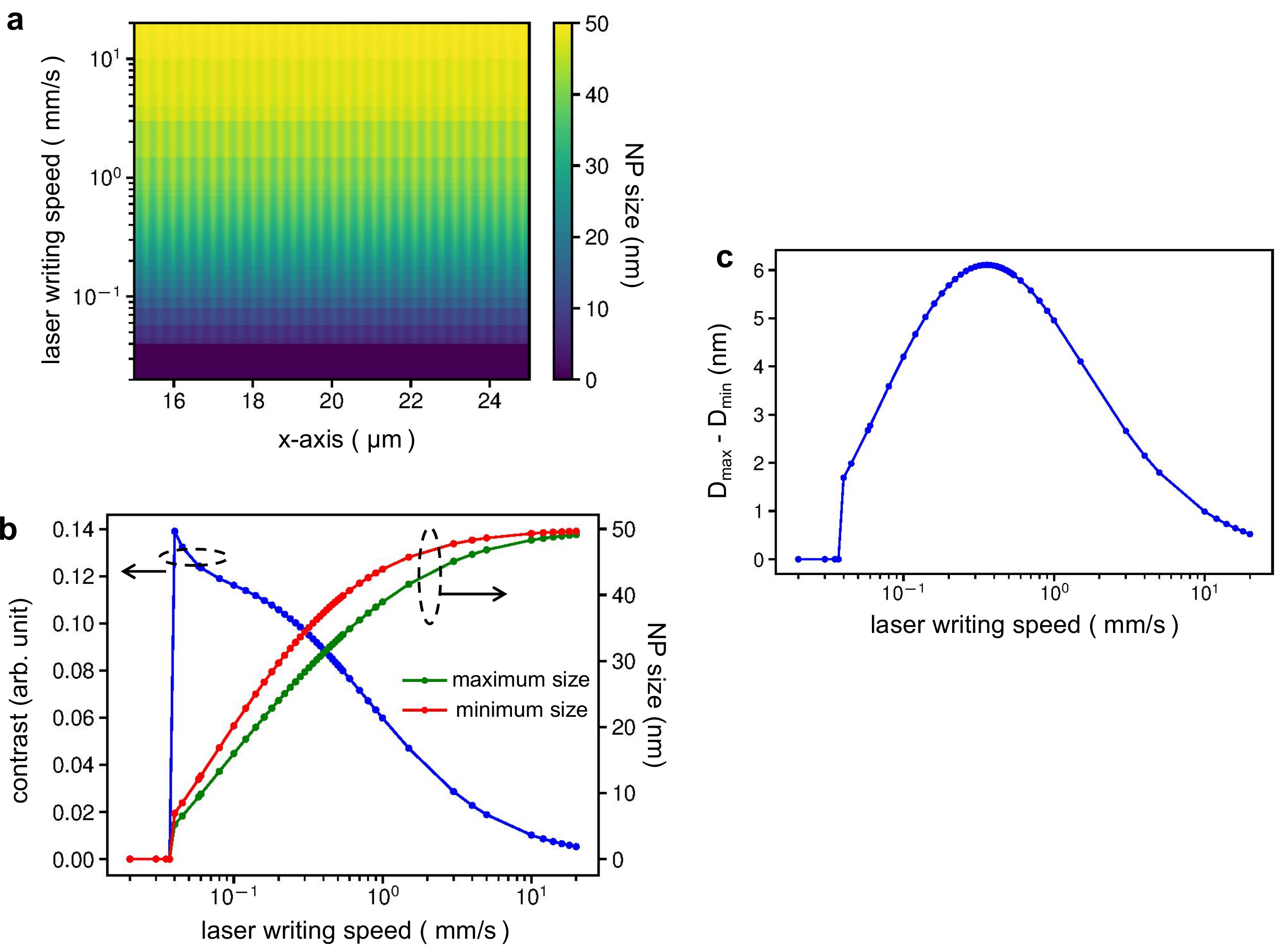}
    \caption{Calculated map of Ag NP size vs writing speed and x-axis (a); the maximum, minimum and contrast of NP size for various speed (b); the NP size difference at the peak and in the valley (c).}
\label{fgr:chapter5_selfOrg_fig5}
\end{figure}

The size of Ag NP after photo-oxidation modulation is calculated and shown in Fig. \ref{fgr:chapter5_selfOrg_fig5} (a)  for various speeds. It is clear that the high contrast grooves only exist if the writing speed is neither too slow (<40 $\mu m/s$) nor too fast (>30 $mm/s$). At low writing speed,  the Ag NP tend to dissolve completely because of the strong size-reduction by photo-oxidation. For laser writing at a high speed, the standing wave modulation is insignificant because the oxidation time is limited and is too short. The maximum and minimum sizes and their difference for the grooves obtained at various writing speeds are shown in Fig. \ref{fgr:chapter5_selfOrg_fig5} (b-c). The contrast is defined as $\frac{D_{max}-D_{min}}{D_{max}+D_{min}}$. The abrupt rising followed by gradual decrements of the contrast value show the corresponding behavior of the standing wave modulation, which is writing speed dependent. The self-organization of Ag nano-gratings by continuous-wave laser was shown to take places only in a specific speed region ranging from $150 \pm 100 \mu m/s$ to $3 \pm 2 mm/s$ \cite{destouches2014self,liu2016selfthesis}. Here, the simulations present similar speed range. Nevertheless, the modeling is different from the experiments by Ref. \cite{destouches2014self}. For example, large Ag NPs (> 35 nm in diameter) were everywhere after the standing wave modulation by simulations; in contrast, these large NPs were aligned periodically and spaced in chains that formed 1D nanoparticle gratings \cite{destouches2014self}. Indeed, NPs moved by external forces to diffuse and finally formed the 1D pattern was not considered by this model. To the best of our knowledge, the self-organization process remains unclear till now. Possible origins may due to the electomagnetic force \cite{eurenius2008grating}, the Soret effect \cite{braibanti2008does}, and the electrostatic force. Though the optical tweezers \cite{ashkin1970acceleration} are commonly used for trapping particles in a liquid, whether the electromagnetic force can move Ag NPs in a mesoporous material by a low intensity (e.g. continuous-wave) laser is, however, questionable and should be accounted for only in for small easily polarised species. 

Based on a simplified model, photo-oxidation modulation can lead to the formation of Ag nano-gratings.  In addition, it is possible that NP motion and trapping can also play a role in the self-assembled grating formation.

\subsection{Species migration}

As far as species migration is concerned, metallic species generally tend to move in mesoporous matrices subjected to laser radiation. The laser-triggered gradients of concentration, $C$, temperature, $T$, and electrostatic force, $\boldsymbol{F}$, are able to induce ion and cluster migration as follows:

\begin{equation}
    \boldsymbol{J} = -D \nabla C - C D_T \nabla T + m C \boldsymbol{F}+C\boldsymbol{U}
    \label{eqn:chap6_soret_eq2}
\end{equation}

\noindent where D is diffusion coefficient of the Ag species; $D_T$ is the thermal diffusion coefficient; $m =D/kT$; $k$ is Boltzmann constant; $\boldsymbol{U}$ is the velocity of a convective flow; and $T(r)$ is the thermal field induced by the laser. For different size of NPs, Soret coefficient is calculated as a function of temperature as follows: \cite{braibanti2008does}

\begin{equation}
\begin{aligned}
    S_T (T,R) &= D_T / D \\
    &= S^{\infty}_{T}(R) [1 - \exp(\frac{T_{sc}(R) - T}{T_0(R)})]
    \label{eqn:chap6_soret_eq3}
\end{aligned}
\end{equation}

\noindent where $T_0$ is a fitting parameter, $R$ is the nanoparticle radius, and $T_{sc}(R)$ is the specific temperature, at which $S_T$ changes its sign. 

When thermal gradient prevails, ion and small cluster drift velocity depends on their concentration,  $C$, and on the size-dependent thermophoretic mobility, or thermal diffusion coefficient ($D_T$). Depending on the sign of $D_T$, the drift can be realized toward either the coldest or the hottest side, leading to a steady state concentration. 

Thus, the presented model can be extended by including additional effects affecting  Ag NPs growth and thermal diffusion to simulate the self-assembly of nanogratings. Optical properties of these structures is also a subject of additional studies that should be performed in future. This topic is too large is out of scope of the present thesis.

\newpage
\chapter{Appendix}

\section{Appendix A: deep neural networks for predicting nanocrystal size of anatase TiO$_2$}

Deducing the TiO$_2$ nanocrystal size based on Raman spectra is presented in this section. Traditionally, the inverse problem without human's handful adjustments of parameters can be done by the optimization algorithms. However, fitting of the Raman spectra requires selection method, patience and time. Alternatively, the problem can be solved without fitting algorithm through deep neural networks. Deep learning has attracted widespread attentions and increasing interests due to its great success in computer vision \cite{krizhevsky2012imagenet}, speech recognition \cite{hinton2012deep}, and gaming \cite{silver2016mastering}. Very recently, because of its capabilities in multi-parameters fitting on a high-dimensional space, deep learning algorithms have started to be applied in the field of researches of material design \cite{sanchez2018inverse, carrasquilla2017machine,raccuglia2016machine, yao2019intelligent}, microscopy and spectroscopy \cite{chen2014deep, wu2017air,esteva2017dermatologist, rivenson2017deep, ota2018ghost, rivenson2018phase}, and nano- photonic design \cite{liu2018training, ma2018deep,wiecha2019pushing, xu2019deep, malkiel2018plasmonic}. The algorithms of deep neural networks are also explored to be applied on photonic platforms  \cite{lin2018all, shen2017deep, mengu2019analysis} to boost the calculations aiming at near light speed computing.

\begin{figure}[ht!]
 \centering
    \includegraphics[width=0.8\textwidth]{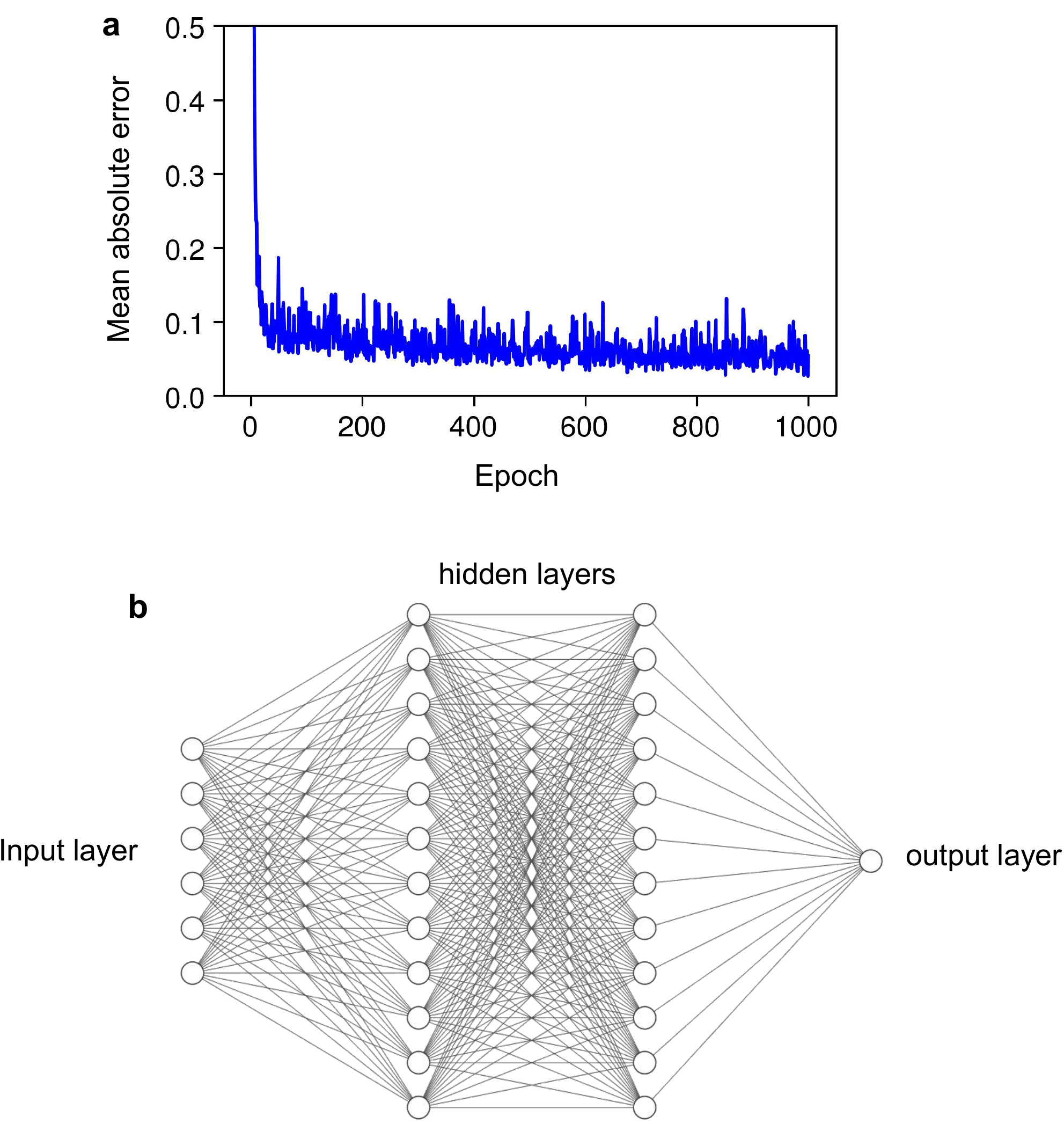}
    \caption{Mean absolute error during training (a). Fully connected structure  obtained by using Raman spectra as the input, while the output is nanocrystal size (b).}
\label{fgr:raman_ml_01}
\end{figure}

Among the deep neural networks for inverse design problem, the significant challenges in training the network based on non-unique mapped data sets are inevitable. This is fundamental for many physical problems in photonic design, and curve fittings. As a matter of fact, the non-unique spectrum-to-crystal size by the phonon confinement model is unfortunately the case. Certainly, one can solve this problem by pealing off the training data into a new data set, so that each data inside it is distinct. However, it requires time and patience, and turned out be inefficient \cite{liu2018training,kabir2008neural}. 

To solve this problem, D. Liu et al. proposed a method based on a tandem network structure\cite{liu2018training}. The cascaded network consists of an inverse training network and a forward training network. The idea is to construct a network that can generate data in a distinct space based on the separated parts of networks. The process can mainly be divided as the following steps: firstly, forward training the second part network; the forward process is efficient because the mapping is unique in this direction; secondly, construct the first part of the network and feed the results into the previously trained network; the data set for this step can be very large, because the input are the same as the output; during the training, the weights inside the pre-trained network by the first step are set to be constant; after the training, the network automatically selects the weights so that the generated data locates in a distinct space.

Inspired by this idea, the nanocrystal of anatase TiO$_2$ are estimated using a deep neural network in the fitting Raman spectrum. The data set is generated based on the phonon confinement model (Eq. \ref{eqn:phonon-confinement_p1}). Before diving into the cascading network, the first attempt is to train an inverse-fitting network directly. Figure \ref{fgr:raman_ml_01}(a) shows the mean absolute error during training. The corresponding network structure is shown in Figure \ref{fgr:raman_ml_01}(b). It is obvious that the network is very hard to be trained as the mean absolute error oscillates even after 1000 epoches. The network used here has two hidden layers of 100 unit cells inside each layer. The input layer has 200 cells and the output is one cell. The training is still very hard even for deeper networks of 6 layers of 500 cells inside each layer.

\begin{figure}[ht!]
 \centering
    \includegraphics[width=0.95\textwidth]{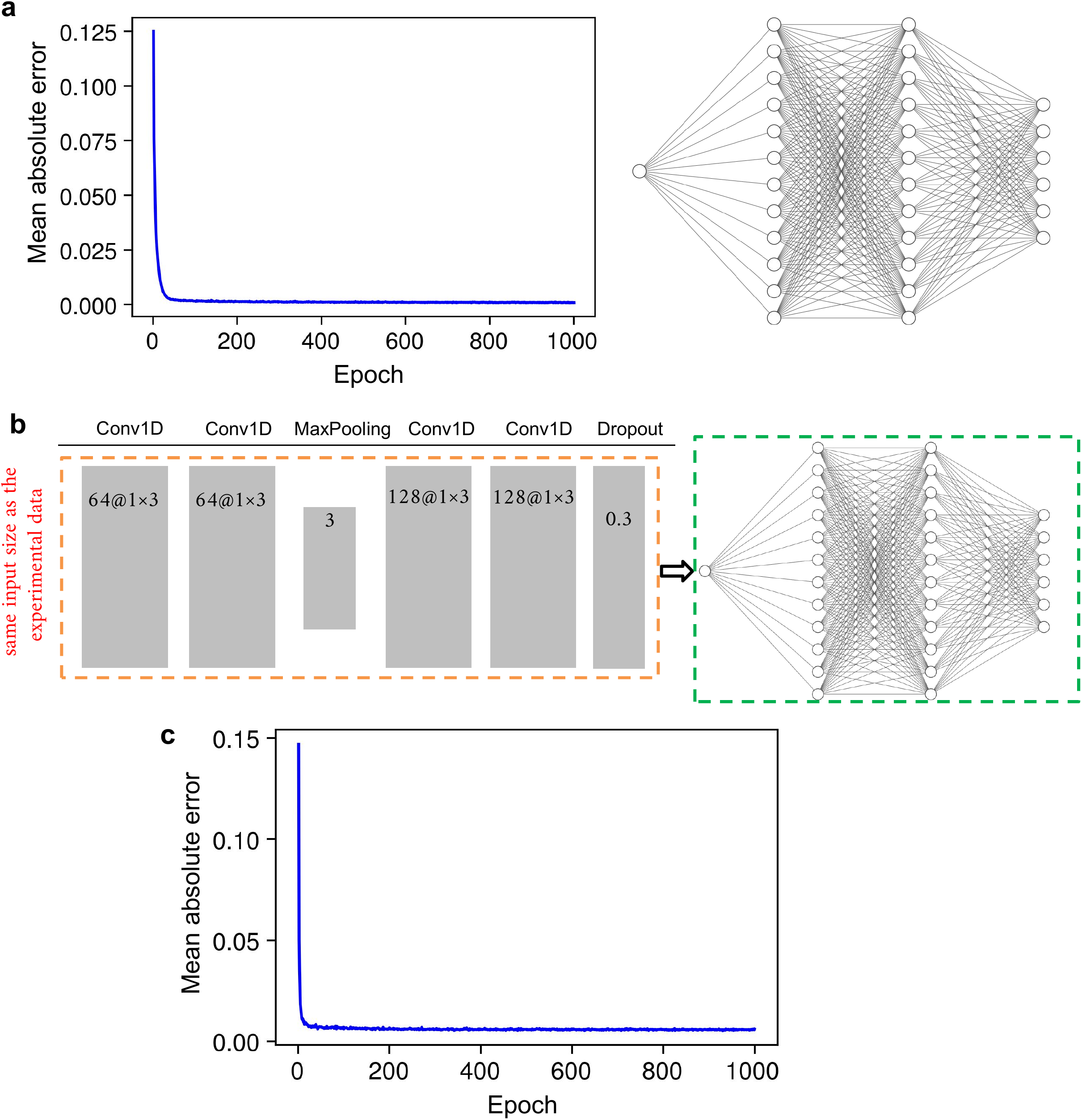}
    \caption{Training efficiency for the structure using size as the input and Raman spectrum as the output (a).The generating network (the dashed red rectangle) using discrete experimental data as input and the output is fed into the pre-trained network by step (a) (dashed green rectangle). Mean absolute error vs epoch for network shown figure (b).}
\label{fgr:raman_ml_02}
\end{figure}

\begin{figure}[ht!]
 \centering
    \includegraphics[width=\textwidth]{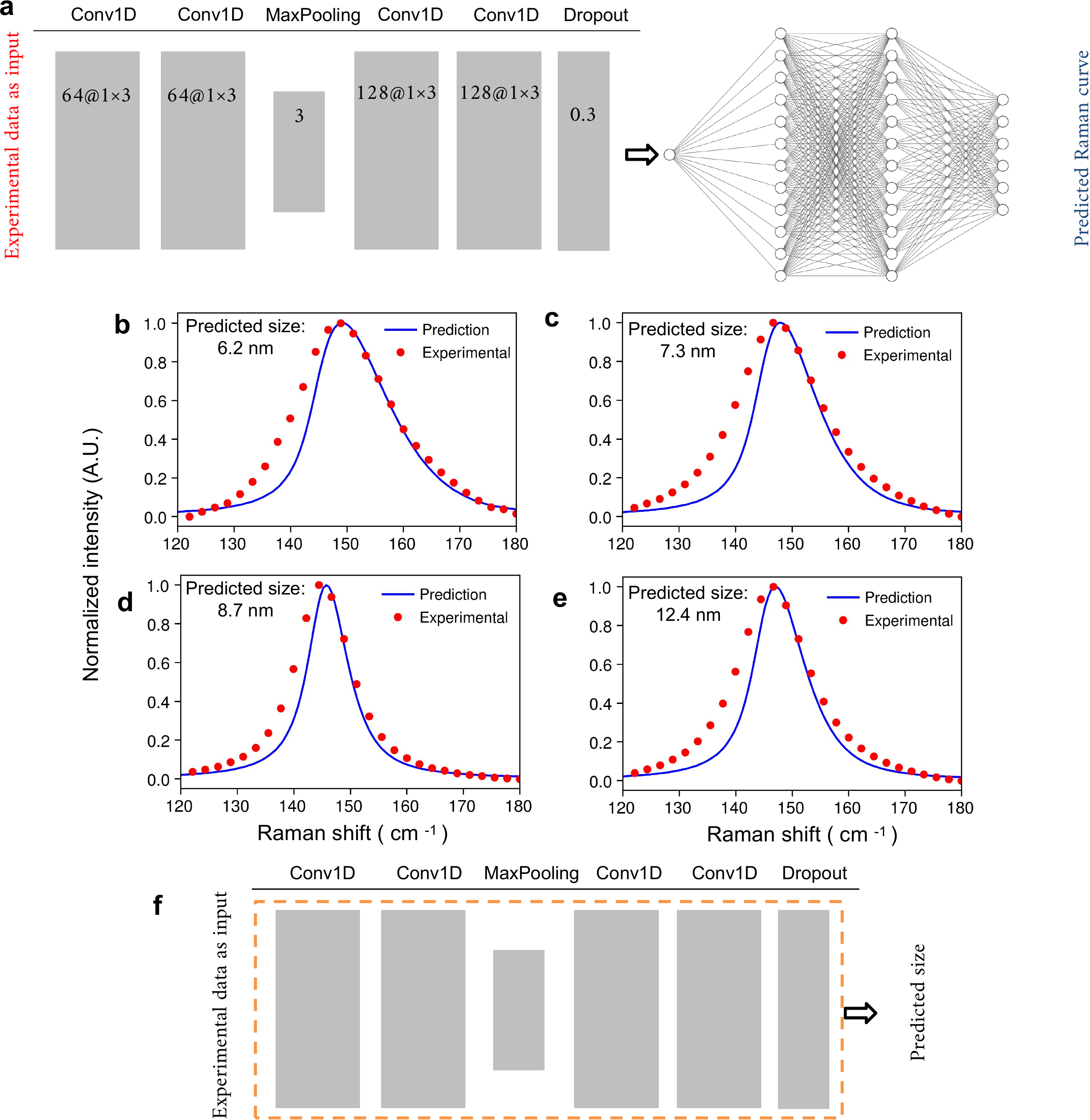}
    \caption{The network after training is then used for prediction (a). The predicted Raman spectra for experiments calcined at 400, 500, 600 and 700 $^{\circ} C$ are shown in figure (b), (c), (d), and (e) respectively. Experimental data (samples without seeds) are taken from Ref. \cite{sharma2019crystal}. The yellow dashed rectangle shows the network part for predicting the nanocrystal size (f).}
\label{fgr:raman_ml_03}
\end{figure}

On the contrary, the forward training is very efficient. Figure \ref{fgr:raman_ml_02}(a) shows the mean absolute error decrease rapidly and the vibrations is very small. The forward training network has identical structure to the inverse network used above, which only reverse the input and output layer. In the next step, a new network is designed and shown in Figure \ref{fgr:raman_ml_02}(b) (red dashed rectangle) having 6 hidden layers. The convolution layers are introduced for identifying Raman curve features. For example, the first hidden layer has 64 convolution cores of 3 pixels each. They are used to identify different features. The maxpooling layer of core of 3 pixels is to minimize the data size. The dropout layer is used to avoid over-fitting. The input layer has 27 cells which is the size of the experimental Raman spectrum. The output of this part network is then fed into the pre-trained forward network (shown by green dashed rectangle in Figure \ref{fgr:raman_ml_02}(b)).

After being successfully connected, the total network becomes a new deep network whose output layer has 200 cells representing the Raman spectrum in the range of 100 to 200 $cm^{-1}$. As discussed above, the data set used for training can be very large since the input and output are Raman spectra. For this sake, the Raman spectra are generated by the phonon confinement model. The input spectrum are selected according to the experimental Raman shift. The training by this network is then very efficient as shown by Figure \ref{fgr:raman_ml_02}(c). 

The whole trained network (Figure \ref{fgr:raman_ml_03}(a)) can be used for Raman spectrum generation and fitting. Splitting the first part of that network, it can be used for predicting the nano crystal size of TiO$_2$. Figure \ref{fgr:raman_ml_03}((b)-(e)) shows the Raman curve comparisons of prediction and Experiment. The nano crystal size is predicted by the network shown by Figure \ref{fgr:raman_ml_03}(f). The good fittings indicates the inverse training is successful as indicated by the mean absolute error in Figure \ref{fgr:raman_ml_02}(c). Nevertheless, in contrast to the theoretical model, the experimental spectra are more expanded in the left part of the curve (for instance, in \ref{fgr:raman_ml_03}(b)). In fact, the differences comes from the phonon confinement model itself: the uncertainty of the dispersion relation of the anatase TiO$_2$, the assumption of isotropic dispersion relations, and the absence of consideration of surface effects for very small nanocrystals \cite{sharma2019crystal,zhang2000raman}.

\section{Appendix B: the timestep variations during laser writing  in simulations}

\begin{figure}[ht!]
 \centering
    \includegraphics[width=0.95\textwidth]{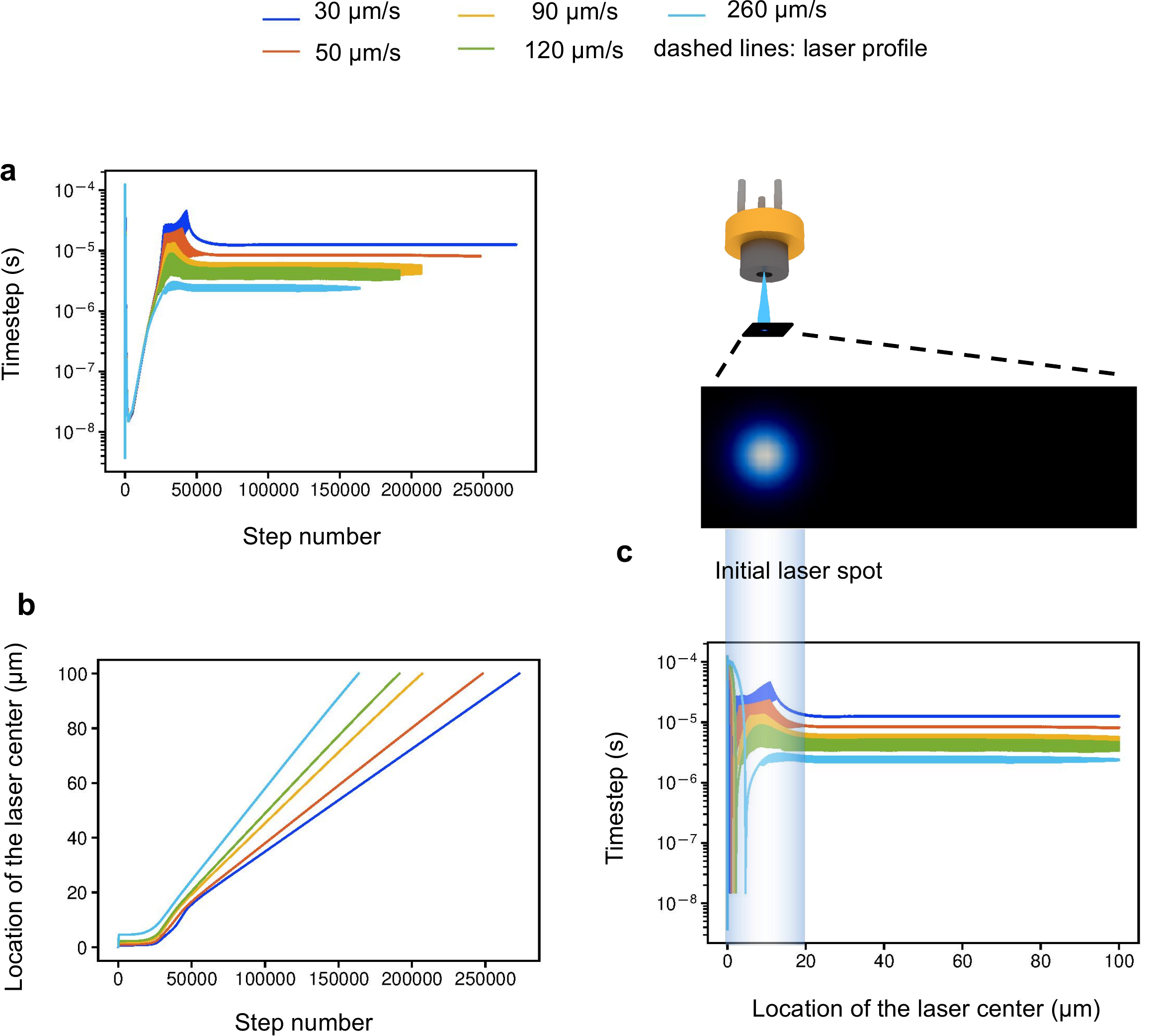}
    \caption{Adaptive timestep (a) and the location of laser center (b) as functions of step number during the calculation. The relation of timestep and location of laser center is plotted in (c). The initial laser spot size is also depicted. The dramatic variations of timestep is obviously seen to happen inside the initial laser spot, which corresponds to the fast temperature change before a steady state.}
\label{fgr:chapter5_appd}
\end{figure}

The Bulirsch-Stoer algorithm provides the forecasting timestep that is very useful in calculations. As discussed in the previous sections, the time step varies according to the reaction speed. Thus, the timestep acts as the benchmark of the whole system. Figure \ref{fgr:chapter5_appd}(a) shows the relation of adaptive timestep with step number in calculations. The initial timestep is set to 0.1 ns. The timestep changes dramatically in the first few hundreds of steps, which slowly changes as the step marches forward before the steady value is reached. The same trend are observed throughout simulations at different writing speeds. The timestep at steady-state is smaller and the calculation stops earlier as the writing speed is higher. To identify the major role to timestep, the laser center locations are calculated as $V_S \times t$ and shown in Figure \ref{fgr:chapter5_appd}(b). The variations of timestep only appear before the laser moves 20 $\mu m$ in the writing direction. To find out the locations of non- steady-states, the relation of timestep and laser center locations are plotted in Figure \ref{fgr:chapter5_appd}(c). The insert map shows the laser spot. It is obvious that the timestep changes very quickly inside the region of size less than the laser spot(e.g. 20$\mu m$). As one may record from Figure \ref{fgr:chapter5_tempRise_p1} that the temperature changes at the same time, the temperature seems to be the major role in affecting the timestep. This is possible because temperature has great influences on the physicochemical process in Ag NP's growth.

\bibliographystyle{unsrt}
\bibliography{main}

\newpage
\chapter{Publications}

\section{Publications in peer-reviewed journals}
\begin{itemize}

\item[1.] H. Ma, R.A. Zakoldaev, A. Rudenko, M. Sergeev, V.P. Vadim, T.E.  Itina: Well-controlled femtosecond laser inscription of periodic void structures in porous glass for photonic applications. \textit{Optics Express, Vol. 26 (2017): pp.33261-33270.}

\item[2.] A. Rudenko, H. Ma, V.P. Vadim, J.-P. Colombier, T.E. Itina: On the role of nanopore formation and evolution in multi-pulse laser nanostructuring of glasses. \textit{Applied Physics A, Vol. 124 (2018): pp.63.}
     
\item[3.] N. Sharma, H. Ma, T. Bottein, M. Bugnet, F. Vocanson, D. Grosso, T.E. Itina, Y. Ouerdane, N. Destouches: Crystal Growth in Mesoporous TiO$_2$ Optical Thin Films. \textit{The Journal of Physical Chemistry C, Vol. 123 (2019): pp.6070-6079.}

\item[4.] H. Ma, S. Bakhti, A. Rudenko, F. Vocanson, D.S. Slaughter, N. Destouches, T.E. Itina: Laser-Generated Ag Nanoparticles in Mesoporous TiO$_2$ Films: Formation Processes and Modeling-Based Size Prediction \textit{The Journal of Physical Chemistry C, Vol. 123 (2019): pp. 25898-25907, DOI:10.1021/acs.jpcc.9b05561.}

\item[5.] T.E. Itina, R.A. Zakoldaev, M. Sergeev, H. Ma, S. Kudryashov, O. Medvedev, V.P. Veiko: Ultra-short laser-induced high aspect ratio densification in porous glass. \textit{Optical Materials Express, Vol. 9 (2019): pp. 4379-4389}

\item[6.] Y. Andreeva, V. Koval, M. Sergeev, V.P. Veiko, N. Destouches, F. Vocanson, H. Ma, A. Loshachenko, T.E. Itina: Picosecond laser writing of Ag-SiO$_2$ nanocomposite nanogratings for optical filtering. \textit{Optics and Lasers in Engineering, Vol. 124 (2020): pp.105840.}

\end{itemize}

\section{Participation in conferences}
\begin{itemize}

\item[1.]  H. Ma, S. Bakhti, Y. Andreeva, A. Rudenko, F. Vocanson, D.S. Slaughter, N. Destouches, T.E. Itina. On the mechanisms of metal nanoparticle formation in CW laser-irradiated thin porous films. \textit{European Materials Research Society (E-MRS 2019) Spring Meeting, Symposium V Laser interactions with materials: from fundamentals to applications, May 2019, Nice, France} (oral talk)

\item[2.] H. Ma, S. Bakhti, Z. Liu, G. Vitrant, L. Saviot, D.S. Slaughter, T.E. Itina, Nathalie Destouches. Increased heating of metallic nanoparticles for lower deposited light energy: measurement and explanation of a counter-intuitive behavior. \textit{C'Nano 2018 The nanoscience meeting, Dec. 2018, Toulon, France.} (oral talk, co-author)

\item[3.] Y. Andreeva, M. Sergeev, H. Ma, N. Sharma, N. Destouches, T.E. Itina, Vadim Veiko: Experimental and numerical study of laser-induced metal nanoparticle formation in thin porous films. \textit{International conference of Modern Nanotechnologies and Nanophotonics for Science and Industry (MNNSI), Nov. 2018, Suzdal, Russia.} (oral talk, co-author)

\item[4.] H. Ma, S. Bakhti, Z. Liu, A. Rudenko, D.S. Slaughter, N. Destouches, T.E. Itina. A perspective of the laser induced temperature increase on silver nanoparticles growth in titaniun dioxide thin films. \textit{The International Workshop on Metallic Nano-Objects (MNO), Nov. 2018, Lyon, France.} (oral talk) 

\item[5.] T.E. Itina, A. Rudenko, J-P. Colombier, H. Ma, R.A. Zakoldaev, M.M. Sergeev, V.P. Veiko: Multi-physical modeling of laser nano-and micro-structuring of glasses. \textit{LPM2018-Laser Precision Microfabrication Symposium 2018, Jun. 2018, Edinburgh, United Kingdom} (invited talk, co-author)

\item[6.] T.E. Itina, J-P. Colombier, H. Ma, M.M. Sergeev, R.A. Zakoldaev, V.P. Veiko. Ultra-Short Laser Structuring of Optical Materials in Volume. \textit{International High Power Laser Ablation Symposium (HPLA 2018), Mar. 2018, Santa Fe, United States.} (oral talk, co-author) 

\item[7.] T.E. Itina, H. Ma, A. Rudenko, S. Mottin, V.P. Veiko, M.M. Sergeev, R.A. Zakoldaev. Laser assisted periodic nanostructure formation in dielectric materials: formation mechanisms. \textit{ALT'17, Sep. 2017, Busan, South Korea.} (invited talk)

\item[8.] A. Rudenko, H. Ma, T.E. Itina, V.P. Veiko. Ultra-short Laser Interactions for Advanced Photonic Technologies. \textit{Progress In Electromagnetics Research Symposium (PIERS 2017), May 2017, St Petersburg, Russia.} (oral talk, co-author)

\item[9.] T. E. Itina, A. Rudenko, J.-P. Colombier, H. Ma. Ultra-short laser structuring of glasses: Predictive modeling insights. \textit{$7^{th}$ European Conference on Applications of Femtosecond Lasers in Materials Science, FemtoMat 2017, 20-22 March, Mauterndorf Castle, Salzburg, Austria.} (invited talk, co-author)

\end{itemize}

\newpage\null\thispagestyle{empty}\newpage\null\thispagestyle{empty}\newpage

\end{document}